\documentclass[a5paper]{article}
\usepackage[utf8]{inputenc}

\usepackage{amsmath}
\usepackage{amsfonts}
\usepackage{amssymb}
\usepackage{mathtools}

\usepackage{threeparttable}
\usepackage{multirow}
\usepackage{dcolumn}
\usepackage{booktabs}

\usepackage{graphicx}
\usepackage{float}
\usepackage[subrefformat=parens,labelformat=parens,font=small]{subfig}

\usepackage{comment}
\usepackage[left=1.50cm, right=1.50cm, top=2.00cm, bottom=2.00cm]{geometry}

\usepackage[breaklinks=true]{hyperref}
\usepackage{breakcites}
\hypersetup{
	colorlinks=true, 
	linktoc=all,     
	linkcolor=blue,  
}

\renewcommand{\vec}[1]{\boldsymbol{#1}}

\author{Yajun Zhu \footnote{\href{mailto:mazhuyajun@ust.hk}{mazhuyajun@ust.hk}} \ and Kun Xu \footnote{\href{mailto:makxu@ust.hk}{makxu@ust.hk}} \\
\textit{\small Department of Mathematics, The Hong Kong University of Science and Technology}}
\date{}

\title{The first decade of unified gas kinetic scheme}

\begin{document}

\maketitle

\tableofcontents

\clearpage

\section{Introduction}

The unified gas kinetic scheme (UGKS) developed by Xu et al. \cite{xu2010ugks,xu2015book} is a multiscale method based on the direct modeling of flow physics on the numerical mesh size and time step scales for the continuum and rarefied gas simulations.
Considering the fact that description of flow behavior depends closely on the scales of observation, while all theoretical equations, such as the Boltzmann equation or the Navier--Stokes (NS) equations, are constructed and valid only on their modeling scales, the methodology of direct modeling targets on constructing the corresponding flow physics directly on the numerical scales for a better description of multiscale transport process with high efficiency.

The flow physics in the rarefied regime is the free transport and collision of microscopic particles, which can be well described by the Boltzmann equation; while in the hydrodynamic regime, the phenomena are mainly macroscopic wave interaction, for which it would be inappropriate and low efficient to directly implement the Boltzmann dynamics.
The Boltzmann equation is the fundamental governing equation of rarefied gas dynamics, which is usually regarded to be valid for all flow regimes. 
But, it only means that all regimes physics is to be recovered by resolving the physics on the kinetic scale and accumulating its effect to hydrodynamic level. 
The kinetic scale for the modeling of the Boltzmann equation is the particle mean free path and the time between successive collisions (collision time), which has to be matched by the resolution in the numerical discretization.
The NS equations are derived from the evolution of fluid elements with continuum mechanics assumption in the hydrodynamic regime, modeling the viscous shearing and heat conduction effect on the macroscopic scale, which can be derived from the coarse-graining process of the Boltzmann equation.
The NS equations are widely used in the computational fluid dynamic (CFD) for numerical simulations, and achieve great success in industrial applications, including aeronautics and aerospace, weather forecast, biological engineering, etc.
However, the extension to transition regimes is not so successful, because the generality of the multiscale coarse-graining methodology is limited due to that the non-equilibrium flow phenomena cannot be fully described through a few constitutive relations with a limited number of degrees of freedom with the variation of scales. 
Moreover, it is difficult to determine the specific scale of flow physics in the transition regime, not even know the appropriate flow variables in such a scale.

In the computations, an underlying scale is the numerical scale of discretization, such as the mesh size and the time step, which are the preset scales we would like to use to investigate the problems.
Therefore, direct modeling on the numerical scales targets on describing the flow evolution directly on the observational scale.
By this way, the modeling scale and the discretization scale are the same, and governing equations corresponding to the mesh size or time step have to be constructed.
The methodology of direct modeling on discretization scale makes it possible to construct a truly multiscale method with high efficiency.

In the UGKS, the key ingredients are to follow the basic conservation laws of macroscopic flow variables and the distribution function of microscopic particles in a discrete control volume, and to construct a multiscale flux function by taking into account the contribution from both particles' transport and collision in the modeling scale.
On one hand, the conservations are the fundamental laws and are valid in all scales; on the other hand, a time-dependent flux function is constructed using the integral solution of the kinetic model, which presents the local flow evolution on the numerical scales with the account of the numerical resolution and local physical state.
With the adaptive flux function, the UGKS is capable of simulating multiscale flow dynamics with large variation of cell Knudsen number in a high efficiency, and it does not require the mesh size and time step to be smaller than the mean free path and particle collision time with the kinetic scale resolution.
With the discrete velocity points, the UGKS could recover the NS solutions in the hydrodynamic limit and provide highly non-equilibrium rarefied solution in the kinetic scale.

In the past decade, the UGKS \cite{xu2010ugks,huang2012multidimension,xu2015book} gets a fast development and extension.
The Shakhov model \cite{shakhov1968model} was introduced to replace the Bhatnagar--Gross--Krook (BGK) model \cite{bhatnagar1954model} in the original UGKS to achieve a more flexible Prandtl number \cite{xu2011shakhov}.
Based on the Rykov model \cite{rykov1975model}, the rotational and vibrational degrees of freedom are included in the UGKS \cite{liu2014diatomic,zhang2015thesis,wang2017diatomic} for better description of diatomic gases, especially for high speed flows.
With the similar consideration of the UGKS, Guo et al. \cite{guo2013dugks,guo2015dugks} developed the discrete unified gas kinetic scheme (DUGKS), which adopts the discrete solution of the kinetic equation along characteristic line to couple the particles' free motion and collision.
The DUGKS has the capability for multiscale flow simulations in all Knudsen regimes as well, and has a simple flux function.
The UGKS and DUGKS have been validated by many numerical cases of neutral gas flows, including micro flow \cite{wang2012sound,wang2014slider,wang2016onsager,huang2013micro,liu2015nano,chen2012radiometer} and high-speed rarefied flow \cite{jiang2016thesis,jiang2019aia}.
The direct modeling methodology provides a framework for multiscale modeling of transport process, and has been applied in many fields, such as radiative transfer \cite{sun2015radiative,sun2015frequency}, phonon transport \cite{guo2016phonon}, plasma \cite{liu2017plasma}, neutron transport \cite{shuang2019parallel,tan2020neutron}, multicomponent and multiphase flow \cite{xiao2019multicomponent,yang2019phase}, granular flow \cite{liu2019granular,wang2018granular,wang2019monodisperse}.
Several techniques have been developed and implemented to further improve the UGKS, such as unstructured mesh computation \cite{zhu2016unstructured,sun2017unstructured}, moving grids \cite{wang2019ale,chen2012mesh}, velocity space adaptation \cite{chen2012mesh,chen2019conserved}, memory reduction \cite{chen2017memory,yang2018memory}, implicit algorithms \cite{zhu2016iugks,yuan2020conservative,sun2017iugks,sun2018iugks}, and further simplification and modification \cite{chen2016simplification,liu2012modified}.
The UGKS is becoming an important tool for solving multiscale problems, and shows great potentials in many fields of engineering applications.

\section{Unified gas kinetic scheme}\label{sec:ugks}
In this section, the construction of the UGKS based on the BGK model \cite{bhatnagar1954model} will be introduced in details.

In rarefied gas dynamics, gas distribution function $f(u,v,w,x,y,z,t)$ is used to describe the statistical behavior of microscopic gas particles.
It denotes the number density (or mass density) of the particles around the microscopic velocity $\vec{u} = (u, v, w)^T$ and the physical location $\vec{x} = (x,y,z)^T$ at time $t$.
All macroscopic flow variables can be obtained by taking moments of the gas distribution function.
For example, the conservative flow variables
$\vec{w} = (\rho, \rho U, \rho V, \rho W, \rho E)^T$, i.e., the densities of mass, momentum and energy are computed by
\begin{equation}\label{eq:conserved_variable}
\vec{w} (x,y,z, t) = \int{f (x,y,z, u, v, w, t) \vec{\psi}(u, v, w) d\Xi},
\end{equation}
where $\vec{\psi}(u,v,w) = (1, u, v, w, \frac{1}{2} (u^2 + v^2 + w^2))^T$ is the collision invariants, and $d\Xi = dudvdw$.
The shear stress $\vec{P}$ and heat flux $\vec{q}$ are calculated from
\begin{equation}\label{eq:stress_tensor}
P_{\alpha \beta} = \int {c_\alpha c_\beta f d\Xi},
\end{equation}
and
\begin{equation}\label{eq:heat_flux}
q_{\alpha} = \int {c_\alpha \frac{|\vec{c}|^2}{2} f d\Xi} ,
\end{equation}
where $\vec{c} = (u - U, v - V, w- W)^T$ denotes the vector of peculiar velocity.

\subsection{Direct modeling on numerical discretization scale}
In order to capture the non-equilibrium physics, discrete distribution function is usually employed in the kinetic solvers \cite{broadwell1964dvm,bobylev1995rgd,yang1995rarefied,aristov2012direct}, where the velocity space is discretized by a set of velocity points.

\begin{figure}[htbp]
\centering
\subfloat[\label{fig:physical_space}]
{\includegraphics[width=0.45\textwidth]{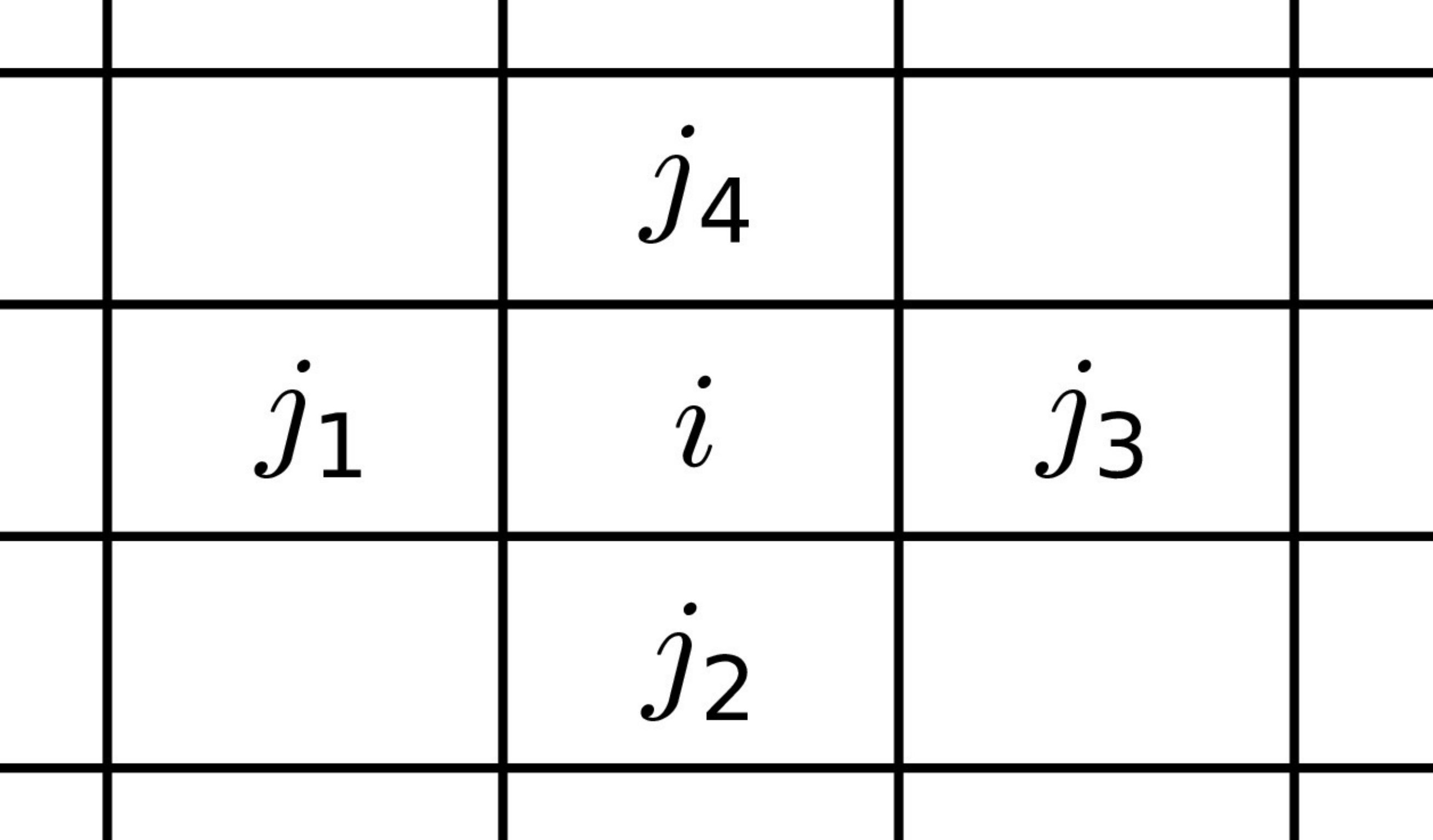}} \hspace{2em}
\subfloat[\label{fig:velocity_space}]
{\includegraphics[width=0.45\textwidth]{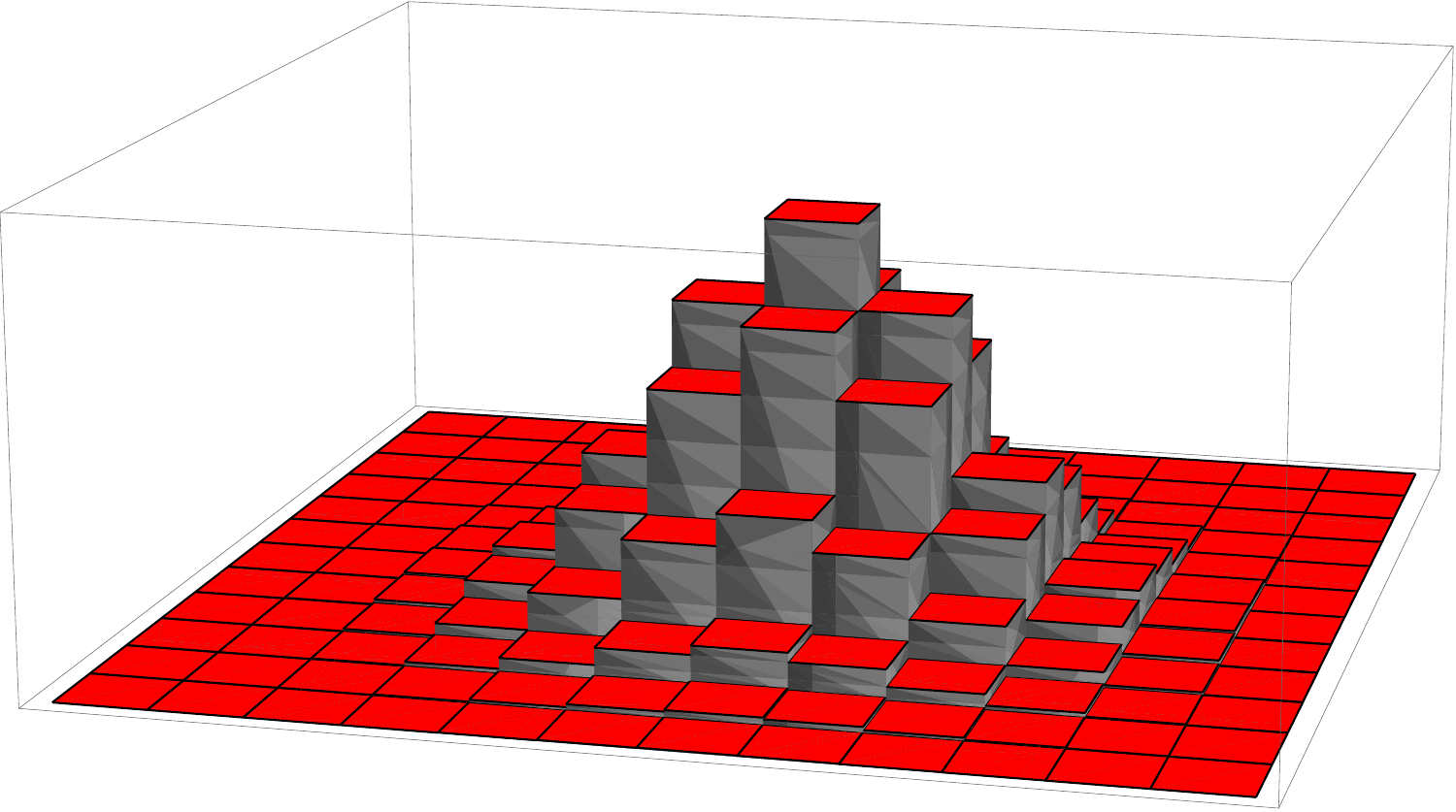}}
\caption{\label{fig:space_discretization}Physical space and velocity space for two-dimensional cases. (a) physical space, (b) velocity space.}
\end{figure}

As shown in Fig.~\ref{fig:space_discretization}, in the framework of finite volume method, for a discrete control volume $(\delta \vec{x}_i, \delta \vec{u}_k)$ in the phase space, the averaged gas distribution function is
\begin{equation}\label{eq:cell_averaged_f}
f_{i,k}(t) = \frac{1}{V_i \mathcal{V}_k} \int_{\delta \vec{x}_i} \int_{\delta \vec{u}_k} {f(x, y, z, u, v, w,t) du dv dw dx dy dz},
\end{equation}
where $V_i$ is the volume of cell $i$ in physical space, $\mathcal{V}_k$ is the volume of cell $k$ in velocity space or the integration weight at the discrete velocity point $\vec{u}_k$.
During a discrete time step $\Delta t = t^{n+1} - t^n$, the governing equation for evolution of cell distribution function is
\begin{equation}\label{eq:micro_conservation}
f_{i,k}^{n+1} = f_{i,k}^n - \frac{\Delta t}{V_i} \sum_{j \in N(i)} {u_{n,ij,k} {\bar f}_{ij,k} S_{ij}} + \int_{0}^{\Delta t} {\mathcal{J}_k (f_i,f_i) dt},
\end{equation}
where $N(i)$ denotes the set of the face-bordered neighbors of cell $i$, $j$ is one of the neighboring cells, and the interface between cells $i$ and $j$ is denoted by $ij$.
$u_{n,ij,k} = \vec{u}_k \cdot \vec{n}_{ij}$ denotes the normal velocity of a particle across the interface $ij$, where $\vec{n}_{ij}$ is the outward normal vector of face $ij$, and $\vec{u}_k$ denotes the particle velocity.
$S_{ij}$ is the area of cell interface $ij$, and $\bar{f}_{ij,k}$ is the time-averaged distribution function at cell interface $ij$ over time step $\Delta t$.
$\mathcal{J}_k(f,f)$ represents particles' collision.
It can be seen that Eq.~\eqref{eq:micro_conservation} describes the evolution of the distribution function due to particles moving across cell interfaces and interacting with each other, which is the conservation law of gas particles in the control volume.

By taking the moments of the microscopic equation, the macroscopic governing equations on the numerical scale are
\begin{equation}\label{eq:macro_conservation}
\vec{w}_i^{n+1} = \vec{w}_i^n - \frac{\Delta t}{V_i}\sum_{j \in N(i)}{\vec{F}_{ij} S_{ij}},
\end{equation}
where $\vec{F}_{ij}$ denotes the time-averaged flux of conservative flow variables across face $ij$, which is computed by taking moments of microscopic fluxes $\mathcal{F}_{ij,k}$
\begin{equation}\label{eq:macro_flux_from_micro}
\vec{F}_{ij} = \sum_k {\mathcal{F}_{ij,k}\vec{\psi}_k \mathcal{V}_k} = \sum_k{u_{n,ij,k} {\bar f}_{ij,k} \vec{\psi}_k \mathcal{V}_k }.
\end{equation}
The collision term satisfies the compatibility condition
\begin{equation}\label{eq:compatibility_condition}
\sum_k {\mathcal{J}_k (f_i,f_i) \vec{\psi}_k \mathcal{V}_k} = \vec{0}.
\end{equation}
Eq.~\eqref{eq:macro_conservation} directly gives the conservation of mass, momentum and energy on the discretization scale.

\begin{figure}[htbp]
	\centering
	\subfloat[]
	{\includegraphics[width=0.45\textwidth]{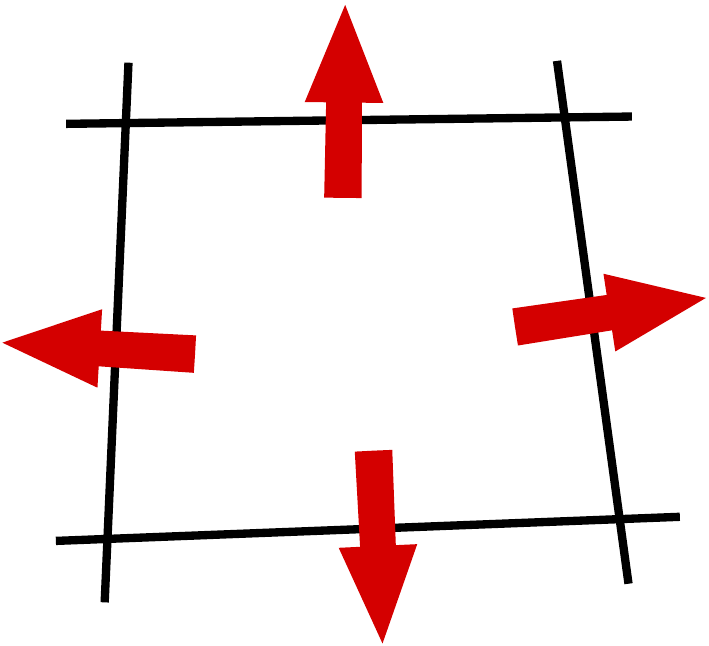}} \hspace{2em}
	\subfloat[]
	{\includegraphics[width=0.45\textwidth]{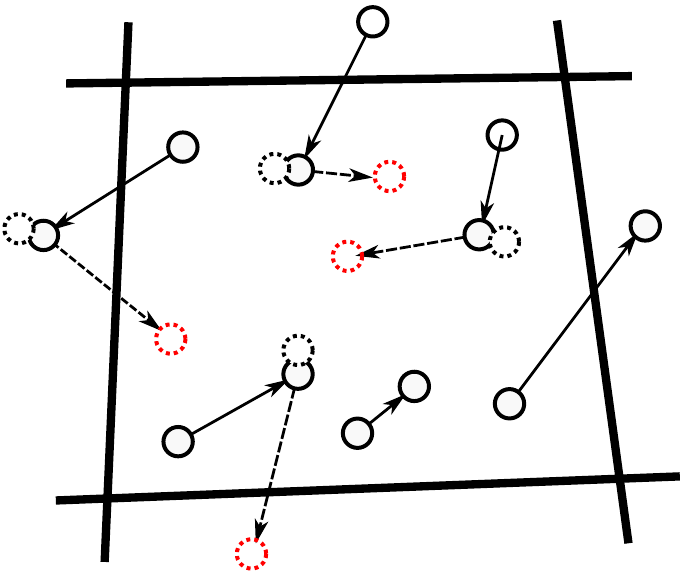}}
	\caption{\label{fig:fvm_evolution}Direct modeling on discretization scale. (a) Conservation on macroscopic level, and (b) conservation on microscopic level.}
\end{figure}

It should be emphasized again that as shown in Fig.~\ref{fig:fvm_evolution}, Eqs.~\eqref{eq:micro_conservation} and \eqref{eq:macro_conservation} describe the conservation of gas distribution function and macroscopic flow variables, which is supposed to be valid at arbitrary discretization scale regardless of variations of mesh size and time step.
The external force is ignored, and the UGKS with consideration of external force can be found in the studies of Xiao et al. \cite{xiao2017gravity,xiao2018force}.
Following the basic conservation laws is the fundamental step in the construction of a multiscale UGKS on the numerical scales.
With these conservation laws, we can see from Eq.~\eqref{eq:macro_conservation} that the evolution mainly depends on the microscopic flux $\mathcal{F}_{ij,k}$ across cell interfaces.
In the UGKS, a multiscale flux constructed from the integral solution of the BGK model equation \cite{bhatnagar1954model} is the key to provide an appropriate description of the local dynamic evolution.
In other words, the time evolution solution of the gas distribution function is directly modeled and used in the UGKS.

\subsection{Scale-adaptive flux function}\label{sec:flux_function}
The BGK kinetic model is written as
\begin{equation}\label{eq:bgk_equation}
\frac{\partial f}{\partial t}
+ u \frac{\partial f}{\partial x}
+ v \frac{\partial f}{\partial y}
+ w \frac{\partial f}{\partial z}
= \frac{g-f}{\tau},
\end{equation}
where the collision term is approximated by a relaxation process of the distribution function approaching to a local equilibrium state.
The equilibrium state $g$ is a Maxwellian distribution
\begin{equation}\label{eq:maxwellian_distribution}
g_M =  \rho \left(\frac{m_0}{2\pi k_B T}\right)^{\frac{3}{2}} e^{-\lambda\left[(u -U)^2 + (v - V)^2 + (w-W)^2\right]},
\end{equation}
where $m_0$ denotes molecular mass of gas particle, $k_B$ is the Boltzmann constant.
$\lambda$ is related to temperature $T$ by
\begin{equation}
	\lambda = \frac{m_0}{2 k_B T}.
\end{equation}
In the BGK model (\ref{eq:bgk_equation}), $\tau$ is the relaxation time or particles' mean collision time, which denotes the averaged time interval between two collisions.
It can be computed by
\begin{equation}\label{eq:tau_computation}
\tau = \frac{\mu}{p},
\end{equation}
where $\mu$ is dynamic viscosity and $p$ the static pressure.
The dynamic viscosity can be computed by Sutherland law \cite{sutherland1893viscosity}
\begin{equation}\label{eq:sutherland}
\mu = \mu_0 \left(\frac{T}{T_0}\right)^{\frac{3}{2}} \frac{T_0 + S}{T + S},
\end{equation}
or the power law
\begin{equation}\label{eq:dynamic_viscosity}
\mu = \mu_0 \left(\frac{T}{T_0}\right)^\omega,
\end{equation}
where $\mu_0$ is the reference dynamic viscosity at the reference temperature $T_0$, $\omega$ is the temperature index which is related to the molecular gas model, and $S$ denotes a model parameter of a constant temperature.

For a specific physical location $\vec{x}_0$, the analytic solution of the kinetic model equation along the characteristic line $\vec{x}^\prime = \vec{x}_0 + \vec{u} (t^\prime - t_0)$ gives
\begin{equation}\label{eq:integral_solution}
f(\vec{x}_0, t) = \frac{1}{\tau} \int_{t_0}^{t} {g(\vec{x}^{\prime}, t^\prime) e^{-(t-t^\prime)/\tau} dt^\prime} + e^{-(t-t_0)/\tau} f_0(\vec{x}_0 - \vec{u} (t-t_0)),
\end{equation}
where $f_0(\vec{x})$ is the initial distribution function around $\vec{x}_0$ at the beginning of each time step $t_0$.
$g(\vec{x}, t)$ represents the equilibrium state in time and space.
For convenience, assume the initial time $t_0 = 0$ and the investigated location $\vec{x}_0 = \vec{0}$.
From Eq.~\eqref{eq:integral_solution} it can be found that provided with the initial distribution function $f_0(\vec{x})$ and the equilibrium state $g(\vec{x}, t)$ around a cell interface, the time-dependent distribution  function can be obtained at the cell interface.

For a second order method, the initial distribution function and the equilibrium state at the cell interface $\vec{x}_0 = \vec{x}_{ij}$ can be expanded as
\begin{equation}\label{eq:f_expansion}
f_0(\vec{x}) = f_0 + f_{\vec{x}} \cdot \vec{x},
\end{equation}
and
\begin{equation}\label{eq:g_expansion}
g(\vec{x},t) = g_0  + g_{\vec{x}} \cdot \vec{x} + g_t t,
\end{equation}
where $f_{\vec{x}}$ is the spatial derivative of initial distribution function, and $g_{\vec{x}}$ and $g_t$ are the spatial and temporal derivatives of the equilibrium state, respectively.
In Eqs.~\eqref{eq:integral_solution}, \eqref{eq:f_expansion} and \eqref{eq:g_expansion}, the independent variable $\vec{u}$ is ignored, and in the following description, we assume using a discrete distribution function at velocity point $\vec{u}_k$, and the subscript $k$ will be ignored without specific statement.
Substituting Eqs.~\eqref{eq:f_expansion} and \eqref{eq:g_expansion} into Eq.~\eqref{eq:integral_solution}, the time-dependent distribution function at cell interface $ij$ can be obtained
\begin{equation}\label{eq:fij_expression}
f_{ij}(t) = c_1 f_0 + c_2 f_{\vec{x}} \cdot\vec{u}
+ 	c_3 g_0 + c_4 g_{\vec{x}} \cdot \vec{u} + c_5 g_t ,
\end{equation}
where
\begin{equation}\label{eq:fij_coefficients}
\begin{aligned}
c_1 &=  e^{-t/\tau}, \\
c_2 &= -t e^{-t/\tau},\\
c_3 &=  1 - e^{-t/\tau}, \\
c_4 &= t e^{-t/\tau} - \tau (1-e^{-t/\tau}), \\
c_5 &= t -  \tau (1-e^{-t/\tau}).		
\end{aligned}
\end{equation}
Integrating over a time step $\Delta t$, the time-averaged flux is
\begin{equation}\label{eq:fij_flux}
\begin{aligned}
\mathcal{F}_{ij}
&= u_{n,ij} \left[\left( C_1 f_0 + C_2 f_{\vec{x}} \cdot\vec{u} \right)
+ \left( C_3 g_0 + C_4 g_{\vec{x}} \cdot \vec{u} + C_5 g_t \right) \right] \\
&= u_{n,ij} (\mathcal{F}_{ij}^{fr} + \mathcal{F}_{ij}^{eq}),
\end{aligned}
\end{equation}
where $\mathcal{F}_{ij}^{fr}$ and $\mathcal{F}_{ij}^{eq}$ denote the fluxes contributed from the initial distribution function and equilibrium state, respectively.
The coefficients are
\begin{equation}\label{eq:fij_flux_coefficients}
\begin{aligned}
C_1 &=  \frac{\tau}{\Delta t} (1 - e^{-\Delta t/\tau}), \\
C_2 &=  \Delta t \left[\frac{\tau}{\Delta t}  e^{-\Delta t/\tau} - \frac{\tau^2}{\Delta t^2} (1 - e^{-\Delta t/\tau}) \right], \\
C_3 &=  1 - \frac{\tau}{\Delta t} (1 - e^{-\Delta t/\tau}), \\
C_4 &= \Delta t \left[ \frac{2\tau^2}{\Delta t^2} (1 - e^{-\Delta t/\tau}) - \frac{\tau}{\Delta t} - \frac{\tau}{\Delta t} e^{-\Delta t/\tau} \right],\\
C_5 &= \Delta t \left[ \frac{1}{2}  - \frac{\tau}{\Delta t}   + \frac{\tau^2}{\Delta t^2} (1 - e^{-\Delta t/\tau}) \right].
\end{aligned}
\end{equation}

\begin{figure}[htbp]
\centering
\includegraphics[width=0.5\textwidth]{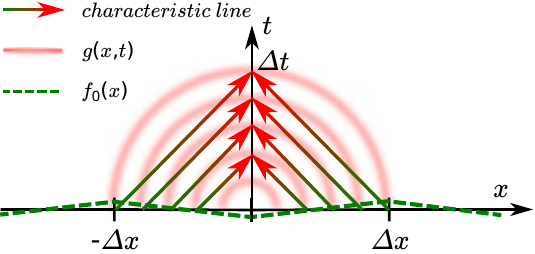}
\caption{\label{fig:flow_physics}Flow physics on the numerical discretization scale $(\Delta \vec{x}, \Delta t)$ described by the integral solution of the kinetic model.}
\end{figure}

As shown in Fig.~\ref{fig:flow_physics}, the integral solution describes the evolution of the distribution function from the initial state to the local equilibrium with the accumulation of particles' collision.
When the discrete time scale is much larger than particles' mean collision time $\Delta t \gg \tau$, the local equilibrium state dominates in the flux function, and the flow behavior performs as macroscopic wave spreading and interaction; when the time scale is small $\Delta t < \tau$, the initial distribution function is dominant and the flow physics is the free transport of microscopic particles.
It can be found that the integral solution not only couples the particles' free transport and collision processes, but also provides a transition from kinetic scale to hydrodynamic scale, where the observed local physics is determined by the cell Knudsen number, i.e., $\tau / \Delta t$.
For a well-resolved solution, the flux function is able to provide the flow physics on the corresponding modeling scale of the numerical time step.
The scale adaptation through the ratio of $\tau$ and $\Delta t$ is the key to construct a truly multiscale numerical method.
In the traditional numerical partial differential equation (PDE) approach, usually it requires the mesh size and the time step small enough to match the physical modeling scale in the construction of the governing equations, while the UGKS has no specific requirement for the numerical discretization, except for the stability condition and necessary spatial resolution for well-resolving local flow structure.
Owing to the adaptive multiscale flux function and direct modeling concept, the UGKS is supposed to have better efficiency in the numerical simulation of multiscale problems with large variation of Knudsen numbers.

It should be pointed out that there are several assumptions for derivation of Eqs.~\eqref{eq:fij_expression} and \eqref{eq:fij_flux}, which demonstrate the range of its validity.
First, the collision time $\tau$ is assumed to be a local constant, so the integral solution only describes the local physical evolution within a cell size $\Delta x$ and the corresponding time step $\Delta t$ as shown in Fig.~\ref{fig:flow_physics}.
Second, the distribution functions are in the form of Eqs.~\eqref{eq:f_expansion} and \eqref{eq:g_expansion}, which implies that the flow structure in a cell can be well resolved by a linear function.
In another words, even with the so-called scale adaptive property, for the cases with very complicated flow structure unresolved by the mesh size and time step, the integral solution is unable to provide correct time evolution without further numerical modeling of the local physics on a coarse mesh and large time step.

\subsection{Numerical algorithm}
The main ideas to construct the UGKS, i.e., direct modeling on the numerical scales and multiscale flux function, have been introduced in previous section.
The detailed algorithm for implementing the UGKS will be presented in the following description.

\subsubsection{Initial distribution function}\label{sec:initial_condition}

For CFD computation, the initial macroscopic flow variables at the beginning step are given, such as
\begin{equation}
\vec{w} =
\begin{pmatrix}
\rho \\
\rho U\\
\rho V \\
\rho W \\
\frac{1}{2}\rho (U^2 + V^2 + W^2) + \frac{3}{2} \rho \frac{k_B}{m_0} T
\end{pmatrix},
\end{equation}
from which the distribution function can be initialized by the corresponding Maxwellian distribution computed from the given macroscopic flow variables.
The relaxation time $\tau$ is determined by Eqs.~\eqref{eq:tau_computation} and \eqref{eq:dynamic_viscosity}.
If the variable soft sphere (VSS) model is adopted, the Knudsen number determined by the initial condition is
\begin{equation}\label{eq:knudsen_number}
{ Kn} = \frac{l_{mfp}}{L} = \frac{4\alpha(5-2\omega)(7-2\omega)}{5(\alpha+1)(\alpha+2)} \sqrt{\frac{m_0}{2 \pi k_B T}} \frac{\mu}{\rho L},
\end{equation}
where $l_{mfp}$ is particles' mean free path, denoting the averaged traveling distance between two collisions.
$L$ is the reference length.
For variable hard sphere (VHS) model, $\alpha = 1$; and for hard sphere (HS) model, $\alpha=1$ and $\omega=0.5$.
From Eq.~\eqref{eq:knudsen_number}, the relation among the Mach number, the Reynolds number, and the Knudsen number is given by 
\begin{equation}\label{eq:kn_re_ma}
{ Kn} \cdot { Re} = \frac{4\alpha(5-2\omega)(7-2\omega)}{5(\alpha+1)(\alpha+2)} \frac{\gamma}{2\pi} { Ma},
\end{equation}
where $\gamma$ is the heat capacity ratio, which is $1.4$ for diatomic gas and $5/3$ for monatomic gas.

Considering a higher order accuracy, spatial reconstruction is required for the flow variables to improve the numerical resolution.
The reconstructed initial distribution function within a control volume has the form of
\begin{equation}\label{eq:reconstructed_f0}
f_i(\vec{x}) = f_i + f_{\vec{x}}^i \cdot (\vec{x} - \vec{x}_i),
\end{equation}
where $f_{\vec{x}}^i$ is the spatial derivative of the distribution function in cell $i$.
On a structured mesh, the spatial reconstruction can be carried out through direction by direction.
For example, the spatial gradient along $x$-direction can be computed by
\begin{equation}
	f^i_x = \left( {\text{sign}}(s_1) + {\text{sign}}(s_2) \right) \frac{|s_1| |s_2|}{|s_1| + |s_2|} ,
\end{equation}
with van Leer limiter \cite{van1977towards}, where $s_1 = (f_i - f_{i+1}) / (x_i - x_{i+1})$ and $s_2 = (f_i - f_{i-1}) / (x_i - x_{i-1})$.
For unstructured mesh, the least-square method is commonly used in the CFD computation, together with the slope limiters, such as Barth--Jespersen limiter \cite{barth1989design} and Venkatakrishnan limiter \cite{venkatakrishnan1995convergence}.

\subsubsection{Flux evaluation}\label{sec:flux_evaluation}

\begin{figure}[htbp]
	\centering
	\subfloat[]{\includegraphics[width=0.3\textwidth]{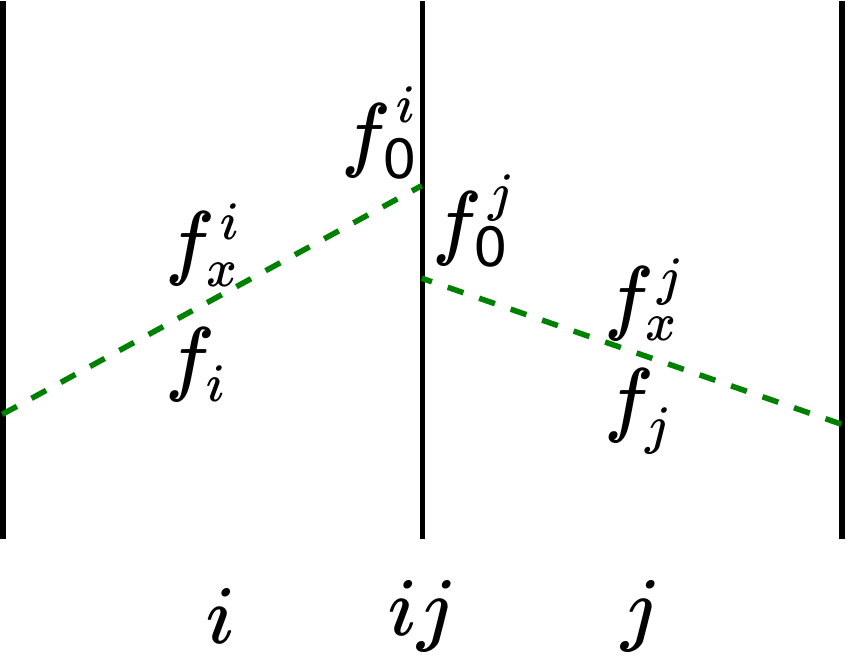}} \hspace{2em}
	\subfloat[]{\includegraphics[width=0.3\textwidth]{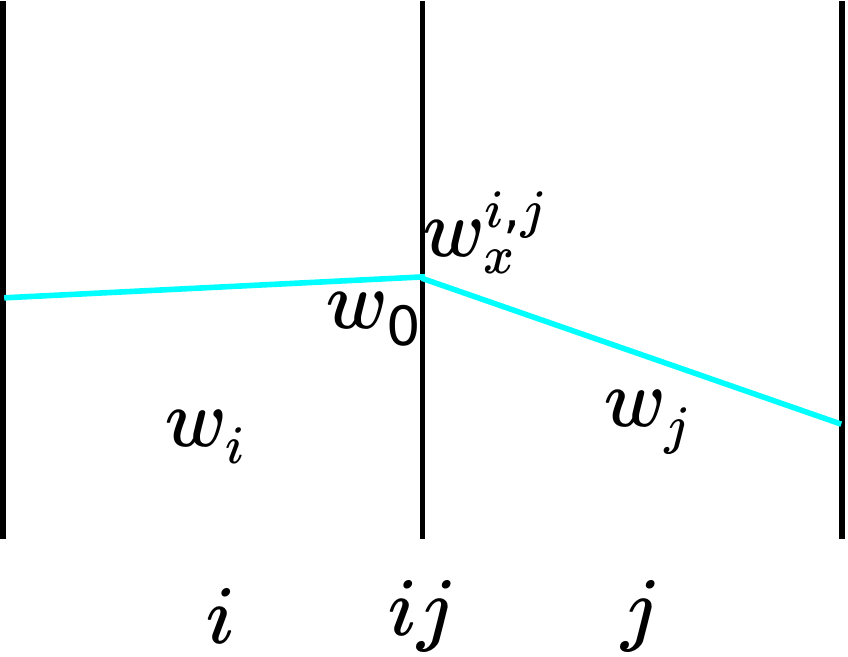}}
	\caption{\label{fig:interface_distribution}Distribution function around a cell interface. (a) Initial distribution function $f_0(\vec{x})$, and (b) the equilibrium state $g(\vec{x})$.}
\end{figure}
As shown in Fig.~\ref{fig:interface_distribution}, from the reconstructed distribution function in the cell, the initial distribution function at the cell interface can be obtained by
\begin{equation}\label{eq:interface_f0}
f_0(\vec{x}) =
\begin{cases}
f_0^i + f_{\vec{x}}^i \cdot \vec{x}, \quad  \vec{x} \cdot \vec{n}_{ij} < 0 \\
f_0^j + f_{\vec{x}}^j \cdot \vec{x}, \quad  \vec{x} \cdot \vec{n}_{ij} > 0 \\
\end{cases}
\end{equation}
where $f_0^i$ and $f_0^j$ are the interpolated values at the left and right sides of the interface.
The conservative variables are computed from the initial gas distribution function
\begin{equation}\label{eq:collided_w0}
\vec{w}_0
=
\sum_{\vec{u}_k \cdot \vec{n}_{ij} > 0} {f_{0,k}^i \vec{\psi}_k \mathcal{V}_k }
+
\sum_{\vec{u}_k \cdot \vec{n}_{ij} < 0} {f_{0,k}^j \vec{\psi}_k \mathcal{V}_k },
\end{equation}
and the spatial gradients of the macroscopic variables are obtained by local reconstruction
\begin{equation}\label{eq:interface_wx}
\vec{w}_{\vec{x}} =
\left(\vec{w}_x^{i,j}, \vec{w}_y, \vec{w}_z\right)^T,
\end{equation}
where two slopes $\vec{w}_x^{i,j}$ are employed for the spatial gradients along the normal direction of cell interface, while one slope is used for the other tangential directions \cite{xu2001gks,xu2005multidimensional}.
The spatial derivative of the equilibrium state $g_{\vec{x}}$ is computed from the gradient of macroscopic flow variables $\vec{w}_{\vec{x}}$ by the chain rule
\begin{equation}
	g_{\vec{x}} = \frac{\partial g}{\partial \rho} \frac{\partial \rho}{\partial \vec{x}} +
	\frac{\partial g}{\partial \rho U} \frac{\partial \rho U}{\partial \vec{x}} +
		\frac{\partial g}{\partial \rho V} \frac{\partial \rho V}{\partial \vec{x}} +
			\frac{\partial g}{\partial \rho W} \frac{\partial \rho W}{\partial \vec{x}} +
	\frac{\partial g}{\partial \rho E} \frac{\partial \rho E}{\partial \vec{x}},
\end{equation}
which is equivalent to the computation by using a Taylor expansion \cite{xu2001gks}.
For example, the gradient of equilibrium state in $x$-direction can be written as
\begin{equation}\label{eq:gx_expansion}
	\frac{\partial g}{\partial x} = g_M \left[a_0 + a_1 u + a_2 v + a_3 w + \frac{a_4}{2} (u^2 + v^2 + w^2)\right],
\end{equation}
where
\begin{equation}\label{eq:gx_coef}
	\begin{aligned}
	a_4 &= \frac{4\lambda^2}{3} \left(R_0 - 2 U R_1 - 2 V R_2 - 2 W R_3\right), \\
	a_3 & = 2\lambda R_3 - W a_4, \\
	a_2 & = 2\lambda R_2 - V a_4, \\
	a_1 & = 2\lambda R_1 - U a_4,\\
	a_0 & = \frac{\partial \rho}{\partial x} - U a_1 - V a_2 - W a_3 - \frac{1}{2} (U^2 + V^2 + W^2 + \frac{3}{2\lambda}) a_4, \\
	\end{aligned}
\end{equation}
and the related parameters are
\begin{equation}\label{eq:gx_coef_calc}
	\begin{aligned}
	R_0 &= 2 \frac{\partial \rho E}{\partial x} - \left(U^2 + V^2 + W^2 + \frac{3}{2\lambda}\right) \frac{\partial \rho}{\partial x},\\
	R_1 &= \frac{\partial \rho U}{\partial x} - U \frac{\partial \rho}{\partial x}, \\
	R_2 &= \frac{\partial \rho V}{\partial x} - V \frac{\partial \rho}{\partial x},\\
	R_3 &= \frac{\partial \rho W}{\partial x} - W \frac{\partial \rho}{\partial x}.
	\end{aligned}
\end{equation}
The temporal gradient of the conservative variables is obtained by the compatibility condition at $t=0$
\begin{equation}
\frac{d}{dt} \int {(g-f) \vec{\psi} d\Xi} = \vec{0},
\end{equation}
which gives
\begin{equation}\label{eq:interface_wt_a}	
\vec{w}_t
=
\int {g_t \vec{\psi} d\Xi}
=
\int {f_t \vec{\psi} d\Xi}
=
-\sum_k {\vec{u}_k f_{\vec{x},k} \vec{\psi}_k \mathcal{V}_k},
\end{equation}
or by the compatibility condition over a whole time step \cite{xu2001gks,liu2012modified}
\begin{equation}
\int_{0}^{\Delta t} {(g-f) \vec{\psi} d\Xi} = \vec{0},
\end{equation}
or by the Lax--Wendroff method \cite{xu2010ugks}
\begin{equation}\label{eq:interface_wt_b}	
\vec{w}_t = \int {g_t \vec{\psi} d\Xi} = -\int {(\vec{u} \cdot g_{\vec{x}}) \vec{\psi} d\Xi}.
\end{equation}
All these three mentioned methods can maintain the second order accuracy.
From the temporal gradient of macroscopic flow variables $\vec{w}_t$, the temporal gradient of equilibrium distribution function can be computed by the chain rule
\begin{equation}
g_t = \frac{\partial g}{\partial \rho} \frac{\partial \rho}{\partial t} +
\frac{\partial g}{\partial \rho U} \frac{\partial \rho U}{\partial t} +
\frac{\partial g}{\partial \rho V} \frac{\partial \rho V}{\partial t} +
\frac{\partial g}{\partial \rho W} \frac{\partial \rho W}{\partial t} +
\frac{\partial g}{\partial \rho E} \frac{\partial \rho E}{\partial t}.
\end{equation}
Similarly, $g_t$ can be calculated as well by Eqs.~\eqref{eq:gx_expansion}, \eqref{eq:gx_coef} and \eqref{eq:gx_coef_calc}, in which $\partial /\partial x$ needs to be replaced by $\partial / \partial t$.

So far, we have the initial distribution function $f_0(\vec{x})$ and the equilibrium state $g(\vec{x},t)$ around the cell interface, and the microscopic flux $\mathcal{F}_{ij}$ and macroscopic flux $\vec{F}_{ij}$ can be obtained from Eqs.~\eqref{eq:fij_flux} and \eqref{eq:macro_flux_from_micro}.
In section \ref{sec:flux_function}, it is pointed out that the integral solution is valid when the mesh size could resolve the flow structure.
For the cases that there are discontinuities in the flow field, such as shocks, the mesh size is usually not fine enough to resolve the flow structure around the discontinuities in the hydrodynamic regime.
At this time, a shock capturing scheme is preferred.
Therefore, extra numerical dissipation has to be added in the local region near discontinuities by changing the relaxation time $\tau$.
Specifically, the relaxation time $\tau$ at the cell interface for flux evaluation is computed by
\begin{equation}\label{eq:numerical_tau}
\tau = \frac{\mu}{p} + |\frac{p^i - p^j}{p^i+p^j}| \Delta t,
\end{equation}
where $p^i$ and $p^j$ are the interpolated pressure at the left and right sides of cell interface to distinguish a discontinuity.

\subsubsection{Solution update}
In the UGKS, the macroscopic flow variables and the gas distribution function are updated simultaneously.
With the macroscopic flux across cell interfaces, the conservative variables $\vec{w}^{n+1}$ at $t^{n+1}$ can be updated by conservation law \eqref{eq:macro_conservation}.
For the collision term, the UGKS adopts an implicit treatment using the trapezoidal rule
\begin{equation}
\mathcal{J}(f,f) = \frac{1}{2}\left[\frac{g_i^{n+1} -f_i^{n+1}}{\tau_i^{n+1}} + \frac{g_i^n -f_i^n}{\tau_i^n}\right],
\end{equation}
where $g_i^{n+1}$ and $\tau_i^{n+1}$ are obtained from the updated macroscopic variables $\vec{w}_i^{n+1}$.
From Eq.~\eqref{eq:micro_conservation}, the updated distribution function is
\begin{equation}\label{eq:micro_updation}
f_i^{n+1} = f_i^n + \frac{2\tau_i^{n+1}}{\Delta t + 2\tau_i^{n+1}}
\left[- \frac{\Delta t}{V_i} \sum_{j \in N(i)} {\mathcal{F}_{ij} S_{ij}} + \frac{\Delta t}{2}\left(\frac{g_i^{n+1} -f_i^{n}}{\tau^{n+1}} + \frac{g_i^n -f_i^n}{\tau^n}\right)\right].
\end{equation}

\subsubsection{Boundary treatment}\label{sec:ugks_bc}
\begin{figure}[htbp]
	\centering
	\subfloat[\label{fig:bc_a}]
	{\includegraphics[width=0.3\textwidth]{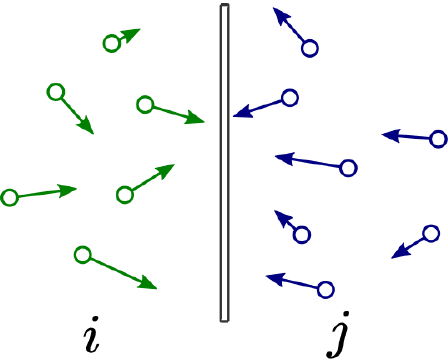}}\hspace{2em}
	\subfloat[\label{fig:bc_b}]
	{\includegraphics[width=0.4\textwidth]{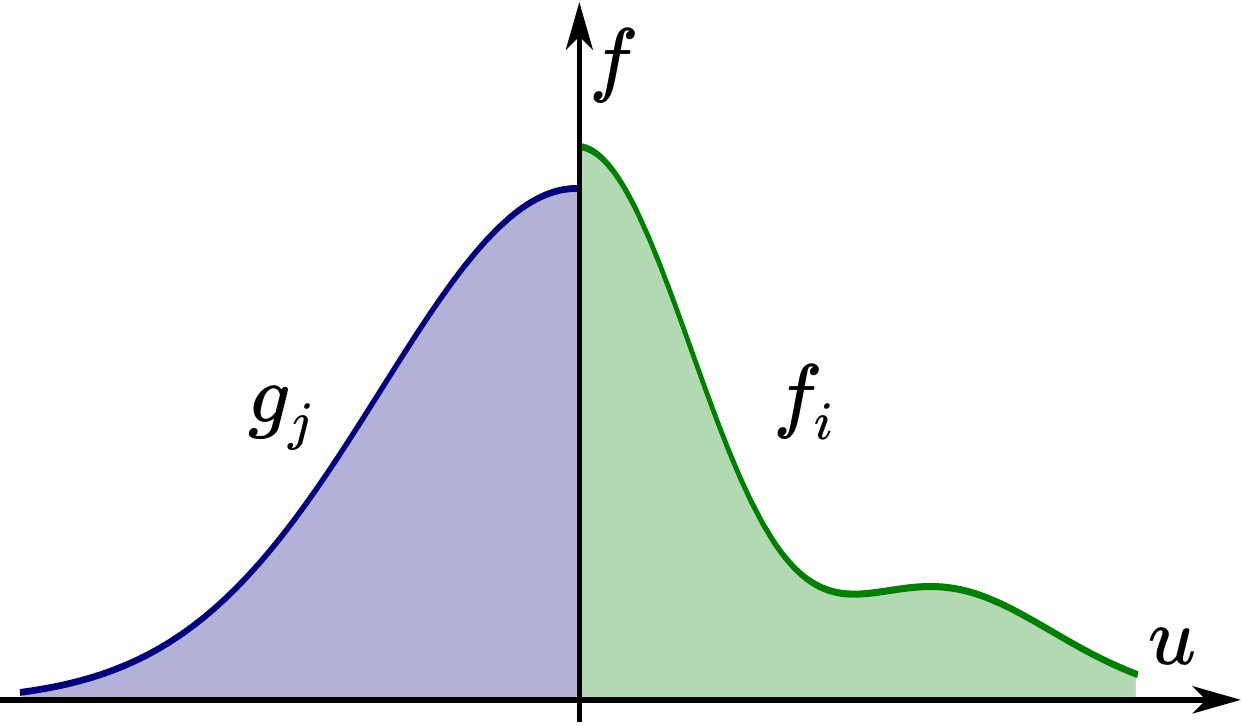}}
	\caption{\label{fig:boundary_condition}Illustration for boundary treatment. (a) Inner cell $i$ and ghost cell $j$ around cell interface, (b) distribution function at cell interface.}
\end{figure}

In the CFD computations, ghost cell is commonly used to implement the boundary condition.
As shown in Fig.~\subref*{fig:bc_a}, assume that the inner cell is on the left side of boundary interface, denoted by $i$, and the corresponding ghost cell is on the right side, denoted by $j$.
For kinetic solvers based on the discrete velocity method, the distribution functions on both sides of the interface need to be constructed, as illustrated in Fig.~\subref*{fig:bc_b}.

\noindent
\textbf{(a) Isothermal wall boundary condition}

For isothermal wall with complete accommodation, the diffusive reflection condition is conducted at the boundary, where the distribution function on the ghost cell side is a Maxwellian distribution.
This boundary is commonly employed in the rarefied gas simulation, which can provide slip velocity at the wall surface automatically.
In order to recover the non-slip boundary in the continuum limit, a multiscale boundary was proposed by Chen et al. \cite{chen2015comparative} through using the local integral solution again.
With the given wall temperature $T_w$ and velocity $\vec{U}_w$ (only tangential velocity is considered here), the only unknown for the reflected Maxwellian distribution $g_w$ is the wall density, which can be computed from the non-penetration condition
\begin{equation}\label{eq:nonpenetration}
F_{\rho}^{in} + F_{\rho}^{w} = 0,
\end{equation}
where the incident mass from inner cell is computed by
\begin{equation}
F_{\rho}^{in} = \sum_{u_{n,ij,k} > 0}{u_{n,ij,k} \bar{f}_{i,k}  \mathcal{V}_k},
\end{equation}
and the reflected mass is
\begin{equation}
F_{\rho}^{w}=
\rho_w \sum_{u_{n,ij,k} < 0}{u_{n,ij,k} \hat{g}_{w,k} \mathcal{V}_k}.
\end{equation}
$\hat{g}_{w}$ is the Maxwellian distribution for unit density, i.e.,
\begin{equation}
\hat{g}_{w} =\left(\frac{m_0}{2\pi k_B T_w}\right)^{\frac{3}{2}} {\exp}\left[{-\frac{m_0 (\vec{u}-\vec{U}_w)^2}{2 k_B T_w}}\right].
\end{equation}
$\bar{f}_{i,k}$ denotes the averaged distribution function constructed from Eq.~\eqref{eq:fij_expression}, where the initial distribution function is extrapolated from the inner cells, and the required macroscopic flow variables at cell interface are determined from the no slip boundary condition with a zero gradient pressure, i.e., $T = T_w$, $\vec{U} = \vec{U}_w$ and $p = p_i$.

From non-penetration condition \eqref{eq:nonpenetration}, the reflected wall density $\rho_w$ can be determined, and the microscopic flux across the boundary will be
\begin{equation}
\mathcal{F}_w =
\begin{cases}
u_{n,ij} \bar{f}_i,       & \quad u_{n,ij} > 0\\
u_{n,ij} \rho_w \hat{g}_w, &\quad  u_{n,ij} < 0
\end{cases}
\end{equation}
and the macroscopic fluxes are
\begin{equation}
\vec{F}_w =
\sum_{u_{n,ij,k} > 0}{u_{n,ij,k}\bar{f}_{i,k} \vec{\psi}_k \mathcal{V}_k}
+
\rho_w \sum_{u_{n,ij,k} < 0}{u_{n,ij,k} \hat{g}_{w,k} \vec{\psi}_k \mathcal{V}_k}.
\end{equation}

\noindent
\textbf{(b) Inlet and outlet boundary condition}

There are many inlet and outlet boundary conditions for CFD simulations in the continuum regime, such as the farfield boundary that all flow variables in the freestream are specified, subsonic inlet consisting of the specification of total pressure and total temperature, subsonic outlet with a predefined static pressure, supersonic inlet and outlet.
For NS solvers, the macroscopic flow variables at the cell interface can be determined by the boundary condition and flow variables in the inner cell.
Usually the characteristic variable method with Riemann invariants is adopted.
Details can be found in the CFD textbook \cite{blazek2015cfd}.

In the UGKS, one convenient way for conducting the inlet boundary condition is to set the macroscopic state of ghost cell with the flow variables determined by the boundary conditions using characteristic variable method, and to initialize the distribution function in the ghost cell by the corresponding Maxwellian equilibrium.
While for outlet boundaries, the macroscopic variables and distribution function should be interpolated from the interior domain for supersonic flow.
Specifically, the states in the ghost cell are in the form of
\begin{equation}\label{eq:ghost_state_macro}
\vec{w}_j = \mathcal{B}(\vec{w}_i, \vec{w}_{bc}),
\end{equation}
and
\begin{equation}\label{eq:ghost_state_micro}
f_j = \mathcal{B}(f_i, f_{bc}),
\end{equation}
where $\mathcal{B}$ is the operator for calculating boundary states, and $\vec{w}_{bc}$ and $f_{bc}$ are the prescribed variables to specify the boundary.
With the ghost states, fluxes across the inlet and outlet boundaries can be computed as same as that for inner interface.

The boundary condition based on the Riemann invariant is only valid in the hydrodynamic regime. A semi-empirical boundary condition was proposed to include the rarefied effect \cite{chen2012mesh}, where a weight function $\alpha = e^{-1/{ Kn}}$ constructed from the global Knudsen number is used to modify the boundary condition as $\vec{w}_j = \alpha \vec{w}_{\infty} + (1-\alpha)\mathcal{B}(\vec{w}_i, \vec{w}_{\infty})$.
This modification provides a smooth transition connecting the two limiting cases of free molecular and continuum flows.

\noindent
\textbf{(c) Symmetric plane}

For a symmetric boundary, both the macroscopic and microscopic flow variables are mirrored with respect to the boundary, i.e.,
\begin{equation}
(\rho_j,  U_j, V_j, W_j, T_j)^T =
(\rho_i, -U_i, V_i, W_i, T_i)^T,
\end{equation}
and
\begin{equation}
f_j(u, v, w) = f_i (-u, v, w).
\end{equation}
Then the boundary calculation is the same as the interface in interior domain.

\subsubsection{Flow chart}
\begin{figure}[hbtp]
	\centering
	\includegraphics[width=0.9\textwidth]{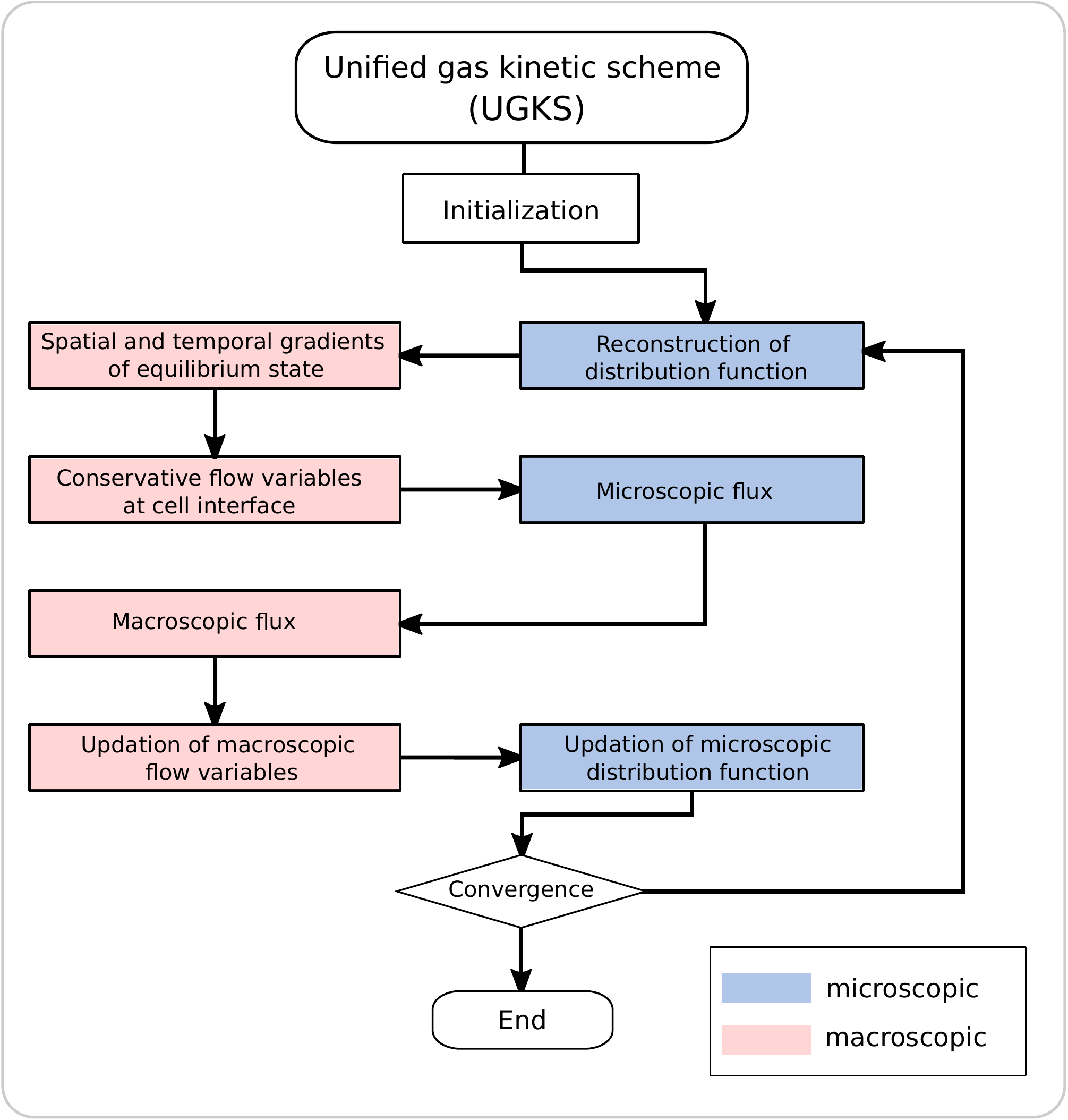}
	\caption{\label{fig:ugks_flowchart}Flow chart of UGKS.}
\end{figure}

The flow chart for the UGKS computation is drawn in Fig.~\ref{fig:ugks_flowchart}.
The procedures to evolve the flow field by the UGKS from $t^n$ to $t^{n+1}$ can be summarized as
\begin{description}
	\item[Step 1] Reconstruction of distribution function \\
	Get the spatial distribution of microscopic variable in the cell $f_i(\vec{x})$ in Eq.~\eqref{eq:reconstructed_f0}, and obtain the initial distribution function at the cell interface $f_0(\vec{x})$ by spatial interpolation, see Eq.~\eqref{eq:interface_f0}.
	\item[Step 2] Reconstruction of macroscopic flow variables \\
	Compute the conservative variables $\vec{w}_0$ from compatibility condition \eqref{eq:collided_w0}, and obtain the spatial derivatives $\vec{w}_{\vec{x}}$ by local reconstruction, then compute the temporal gradient of the conservative flow variables $\vec{w}_t$ from Eq.~\eqref{eq:interface_wt_b}, and correspondingly the spatial and temporal gradients of the equilibrium distribution can be obtained by chain rules.
	\item[Step 3] Flux evaluation \\
	 Compute the microscopic flux across the cell interfaces and boundaries $\mathcal{F}_{ij}$ from Eq.~\eqref{eq:fij_flux}, and get the macroscopic flux $\vec{F}_{ij}$ by Eq.~\eqref{eq:macro_flux_from_micro}.
	\item[Step 4] Solution update \\
	Update the conservative flow variables $\vec{w}^{n+1}$ based on conservation law \eqref{eq:macro_conservation}, and get the corresponding equilibrium state $g^{n+1}$, and then update the $f^{n+1}$ by Eq.~\eqref{eq:micro_updation}.
\end{description}

The UGKS employs the integral solution of the kinetic model over a finite time step with the accumulative effect of particles collision during the transport process, and evolves both the macroscopic flow variables and microscopic distribution function, which makes it possible to use an implicit treatment of the collision term to overcome its stiffness in the continuum regimes.
With the Chapman--Enskog expansion \cite{chapman1990mathematical}, the asymptotic property \cite{dimarco2011exponential,filbet2010class} has been analyzed for the NS limit \cite{liu2016boltzmann,liu2020linear}.
Unlike kinetic methods using splitting treatment of particles' transport and collision, the UGKS removes the constraint that the mesh size and the time step have to be smaller than the mean free path and mean collision time in order to capture NS solutions.
The UGKS follows the conservation law in the framework of finite volume method, and provides detailed flow physics evolution through time-dependent solution of the distribution function on the numerical discretization scale.
Therefore, the UGKS is a multiscale method which is capable of capturing the flow physics in all Knudsen number regimes, and it is supposed to be more efficient due to its scale-adaptive property.

\subsection{Numerical simulations}
The UGKS has been applied in the simulations of non-equilibrium flows, mainly for the microflow and the high-speed flows in rarefied regimes.
Some examples will be given in the following to illustrate the capability of the UGKS for micro flows and supersonic flows.

\subsubsection{Couette flow}

The Couette flow is a standard simple test for the whole flow regime.
It is a steady flow that is driven by the surface shearing of two infinite and parallel plates moving oppositely along their own planes.
The Knudsen number is defined as ${ Kn} =l_{HS} / h$, where $l_{HS}$ is the mean free path based on hard sphere model, and $h$ is the distance between the plates.

\begin{figure}[htbp]
\centering
\subfloat[\label{fig:ugks_couette_u}]
{\includegraphics[width=0.4\textwidth]{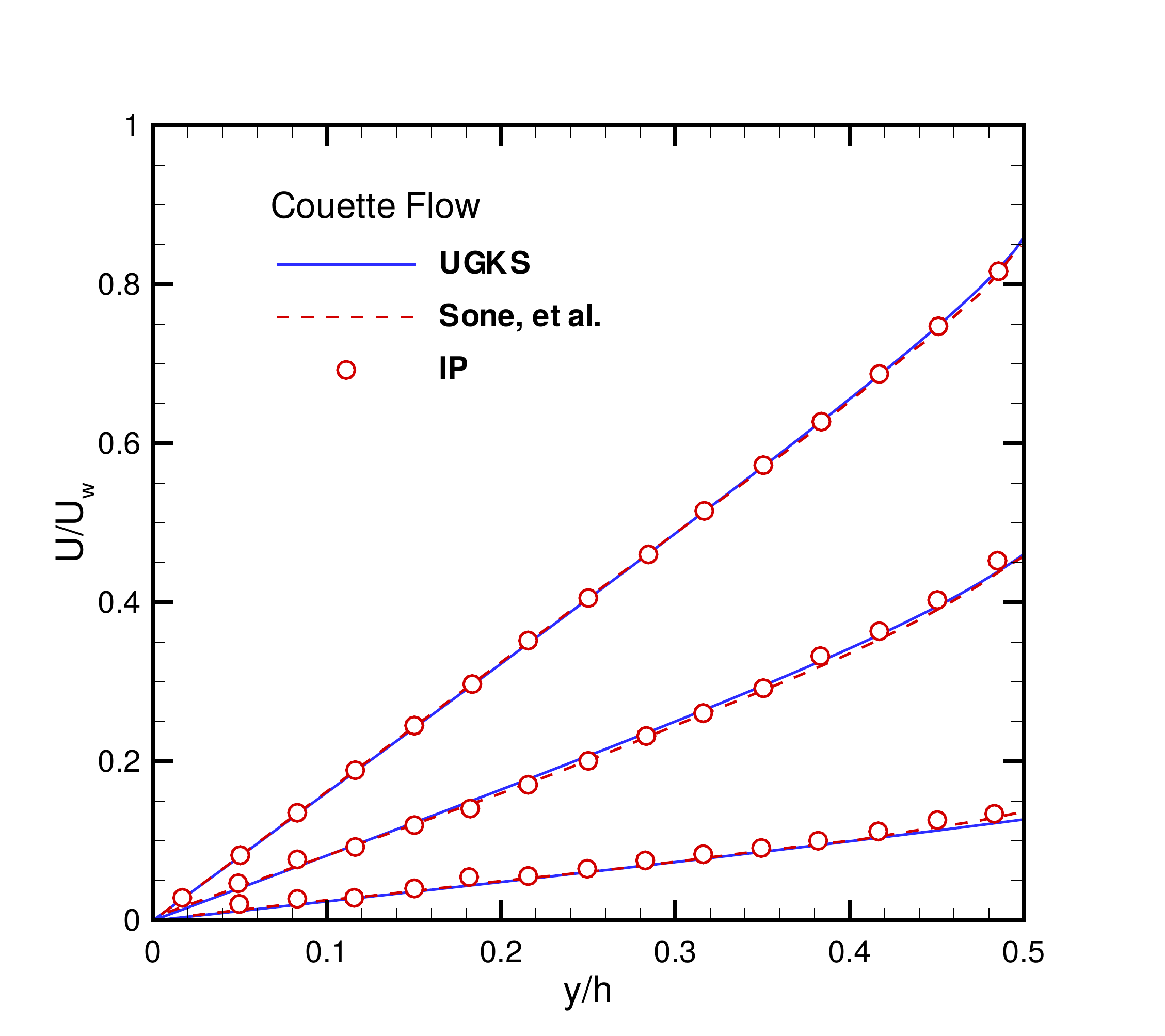}}
\subfloat[\label{fig:ugks_couette_tau}]
{\includegraphics[width=0.4\textwidth]{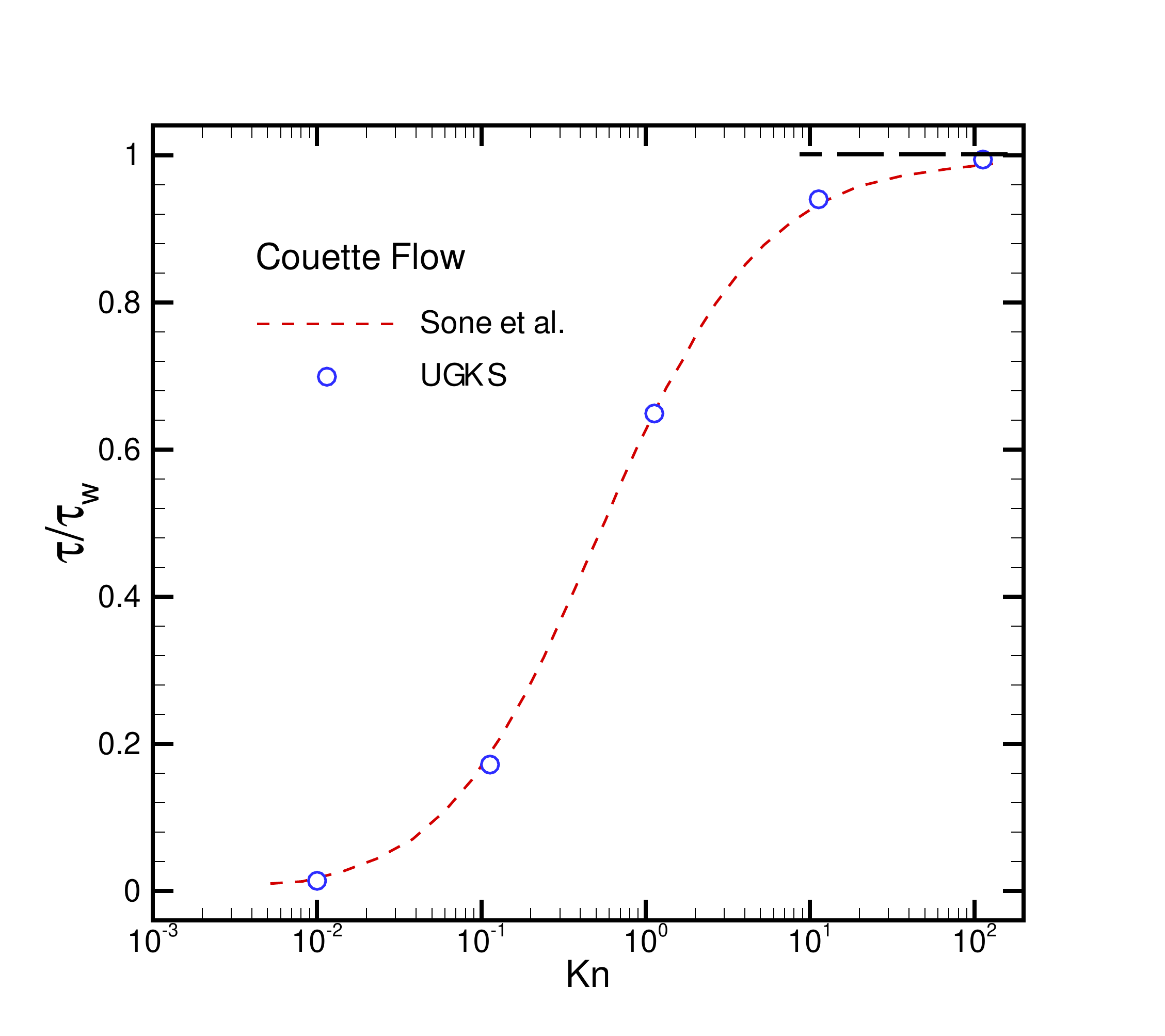}}
\caption{Couette flows \cite{huang2013micro}. (a) Comparison of velocity profiles at $Kn=0.2/\sqrt{\pi},2/\sqrt{\pi}, 20/\sqrt{\pi}$ between the information preserving (IP) method \cite{fan2001statistical}, linearized Boltzmann equation \cite{sone1990lbe} and UGKS; (b) relation of drag versus Knudsen number. }
\end{figure}

\begin{figure}[htbp]
\centering
\subfloat[\label{fig:ugks_tcouette_temperature}]
{\includegraphics[width=0.4\textwidth]{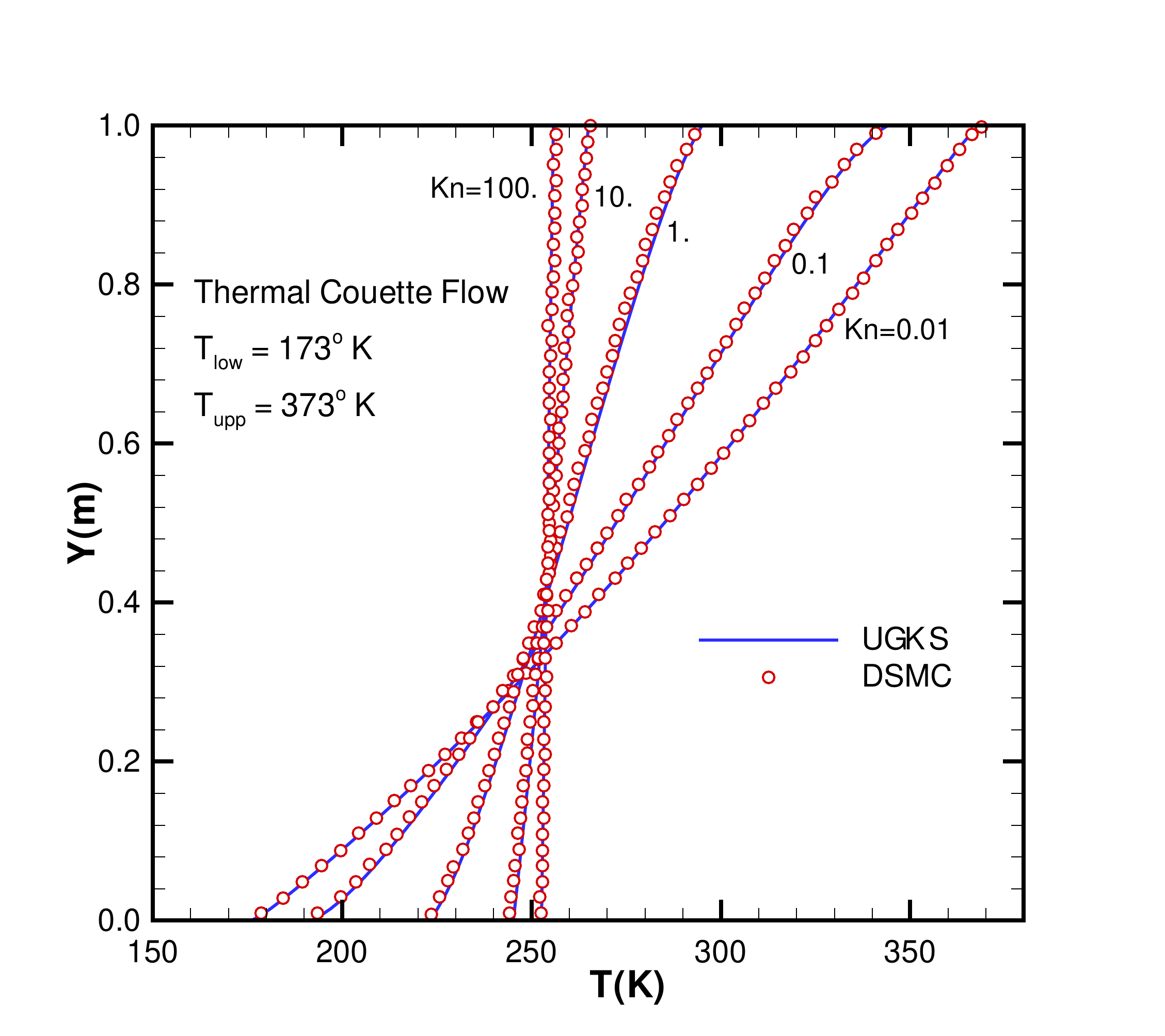}}
\subfloat[\label{fig:ugks_tcouette_heatflux}]
{\includegraphics[width=0.4\textwidth]{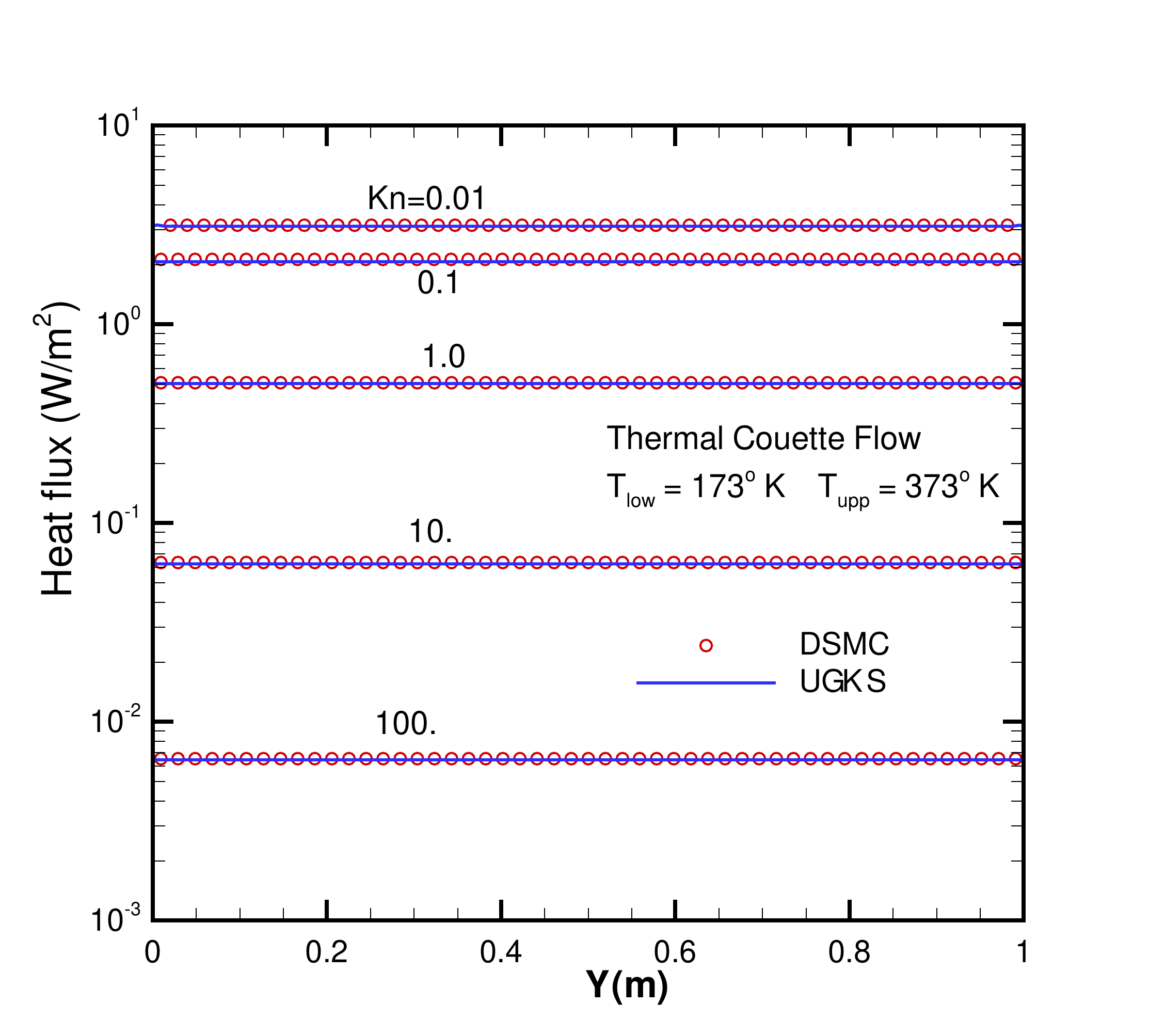}}
\caption{\label{fig:ugks_tcouette}Thermal Couette flow at ${ Kn} = 0.01, 0.1, 1, 10$, and
$100$ calculated by DSMC and UGKS \cite{huang2013micro}. (a) Temperature profiles, (b) heat transfer profiles. }
\end{figure}

In the transition regime, three Knudsen numbers are considered, i.e., $0.2/\sqrt{\pi}$, $2/\sqrt{\pi}$, and $20/\sqrt{\pi} $.
To resolve the flow fields well, $100$ cells are employed in the current calculation for all three cases.
Figure \subref*{fig:ugks_couette_u} compares the velocity profiles given by UGKS, the linearized Boltzmann equation \cite{sone1990lbe}, and IP-DSMC results \cite{fan2001statistical}.
All solutions have excellent agreement. 
Figure \subref*{fig:ugks_couette_tau} also compares the relation of the surface shear stress versus the Knudsen number given by various methods.
The normalization factor is the collisionless solution \cite{fan2001statistical}.
Both solutions agree nicely with each other in the whole flow regime.
The above test is basically an isothermal one.

Simple heat conduction problem in rarefied gas is also a valuable case to test the capability to capture thermal effect.
This consists of two stationary parallel surfaces maintained at different temperatures.
The same problems have been studied in  \cite{sun2002direct,masters2007octant}.
The up and down surfaces are maintained at temperature of $173 { K}$ and $373{ K}$ separately with a $1{ m}$ gap between them and the intervening space is filled with argon gas at various densities to have the corresponding Knudsen numbers ${ Kn} = 0.01$, $0.1$, $1$, $10$, and $100$.
The one dimensional computational domain is discretized with $100$ cells in the physical space and $28 \times 28$ grid points in the velocity space.
Figure \ref{fig:ugks_tcouette} presents the temperature profiles and heat flux results from the unified scheme and the benchmark DSMC solution.
There is an excellent agreement between UGKS and DSMC solutions in the whole range of Knudsen numbers.

\subsubsection{Micro cavity flow}

\begin{figure}[htbp]
\centering
\subfloat[]{
\includegraphics[width=0.45\textwidth]{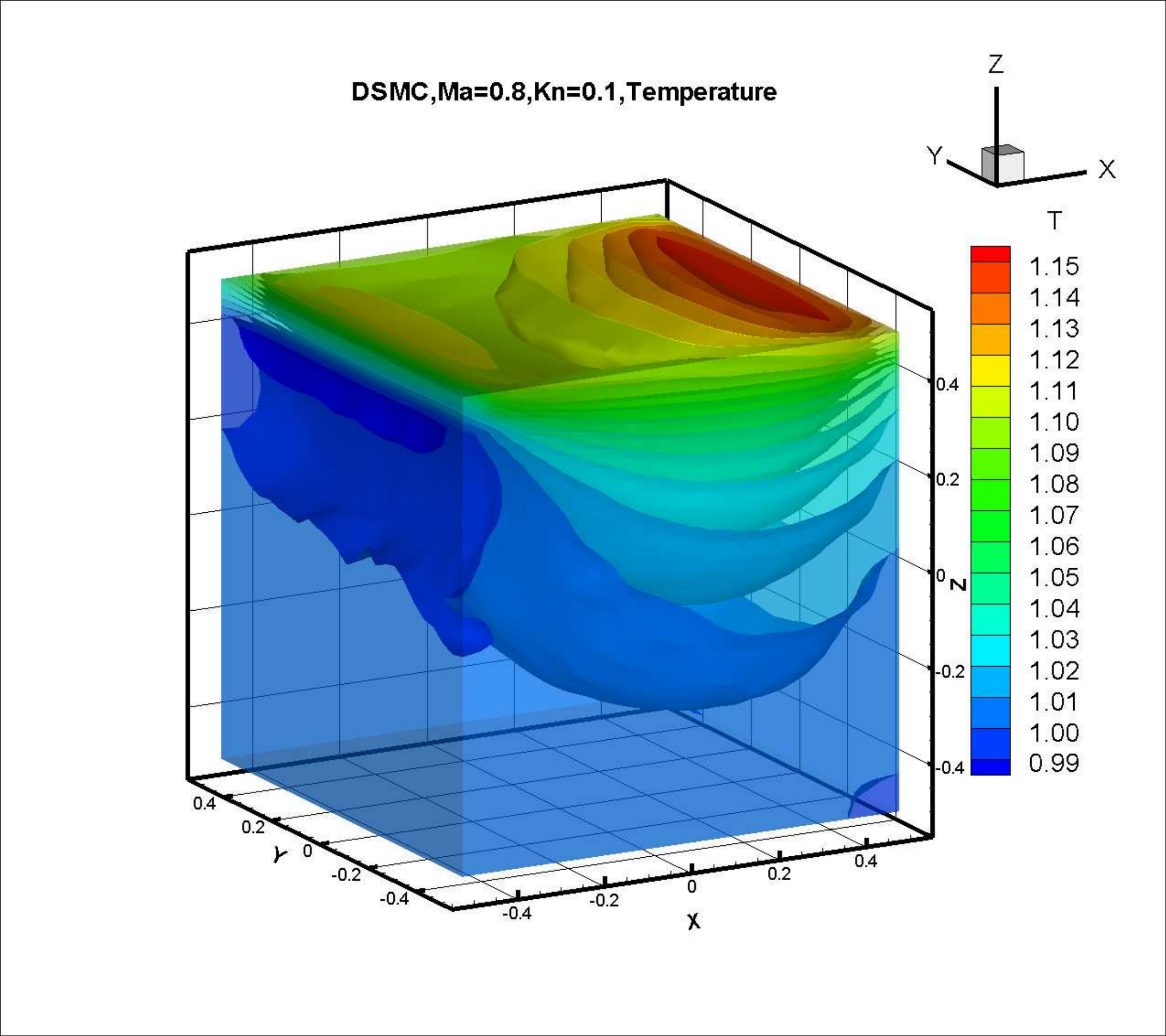}}\hspace{1em}
\subfloat[]{\includegraphics[width=0.45\textwidth]{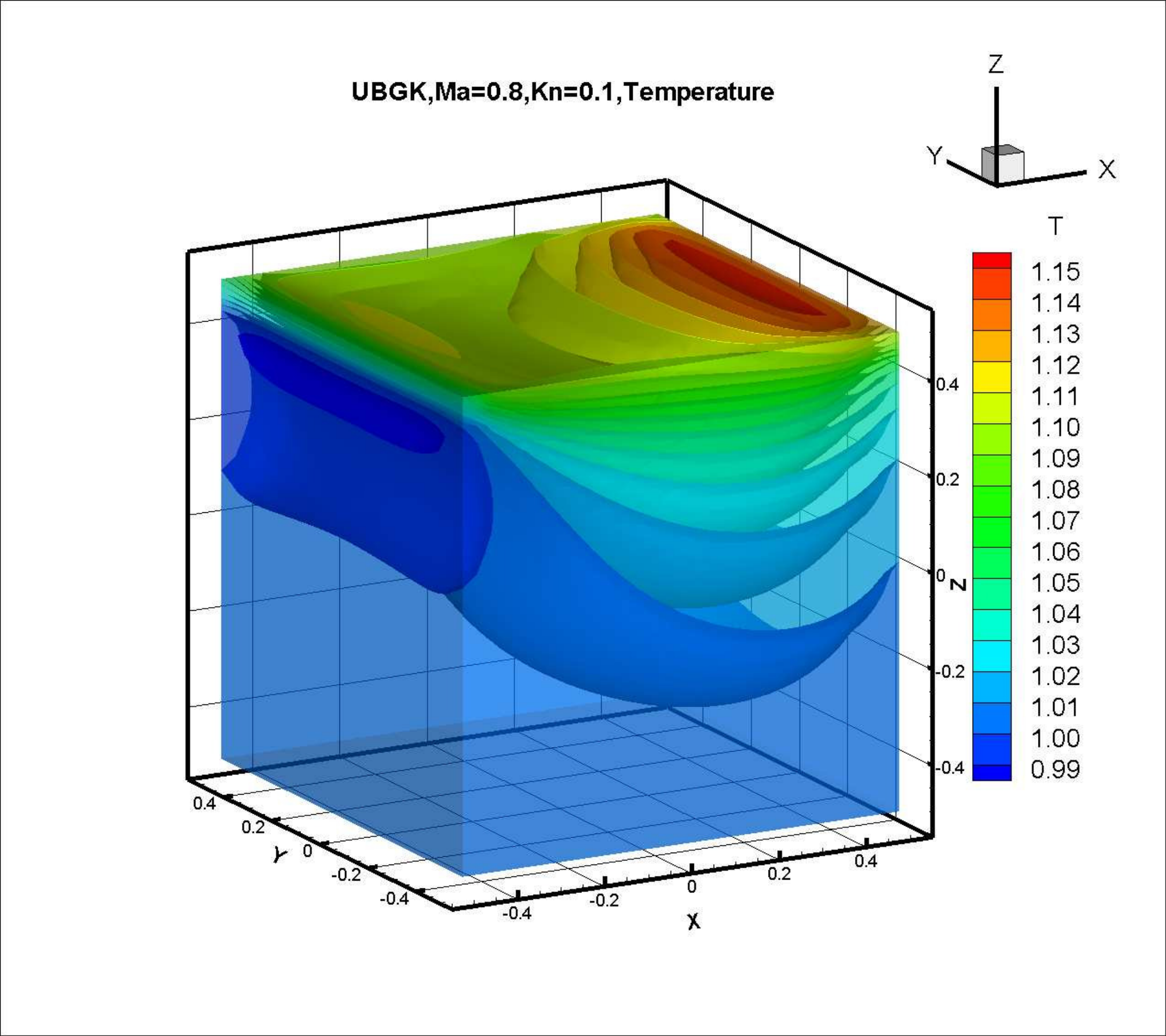}}
\caption{\label{fig:ugks_3d_cavity}Temperature distribution in 3D cavity simulation at Kn=0.1 and Ma=0.8 \cite{huang2012multidimension}. (a) DSMC data and (b) UGKS solutions.}
\end{figure}

\begin{figure}[htbp]
\centering
\subfloat[]{
\includegraphics[width=0.48\textwidth]{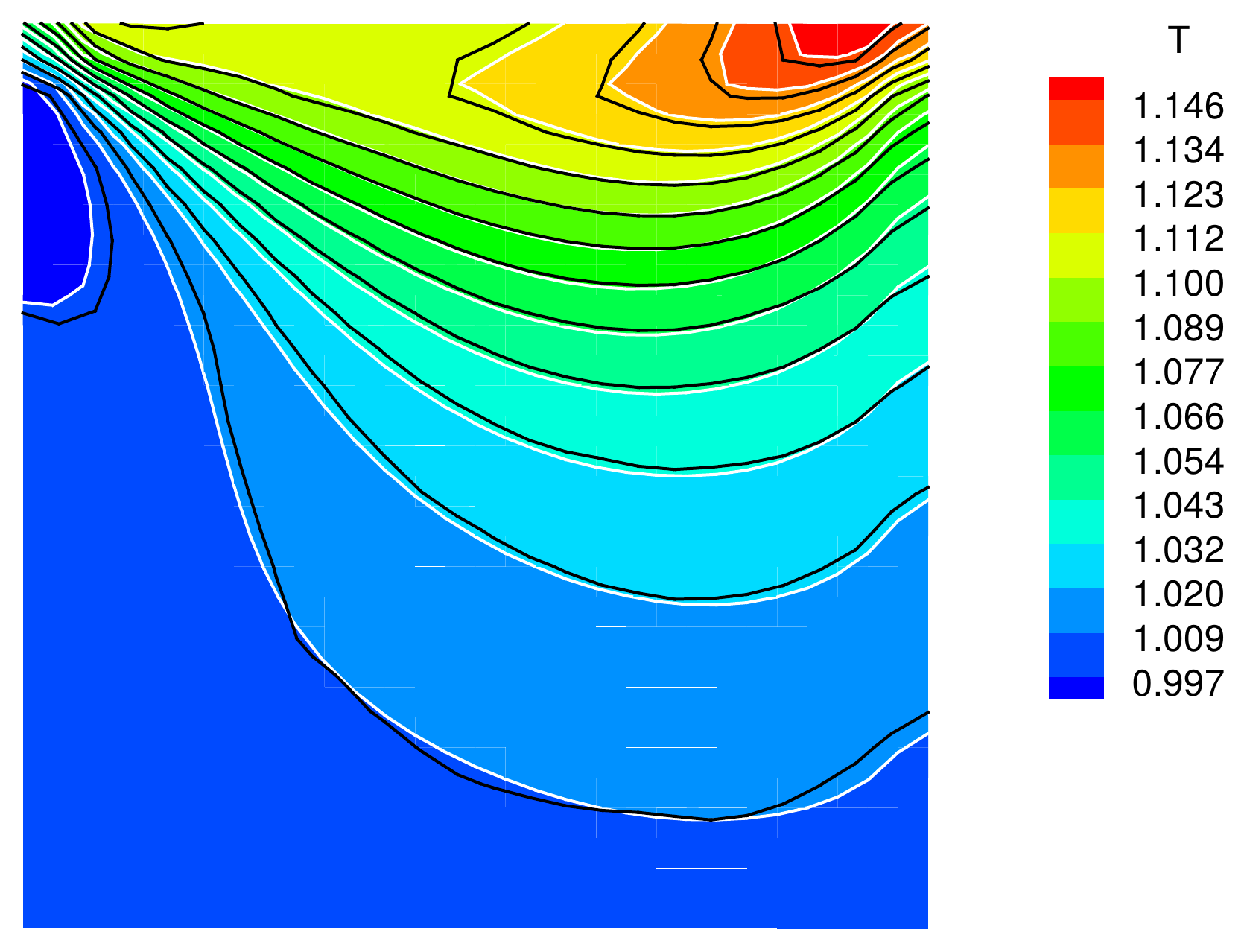}}
\subfloat[]{
\includegraphics[width=0.48\textwidth]{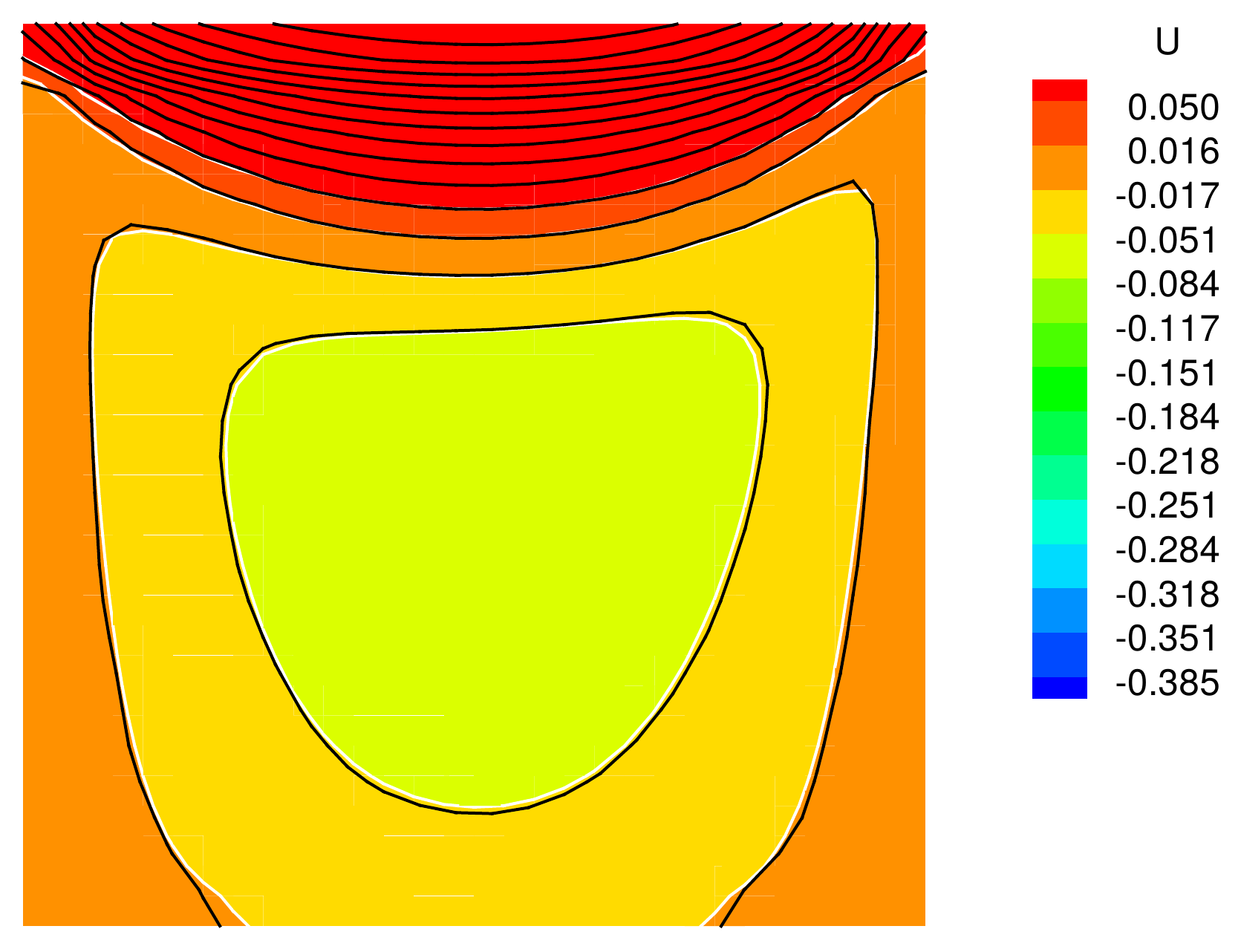}} \\
\subfloat[]{
\includegraphics[width=0.48\textwidth]{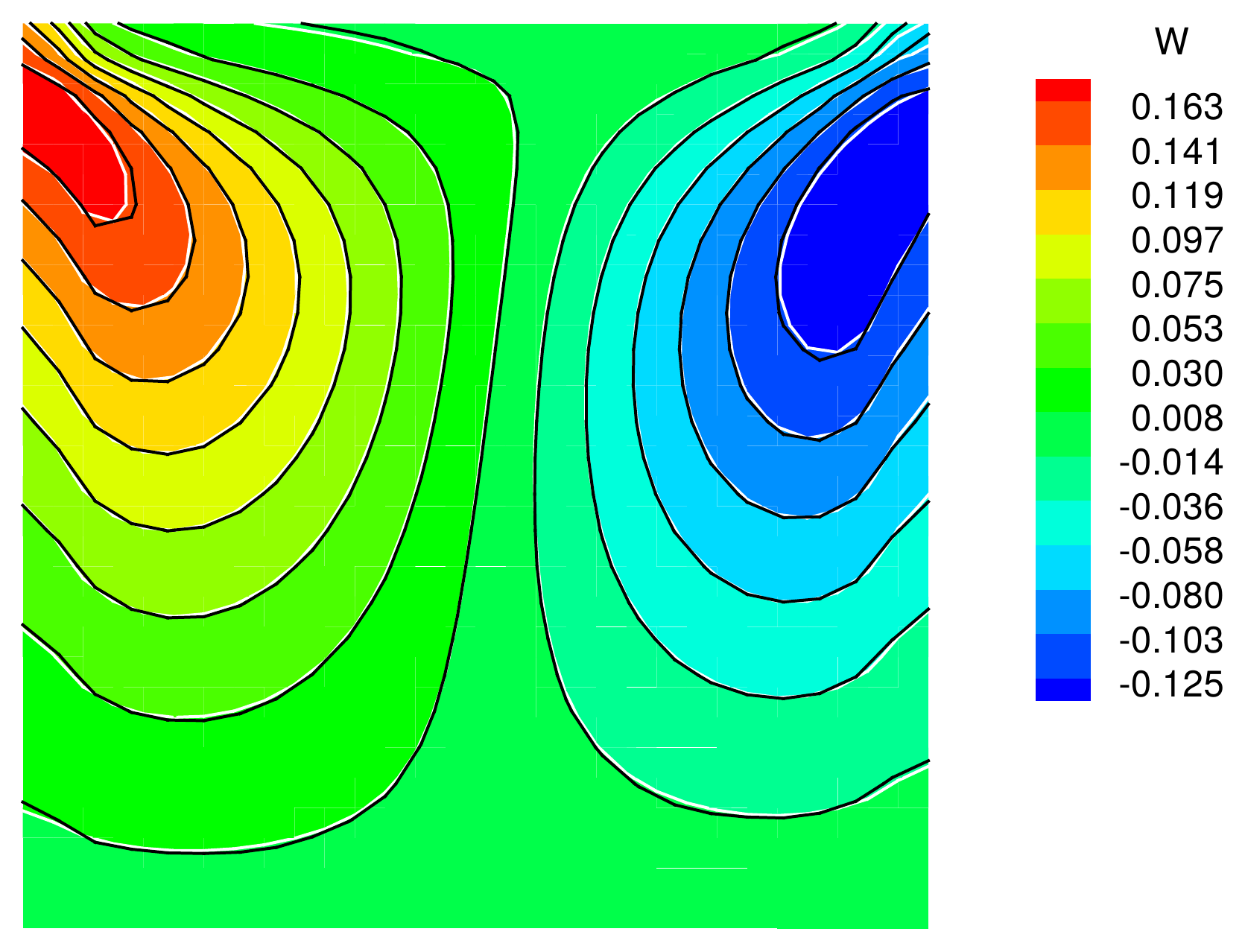}}
\subfloat[]{
\includegraphics[width=0.48\textwidth]{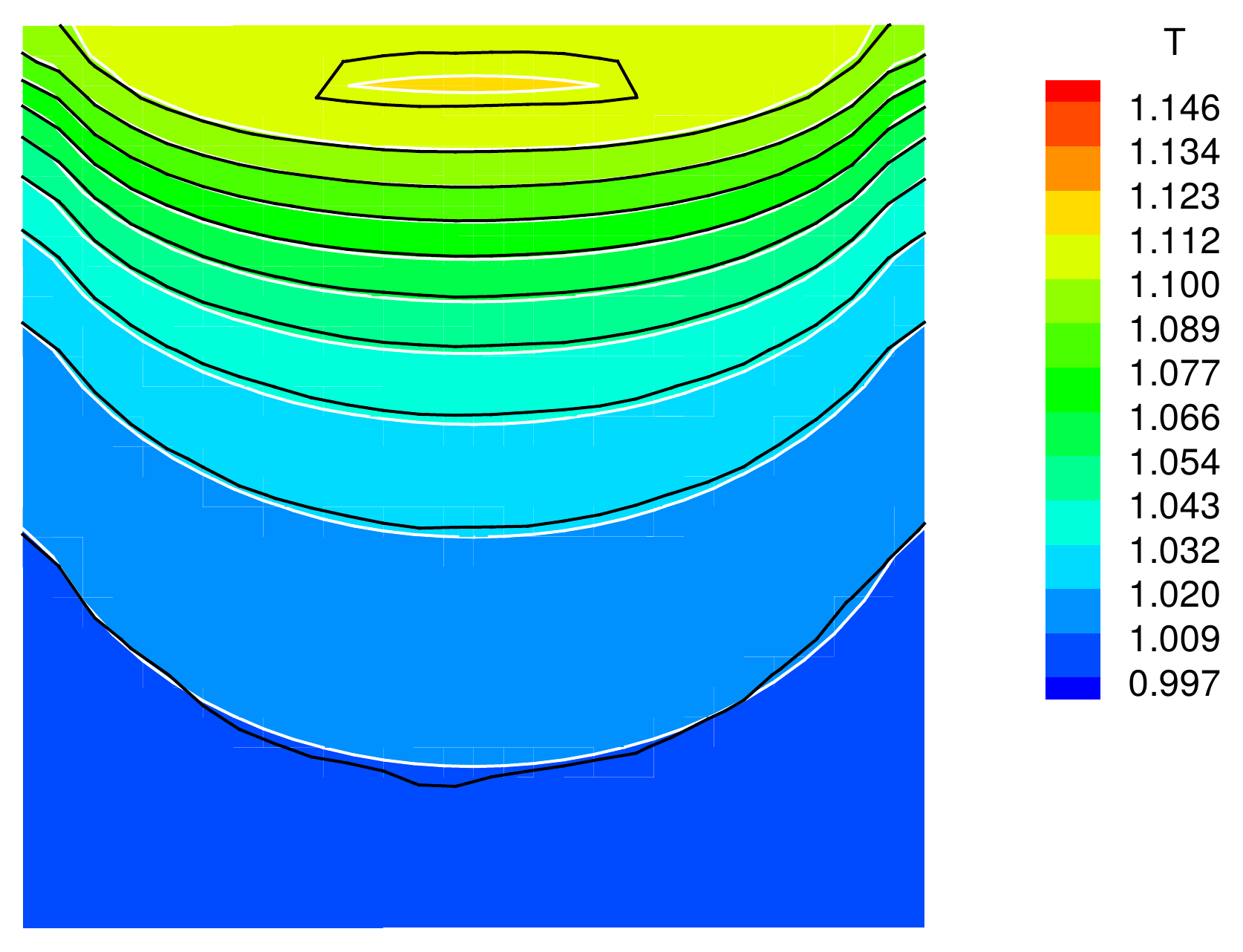}}
\caption{\label{fig:3d_cavity_cut}Flow distributions in different cut-planes in 3D simulation \cite{huang2012multidimension}.
 Black lines: DSMC, white lines: UGKS.  (a) Temperature in symmetric $zx$-plane, (b) $x$-component velocity in symmetric $zx$-plane, (c) $z$-component velocity in symmetric $zx$-plane, and (d) temperature in symmetric $zy$-plane.}
\end{figure}

For the lid-driven cavity flows, the gaseous medium is assumed to consist of monatomic molecules corresponding to that of argon with mass, $m_0 =6.63 \times 10^{-26} { kg}$.
In the DSMC solution, the variable hard sphere (VHS) collision model has been used, with a reference particle diameter of $d = 4.17 \times 10^{-10} { m}$.
The wall temperature is kept the same as the reference temperature, i.e., $T_w = T_0 = 273{ K}$.
In the study, the wall velocity is kept fixed, which is set according to a Mach number 0.8.
The Knudsen number variation is achieved by varying the density.
Maxwell model is used to represent surface accommodation, where only the case with full wall accommodation is presented.
Figure \ref{fig:ugks_3d_cavity} and \ref{fig:3d_cavity_cut} show the three dimensional temperature contours and flow distribution in different cut-planes.
Excellent match between the DSMC data and the UGKS solutions has been obtained.

\subsubsection{Slit flow}

\begin{figure}[htbp]
\centering
\subfloat[\label{fig:slit_mesh}]{\includegraphics[width=0.46\textwidth]{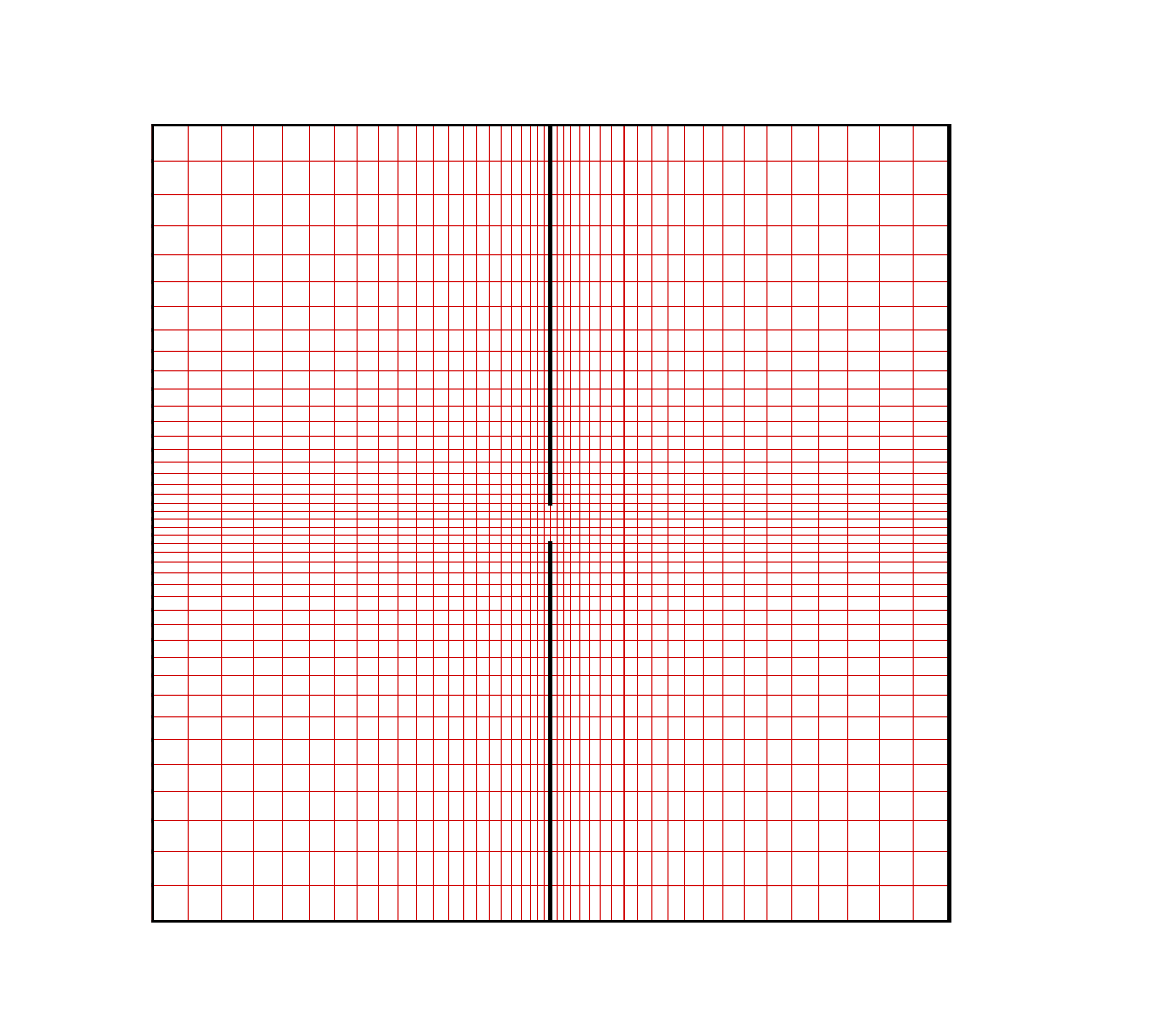}} \hspace{1em}
\subfloat[]{\includegraphics[width=0.46\textwidth]{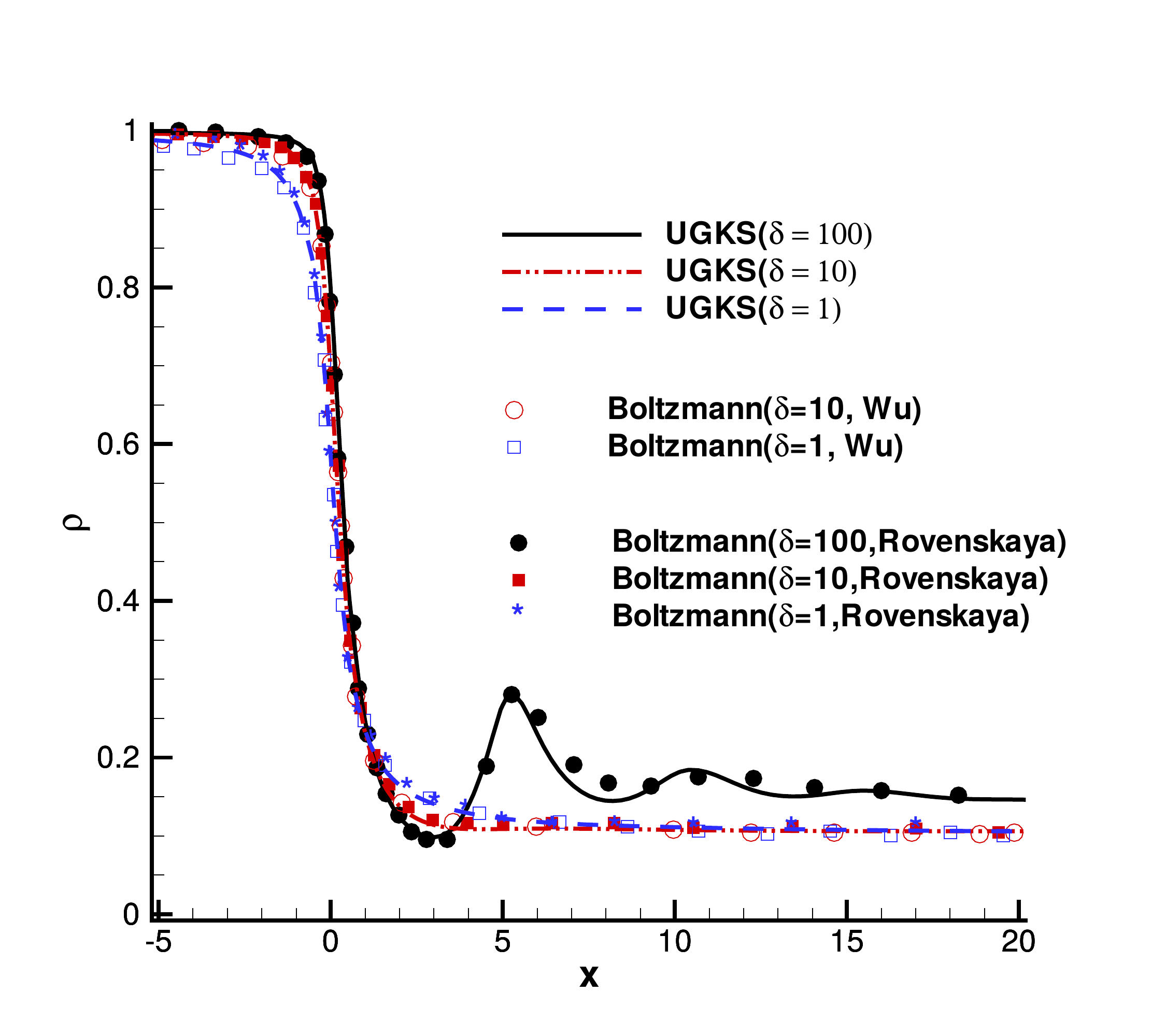}} \\
\subfloat[]{\includegraphics[width=0.46\textwidth]{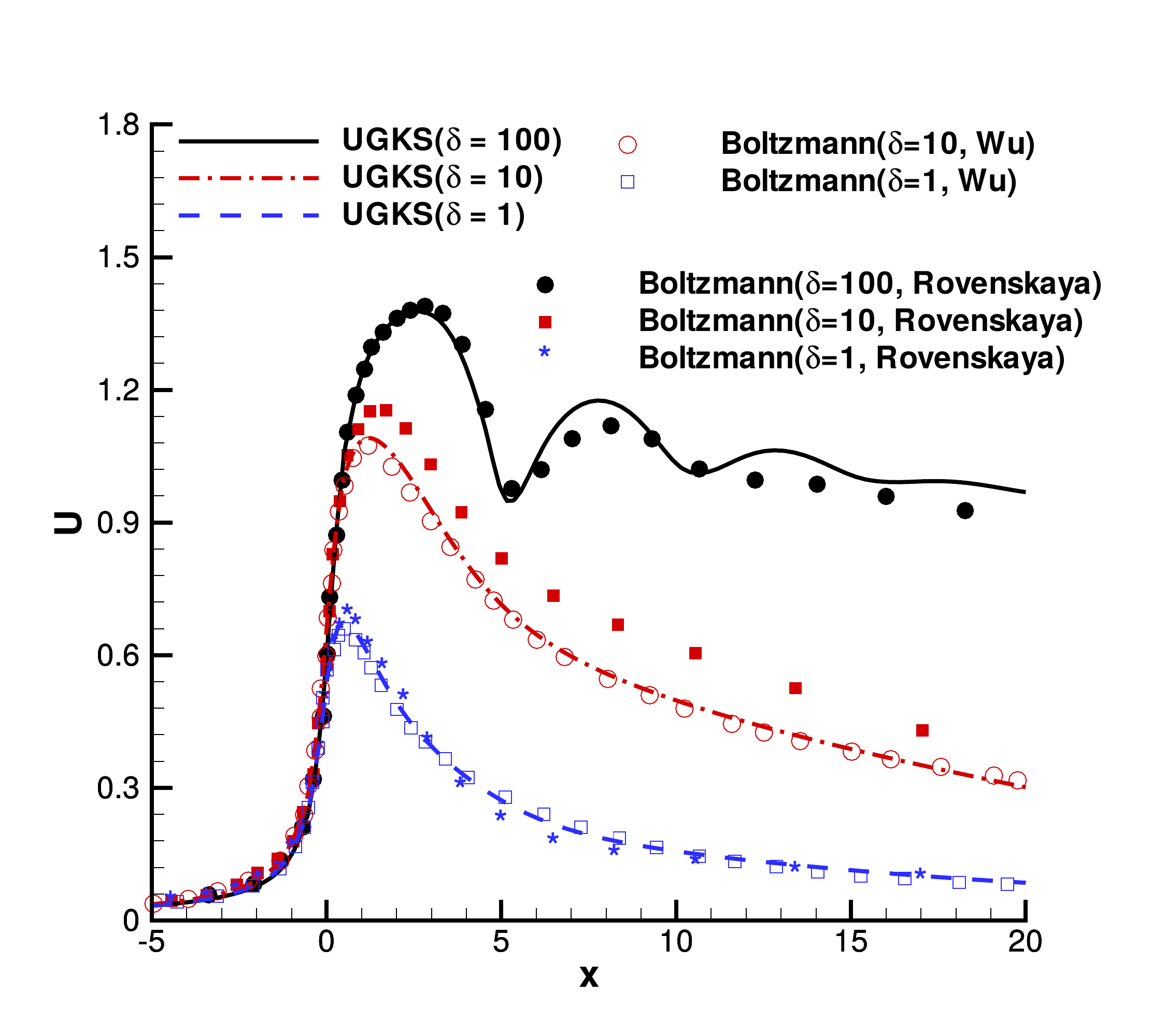}} \hspace{1em}
\subfloat[]{\includegraphics[width=0.46\textwidth]{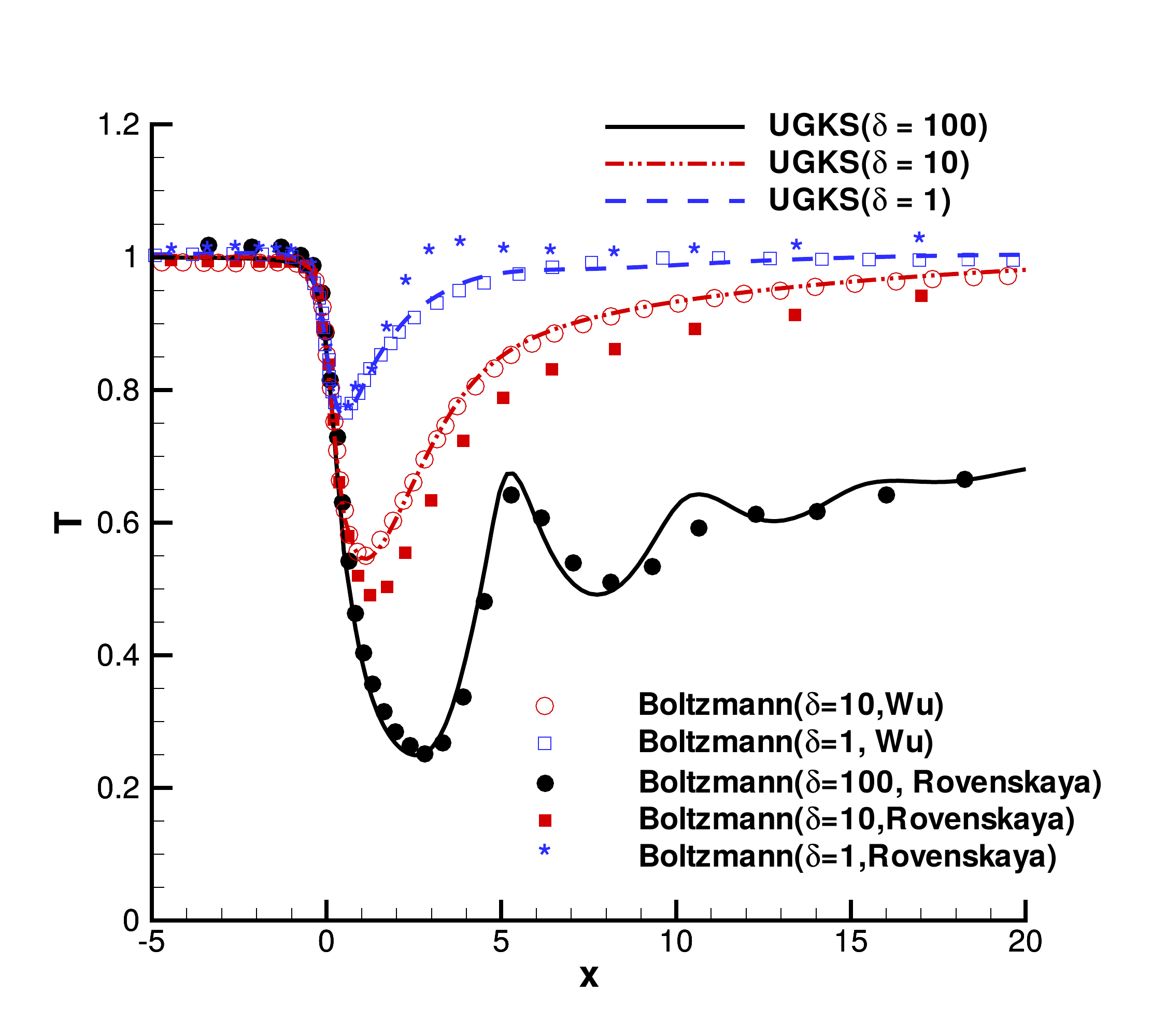}}
\caption{\label{fig:slit} Gas flow through a slit \cite{yu2013thesis}. (a) Mesh distribution, (b) density, (c) velocity, and (d) temperature along the central line.}
\end{figure}

The two-dimensional flow through a slit, placed between two reservoirs, has been simulated to study the UGKS solution compared with the full Boltzmann solution.
The pressure ratio between two reservoirs is $10$, and different rarefaction parameters are considered.
The simulating gas is monatomic argon, which is separated in two reservoirs with equilibrium temperatures $T_0$ and $T_1$, and pressures $p_0$ and $p_1$ ($p_1 < p_0$) respectively.
The computational domain is illustrated in Fig.~\subref*{fig:slit_mesh}.
The velocity distribution function in each reservoir is assumed to be the Maxwellian distribution function corresponding to the appropriate reservoir pressures and temperature.
The slit height is equal to $H$ in the $y$-direction, and the size of the computational domain is $L$, and the ratio is $H/L = 0.01$.
The gas flow through the slit between two reservoirs is determined by the rarefaction parameter $\delta$ defined as
\begin{equation}
\delta = \frac{p_0 H}{\mu_0} \sqrt{\frac{m_0}{2 k_B T_0}}.
\end{equation}
The Knudsen number is on the order of ${ Kn} = 1/\delta$.

The numerical calculations are carried out for the rarefaction parameter ranging from $\delta =1$, $10$ (transition regime) and $100$ (hydrodynamic regime).
The UGKS uses a total $88,400$ mesh points in the physical space, which is about $1/4$ of the spatial mesh points used by the full Boltzmann solver in \cite{rovenskaya2013comparison}, the velocity space is $28 \times 28$ mesh points.
The axial distributions of the number density, $x$-component of the bulk velocity and the temperature for $\delta = 1$, $10 $, and $100$ obtained from the UGKS and the reference full Boltzmann solutions are shown in Fig.~\ref{fig:slit}.
In all three cases, the density variations are qualitatively the same.
The axial velocity $U$ is close to zero at the distance $-5H$ and increases considerably around the slit.
The decreasing of the velocity is not as fast as for the density.
The flow acceleration in the slit is more important for the smaller pressure ratios.
The temperature variation along the axis depends essentially on the pressure ratio.
It is seen that the flow distributions of the full Boltzmann solution of Rovenskaya et al. are different from those obtained by the UGKS, especially for the velocity and temperature at $\delta = 10$.
Another full Boltzmann solver \cite{wu2013deterministic} is used to test the same case.
The results from the new full Boltzmann solver are presented in Fig.~\ref{fig:slit} as well.
Surprisingly, there is a perfect match between UGKS and Wu's Boltzmann solution.
For the slit case, the UGKS is as accurate as the full Boltzmann solver.

\subsubsection{High-speed moving ellipse}

\begin{figure}[htbp]
\centering
\subfloat[]{
\includegraphics[width=0.3\textwidth]{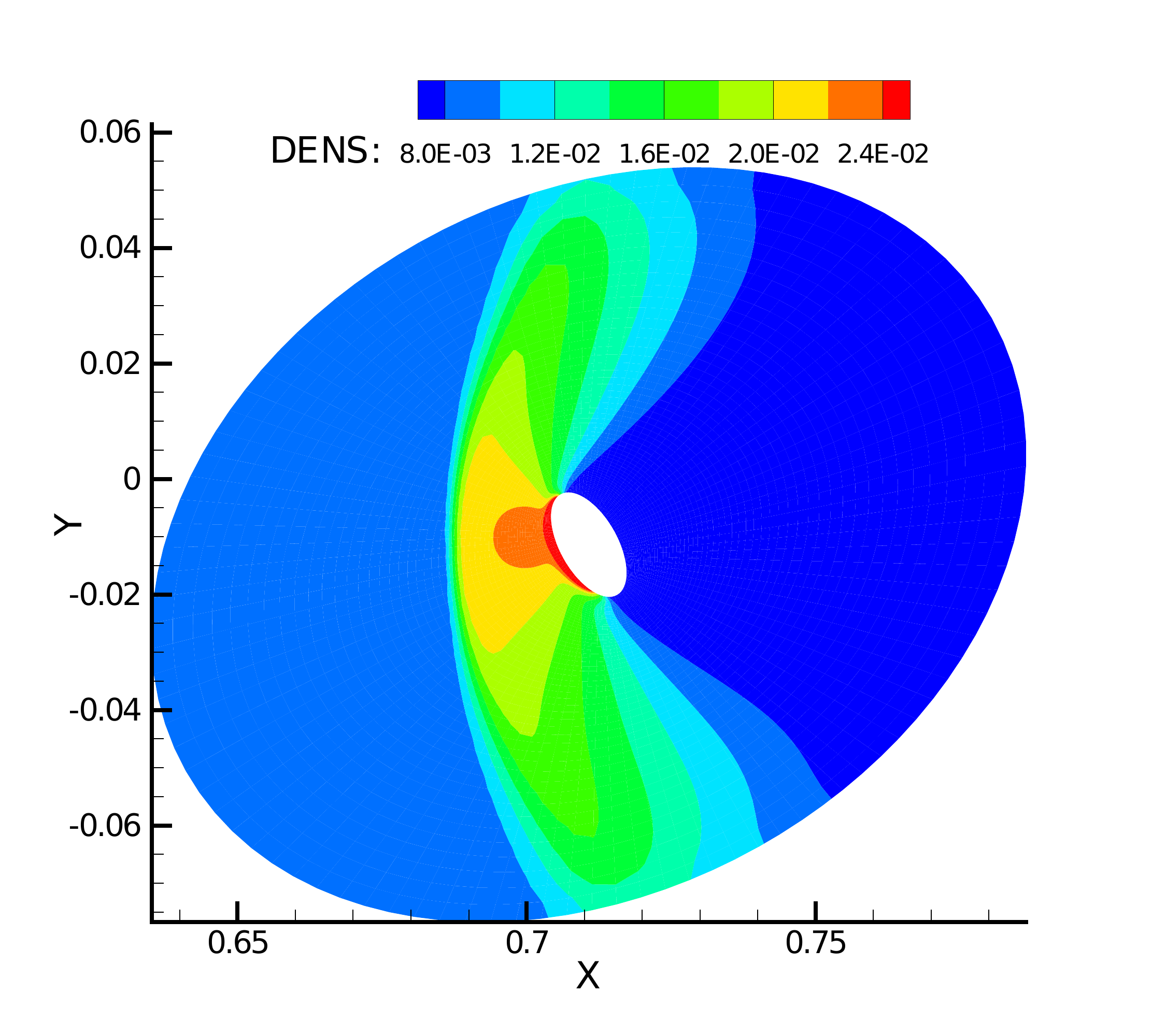}}
\subfloat[]{
\includegraphics[width=0.3\textwidth]{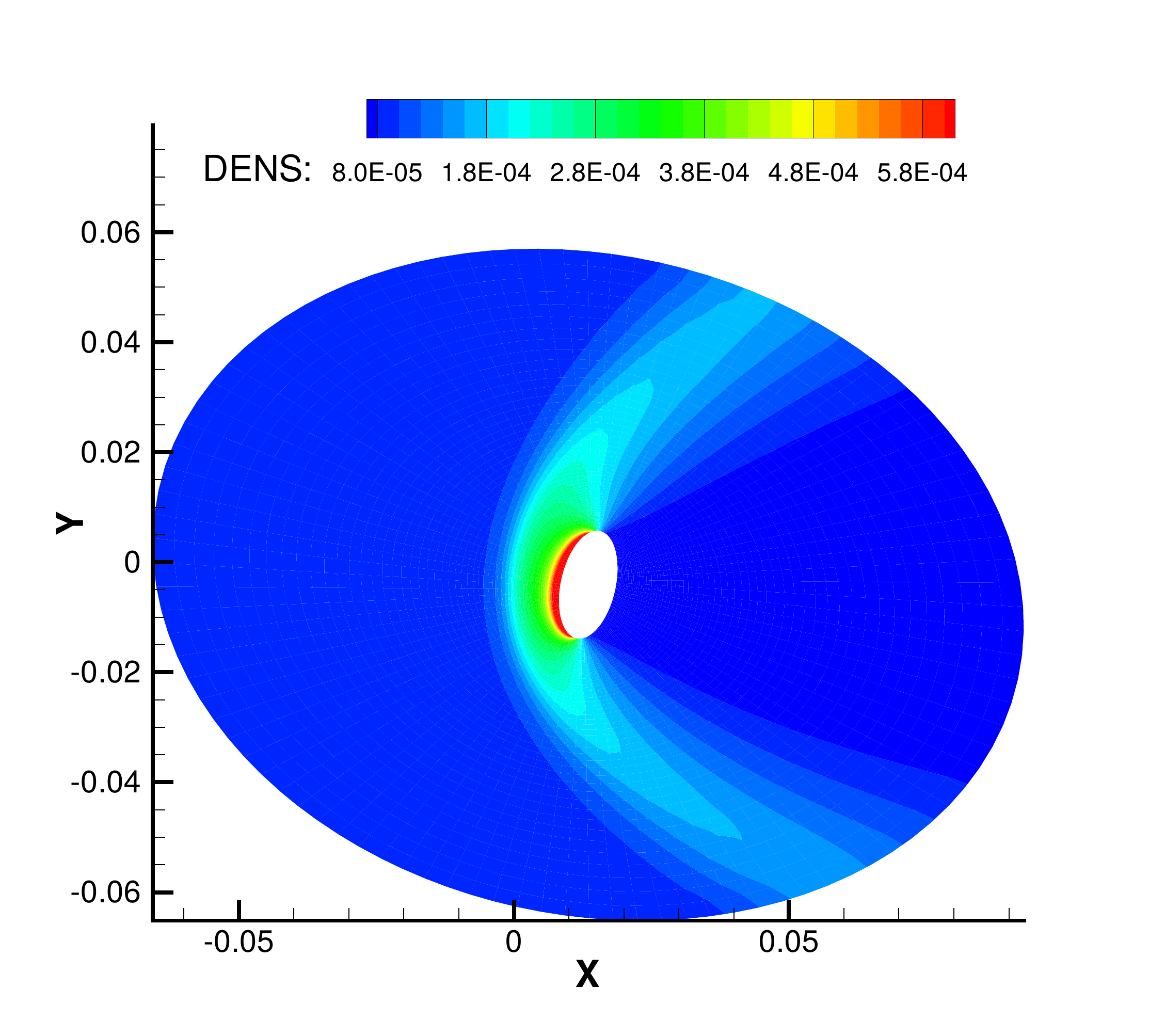}}
\subfloat[]{
\includegraphics[width=0.3\textwidth]{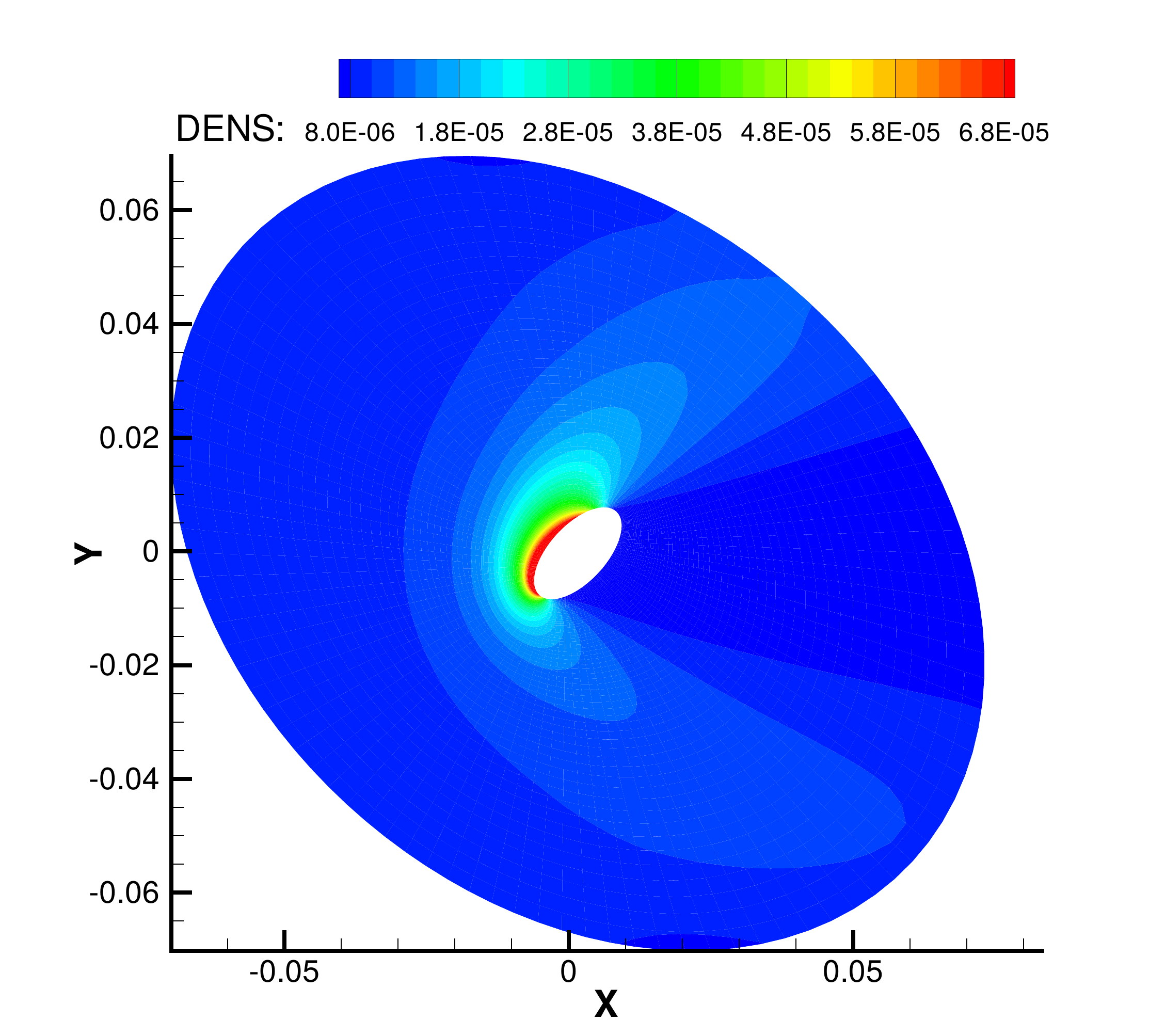}}
\caption{\label{fig:ellipse_density}Density field for shock driven ellipse \cite{chen2012mesh}. (a) ${ Kn}=0.01$, (b) ${ Kn}=0.1$, and (c) ${ Kn}=1$.}
\end{figure}

A freely moving ellipse rests initially in a flow with velocity of $1538.73{ m/s}$, temperature of $273{ K}$, and dynamic viscosity of $2.117\times 10^{-5} { kg/(m \cdot s)}$.
The center of ellipse locates at $(0,0)$ and the angle of attack of the ellipse is $-45^{\circ}$ when the calculation starts.
The incoming flow has a Mach number $5$.
Three cases with different upstream densities, i.e., $8.582\times 10^{-3} { kg/m^3}$, $8.582\times 10^{-5} { kg/m^3}$, and $8.582\times 10^{-6} { kg/m^3}$, are calculated.
The corresponding Knudsen numbers are $0.001$, $0.1$, $1$, respectively.
The long axis of the ellipse is $0.02{ m}$ and the short axis is $0.01{ m}$.
In order to get visible displacement during simulation, the density of ellipse is relatively small, i.e., $1 { kg/m^3}$.
The force and torque on the ellipse are calculated during the flight, which determine the ellipse's flight trajectory and its rotation.

\begin{figure}[htbp]
\centering
\subfloat[]{
\includegraphics[width=0.3\textwidth]{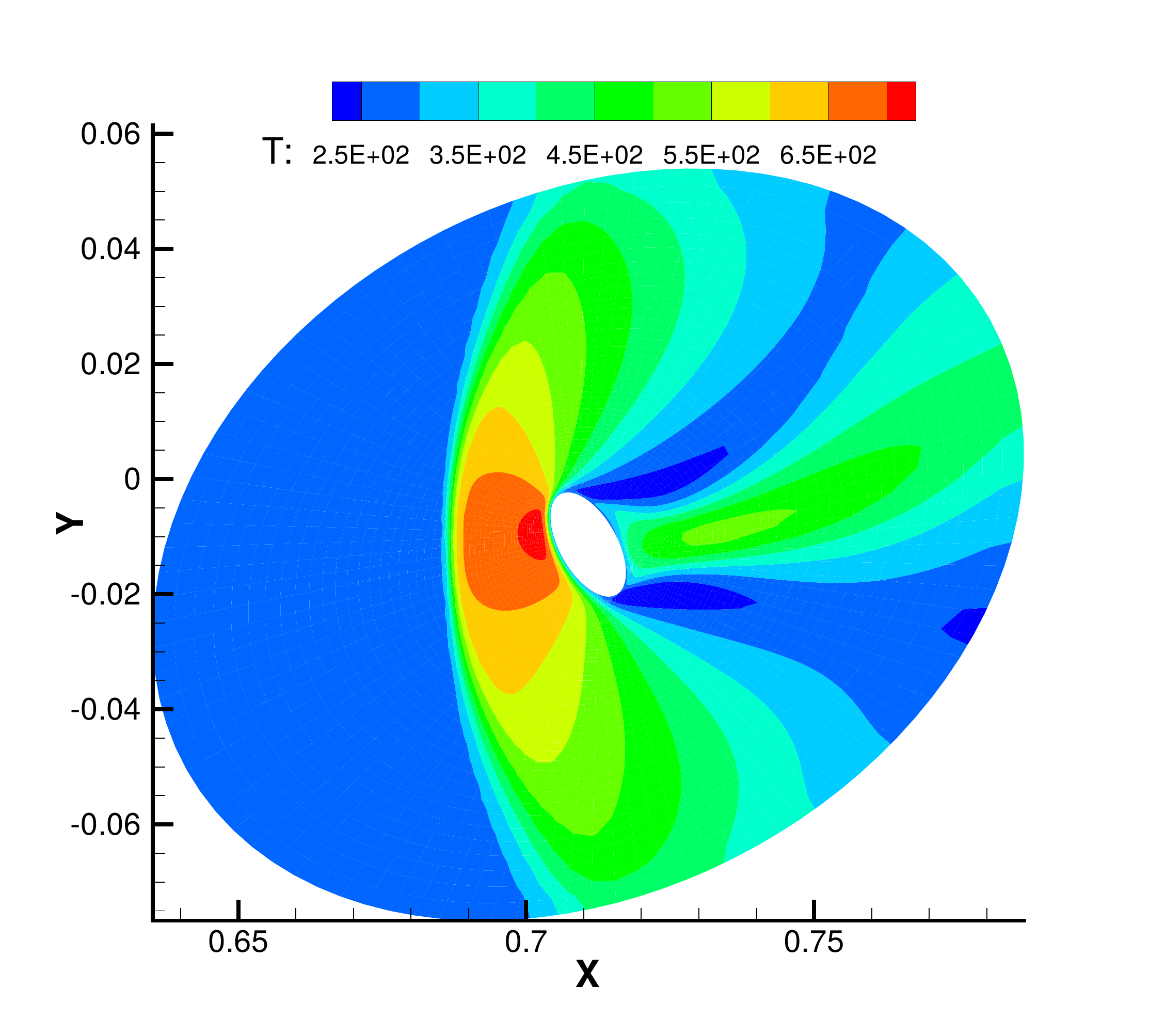}}
\subfloat[]{
\includegraphics[width=0.3\textwidth]{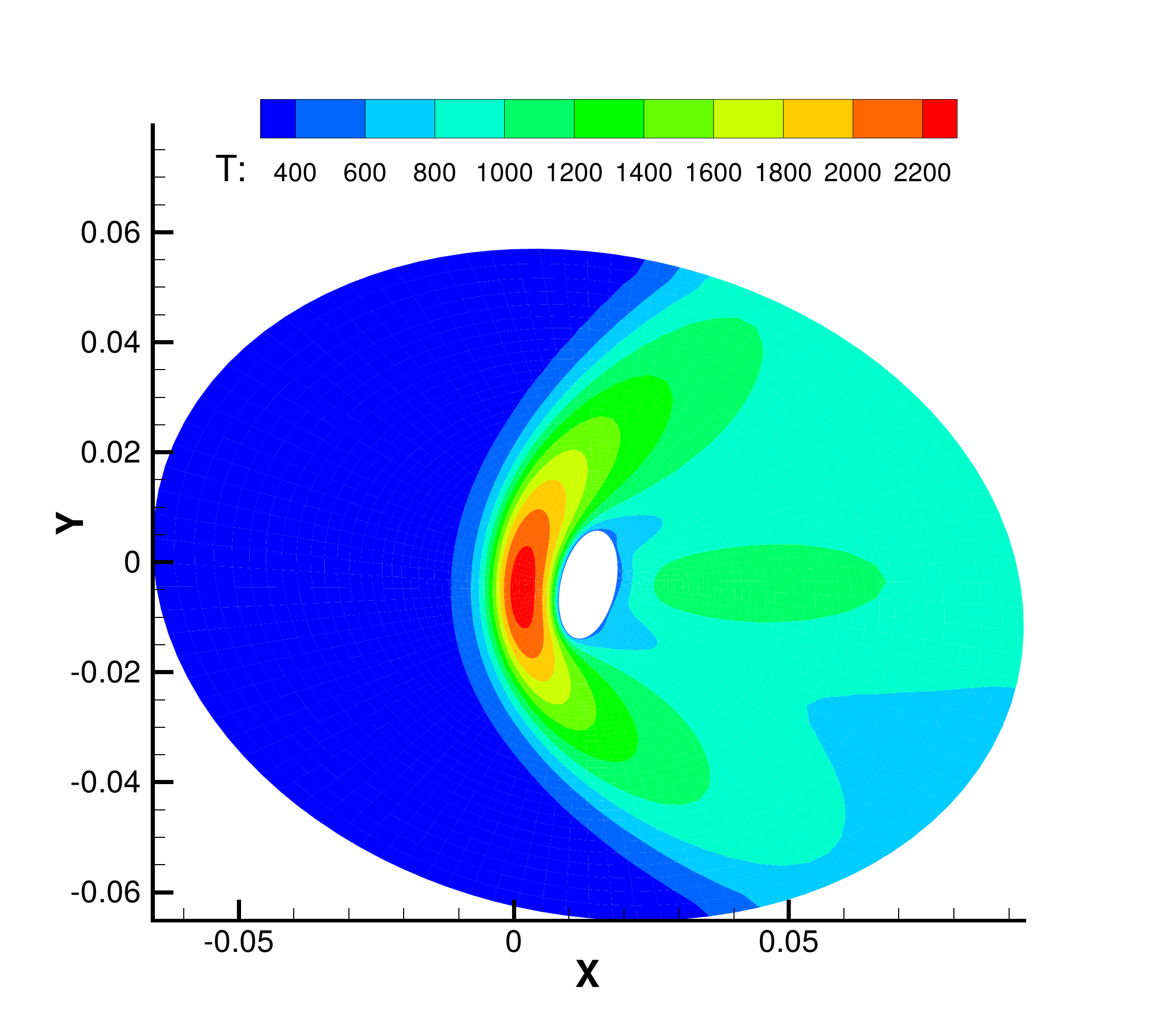}}
\subfloat[]{
\includegraphics[width=0.3\textwidth]{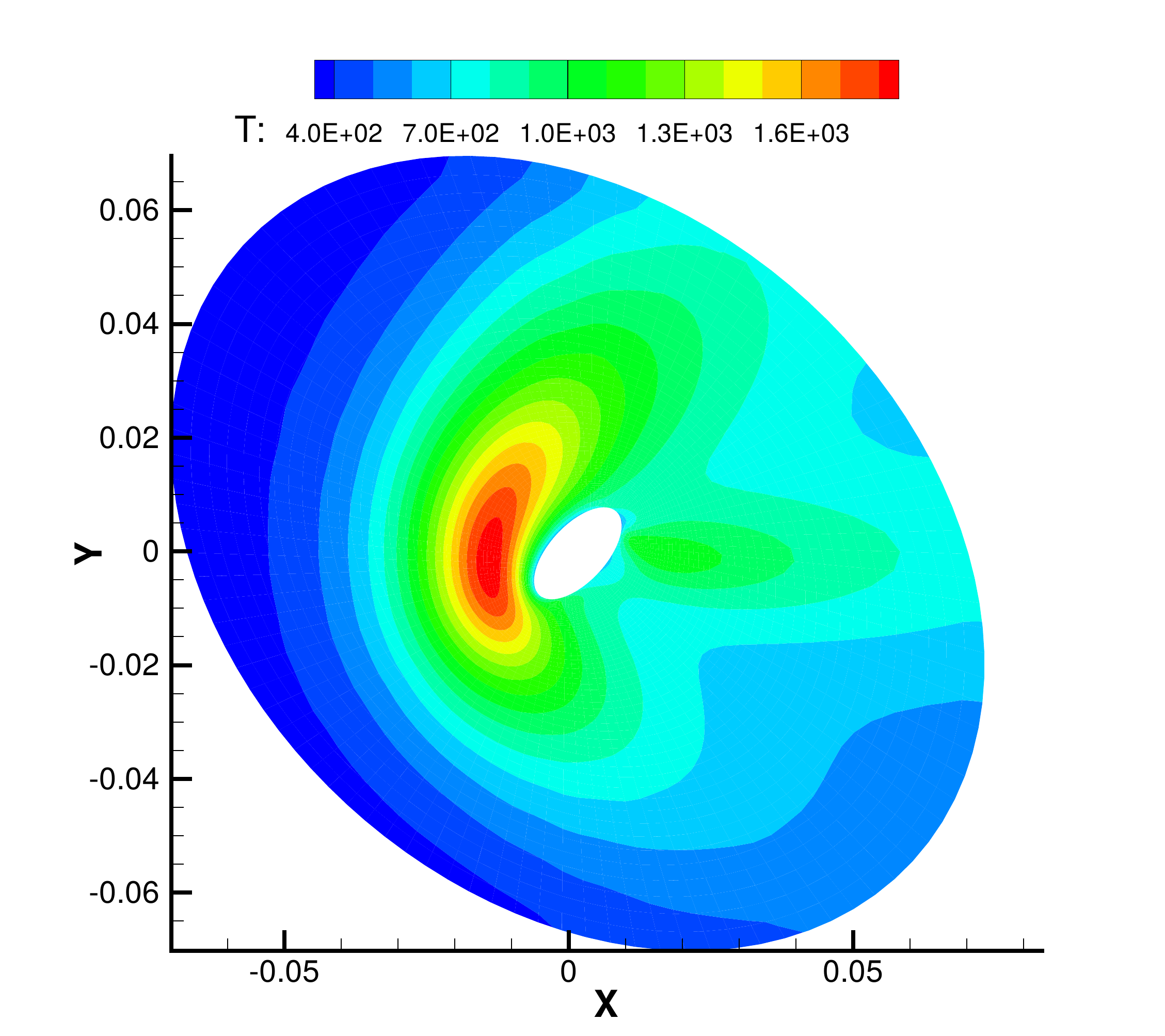}}
\caption{\label{fig:ellipse_temperature}Temperature field for shock driven ellipse \cite{chen2012mesh}. (a) ${ Kn}=0.01$, (b) ${ Kn}=0.1$, and (c) ${ Kn}=1$.}
\end{figure}

Figures \ref{fig:ellipse_density} and \ref{fig:ellipse_temperature} show the density and temperature field at $t = 1.28\times 10^{-3} { s}$.
At ${ Kn} = 0.001$ case, similar to shock capturing scheme for continuum flow, the UGKS with adaptive mesh in the velocity space presents a sharp shock front.
The gas distribution functions are very close to Maxwellian distribution in the whole computational domain expect at the shock front and close to the boundary.
As the Knudsen number increases, the shock thickness gets broaden with a resolved shock structure.
For example, at ${ Kn} = 0.1$, the shock thickness is comparable with the size of the ellipse.

\section{Kinetic model and collision term}
In this section, considering more physical realities, the unified gas kinetic scheme combined with different kinetic models for more accurate solutions will be introduced.

\subsection{Prandtl number correction}

The Prandtl number is defined by
\begin{equation}
{ Pr} = \frac{\mu C_p}{\kappa},
\end{equation}
where $C_p$ is the heat capacity at constant pressure, $\mu$ denotes the dynamic viscosity, and $\kappa$ is the heat conductivity.
For monatomic gas, the Prandtl number is approximately equal to $\frac{2}{3}$.
In the previous section \ref{sec:ugks}, the UGKS based on the BGK model has been introduced.
It is well known that the BGK model \cite{bhatnagar1954model} has a unit Prandtl number, thus it can only provide one correct transport coefficient, i.e., either the dynamic viscosity $\mu$ or the heat conductivity $\kappa$.
In the following, numerical treatment to amend the Prandtl number in the UGKS will be introduced.

\subsubsection{Energy flux modification}
If Eq.~\eqref{eq:tau_computation} is used to determine the relaxation time $\tau$, the viscosity is the same as that in the Navier--Stokes equations in the continuum limit, while the heat flux in the UGKS has to be modified in order to capture the thermal conduction with any realistic Prandtl number.

Similar to the treatment in the gas kinetic scheme (GKS) for solving Navier--Stokes equations \cite{xu2001gks}, the originally developed UGKS \cite{xu2010ugks} amends the macroscopic energy flux across cell interface by
\begin{equation}
	F^{new}_{\rho E} = F_{\rho E} + \left(\frac{1}{ Pr}-1\right) q_{n},
\end{equation}
where $q_{n}$ denotes the normal component of heat flux vector $\vec{q}$ at cell interface, which is computed from the moments of discrete distribution function by
\begin{equation}
	q_{n} = \sum_k {c_{n,ij,k} \bar{f}_{ij,k} \frac{c_k^2}{2} \mathcal{V}_k},
\end{equation}
where $c_{n,ij,k} = \vec{c}_k \cdot \vec{n}_{ij}$ is the normal component of the peculiar velocity.
This modification changes the updated macroscopic variables $\vec{w}^{n+1}$, which will influence the update of distribution function through the implicit equilibrium state $g^{n+1}$ in the collision term .

\subsubsection{Shakhov model}
In order to simulate flow with arbitrary Prandtl number, the UGKS for the Shakhov model \cite{shakhov1968model} was developed in the study \cite{xu2011shakhov}.
The difference between the Shakhov model and the BGK model lies on the equilibrium state.
The equilibrium distribution function in the collision term of the Shakhov model is
\begin{equation}\label{eq:shakhov_model}
 g_s = g_M \left[1+(1-{ Pr}) \vec{c} \cdot \vec{q} \left(\frac{m_0 c^2}{k_B T}-5\right) \middle/ \left(\frac{5 p k_B T}{m_0}\right)\right],
\end{equation}
where $g_M$ is the Maxwellian equilibrium distribution in the BGK model, $p$ and $T$ denote the pressure and temperature, respectively.

On the basis of the UGKS with BGK model, an additional term should be considered during the evaluation of the microscopic fluxes, i.e.,
\begin{equation}
	g_s^{+} = g_M \left[(1-{ Pr}) \vec{c} \cdot \vec{q} \left(\frac{m_0 c^2}{k_B T}-5\right) \middle/ \left(\frac{5 p k_B T}{m_0}\right)\right],
\end{equation}
where the vector of heat flux  $\vec{q}$ is computed by
\begin{equation}
	\vec{q} = \frac{1}{2}\sum_k { \vec{c}_k c_k^2 \bar{f}_k \mathcal{V}_k}.
\end{equation}

\subsubsection{ES-BGK model}
The ellipsoidal statistical BGK (ES-BGK) model proposed by Holway \cite{holway1966es} is another well known kinetic model allowing flexible Prandtl number.
The ES-BGK model replaces the equilibrium state in the BGK model by an anisotropic Gaussian distribution
\begin{equation}
	g_{ES} = \rho \left(\frac{m_0}{2\pi k_B}\right)^{\frac{3}{2}} \frac{1}{\sqrt{{\det}\left|T_{\alpha \beta}\right|}} {\exp}\left[{-\frac{m_0 c_{\alpha} T^{-1}_{\alpha \beta} c_{\beta}}{2 k_B}}\right],
\end{equation}
where $T_{\alpha \beta}$ is the temperature tensor defined by
\begin{equation}\label{eq:temperature_tensor}
	T_{\alpha \beta} =  \frac{1}{ Pr} T \delta_{\alpha \beta} + \left(1-\frac{1}{ Pr}\right) \frac{m_0 P_{\alpha \beta}}{\rho k_B} ,
\end{equation}
where $\delta_{\alpha \beta}$ is the Kronecker delta, $T$ denotes the static temperature, and $P_{\alpha \beta}$ is the stress tensor given in Eq.~\eqref{eq:stress_tensor}.

During the flux calculation, the stress tensor should be computed from the discrete distribution function.
In addition, when using the ES-BGK model, the relaxation time $\tau$ should be computed by
\begin{equation}
	\tau = \frac{\mu}{p { Pr}}.
\end{equation}

\begin{figure}[htbp]
\centering
\subfloat[]{\includegraphics[width=0.44\textwidth]{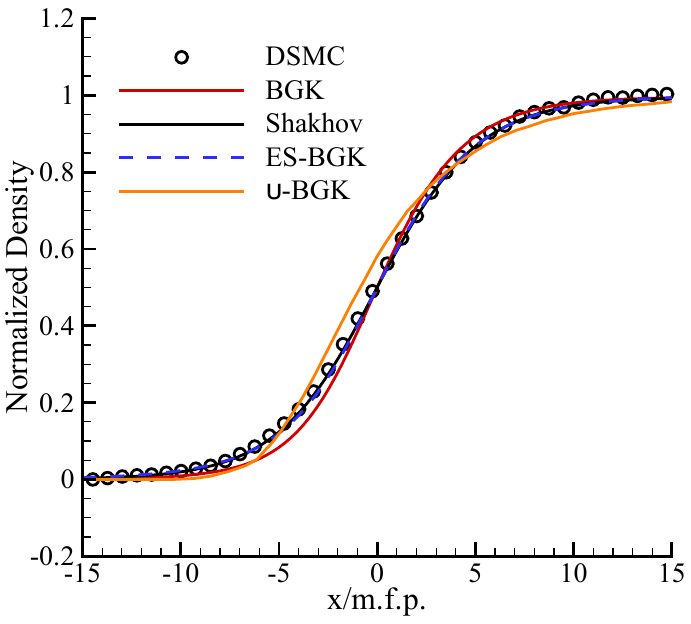}} \hspace{1em}
\subfloat[]{\includegraphics[width=0.44\textwidth]{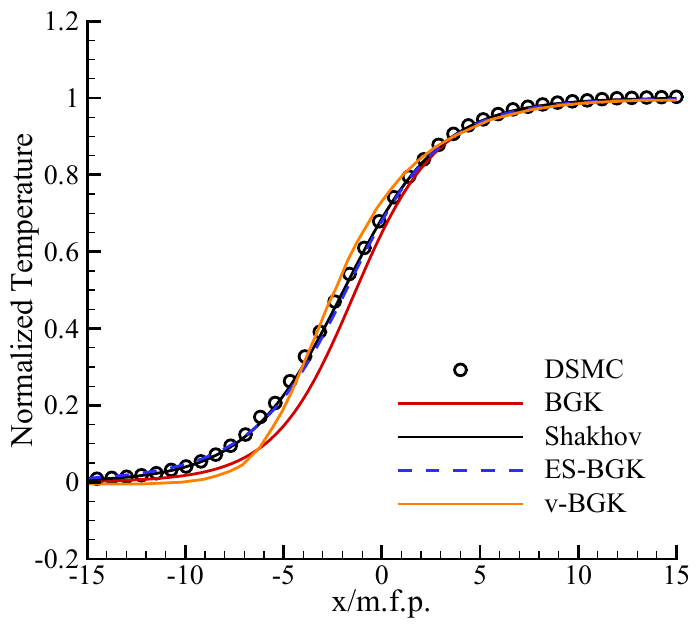}}
\caption{\label{fig:model_shock1.4}Profile of shock structure at ${ Ma}=1.4$ \cite{liu2014model}. (a) Density, and (b) temperature.}
\end{figure}

\begin{figure}[htbp]
\centering
\subfloat[]{\includegraphics[width=0.44\textwidth]{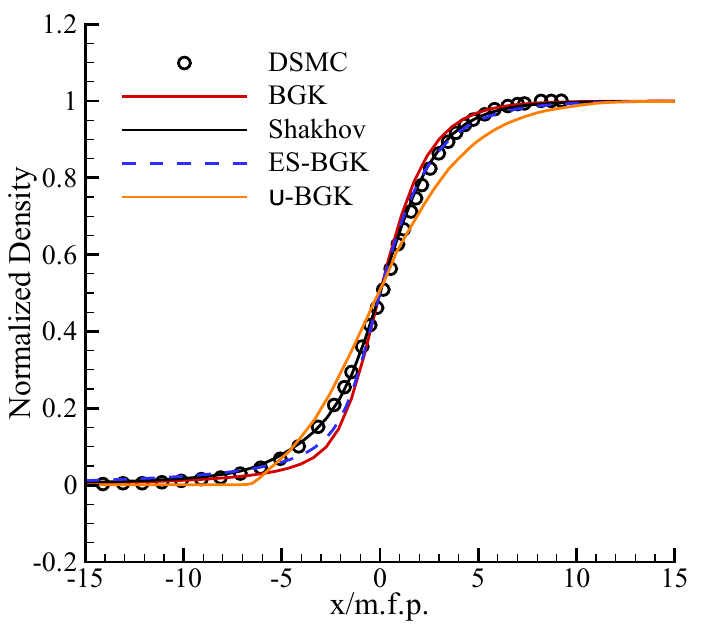}} \hspace{1em}
\subfloat[]{\includegraphics[width=0.44\textwidth]{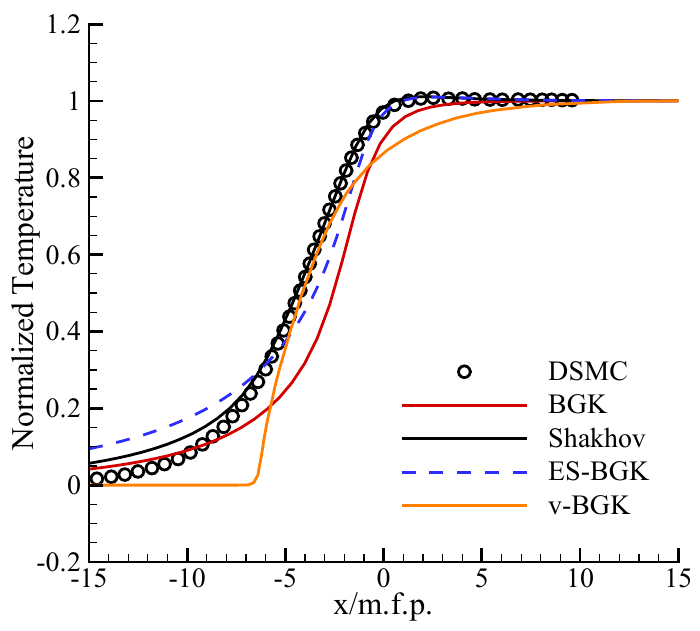}}
\caption{\label{fig:model_shock8}Profile of shock structure at ${ Ma}=8.0$ \cite{liu2014model}. (a) Density, and (b) temperature.}
\end{figure}

The kinetic models to fix the Prandtl number in the UGKS are investigated in \cite{liu2014model}, where the test case of normal shock structure is computed to compare the performance of the Shakhov model, ES-BGK model and a kinetic model with velocity-dependent collision frequency \cite{mieussens2004numerical}.
For normal shock wave, the upstream and downstream conditions are determined by the Rankine--Hugoniot relation.
The mean free path is evaluated by Eq.~\eqref{eq:knudsen_number}, where the parameters are $\alpha = 1.40$ and $\omega=0.81$ for argon gas \cite{shen2006rarefied}.
Figures \ref{fig:model_shock1.4} and \ref{fig:model_shock8} show the normalized density and temperature profiles for the cases at ${ Ma}=1.4$ and $8$.
For the supersonic case with relatively low Mach number, the results predicted by all these kinetic models match well with the benchmark DSMC data.
For the hypersonic case at ${ Ma} = 8$, the density predicted by the Shakhov model agrees well with the DSMC solution, while ES-BGK and BGK models obtain stepper solutions, and the $\nu$-BGK displays a kink.
For the temperature distribution, obvious deviations have been observed for all the mentioned kinetic models.
The Shakhov model obtains better solutions than the others.
An early temperature rise occurs in the upstream, but in other regions the solutions are acceptable.

\begin{figure}[htbp]
\centering
\subfloat[]{\includegraphics[width=0.32\textwidth]{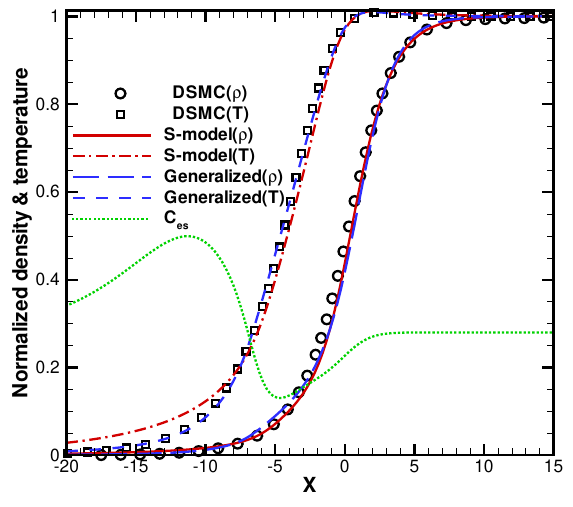}}
\subfloat[]{\includegraphics[width=0.32\textwidth]{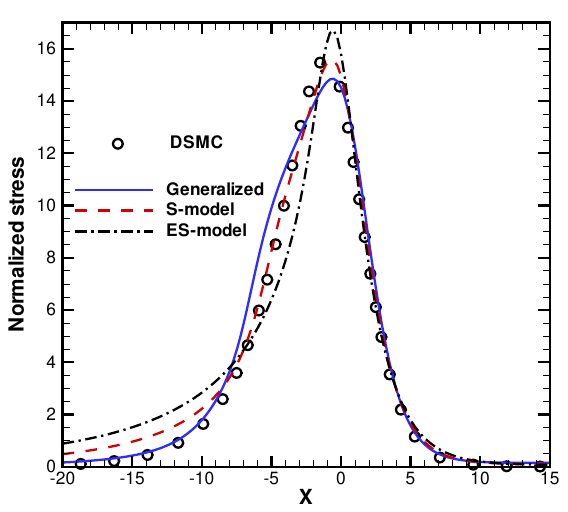}}
\subfloat[]{\includegraphics[width=0.32\textwidth]{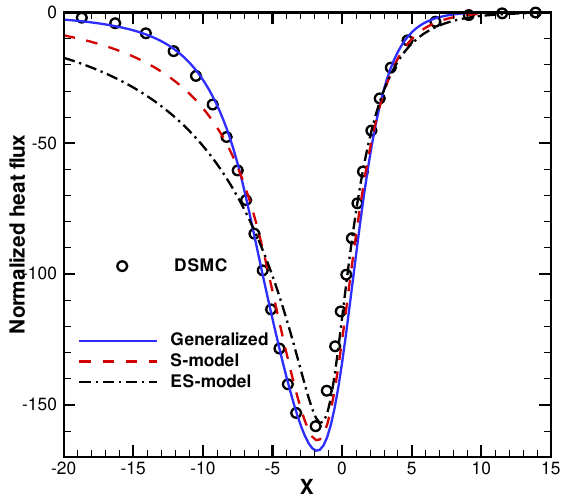}}
\caption{\label{fig:model_shock_general}Profile of shock structure at ${ Ma}=8.0$ \cite{chen2015model}. (a) Density and temperature,  (b) stress, (c) heat flux.}
\end{figure}

Chen et al. \cite{chen2015model} proposed a general kinetic model, which can recover the Shakhov model and the ES-BGK model, and bridge these two models with a free parameter $C_{es}$.
General speaking, the Shakhov model can obtain better results than the ES-BGK model, while ES-BGK can get more accurate solution for the cases driven by temperature gradient.
By adjusting the free parameters, the general kinetic model can make full use of the advantages of these two models.
With a variable $C_{es}$, a good shock structure has been obtained by the general kinetic model in both the upstream and downstream regions, see the results in Fig.~\ref{fig:model_shock_general}.
If high-order moments are evaluated, such as the stress and heat flux, there are still some deviations between the general kinetic model and the DSMC method.
In order to obtain more accurate solution for highly non-equilibrium flows, the kinetic equation with full Boltzmann collision term should be considered.

\subsection{Full Boltzmann collision term}
In rarefied gas dynamics, the fundamental governing equation is the Boltzmann equation \cite{bird1994molecular,shen2006rarefied,cercignani2000rarefied,sharipov2015rarefied}, which can be written as
\begin{equation}\label{eq:boltzmann_equation}
\frac{\partial f}{\partial t} + \vec{u} \cdot \frac{\partial f}{\partial \vec{x}} = \int_{-\infty}^{\infty} \int_{0}^{4\pi} {(f^\prime f_1^\prime - f f_1) c_r \sigma d\Omega d\vec{u}_1},
\end{equation}
where $f=f(\vec{x},\vec{u},t)$, $f_1=f(\vec{x},\vec{u}_1,t)$, $f^\prime=f^\prime(\vec{x},\vec{u},t)$, $f_1^\prime=f^\prime(\vec{x},\vec{u}_1,t)$ denote the distribution functions of particles with microscopic velocities $\vec{u}$ and $\vec{u}_1$ before and after collisions.
$c_r = |\vec{u}-\vec{u}_1|$ represents the relative speed between particles, $\sigma$ is differential cross section, and $\Omega$ denotes solid angle.

For non-equilibrium flow cases that the time step is comparable with the local particle mean collision time, a more accurate UGKS solver can be constructed by integrating the full Boltzmann collision term in the dynamic evolution of flow physics.
Liu et al. \cite{liu2016boltzmann} combined the Boltzmann collision term with the kinetic model in the updating process of gas distribution function of UGKS.
Specifically,
\begin{equation}
	f_{i,k}^{n+1} = f_{i,k}^n - \frac{\Delta t}{V_i} \sum_{j \in N(i)} {u_{n,ij,k} {\bar f}_{ij,k} S_{ij}} + A {\mathcal{J}_k (f_i,f_i)} + B \frac{g_{i,k}^{n+1} - f_{i,k}^{n+1}}{\tau^{n+1}},
\end{equation}
where $\mathcal{J}(f,f)$ denotes the Boltzmann collision term, and $g$ is the equilibrium state in BGK-type kinetic models.
The parameters $A$ and $B$ satisfy $A + B = \Delta t$.
From the study of homogeneous relaxation cases, it is found that numerical solution obtained from the Shakhov model becomes closer to that from the Boltzmann equation after a period of time evolution with collision accumulation.
Therefore, the full Boltzmann collision term is only required in the highly non-equilibrium region with the time step being less than the mean collision time, and the criterion to determine the regions using the full Boltzmann collision term is based on the comparison between the time step $\Delta t$ and a critical time $t_c$.
The critical time is estimated by
\begin{equation}
	t_c = {\min} \left(2, \int{|f - g|d\Xi} / \rho \right) 2 \tau,
\end{equation}
where the deviation of the distribution function from local equilibrium state is evaluated to measure the local non-equilibrium.

Based on the local flow physics and numerical time step, the parameters can be chosen as
\begin{equation}
	(A, B) =
	\begin{cases}
	(\beta^n \Delta t, (1-\beta^n)\Delta t), &\quad \Delta t < t_c^n \\
	(0, \Delta t), &\quad \Delta \ge t_c^n
	\end{cases}
\end{equation}
with
\begin{equation}
	\beta^n =
	\begin{cases}
	1, & \quad \Delta t < 1 / \sup_{\Omega_v} \nu^n \\
	\exp (1-\Delta t \sup_{\Omega_v} \nu^n), &\quad \Delta t \ge 1 / \sup_{\Omega_v} \nu^n
	\end{cases}
\end{equation}
where $\Omega_v$ denotes the computational domain in velocity space and $\nu^n$ is the collision frequency calculated by using a spectral method \cite{mouhot2006fast,wu2013deterministic}.
With these adaptive parameters, the UGKS can give the Boltzmann solutions in the rarefied flow regime and recover the Navier--Stokes solutions efficiently in the continuum regime.

\begin{figure}[htbp]
\centering
\subfloat[]{\includegraphics[width=0.44\textwidth]{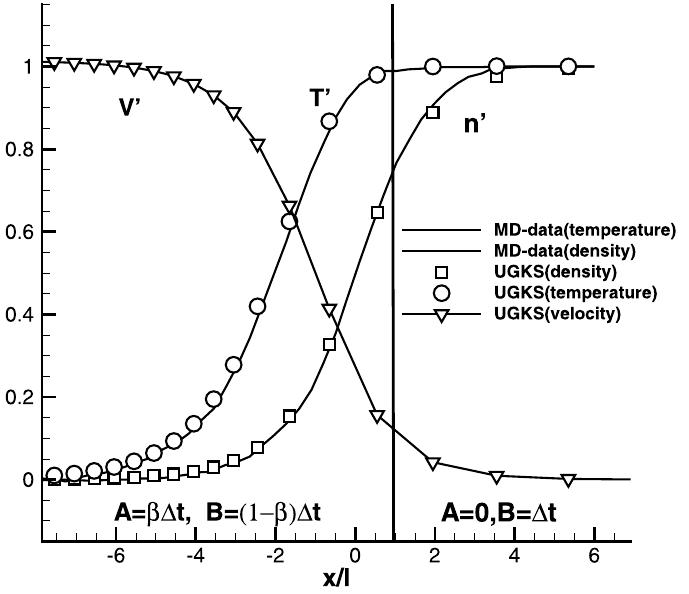}}
\subfloat[]{\includegraphics[width=0.44\textwidth]{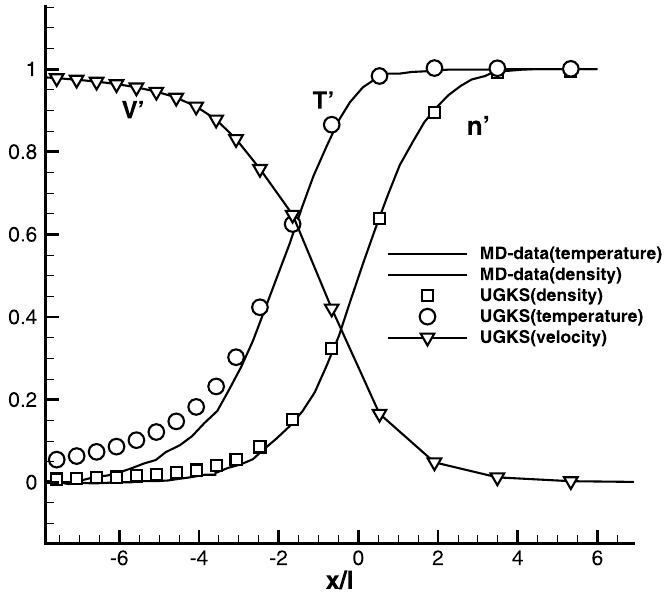}}
\caption{\label{fig:boltzmann_shock} UGKS results of shock structure at ${ Ma}=5$, compared with the MD simulation \cite{liu2016boltzmann}. (a) UGKS combined with full Boltzmann collision term, (b) UGKS with Shakhov model.}
\end{figure}

\begin{figure}[htbp]
\centering
\subfloat[]{\includegraphics[width=0.44\textwidth]{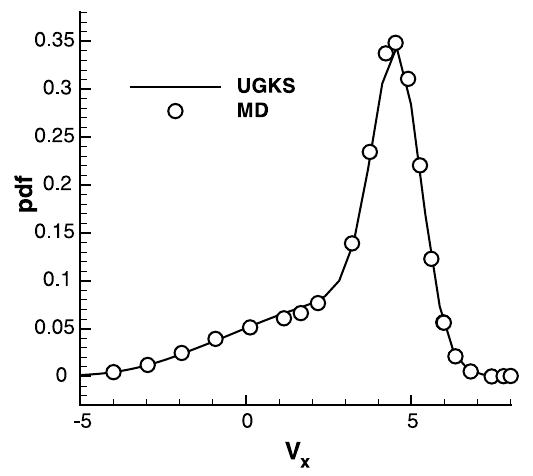}} \hspace{1em}
\subfloat[]{\includegraphics[width=0.44\textwidth]{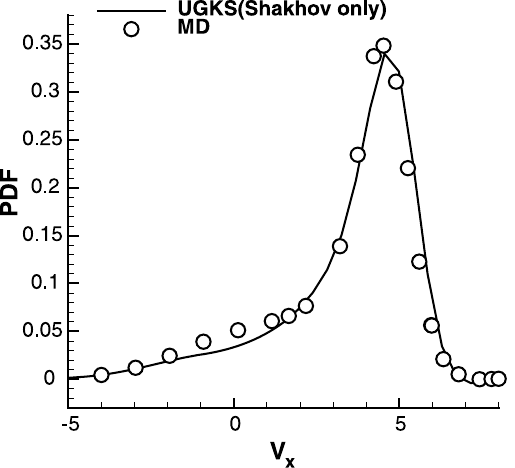}} \\
\subfloat[]{\includegraphics[width=0.44\textwidth]{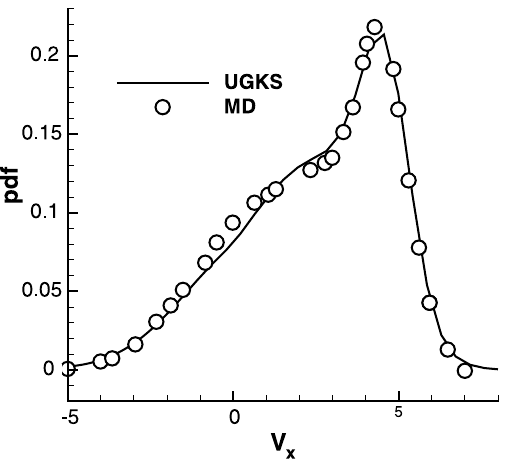}} \hspace{1em}
\subfloat[]{\includegraphics[width=0.44\textwidth]{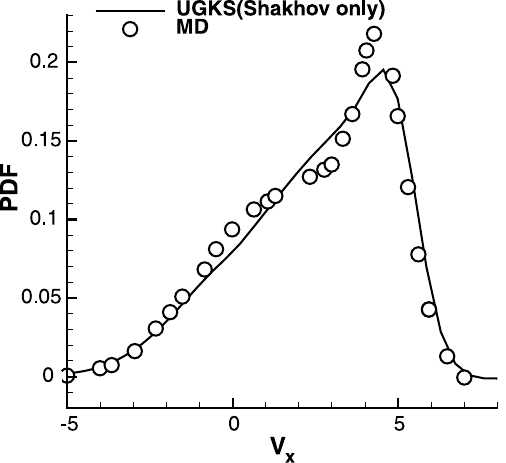}} \\
\subfloat[]{\includegraphics[width=0.44\textwidth]{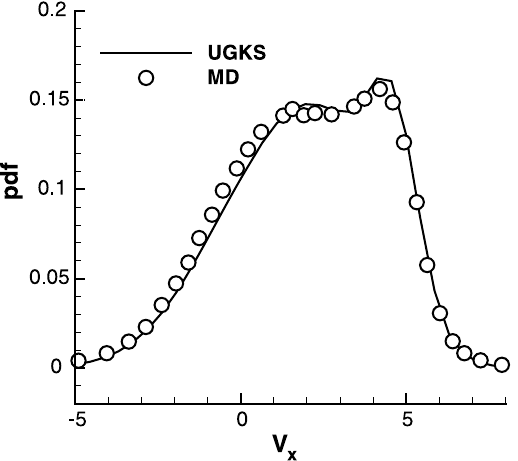}} \hspace{1em}
\subfloat[]{\includegraphics[width=0.44\textwidth]{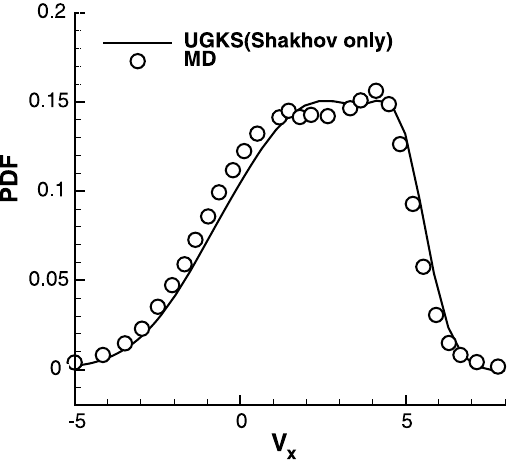}}
\caption{\label{fig:boltzmann_pdf} Distribution function at different location in the shock structure \cite{liu2016boltzmann}.
Left: UGKS solutions combined with Boltzmann collision term. Right: UGKS with Shakhov model.}
\end{figure}

The shock structure at ${ Ma}=5$ is computed by the UGKS with Boltzmann collision term.
As shown in Fig.~\ref{fig:boltzmann_shock},  good agreement with the reference solutions has been obtained, and the early rise of temperature in the upstream in the Shakhov model has been cured by combining the full Boltzmann collision term.
The distribution function inside the shock structure has been plotted in Fig.~\ref{fig:boltzmann_pdf}, where the solutions from molecular dynamic simulation are used as a reference data.
It can be found that the UGKS with full Boltzmann collision term obtains better solutions than that with Shakhov model, but there is no big difference.
Usually, the Shakhov model is good enough for engineering application.

\subsection{Diatomic gas with molecular rotation and vibration}
The UGKS for monatomic gas simulation is presented in the previous description.
For diatomic molecules, for example, ${N_2}$ and ${O_2}$, the characteristic temperature of rotation is around 3 ${K}$ \cite{bird1994molecular}, so the rotational degrees of freedom are already activated in room temperature.
Therefore, in the diatomic gas modeling, the rotational degrees of freedom should be included in the gas distribution function.
In addition, the characteristic temperatures of molecular vibration for ${ N_2}$ and ${ O_2}$ are 3371 ${ K}$ and 2256 ${ K}$ \cite{bird1994molecular} respectively.
For high temperature flows at high Mach numbers, the vibrational degrees of freedom should be taken into account as well.

\subsubsection{Complete relaxation}
For the complete relaxation case, the molecular rotation and vibration are treated as the internal degrees of freedom which have the same temperature as the translational temperature.
The averaged velocities for particles' motion of the internal degrees of freedom are assumed to be zeros.
Specifically, the gas distribution function is $f(x, y, z, u, v, w, \xi, t)$, and the equilibrium state becomes
\begin{equation}
	g(x,y,z, u,v,w,\xi,t)= \rho \left(\frac{\lambda}{\pi}\right)^{\frac{3+K}{2}} e^{-\lambda\left[(u -U)^2 + (v - V)^2 + (w-W)^2\right]} e^{-\lambda \xi^2},
\end{equation}
where $\xi^2 = \xi_1^2 + \xi_2^2 + ... + \xi_K^2$ implies $K$ internal degrees of freedom.
Then the conservative flow variables are computed by
\begin{equation}
	\vec{w} =
	\begin{pmatrix}
	\rho\\
	\rho U \\
	\rho V \\
	\rho W \\
	\rho E	
	\end{pmatrix}
	=
	\int
	\begin{pmatrix}
	1\\
	u\\
	v\\
	w\\
	\frac{1}{2}(u^2+v^2+w^2+\xi^2)
	\end{pmatrix}
	f(u,v,w,\xi)
	du dv dw d\xi,
\end{equation}
where $d\xi = d\xi_1 d\xi_2 ...d\xi_K$.

In order to reduce the computational cost, the reduced distribution functions
\begin{equation}
	\begin{aligned}
	h(x,y,z,u,v,w,t) &= \int f(x,y,z,u,v,w,\xi, t) d\xi, \\
	b(x,y,z,u,v,w,t) &= \int \xi^2 f(x,y,z,u,v,w, \xi, t) d\xi,
	\end{aligned}
\end{equation}
are always employed, and the corresponding equilibrium states are
\begin{equation}
	\begin{aligned}
	h^{eq}(x,y,z,u,v,w,t) &= g_M (x,y,z,u,v,w,t), \\
	b^{eq}(x,y,z,u,v,w,t) &= \frac{K}{2 \lambda} g_M(x,y,z,u,v,w,t).
	\end{aligned}
\end{equation}
The reduced distribution function $h$ and $b$ follow the kinetic equations in the same form of
\begin{equation}
\frac{\partial \Phi}{\partial t}
+ u \frac{\partial \Phi}{\partial x}
+ v \frac{\partial \Phi}{\partial y}
+ w \frac{\partial \Phi}{\partial z}
=
\frac{\Phi^{eq} - \Phi}{\tau}.
\end{equation}
Therefore, the construction of the multiscale flux from the integral solution for $h$ and $b$ has no difference from that of $f$ in previous description.
The only thing that should be noted is that the computation of macroscopic flow variables from the reduced distribution function needs to include the internal degrees of freedom, i.e.,
\begin{equation}
	\vec{w}_i = \sum_k
	\begin{pmatrix}
	1 \\
	u_k \\
	v_k \\
	w_k\\
	\frac{1}{2}(u_k^2 + v_k^2 + w_k^2)	
	\end{pmatrix}
	h_{i,k} \mathcal{V}_k +  \sum_k
	\begin{pmatrix}
	0 \\
	0 \\
	0 \\
	0\\
	\frac{1}{2}
	\end{pmatrix}
	b_{i,k} \mathcal{V}_k,
\end{equation}
and
\begin{equation}
	\vec{F}_{ij} = \sum_k u_{n,ij,k}
	\begin{pmatrix}
	1 \\
	u_k \\
	v_k \\
	w_k\\
	\frac{1}{2}(u_k^2 + v_k^2 + w_k^2)	
	\end{pmatrix}
	\bar{h}_{ij,k} \mathcal{V}_k +  \sum_k u_{n,ij,k}
	\begin{pmatrix}
	0 \\
	0 \\
	0 \\
	0\\
	\frac{1}{2}
	\end{pmatrix}
	\bar{b}_{ij,k} \mathcal{V}_k,
\end{equation}
where $\bar{h}_{ij,k}$ and $\bar{b}_{ij,k}$ are the time averaged distribution function at the cell interface constructed from the integral solution.

By setting the value of $K$, the polyatomic gas with arbitrary number of internal degrees of freedom can be simulated, where the specific heat ratio is related to the number of degrees of freedom by
\begin{equation}
	\gamma = \frac{K+5}{K+3}.
\end{equation}
For monatomic gas, $K=0$ and $\gamma =5/3$; for diatomic gas, $K=2$ and $\gamma=1.4$; and for isothermal case, $K\to \infty$ and $\gamma =1$.
It should be emphasized that this is only the consideration for complete relaxation cases, where the underlying assumption is that the energy change between transitional and internal degrees of freedom is very quick, and the temperatures for translational and internal degrees of freedom are always the same.
By this way, the low-speed supersonic flow can be calculated for the gas with $1< \gamma \le 5/3$.
The method using reduced distribution function to take the internal degrees of freedom into account can be extended to the one- or two-dimensional cases to further reduce the computational cost, where $v$ and $w$ are merged into internal degrees of freedom $\xi$.
Correspondingly, we have $\gamma = (K+3) / (K+1)$  and $\gamma = (K+4) / (K+2)$ for 1D and 2D cases, respectively.
Using the 1D and 2D codes, the polyatomic gases with $1 < \gamma \le  3$ and $ 1 < \gamma \le 2$ can be simulated for not too high Mach number flows.

\subsubsection{Rykov model}
Considering the real gas effect,  the detailed relaxation process of rotational and vibrational degrees of freedom should be modeled.
Liu et al. \cite{liu2014diatomic} employed the Rykov model \cite{rykov1975model} to construct the time evolution solution of the gas distribution function in the UGKS for simulation of diatomic gases.
The rotational degree of freedom is included and modeled by controlling the energy exchange between translational and rotational energy through the relaxation model.
The Rykov kinetic model equation gives
\begin{equation}
	\frac{\partial f}{\partial t}
	+ u \frac{\partial f}{\partial x}
	+ v \frac{\partial f}{\partial y}
	+ w \frac{\partial f}{\partial z} = \frac{g_{tr} - f}{\tau} + \frac{g_{ro} - g_{tr}}{Z_r \tau}.
\end{equation}
The equilibrium states are expressed as
\begin{equation}
	\begin{aligned}
	g_{tr} &=\tilde{g}_{tr} \left[1-\frac{2 m_0 \vec{q}_{tr} \cdot \vec{c}}{15 k_B T_{tr} p_{tr}} \left(\frac{5}{2} - \frac{m_0 c^2}{2 k_B T_{tr}}\right) + (\sigma-1) \frac{m_0 \vec{q}_{ro} \cdot \vec{c}}{k_B T_{tr}} \frac{k_B T_{ro} - \epsilon}{\theta}\right],\\
	g_{ro} &= \tilde{g}_{ro} \left[1-\omega_0\frac{2 m_0 \vec{q}_{tr} \cdot \vec{c}}{15 k_B T p} \left(\frac{5}{2}-\frac{m_0 c^2}{2 k_B T}\right) + (1-\sigma)\omega_1 \frac{m_0 \vec{q}_{ro} \cdot \vec{c}}{k_B T p} \left(1-\frac{\epsilon}{k_B T}\right)\right],
	\end{aligned}
\end{equation}
where
\begin{equation}
\begin{aligned}
\tilde{g}_{tr} &= \rho \left(\frac{m_0}{2 \pi k_B T_{tr}}\right)^{\frac{3}{2}} {\exp}\left[-\frac{m_0 c^2}{2 k_B T_{tr}}\right]
\left(\frac{1}{k_B T_{ro}}\right) {\exp}\left[-\frac{\epsilon}{k_B T_{ro}}\right], \\
\tilde{g}_{ro} &= \rho \left(\frac{m_0}{2 \pi k_B T}\right)^{\frac{3}{2}} {\exp}\left[-\frac{m_0 c^2}{2 k_B T}\right]
\left(\frac{1}{k_B T}\right) {\exp}\left[-\frac{\epsilon}{k_B T}\right],
\end{aligned}
\end{equation}
and
$\theta$ is defined by
\begin{equation}
	\theta = \int {(k_B T_{ro} - \epsilon)^2 f du dv dw d\epsilon}.
\end{equation}
The macroscopic flow variables, such as pressure, temperature and heat flux for translational and rotational degrees are defined by
\begin{equation}
\begin{aligned}
p_{tr} & = \rho \frac{k_B}{m_0} T_{tr} = \frac{1}{3} \int {c^2 f du dv dw d\epsilon}, \\
p_{ro} & = \rho \frac{k_B}{m_0} T_{ro} = \int {\epsilon f du dv dw d\epsilon},\\
\vec{q}_{tr} &= \int {\vec{c} \frac{c^2}{2} f du dv dw d\epsilon },\\
\vec{q}_{ro} &= \int {\vec{c} \epsilon f du dv dw d\epsilon}.
\end{aligned}
\end{equation}
The parameters used in the above model are $\omega_0=0.2354$, $\omega_1 = 0.3049$ and $1/\sigma = 1.55$ for nitrogen gas, and $\omega_0 = 0.5$, $\omega_1=0.286$ and $1/\sigma=1.55$ for oxygen gas.

In order to construct the diatomic UGKS, the Rykov model can be re-written in the same form of the BGK model
\begin{equation}
\frac{\partial f}{\partial t}
+ u \frac{\partial f}{\partial x}
+ v \frac{\partial f}{\partial y}
+ w \frac{\partial f}{\partial z} = \frac{g_{eq} - f}{\tau},
\end{equation}
where the equivalent equilibrium state $g_{eq}$ is
\begin{equation}
	g_{eq} = (1-\frac{1}{Z_r}) g_{tr} + \frac{1}{Z_r} g_{ro}.
\end{equation}
Then the UGKS with Rykov model can be constructed in analogy to that with BGK model.
In the UGKS for diatomic gas simulation, the reduced distribution functions
\begin{equation}
	\begin{aligned}
	G(x, y, z, u, v, w,t) &= \int {f(x, y, z, u, v, w, \epsilon,t) d\epsilon}, \\
	R(x, y, z, u, v, w,t) &= \int {\epsilon f(x, y, z, u, v, w, \epsilon,t) d\epsilon },
	\end{aligned}
\end{equation}
are used to reduce the computational cost.
By using the integral solution along the characteristic line, the time-averaged multiscale flux function for the reduced distribution function, $\bar{G}_{ij}$ and $\bar{R}_{ij}$ across a cell interface $ij$ can be constructed in analogy to Eqs.~\eqref{eq:fij_expression} and \eqref{eq:fij_flux}.
The updation of the macroscopic flow variables can be carried out by
\begin{equation}
\begin{aligned}
\left(
\begin{matrix}
\rho \\ \rho U \\ \rho V \\ \rho W \\ \rho E \\ \rho E_{ro}
\end{matrix}
\right)_i^{n+1}
&=
\left(
\begin{matrix}
\rho \\  \rho U \\ \rho V \\ \rho W \\ \rho E \\ \rho E_{ro}
\end{matrix}
\right)_i^{n} - \frac{\Delta t}{V_i} \sum_{j\in N(i)} \vec{F}_{ij} S_{ij} \\
&+
\Delta t
\left(
\begin{matrix}
0\\ 0\\ 0\\ 0\\ 0\\
\frac{(\rho E_{ro, eq})_i^n - (\rho E_{ro})^n}{2 \tau_i^n} + \frac{(\rho E_{ro, eq})_i^{n+1} - (\rho E_{ro})_i^{n+1}}{2 \tau_i^{n+1}}
\end{matrix}
\right),
\end{aligned}
\end{equation}
where the macroscopic fluxes are
\begin{equation}
	\vec{F}_{ij} =  \sum_k {
			\left[			
			\begin{pmatrix}
			1\\ u_k\\ v_k\\ w_k\\ \frac{u_k^2 + v_k^2+w_k^2}{2}\\ 0
			\end{pmatrix}
			\bar{G}_{ij}
			+
			\begin{pmatrix}
			0\\ 0\\ 0\\ 0\\ 1 \\ 1
			\end{pmatrix}
			\bar{R}_{ij}
			\right] u_{n,ij,k}  \mathcal{V}_k}.
\end{equation}
With the updated macroscopic flow variables, the gas distribution function can be renewed by
\begin{equation}
	\begin{aligned}
	G_i^{n+1} &= G_i^n - \frac{\Delta t}{V_i} \sum_{j \in N(i)} {u_{n,ij} \bar{G}_{ij}  S_{ij}}
	+ \Delta t \left(\frac{G_{eq,i}^n - G_i^n}{2 \tau_i^n} + \frac{G_{eq,i}^{n+1} - G_i^{n+1}}{2 \tau_i^{n+1}}\right),\\
	R_i^{n+1} &= R_i^n - \frac{\Delta t}{V_i} \sum_{j \in N(i)} {u_{n,ij}\bar{R}_{ij}  S_{ij}}
	+ \Delta t \left(\frac{R_{eq,i}^n - R_i^n}{2 \tau_i^n} + \frac{R_{eq,i}^{n+1} - R_i^{n+1}}{2 \tau_i^{n+1}}\right).\\
	\end{aligned}
\end{equation}
In the calculation, the energy relaxation term in the Rykov equation is modeled using a Landau-Teller-Jeans relaxation.
The particle collision time multiplied by rotational collision number defines the relaxation rate for the rotational energy equilibrating with the translational energy.
The rotational collision number $Z_r$ is computed by
\begin{equation}
	Z_r = \frac{Z_r^\ast}{1+(\pi^{3/2}/2) \sqrt{\tilde{T}/T_{tr}} + (\pi + \pi^2/4) (\tilde{T} / T_{tr})},
\end{equation}
where $\tilde{T}$ is the characteristic temperature of intermolecular potential.
For $N_2$ within a temperature range from $30{ K}$ to $3000 { K}$, the values of $Z_r^\ast = 23.0$ and $\tilde{T}=91.5 { K}$ are used.

\begin{figure}[htbp]
\centering
\subfloat[]{\includegraphics[width=0.39\textwidth]{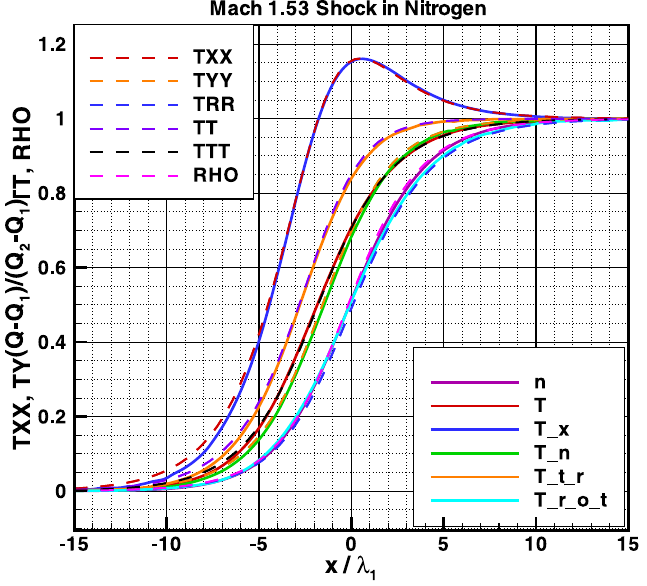}} \hspace{1.5em}
\subfloat[]{\includegraphics[width=0.44\textwidth]{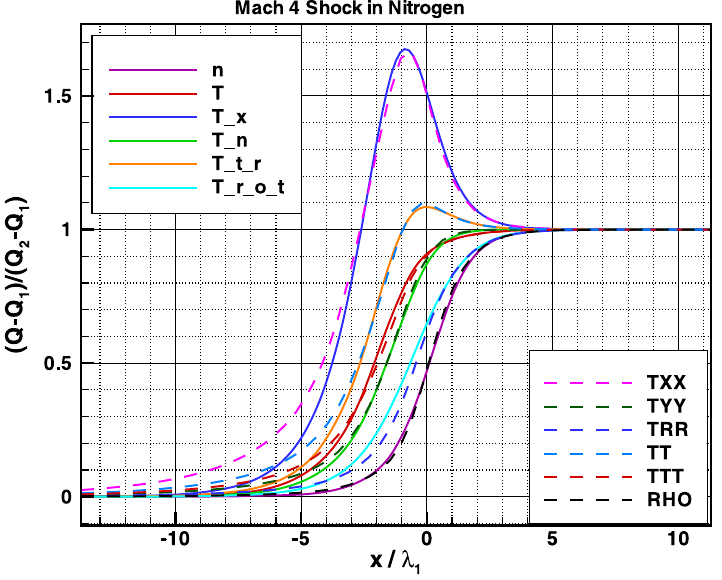}}\\
\subfloat[]{\includegraphics[width=0.44\textwidth]{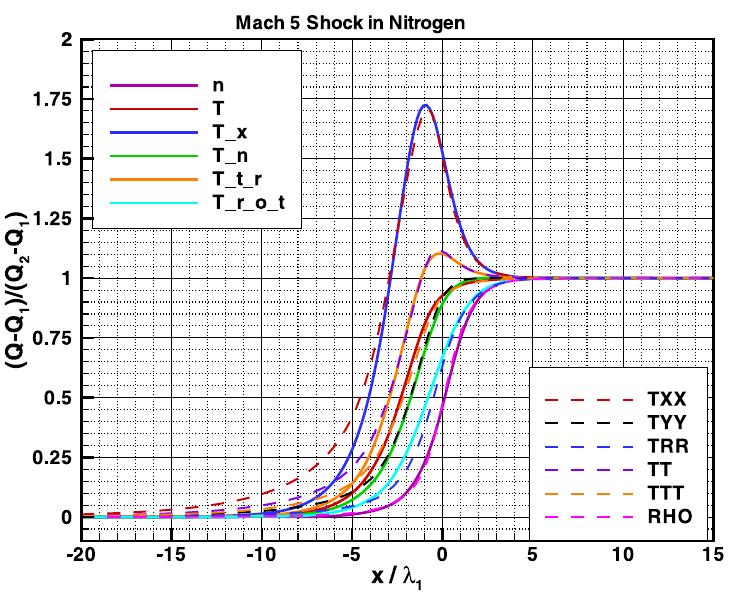}} \hspace{1em}
\subfloat[]{\includegraphics[width=0.43\textwidth]{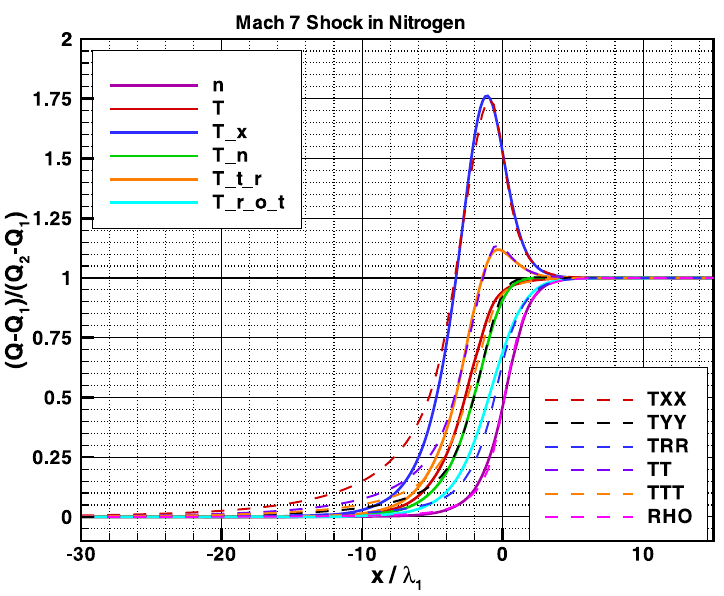}}
\caption{\label{fig:diatomic_shock}Shock structure for nitrogen gas with rotational degrees of freedom \cite{liu2014diatomic}. (a) ${ Ma}=1.53$, (b) ${ Ma} = 4$, (c) ${ Ma}=5$, and (d) ${ Ma}=7$.}
\end{figure}

\begin{figure}[htbp]
\centering
\subfloat[]{\includegraphics[width=0.42\textwidth]{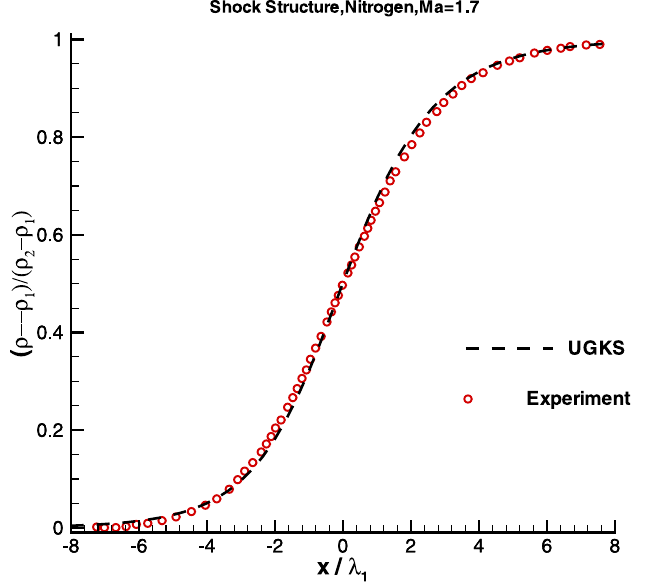}}\hspace{1em}
\subfloat[]{\includegraphics[width=0.42\textwidth]{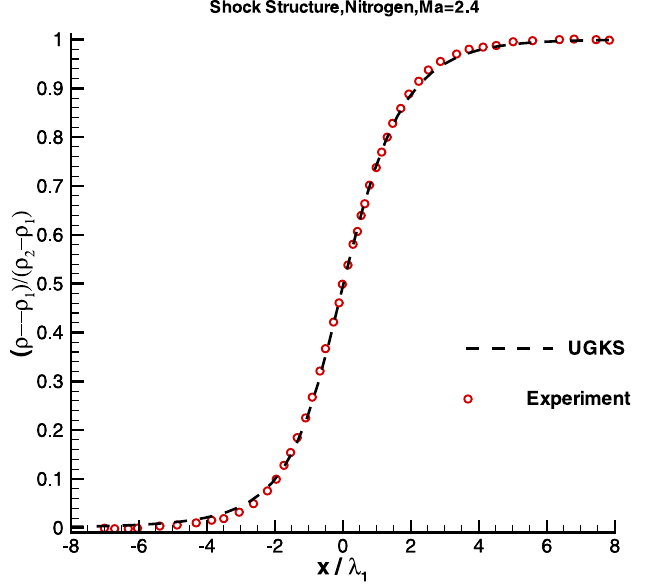}}\\
\subfloat[]{\includegraphics[width=0.42\textwidth]{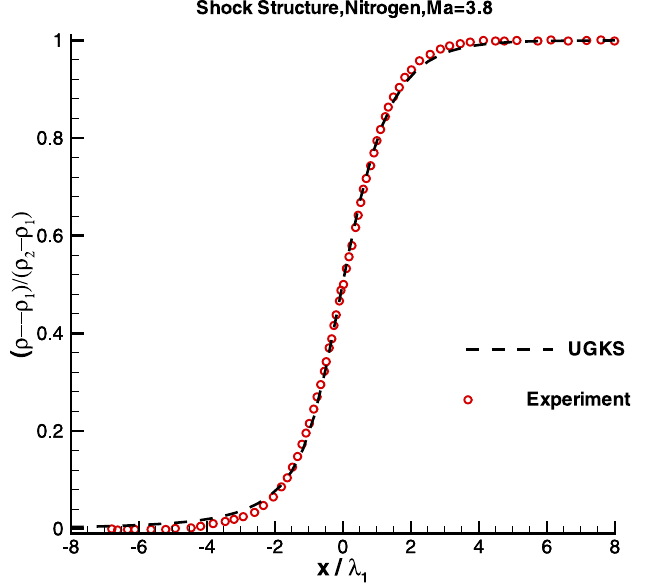}}\hspace{1em}
\subfloat[]{\includegraphics[width=0.42\textwidth]{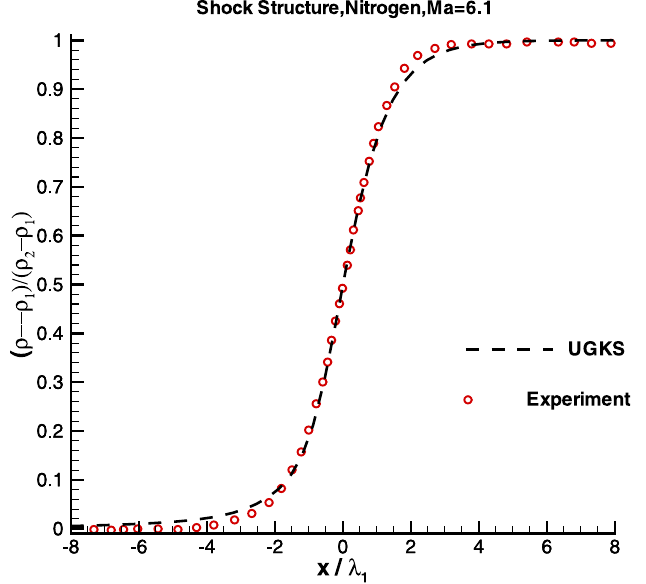}}\\
\subfloat[]{\includegraphics[width=0.42\textwidth]{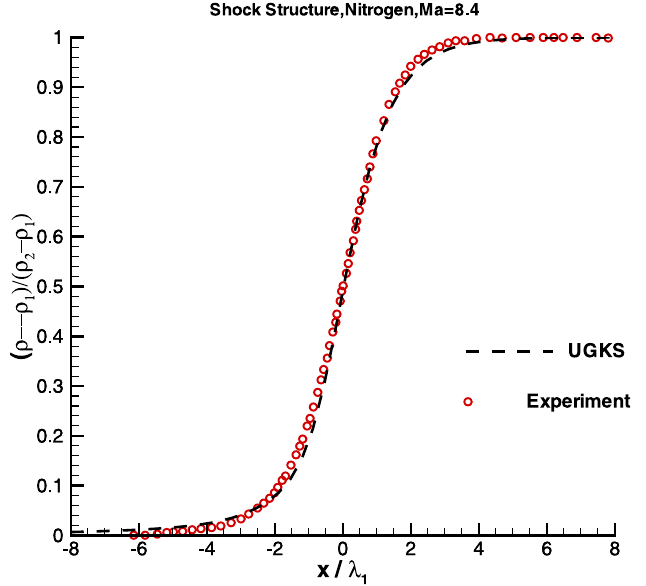}}\hspace{1em}
\subfloat[]{\includegraphics[width=0.42\textwidth]{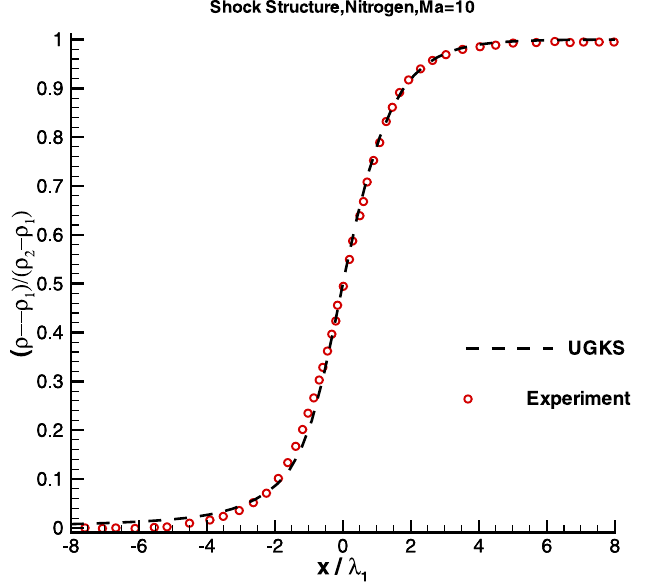}}
\caption{\label{fig:diatomic_shock_exp} Density profiles of the UGKS solutions and experimental measurement at different Mach numbers \cite{liu2014diatomic}. (a) ${ Ma}=1.7$, (b) ${ Ma}=2.4$, (c) ${ Ma}=3.8$, (d) ${ Ma}=6.1$, (e) ${ Ma}=8.4$, and (f) ${ Ma}=10$.}
\end{figure}

The normal shock has been computed for the nitrogen gas at different Mach numbers \cite{liu2014diatomic}.
The collision number is set at a constant of the value $Z_r = 2.4$.
The comparison of the density and temperature between the UGKS and DSMC at ${ Ma}=1.53$, $4.0$, $5.0$, and $7.0$ are given in Fig.~\ref{fig:diatomic_shock}.
Similar to the Shakhov model, the translational temperature in the upstream arises earlier than the data obtained by the DSMC method.
The density profiles at different Mach numbers are illustrated in Fig.~\ref{fig:diatomic_shock_exp}, which agree well with the experimental measurement.

\begin{figure}[htbp]
\centering
\subfloat[]{\includegraphics[width=0.44\textwidth]{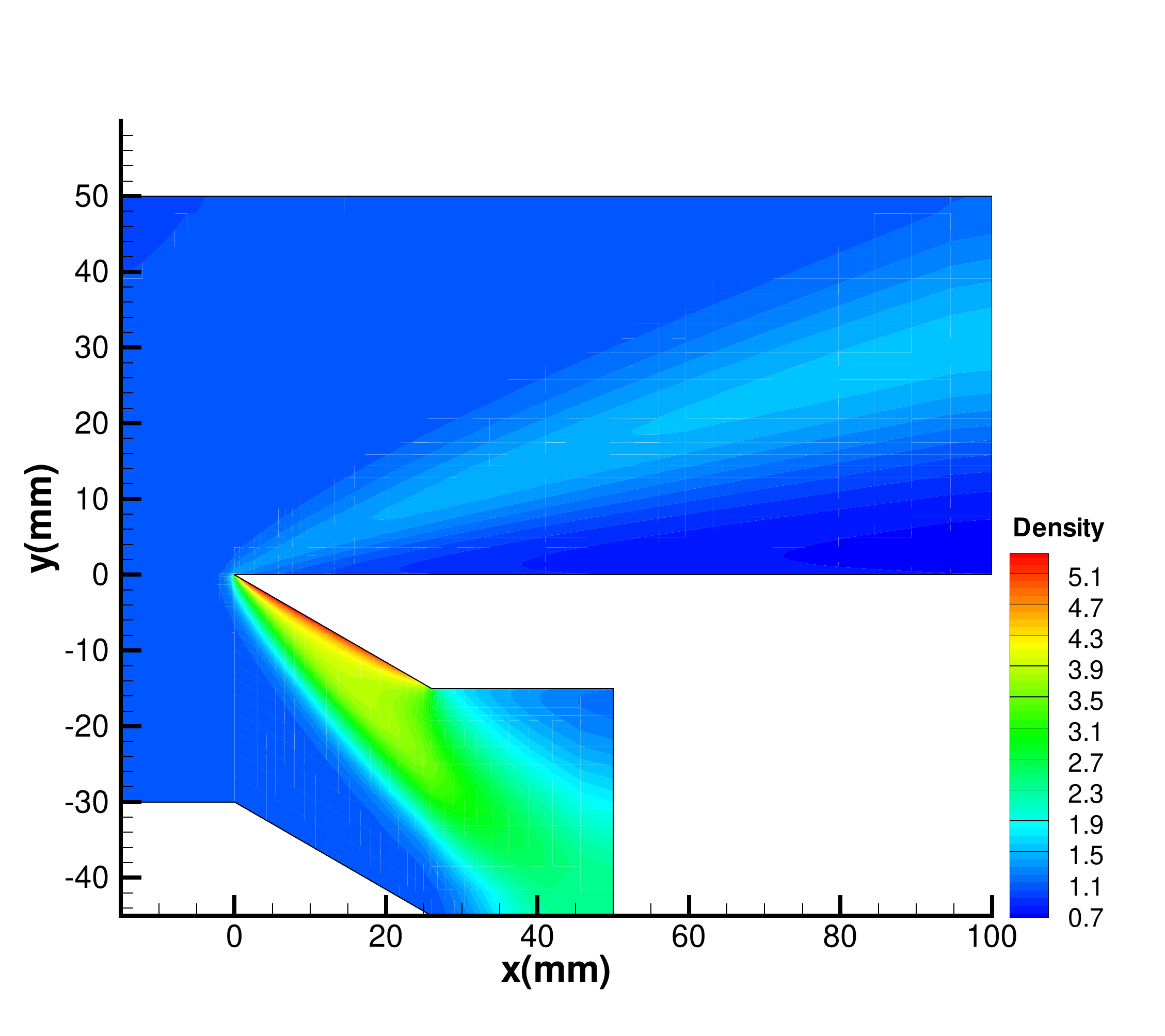}}
\subfloat[]{\includegraphics[width=0.44\textwidth]{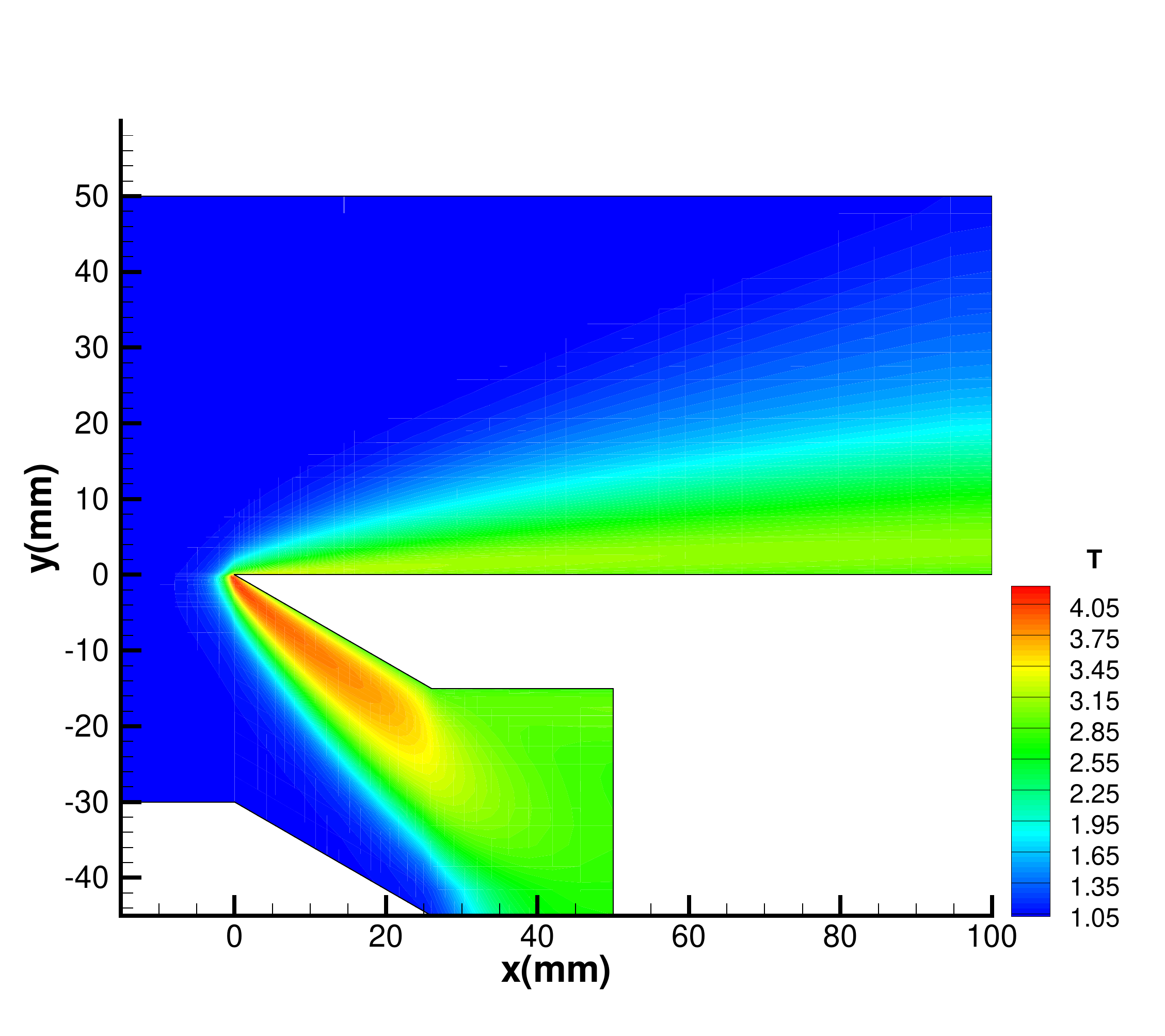}} \\
\subfloat[]{\includegraphics[width=0.44\textwidth]{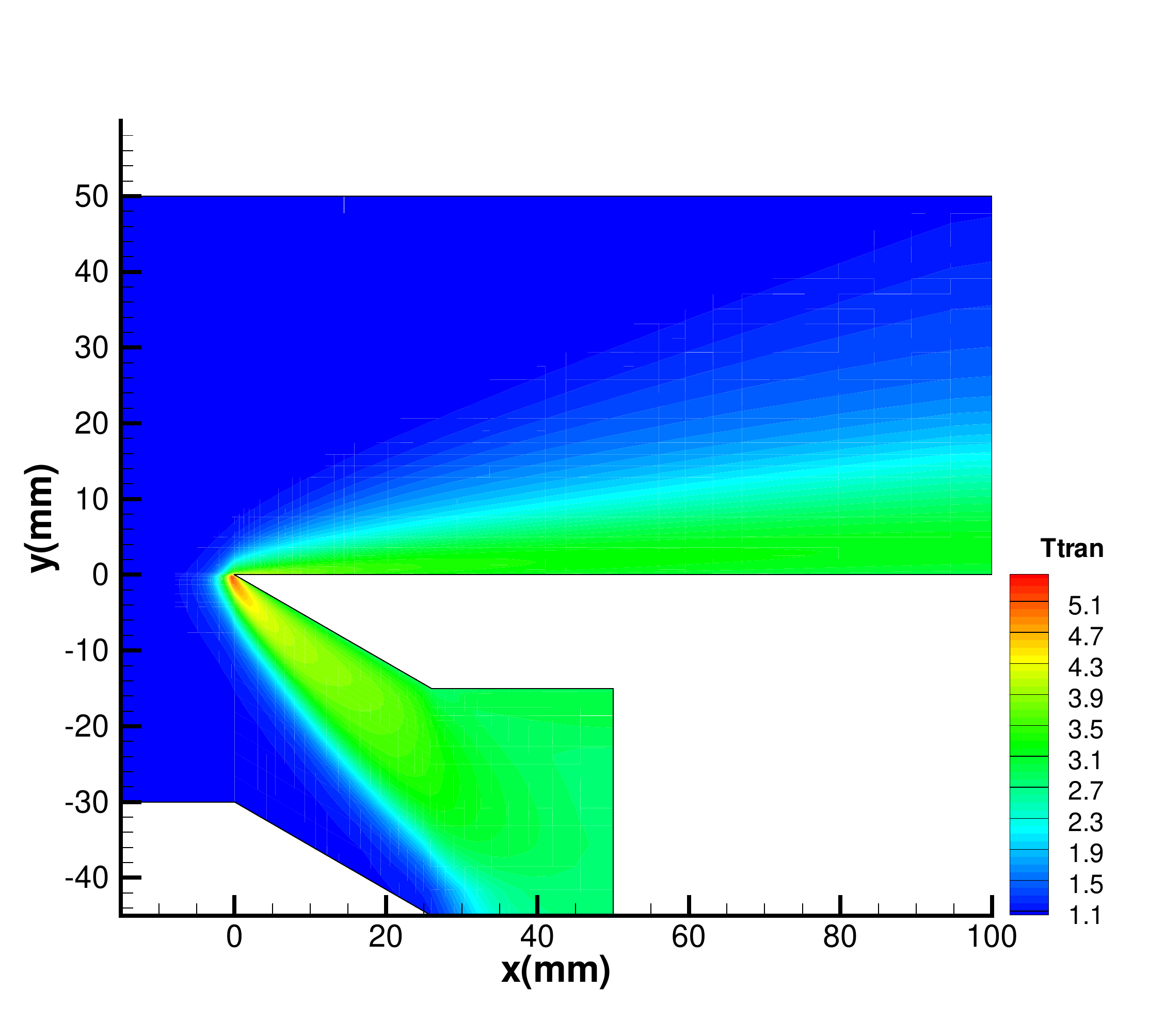}}
\subfloat[]{\includegraphics[width=0.44\textwidth]{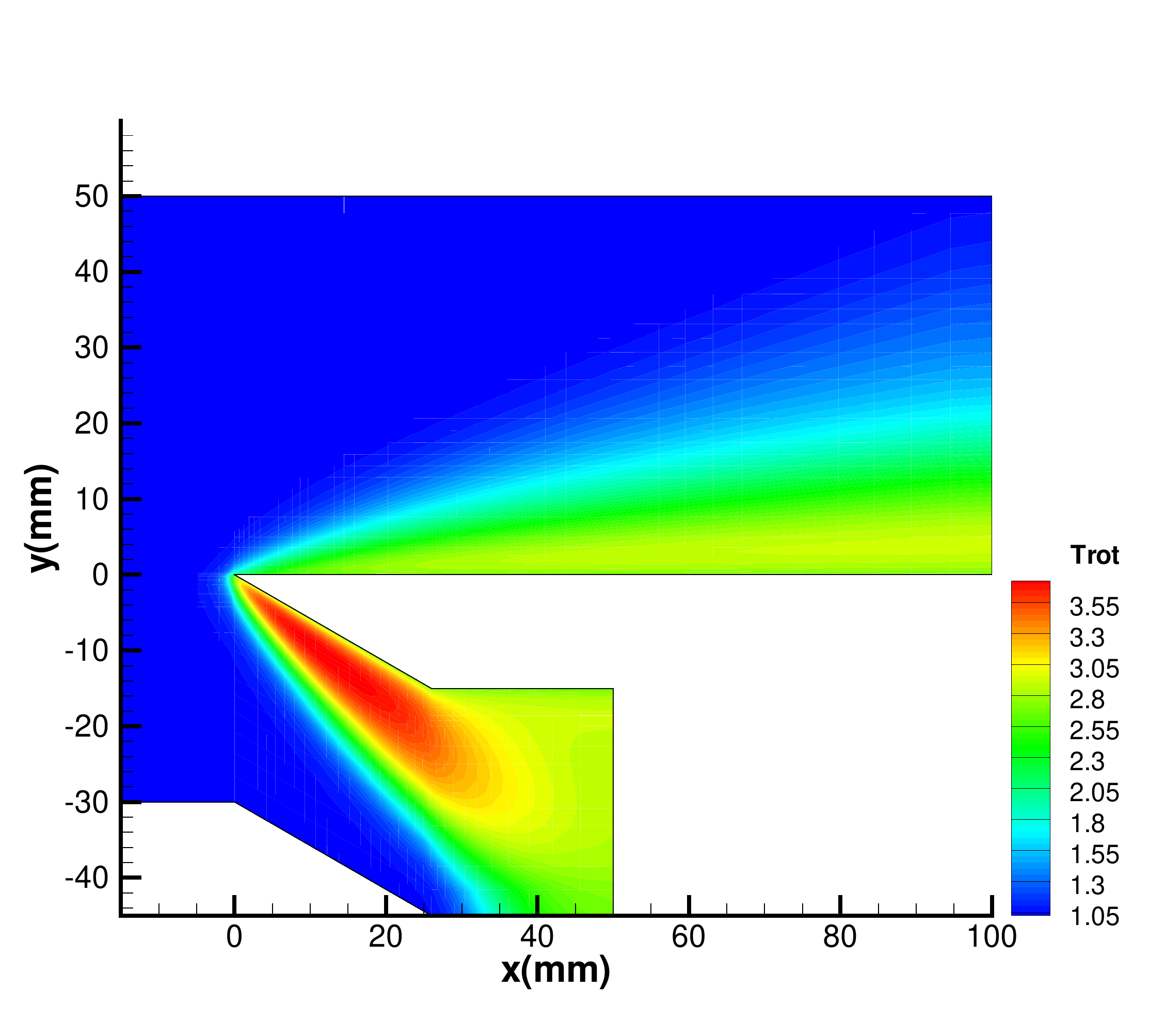}}
\caption{\label{fig:diatomic_plate} High-speed flow around a flat plate \cite{liu2014diatomic}. (a) Density, (b) averaged temperature, (c) translational temperature, and (d) rotational temperature.}
\end{figure}

\begin{figure}[htbp]
\centering
\subfloat[]{\includegraphics[width=0.44\textwidth]{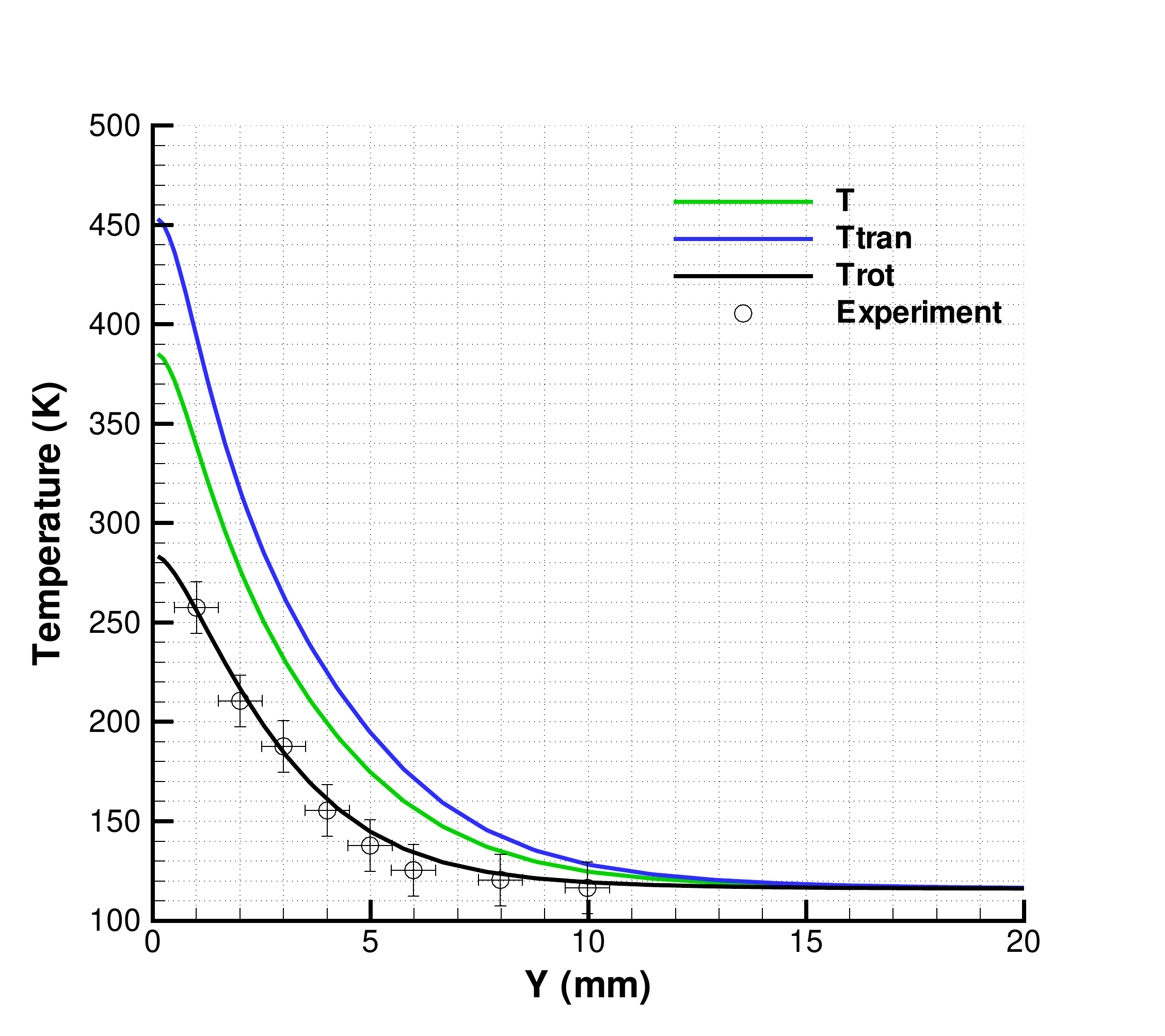}}
\subfloat[]{\includegraphics[width=0.44\textwidth]{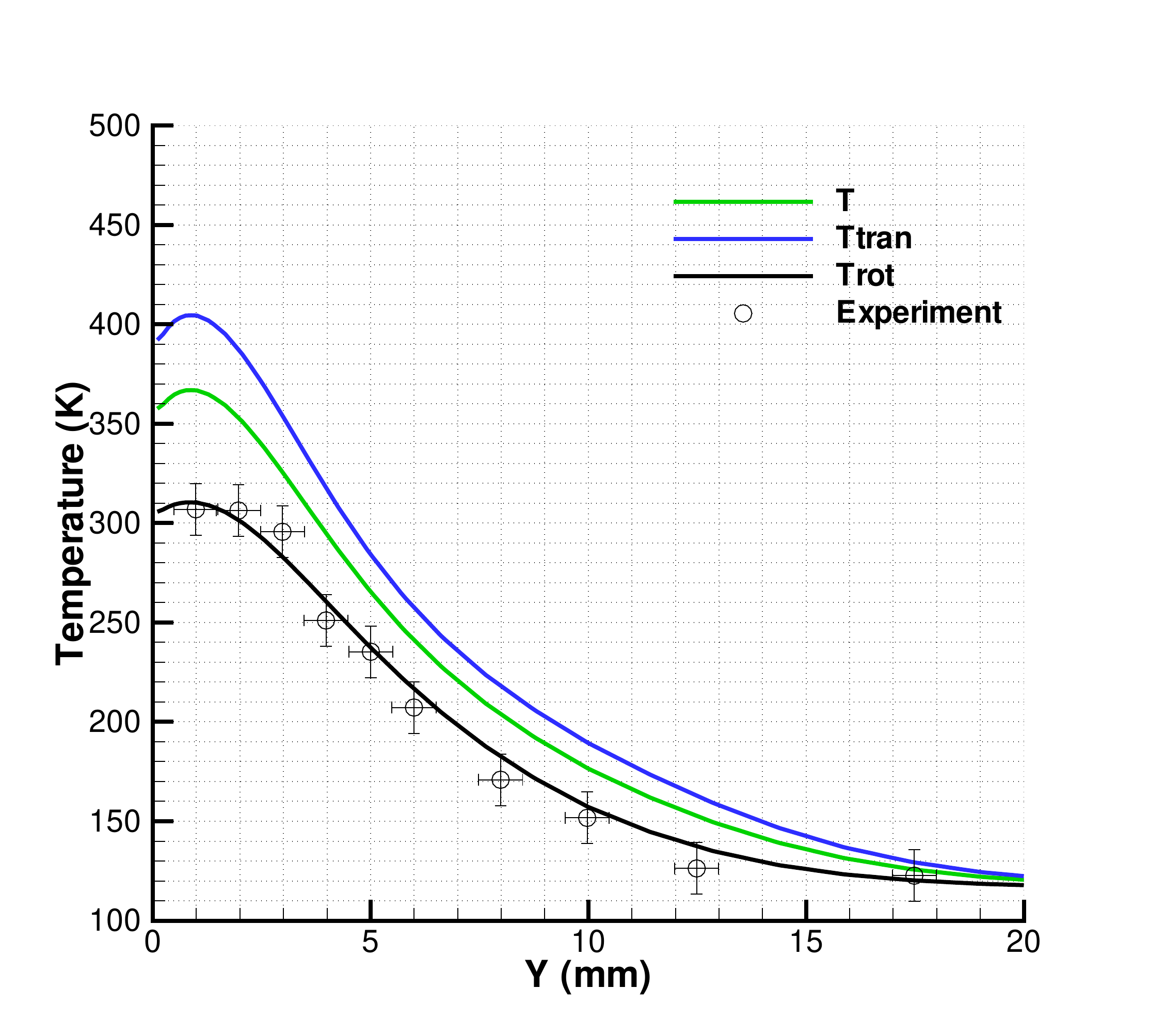}}
\caption{\label{fig:diatomic_plate_exp} Temperature distribution along vertical lines at $x = 5 { mm}$ and $20{ mm}$ \cite{liu2014diatomic}.}
\end{figure}

Following the run34 case in \cite{tsuboi2005experimental}, supersonic flow around a flat plate is computed and compared with the experimental measurement.
The flat plate is made of copper and is cooled by water to preserve a constant wall temperature $290 { K}$.
The stagnation state of the freestream is $983 { Pa}$ and $670 { K}$, and the exit condition is $2.12{ Pa}$ and $116{ K}$.
The exit Mach number is about $4.89$.
In this study, the shock wave and boundary layer at the sharp leading edge are merged.
Highly non-equilibrium between translational and rotational temperatures appears above the flat plate.
Figure \ref{fig:diatomic_plate} presents the density, translational, rotational, and total temperature contours around the sharp leading edge.
The temperature distribution along the vertical line above the flat plate at the location $x=5{ mm}$ and $20 { mm}$ from the leading edge are plotted in Fig.~\ref{fig:diatomic_plate_exp}, where the rotational temperature obtained by the UGKS agree well with the experimental data.

Taking the molecular vibration, a kinetic model \cite{zhang2015thesis,wang2017diatomic} has been proposed to simulate the diatomic gases with activated vibrational degrees of freedom.
The kinetic model equation can be written as
\begin{equation}\label{eq:vib_model}
\frac{\partial f}{\partial t}
+ u \frac{\partial f}{\partial x}
+ v \frac{\partial f}{\partial y}
+ w \frac{\partial f}{\partial z}
=
\frac{g_{t,r,v} - f}{\tau}
+
\frac{g_{tr, v} - g_{t,r,v}}{Z_r\tau}
+
\frac{g_{trv} - g_{tr,v}}{Z_v \tau},
\end{equation}
where $Z_r$ and $Z_v$ denote the rotational and vibrational collision numbers.
The equilibrium states are
\begin{equation}\label{eq:vib_equilibrium}
\begin{aligned}
g_{trv} &= \rho \left(\frac{\lambda}{\pi}\right)^{\frac{3}{2}} e^{-\lambda c^2} \frac{\lambda}{\pi} e^{-\lambda (\xi_1^2 + \xi_2^2)} \frac{4\lambda}{K_v} e^{-\frac{4\lambda}{K_v}\epsilon_v},\\
g_{tr,v} &= \rho \left(\frac{\lambda_{t,r}}{\pi}\right)^{\frac{3}{2}} e^{-\lambda_{t,r} c^2} \frac{\lambda_{t,r}}{\pi} e^{-\lambda_{t,r} (\xi_1^2 + \xi_2^2)} \frac{4\lambda_v}{K_v} e^{-\frac{4\lambda_v}{K_v}\epsilon_v},\\
g_{t,r,v} &= \rho \left(\frac{\lambda_{t}}{\pi}\right)^{\frac{3}{2}} e^{-\lambda_{t} c^2} \frac{\lambda_{r}}{\pi} e^{-\lambda_{r} (\xi_1^2 + \xi_2^2)} \frac{4\lambda_v}{K_v} e^{-\frac{4\lambda_v}{K_v}\epsilon_v}.
\end{aligned}
\end{equation}
Here, $\lambda$, $\lambda_t$, $\lambda_r$, $\lambda_v$ and $\lambda_{t,r}$ are related to the fully relaxed, translational, rotational, vibrational, and partially relaxed temperatures, respectively.
Specifically, we have
\begin{equation}
\begin{aligned}
\lambda &= \frac{(5+K_v)\rho}{4} \Big/
\int_{-\infty}^{+\infty} \int_{-\infty}^{+\infty} \int_{0}^{+\infty} \left[\frac{1}{2} (c^2 + \xi^2) + \epsilon_v \right] f d\epsilon_v d\xi d\Xi, \\
\lambda_t &= \frac{3\rho}{4} \Big/
\int_{-\infty}^{+\infty} \int_{-\infty}^{+\infty} \int_{0}^{+\infty} \frac{1}{2} c^2  f d\epsilon_v d\xi d\Xi, \\
\lambda_r &= \frac{\rho}{2} \Big/
\int_{-\infty}^{+\infty} \int_{-\infty}^{+\infty} \int_{0}^{+\infty} \frac{1}{2} \xi^2 f d\epsilon_v d\xi d\Xi, \\
\lambda_v &= \frac{K_v\rho}{4} \Big/
\int_{-\infty}^{+\infty} \int_{-\infty}^{+\infty} \int_{0}^{+\infty} \epsilon_v  f d\epsilon_v d\xi d\Xi,\\
\lambda_{t,r} &= \frac{5\rho}{4} \Big/
\int_{-\infty}^{+\infty} \int_{-\infty}^{+\infty} \int_{0}^{+\infty} \frac{1}{2} (c^2 + \xi^2)  f d\epsilon_v d\xi d\Xi, \\
\end{aligned}
\end{equation}
where $K_v$ is the number of vibrational degrees of freedom.
This collision model consists of three terms, including the relaxation processes of elastic collision, inelastic collision between molecular translation and rotation, the energy exchange from rotational degrees of freedom to the translational and rotational ones.
Eq.~\eqref{eq:vib_model} is more like a BGK equation with fixed Prandtl number, which can be extended to a Shakhov-like model by using orthogonal polynomials in order to obtain correct relaxation rate of high order moments \cite{zhang2015thesis,wang2017diatomic}.
With the provided kinetic model, the UGKS for diatomic gases with rotational and vibrational degrees of freedom can be constructed.

Different from the translational and rotational degrees of freedom, the vibration degree of freedom $K_v$ is a temperature dependent variable.
According to harmonic oscillator model, the specific vibrational energy associated with a characteristic vibrational temperature $\Theta_v$ is
\begin{equation}
e_v = \frac{k_B}{m_0} \frac{\Theta_v}{e^{2 \lambda k_B \Theta_v / m_0} - 1},
\end{equation}
then according to the equal partition to each degree of freedom, the vibrational degree of freedom $K_v$ can be determined by
\begin{equation}
K_v = \frac{4 k_B}{m_0} \frac{\lambda \Theta_v}{e^{2 k_B \lambda \Theta_v / m_0}-1}.
\end{equation}

\begin{figure}[htbp]
\centering
\subfloat[]{\includegraphics[width=0.44\textwidth]{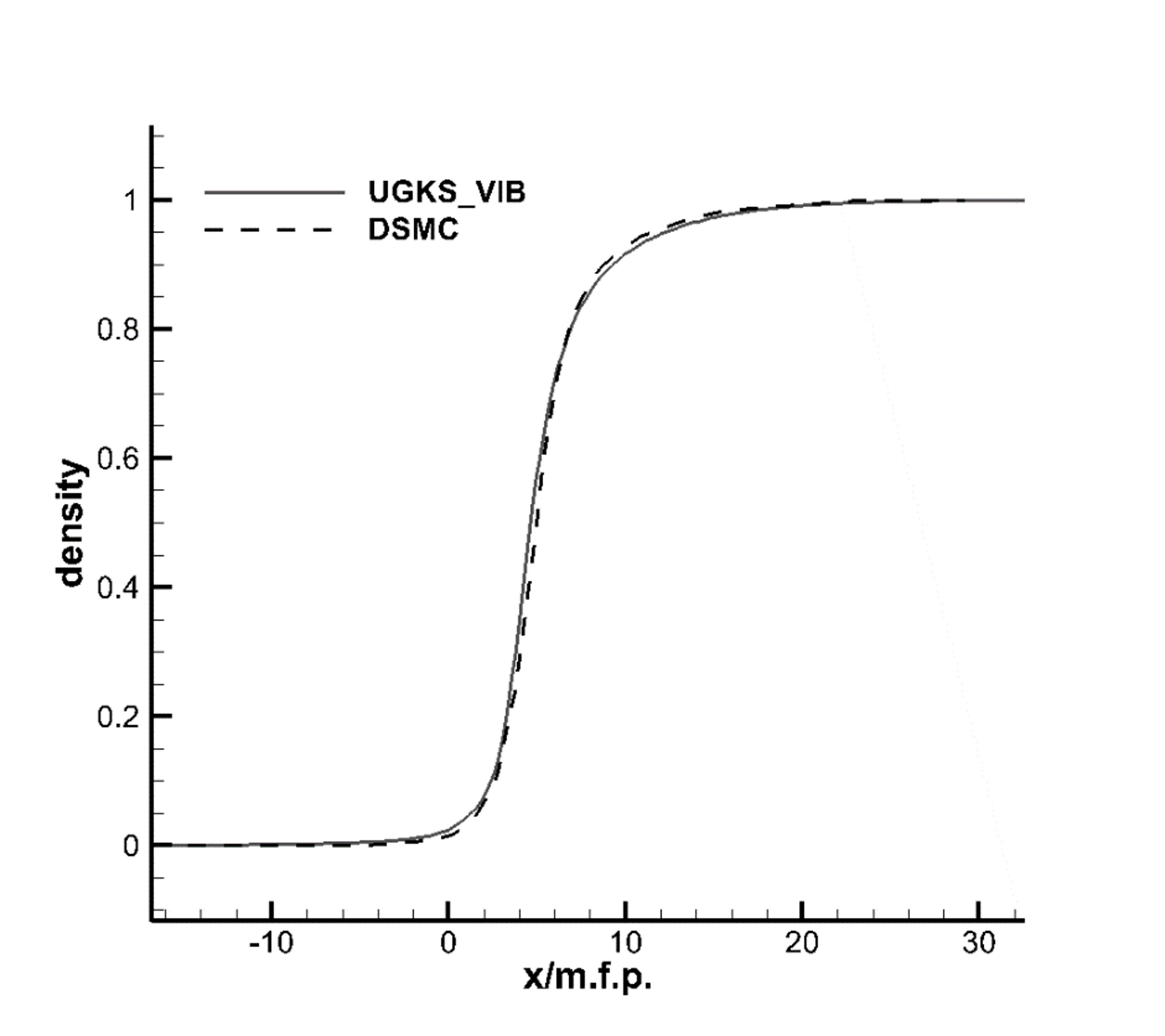}} \hspace{1em}
\subfloat[]{\includegraphics[width=0.44\textwidth]{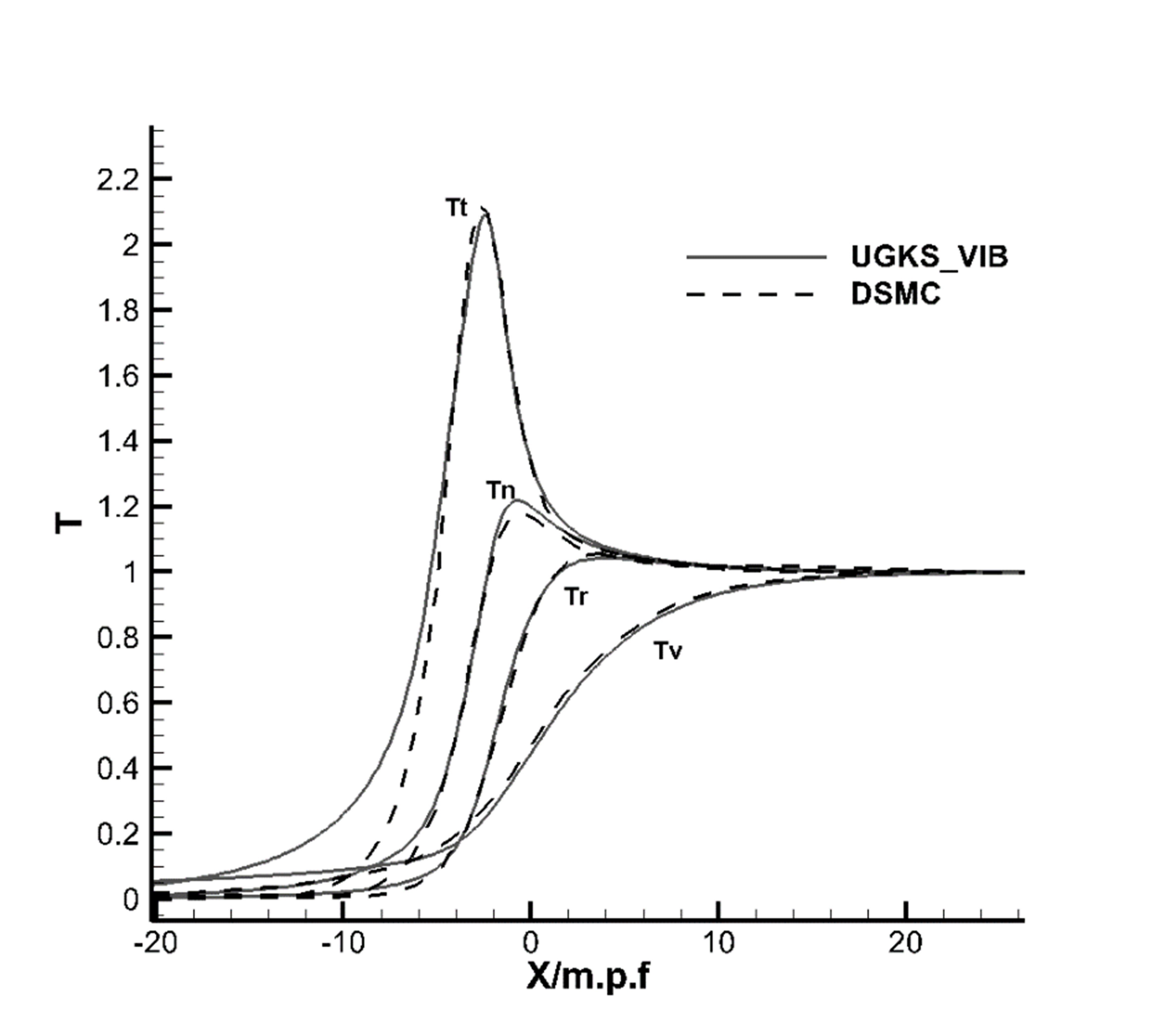}}
\caption{\label{fig:vibration_shock} Normal shock structure of nitrogen gas with vibrational degrees of freedom \cite{zhang2015thesis}. (a) Density, and (b) temperature.}
\end{figure}

The normal shock structure of nitrogen gas at ${ Ma}=10$ is computed in \cite{zhang2015thesis}, where the upstream temperature and density are $226.149 { K}$ and $1.7413\times10^{-2} { kg/m^3}$.
The mean free path is evaluated by Eq.~\eqref{eq:knudsen_number} with $\alpha = 1.36$ and $\omega=0.74$.
The rotational and vibrational collision numbers are chosen as $Z_r = 5$ and $Z_v=35$.
Figure \ref{fig:vibration_shock} shows the density and temperature profiles obtained by the diatomic UGKS and compared with the DSMC data computed by the MONACO code.
The density distribution agrees well with the DSMC results, and similar to the Shakhov model and Rykov model, the translational temperatures along the $x$-direction and other tangential directions are higher than DSMC data in the upstream, and the rotational and vibrational temperature profiles match well with the reference.

\begin{figure}[htbp]
\centering
\subfloat{\includegraphics[width=0.44\textwidth]{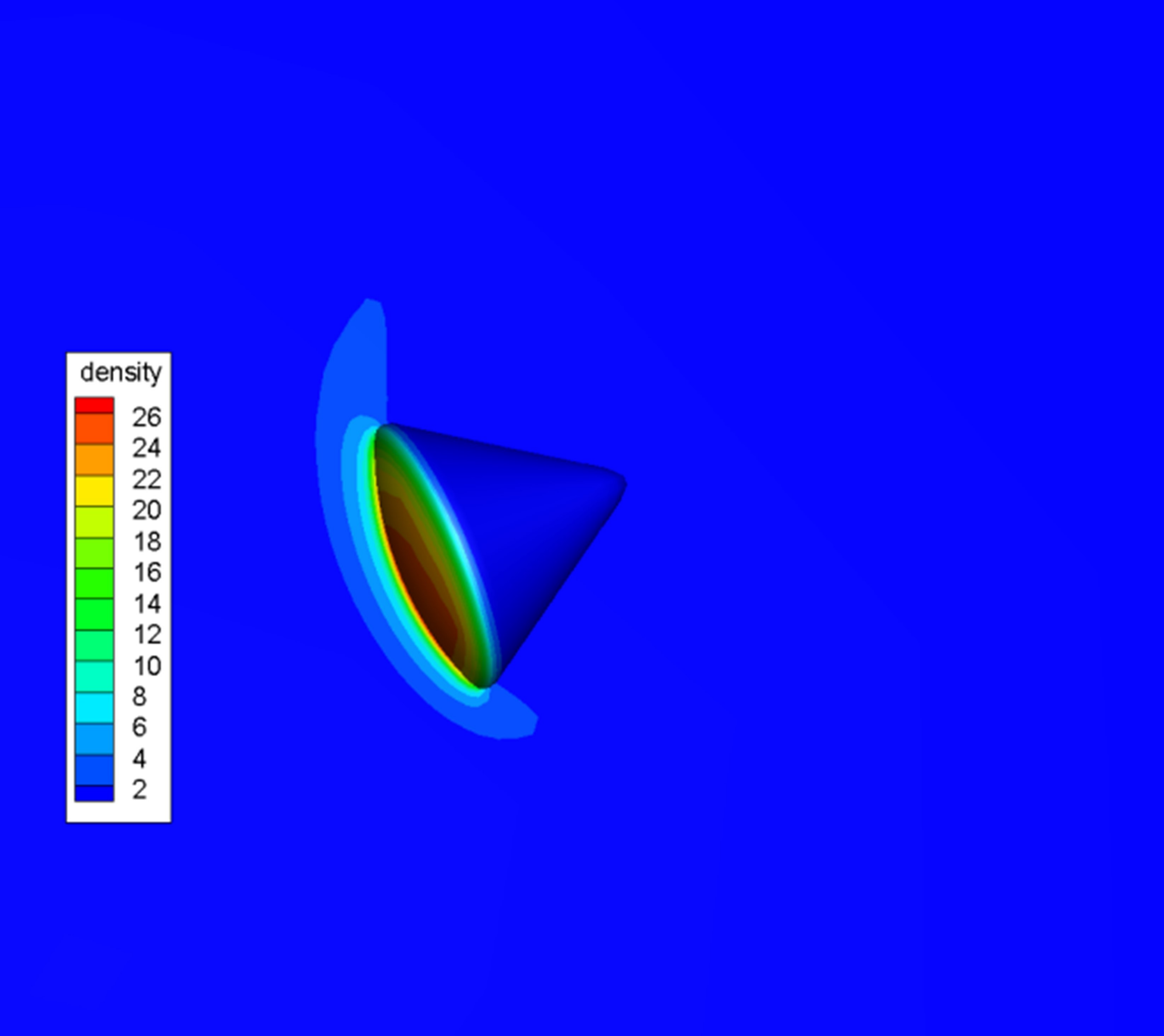}} \hspace{1em}
\subfloat{\includegraphics[width=0.44\textwidth]{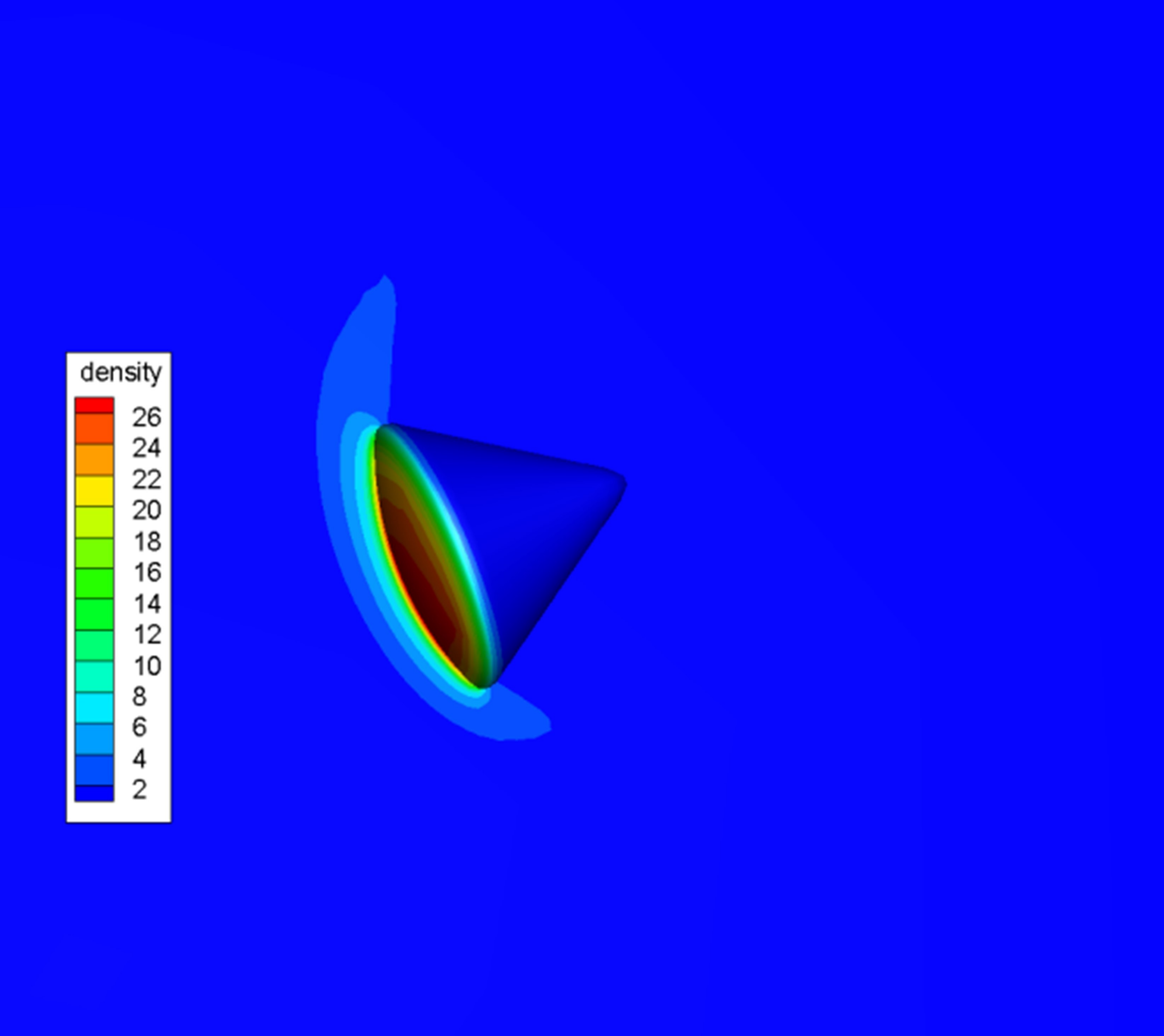}}
\caption{\label{fig:apollo_density} Density distribution of flow around Apollo re-entry vehicle \cite{zhang2015thesis}. (a) Ryhov model, and (b) UGKS with molecular vibration.}
\end{figure}

\begin{figure}[htbp]
\centering
\subfloat{\includegraphics[width=0.44\textwidth]{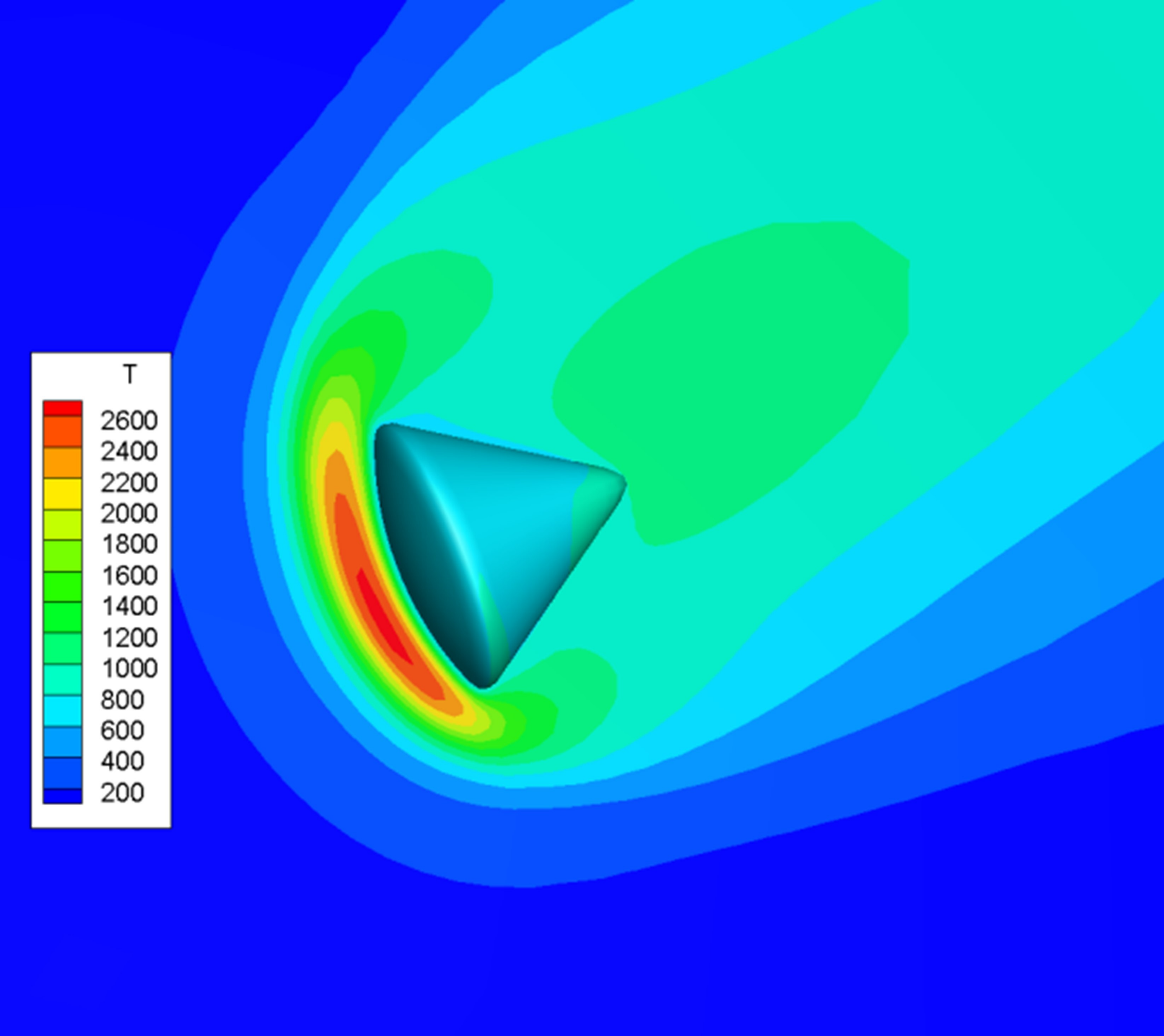}} \hspace{1em}
\subfloat{\includegraphics[width=0.44\textwidth]{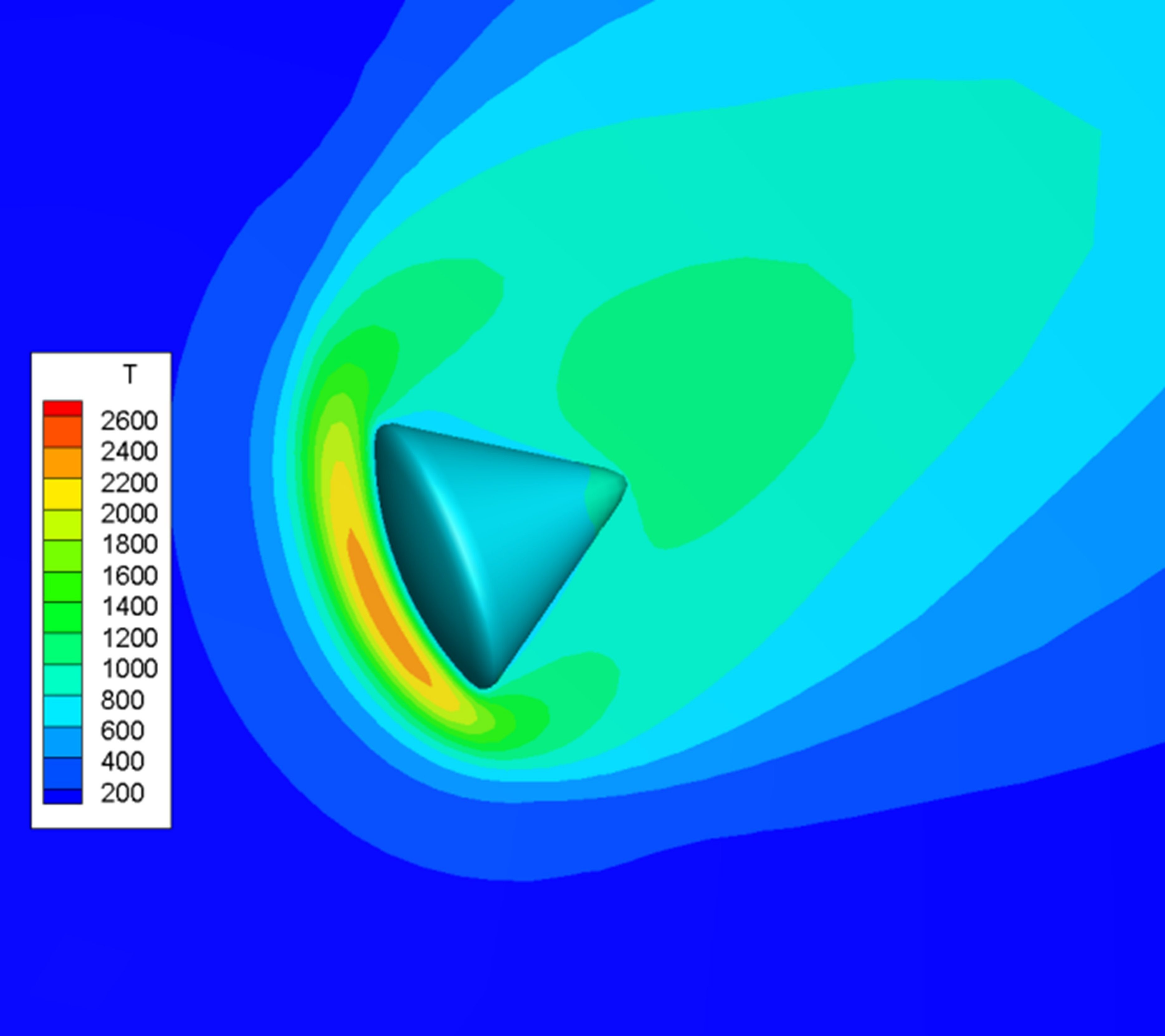}}
\caption{\label{fig:apollo_temperature} Temperature distribution of flow around Apollo re-entry vehicle \cite{zhang2015thesis}. (a) Ryhov model, and (b) UGKS with molecular vibration.}
\end{figure}

\begin{figure}[htbp]
\centering
\subfloat{\includegraphics[width=0.44\textwidth]{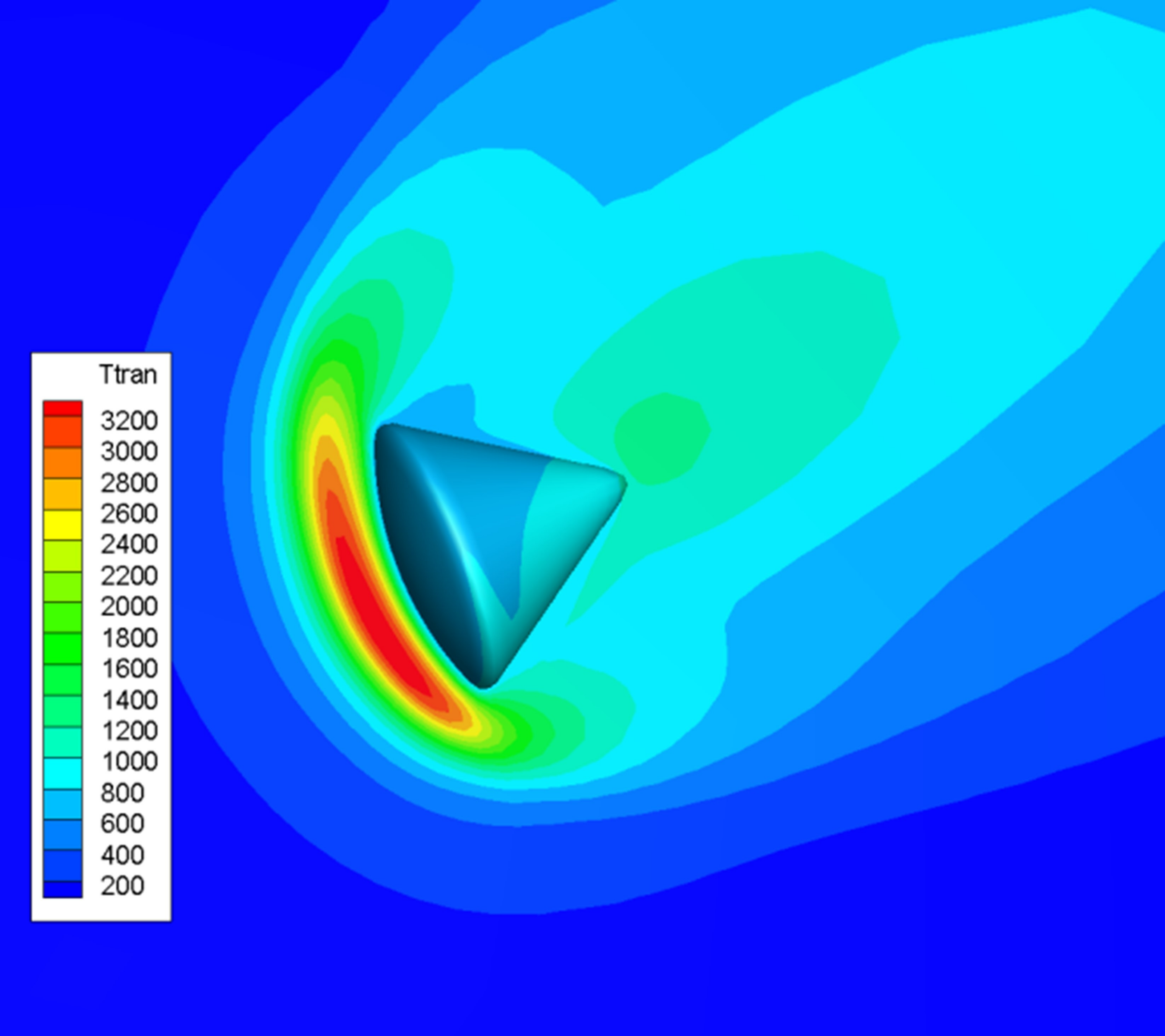}} \hspace{1em}
\subfloat{\includegraphics[width=0.44\textwidth]{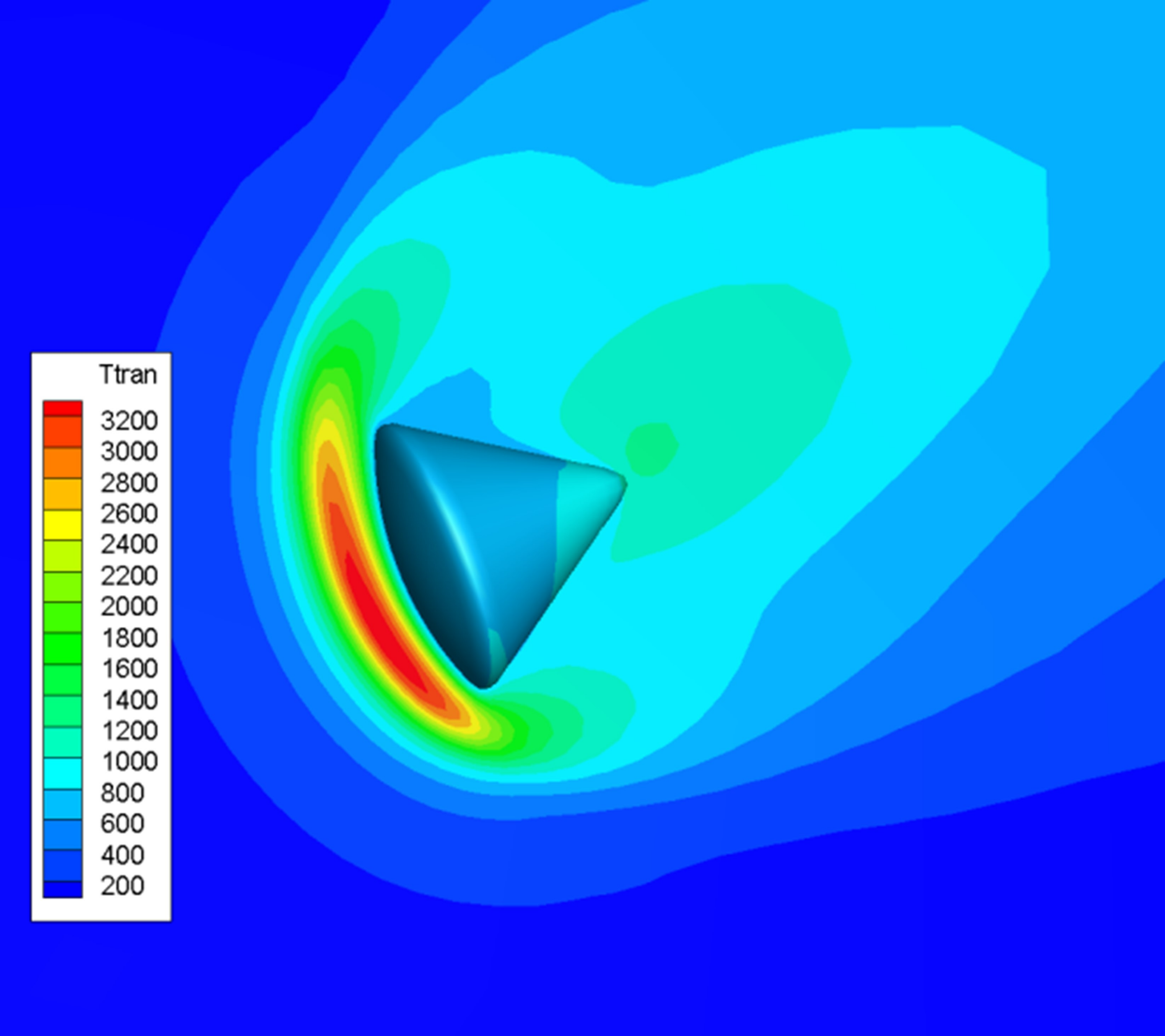}}
\caption{\label{fig:apollo_temperature_tran} Translational temperature distribution of flow around Apollo re-entry vehicle \cite{zhang2015thesis}. (a) Ryhov model, and (b) UGKS with molecular vibration.}
\end{figure}

\begin{figure}[htbp]
\centering
\subfloat{\includegraphics[width=0.44\textwidth]{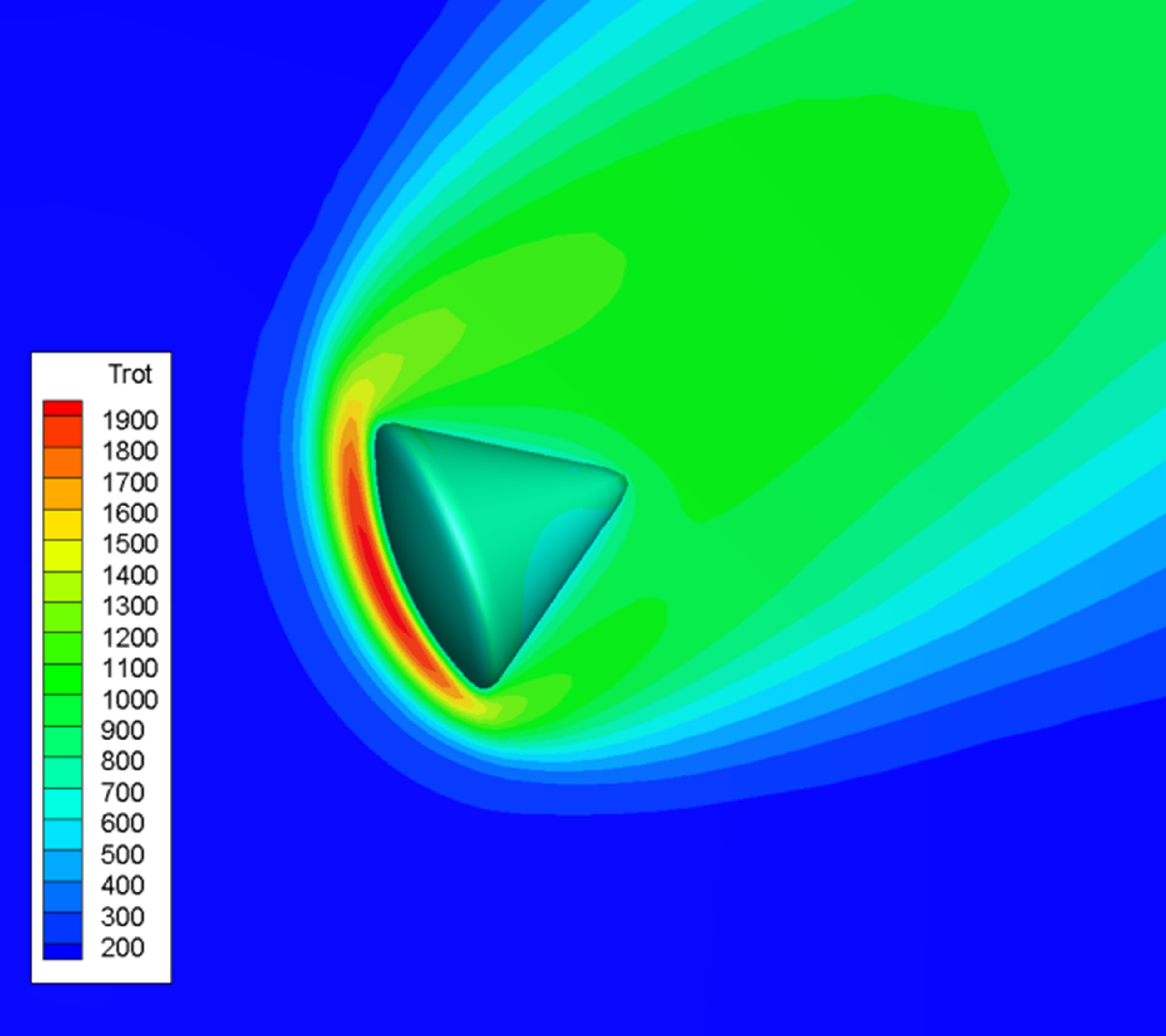}} \hspace{1em}
\subfloat{\includegraphics[width=0.44\textwidth]{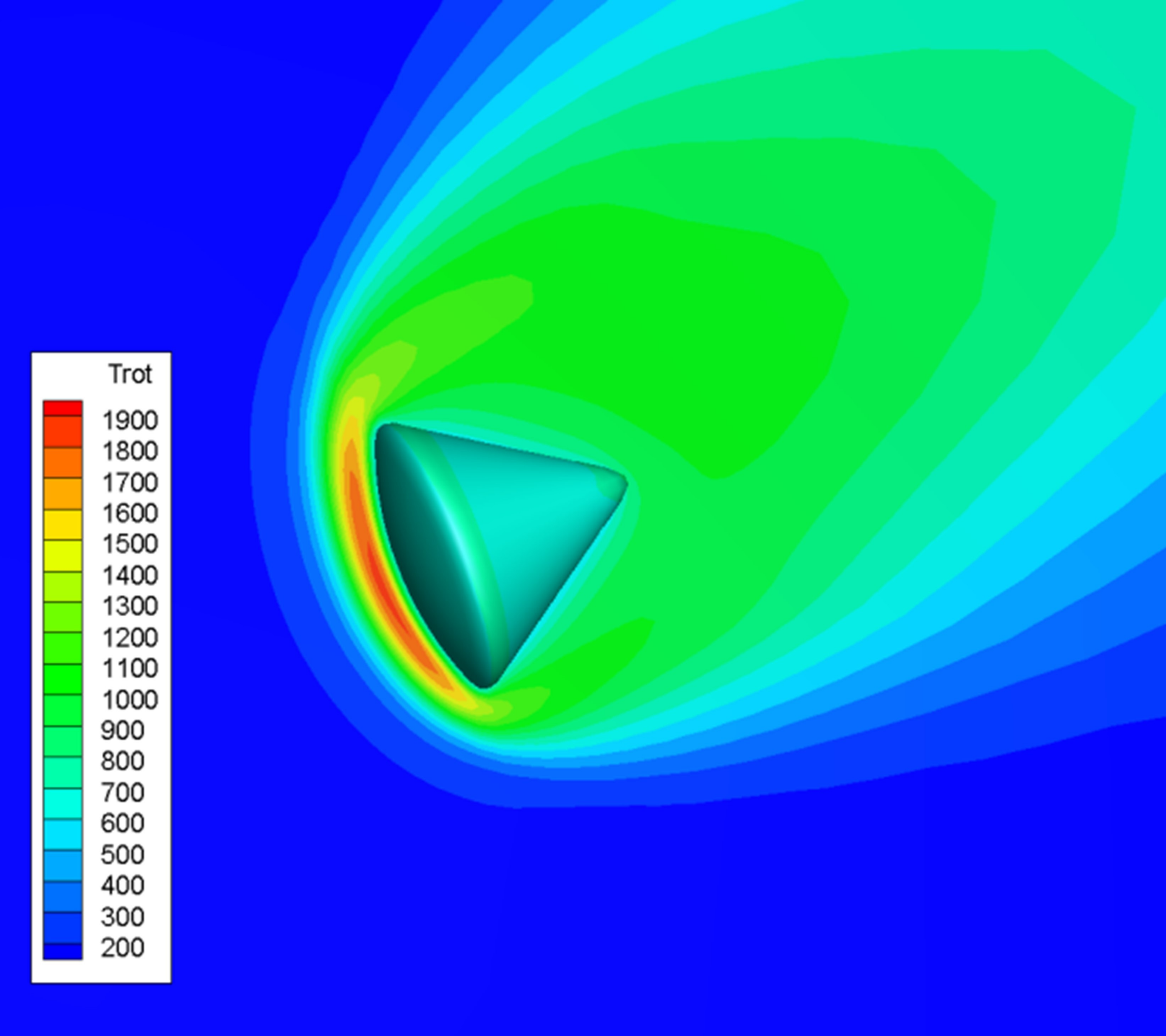}}
\caption{\label{fig:apollo_temperature_rot} Rotational temperature distribution of flow around Apollo re-entry vehicle \cite{zhang2015thesis}. (a) Ryhov model, and (b) UGKS with molecular vibration.}
\end{figure}

\begin{figure}[htbp]
\centering
\includegraphics[width=0.44\textwidth]{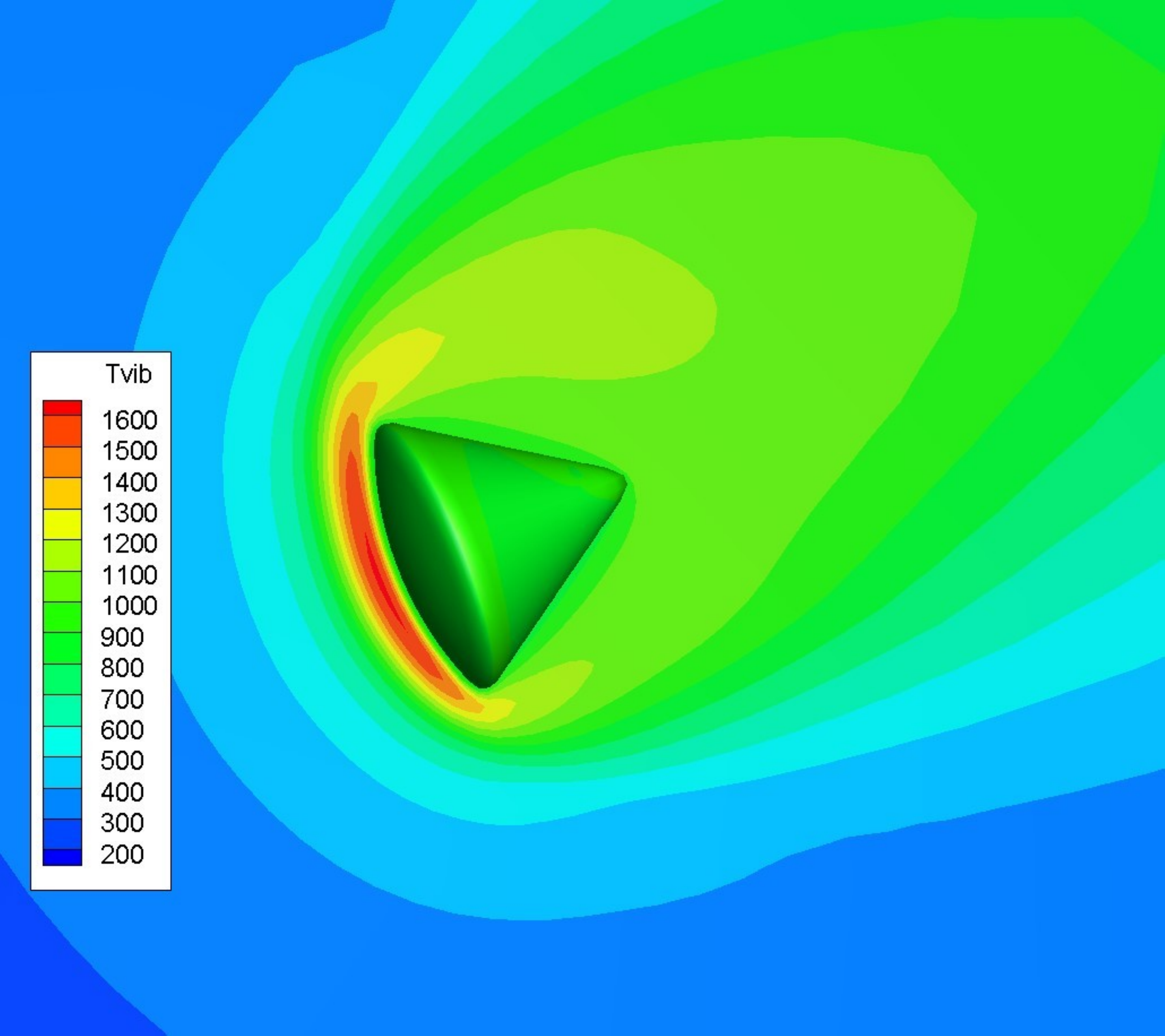}
\caption{\label{fig:apollo_temperature_vib}  Vibrational temperature distribution of flow around Apollo re-entry vehicle \cite{zhang2015thesis}. }
\end{figure}

Flow around the Apollo re-entry capsule is computed at ${ Ma}  = 10.2$, ${ Kn}=0.067$ with a freestream temperature $142.2{ K}$ and a fixed wall temperature $300{ K}$ \cite{zhang2015thesis}.
The nitrogen gas is considered, and the collision numbers adopt $Z_r = 5$ and $Z_v = 10$.
The angle of attack is $170$ degree.
The comparison of flow fields between the diatomic UGKS with Rykov model and the current model with vibrational degree of freedom are presented in Figs~\ref{fig:apollo_density} -- \ref{fig:apollo_temperature_vib}.
It can be found that due to the vibrational degree of freedom, the translational and rotational temperature distributions obtained by the current model are lower than those computed by the Rykov model.
A more realistic solution has been obtained.
\section{Acceleration techniques}
In this section, the acceleration techniques for improving the UGKS in numerical simulations of all flow regimes will be introduced, such as the implicit scheme, multigrid method, parallel computation,  adaptive physical mesh and velocity space, memory reduction techniques, and the novel wave-particle adaptation.

\subsection{Implicit UGKS for unsteady flow}
Implicit method is one of the most commonly used acceleration techniques for solving partial differential equation, which has been widely used in the CFD calculations \cite{jameson1987lower,yoon1988lower,rieger1988solution,jameson1991time,pulliam1993time,tomaro1997implicit}.
In the gas kinetic scheme to solve the Navier--Stokes equations, the implicit methods have been constructed as well \cite{chae2000development,xu2005multidimensional,li2006applications,jiang2012implicit,li2014implicit,tan2017time}.
Although the UGKS is introduced in the previous section as a direct modeling method on the discretization scale, the traditional CFD acceleration techniques can also be applied in the UGKS to obtain higher computational efficiency by transforming the UGKS on discretization scale into a semi-discrete form \cite{zhu2016iugks,zhu2019iugks,zhu2020thesis}.

\subsubsection{Implicit algorithms}\label{sec:iugks_equation}
The semi-discrete governing equations of the macroscopic flow variables and the distribution function are
\begin{equation}\label{eq:semi_macro}
\frac{\partial \vec{w}_i}{\partial t} +  \frac{1}{V_i}\sum_{j \in N(i)}{\vec{F}_{ij} S_{ij}} = \vec{0},
\end{equation}
and
\begin{equation}\label{eq:semi_micro}
\frac{\partial f_i}{\partial t} + \frac{1}{V_i} \sum_{j \in N(i)} {\mathcal{F}_{ij} S_{ij}} = \frac{g_i - f_i}{\tau_i}.
\end{equation}
The governing equations in the semi-discrete form usually describe the instant variation of flow field, which implies a time scale of $t \to 0$.
Specifically, $\vec{F}_{ij}$ and $\mathcal{F}_{ij}$ denote the instant macroscopic and microscopic fluxes across the cell interface $ij$.
However, the UGKS is constructed based on the integral solution of kinetic model, describing the multiscale transport process in a finite time step, and the fluxes $\vec{F}_{ij}$ and $\mathcal{F}_{ij}$ are not only related to the local physical state $\tau $, but also depends on the mesh size determined  time step $\Delta t$, refer to Eqs.~\eqref{eq:macro_flux_from_micro} and \eqref{eq:fij_flux}.
Therefore, the fluxes $\mathcal{F}_{ij}$ and $\vec{F}_{ij}$ in Eqs.~\eqref{eq:semi_macro} and \eqref{eq:semi_micro} should employ the time-averaged fluxes over a finite time step $\Delta t_s$ instead of the instant ones for constructing the implicit UGKS, where the time step $\Delta t_s$ is used to average the time-dependent fluxes in the explicit scheme.
Considering the flow physics in a local region described by the UGKS in Fig.~\ref{fig:flow_physics}, $\Delta t_s$ is determined by the resolution of computational mesh and the maximum of particles' speed
\begin{equation}\label{eq:cfl_condition}
\Delta t_s = {\text{CFL}} \frac{\Delta x}{\max|u|}.
\end{equation}
In order to distinguish the time step $\Delta t_s$ to average fluxes from the following numerical time-marching step $\Delta t$, it should be pointed out that $\Delta t_s$ is only used to evaluate the coefficients before each term of the flux function in Eq.~\eqref{eq:fij_flux_coefficients}.

For unsteady flow evolution in a time step $\Delta t = t^{n+1} - t^n$, the discrete governing equations can be written as
\begin{equation}\label{eq:discrete_macro}
\frac{\vec{w}_i^{n+1} - \vec{w}_i^n}{\Delta t} + \frac{1}{V_i}\sum_{j \in N(i)}{\left[\epsilon \vec{F}_{ij}^{n+1} + (1-\epsilon)\vec{F}_{ij}^{n}\right] S_{ij}} = \vec{0},
\end{equation}
and
\begin{equation}\label{eq:discrete_micro}
\frac{f_i^{n+1} - f_i^n}{\Delta t} + \frac{1}{V_i} \sum_{j \in N(i)} { \left[\epsilon \mathcal{F}_{ij}^{n+1} + (1-\epsilon) \mathcal{F}_{ij}^{n}\right] S_{ij}} =
\epsilon \frac{g_i^{n+1} - f_i^{n+1}}{\tau_i^{n+1}} + (1-\epsilon) \frac{g_i^{n} - f_i^{n}}{\tau_i^{n}},
\end{equation}
where $\epsilon = 0.5$ gives a second-order accurate Crank--Nicolson (CN) scheme, and $\epsilon = 1$ corresponds to the first-order accurate backward Euler method.

Since the equilibrium state $g_i^{n+1}$ has one-to-one correspondence to $\vec{w}_i^{n+1}$ and depends on the distribution function $f_i^{n+1}$, Eqs.~\eqref{eq:discrete_macro} and \eqref{eq:discrete_micro} result in a coupled implicit nonlinear system.
It is very difficult to directly solve the large implicit system coupling macroscopic and microscopic governing equations, and the treatment of implicit equilibrium state $g^{n+1}$ in the collision term is important for the convergent efficiency  in the continuum regime.
In the implicit discrete velocity method \cite{yang1995rarefied}, $g^{n+1}$ is approximated by the explicit term $g^{n}$, which provides a simple way to solve the implicit kinetic equations.
The similar treatment is used in the implicit method of gas kinetic unified algorithm (GKUA) \cite{peng2016igkua} and the earlier implicit UGKS \cite{mao2015igkua}.
Mieussens \cite{mieussens2000discrete,mieussens2000implicit} pointed out that the convergence may slow down considerably if the gain term and loss term are evaluated at different time levels.
The treatment using a lagged equilibrium state will suffer from stiffness problem in the continuum regimes, which will deteriorate the convergence of implicit methods.
Mieussens adopted a linear mapping between the equilibrium state and the distribution function, i.e., $g^{n+1} = g^{n} + \mathcal{M}(f^{n+1} - f^n)$, and implemented a full implicit collision term.
However, $\mathcal{M}$ is a huge matrix with dimensions $N_v \times N_v$, where $N_v$ is the total number of discrete velocity points for each cell, which is unacceptable for practical computations.
Zhu et al. \cite{zhu2016iugks,zhu2016thesis} solved the Eqs.~\eqref{eq:discrete_macro} and \eqref{eq:discrete_micro} in an alternative way and used the most recent solved conservative variables to discretize the equilibrium state in the microscopic implicit equation.
Specifically, for an intermediate solution $\vec{w}_i^{(s)}$ and $f_i^{(s)}$ during the alternative solving process, the governing equation can be written as
\begin{equation}\label{eq:delta_macro}
\frac{1}{\Delta t}\Delta \vec{w}_i^{(s)}
+ \frac{\epsilon}{V_i} \sum_{j \in N(i)}{\Delta \vec{F}_{ij}^{(s)} S_{ij}}
= \vec{R}_i^{(s)},
\end{equation}
and
\begin{equation}\label{eq:delta_micro}
\left(\frac{\epsilon}{\tau_i^{(s+1)}} + \frac{1}{\Delta t}\right) \Delta  f_i^{(s)}
+ \frac{\epsilon}{V_i} \sum_{j \in N(i)} {\Delta \mathcal{F}_{ij}^{(s)} S_{ij}}
= r_i^{(s)},
\end{equation}
where the quantities in the $\Delta$ form is $\Delta Q^{(s)} = Q^{n+1} - Q^{(s)}$, denoting the correction of a specific variable $Q$.
The residuals on the right hand sides of Eqs.~\eqref{eq:delta_macro} and \eqref{eq:delta_micro} are
\begin{equation}\label{eq:residual_macro}
\vec{R}_i^{(s)} = \frac{\vec{w}_i^n - \vec{w}_i^{(s)}}{\Delta t} - \frac{\epsilon}{V_i}\sum_{j \in N(i)}{\vec{F}_{ij}^{(s)} S_{ij}} - \frac{1-\epsilon}{V_i}\sum_{j \in N(i)}{\vec{F}_{ij}^{n}S_{ij}},
\end{equation}
and
\begin{equation}\label{eq:residual_micro}
\begin{aligned}
r_i^{(s)} = \frac{f_i^n - f_i^{(s)}}{\Delta t}
&+ \epsilon \left[\frac{\tilde{g}_i^{(s+1)} - f_i^{(s)}}{\tau_i^{(s)}} - \frac{1}{V_i} \sum_{j \in N(i)} {\mathcal{F}_{ij}^{(s)} S_{ij}}\right] \\
&+ (1-\epsilon) \left[\frac{g_i^{n} - f_i^{n}}{\tau_i^{n}} - \frac{1}{V_i} \sum_{j \in N(i)} {\mathcal{F}_{ij}^{n} S_{ij}}\right].
\end{aligned}
\end{equation}
It should be noted that the fluxes in the residuals are computed the same as that in the explicit UGKS with a time step $\Delta t_s$ by Eqs.~\eqref{eq:fij_flux} and \eqref{eq:macro_flux_from_micro}.
During a numerical time-marching step $\Delta t = t^{n+1} - t^n$, several inner loops are carried out to alternatively solve the coupled implicit system, and the correction for a specific variable is $\Delta Q^{(s)} = Q^{(s+1)} - Q^{(s)}$.
$\tilde{g}_{i}^{(s+1)}$ in Eq.~\eqref{eq:residual_micro} is the equilibrium state computed from the most recently updated conservative variables $\Delta \vec{w}_i^{(s+1)}$ in the inner loops.
After several inner iterations, the flow variables will evolve from $t^n$ to $t^{n+1}$ by $Q^n + \sum_{s} \Delta Q^{(s)} = Q^{(s+1)} \to Q^{n+1}$.

In details, the implicit macroscopic flux on the left hand side of Eq.~\eqref{eq:delta_macro} is approximated by the first-order Euler flux \cite{luo1998fast,luo2001accurate},
\begin{equation}\label{eq:implicit_flux_macro}
\Delta \vec{F}_{ij}^{(s)} = \frac{1}{2} \left[\Delta \vec{T}_i^{(s)} + \Delta \vec{T}_j^{(s)} + \Gamma_{ij} \left(\Delta \vec{w}_i^{(s)} - \Delta \vec{w}_j^{(s)}\right)\right],
\end{equation}
where the Euler equation based flux $\vec{T}$ is
\begin{equation}
\vec{T} =
\begin{pmatrix}
\rho U_n\\
\rho U_n \vec{U} + p \vec{n}_{ij}\\
U_n (\rho E + p)
\end{pmatrix}.
\end{equation}
$U_n = \vec{U} \cdot \vec{n}_{ij}$ is the projected macroscopic velocity along the normal direction of cell interface $ij$.
$\Gamma_{ij}$ denotes the spectral radius of the Jacobian matrix of macroscopic fluxes.
For viscous flow, an additional term $\Gamma_{\nu}$ is required for stability consideration, and then $\Gamma_{ij}$ is computed by
\begin{equation}\label{eq:spectral_radius}
\Gamma_{ij} = (U_n + a_s) + \Gamma_{\nu} = (U_n + a_s)
+ \frac{2 \mu}{\rho |\vec{n}_{ij} \cdot (\vec{x}_j - \vec{x}_i)|},
\end{equation}
where $a_s$ is the speed of sound.
The implicit microscopic flux on the left hand side of Eq.~\eqref{eq:delta_micro} is approximated by first order upwind scheme, i.e.,
\begin{equation}\label{eq:implicit_flux_micro}
\Delta \mathcal{F}_{ij}^{(s)} = \frac{u_{n,ij}}{2}\left[1+{\text{sign}}(u_{n,ij})\right] \Delta f_{i}^{(s)} + \frac{u_{n,ij}}{2}\left[1-{\text{sign}}(u_{n,ij})\right]  \Delta f_{j}^{(s)}.
\end{equation}
Substituting Eqs.~\eqref{eq:implicit_flux_macro} and \eqref{eq:implicit_flux_micro} into Eqs.~\eqref{eq:delta_macro} and \eqref{eq:delta_micro}, the implicit governing equations for inner iterations become
\begin{equation}\label{eq:implicit_macro}
\left(\frac{1}{\Delta t} + \frac{\epsilon}{2 V_i} \sum_{j \in N(i)}{\Gamma_{ij} S_{ij}}\right) \Delta \vec{w}_i^{(s)}
+ \frac{\epsilon}{2 V_i} \sum_{j \in N(i)}{\left(\Delta \vec{T}_{j}^{(s)} - \Gamma_{ij} \Delta \vec{w}_{j}^{(s)}\right) S_{ij}}
= \vec{R}_i^{(s)},
\end{equation}
and
\begin{equation}\label{eq:implicit_micro}
D_i \Delta  f_i^{(s)}
+ \sum_{j \in N(i)} {D_j \Delta  f_j^{(s)}}
= r_i^{(s)},
\end{equation}
where
\begin{equation}
\begin{aligned}
D_i &= \frac{\epsilon}{\tau_i^{(s+1)}} + \frac{1}{\Delta t} + \frac{\epsilon}{2 V_i} \sum_{j \in N(i)} {u_{n,ij} S_{ij} \left[1+{ \text{sign}}(u_{n,ij})\right]}, \\
D_j &= \frac{\epsilon}{2 V_i} u_{n,ij} S_{ij} \left[1-{\text{sign}}(u_{n,ij})\right],
\end{aligned}
\end{equation}
and the variation of Euler flux adopts
\begin{equation}
\Delta \vec{T}_j^{(s)} = \vec{T}(\vec{w}_j^{(s)} + \Delta \vec{w}_j^{(s)}) - \vec{T}(\vec{w}_j^{(s)}),
\end{equation}
to avoid the computation of a Jacobian matrix \cite{sharov1997reordering,luo1998fast}.

The implicit systems formed by Eq.~\eqref{eq:implicit_macro} and Eq.~\eqref{eq:implicit_micro} can be solved in sequence by the numerical algorithms commonly used in CFD computations, such as the lower--upper symmetric Gauss--Seidel (LU-SGS) method \cite{yoon1986multigrid,jameson1987lower,yoon1988lower} and generalized minimal residual (GMRES) method \cite{saad1986gmres}.
In the implicit UGKS (IUGKS), the LU-SGS method or the point relaxation with two spatial sweepings \cite{rogers1995comparison,yuan2002comparison} are usually adopted.

\subsubsection{Boundary condition}

For the IUGKS, there are two parts of computations that require boundary treatment:
(a) the time-averaged fluxes in the residuals on the right hand side of the governing equations;
(b) the correction of flow variables during inner iterations on the left hand side of governing equations.

Since the time averaged fluxes are computed exactly the same as that in the explicit scheme, the boundary treatment becomes the same as that in the explicit scheme, which has been described in Section \ref{sec:ugks_bc}.
While for the correction of flow variables, the boundary condition is implemented by linearization of the explicit boundary condition (\ref{eq:ghost_state_macro}) and (\ref{eq:ghost_state_micro}), i.e.,
\begin{equation}
\Delta \vec{w}_j = \frac{\partial \mathcal{B}(\vec{w}_i, \vec{w}_{bc})}{\partial \vec{w}_i} \Delta \vec{w}_i,
\end{equation}
and
\begin{equation}\label{eq:ghost_state_delta}
\Delta f_j = \frac{\partial g_M(\vec{w}_j)}{\partial \vec{w}_j} \Delta \vec{w}_j.
\end{equation}

For supersonic inlet or far flow field, we have ${\partial g_M(\vec{w}_j)}/{\partial \vec{w}_j} = \vec{0}$ and then $\Delta f_j = 0$, while for subsonic inlet, the correction of distribution function can be set according to $\vec{w}_j$ and $\Delta \vec{w}_j$ in the ghost cell during LU-SGS sweeping.
For the isotermal walls, we have
\begin{equation}
\Delta f_j = \Delta \rho_j \hat{g}_w,
\end{equation}
where the density variation in the ghost cell is computed by the change of inner cell state
\begin{equation}
\Delta \rho_j \int_{u_{n,ij} < 0}{u_{n,ij} \hat{g}_w  d\vec{u}}
= -\sum_{u_{n,ij,k}>0}{u_{n,ij,k}\Delta f_{i,k} \mathcal{V}_k}.
\end{equation}
Straightforwardly, the change of the distribution function in the ghost cell can be set as
$\Delta f_j(u,v,w) = \Delta f_i(-u,v,w)$ and $\Delta f_j(u,v,w) = \Delta f_i(u,v,w)$
for symmetric boundary and extrapolation outlet, respectively.

\subsubsection{Numerical tests}

\noindent
\textbf{(a) Advection of sine wave}

\begin{figure}[htbp]
\centering
\includegraphics[width=0.45\textwidth]{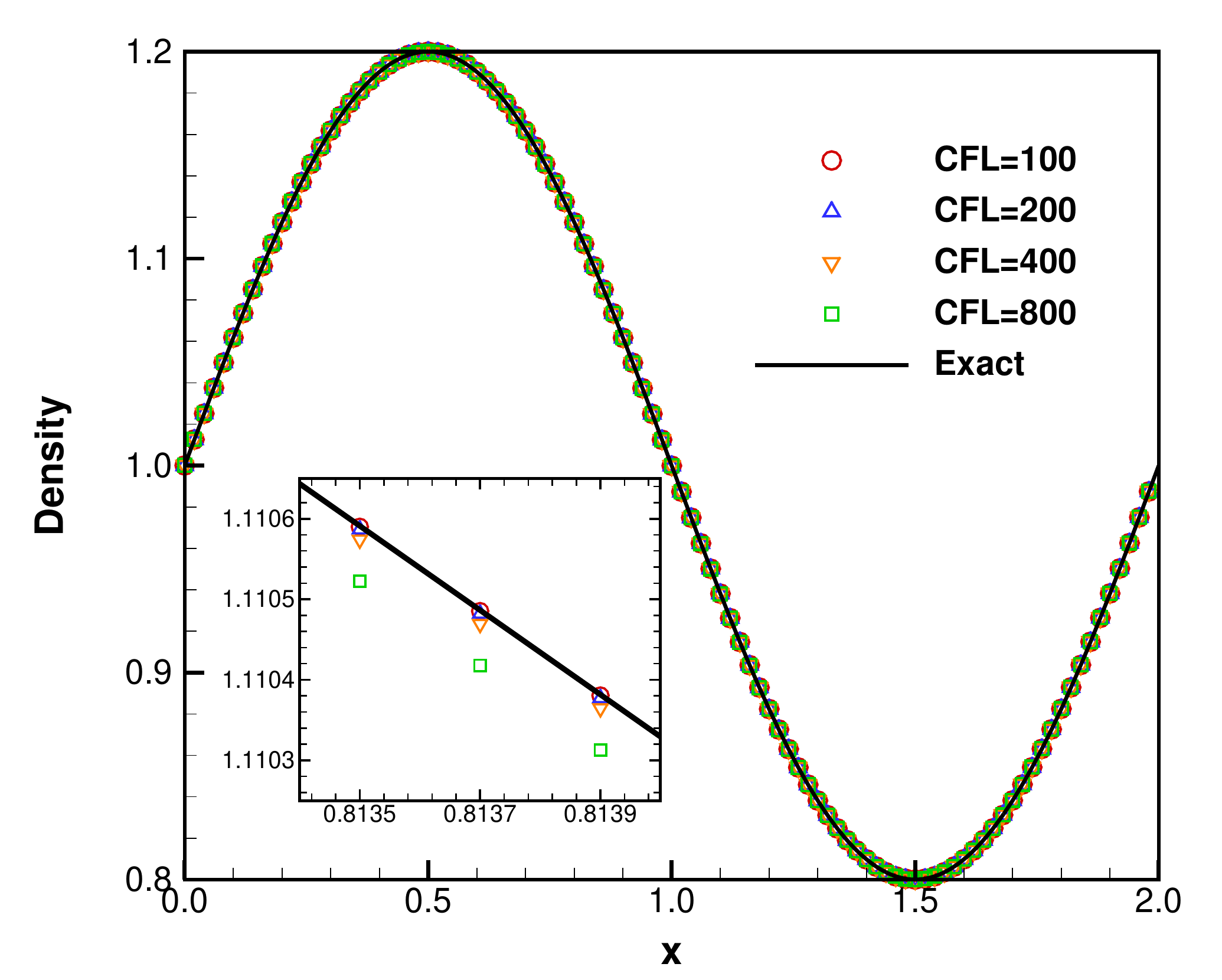}
\caption{\label{fig:sineDensity}Density distribution at time $t=2$ obtained by the IUGKS with different time steps \cite{zhu2019iugks}.}
\end{figure}

For unsteady flow simulations, the one-dimensional case of advection of density perturbation is employed to test the temporal accuracy of the IUGKS.
The initial condition is set as
\begin{equation}
\rho (x) = 1+ 0.2 \sin (\pi x), \quad U(x)=1, \quad p(x) = 1, \quad x \in \left[0,2\right].
\end{equation}
The periodic boundary condition is implemented and it gives an analytic solution
\begin{equation}
	\rho_e(x, t) = 1 + 0.2 \sin \left(\pi(x-t)\right).
\end{equation}

In the Euler limit, the UGKS is supposed to have an error on the order of $\mathcal{O} (\Delta x^2, \Delta t^2)$.
A very fine uniform mesh with $10000$ cells and relatively large time steps are used to capture the convergence order of the IUGKS with respect to time step.
A small mean collision time $\tau = 0.001 \Delta t_s$ is employed to drive the IUGKS to get a continuum inviscid solution.

\begin{figure}[htbp]
\centering
\subfloat[\label{fig:sineNormSecond}]
{\includegraphics[width=0.4\textwidth]{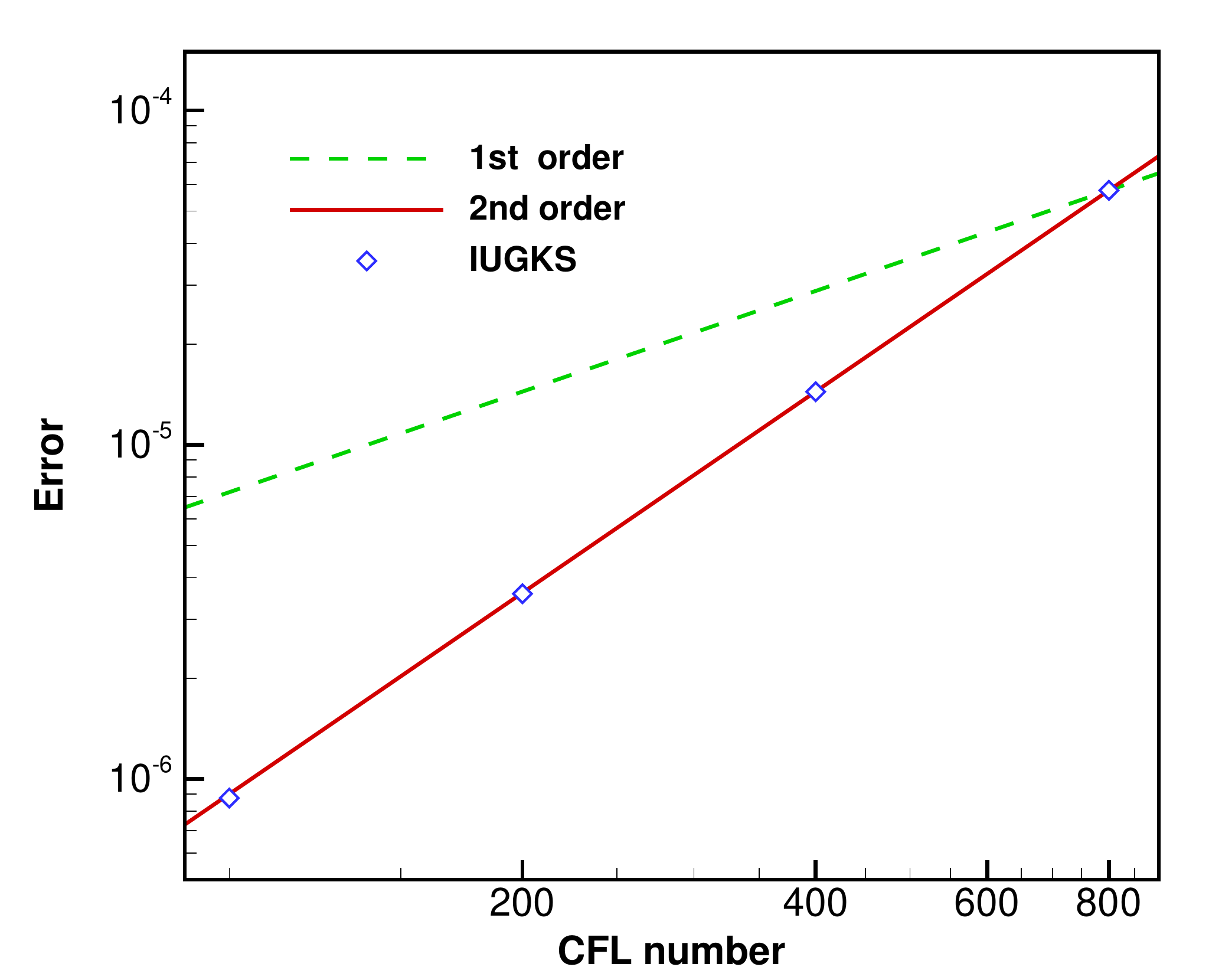}}
\subfloat[\label{fig:sineNormFirst}]
{\includegraphics[width=0.4\textwidth]{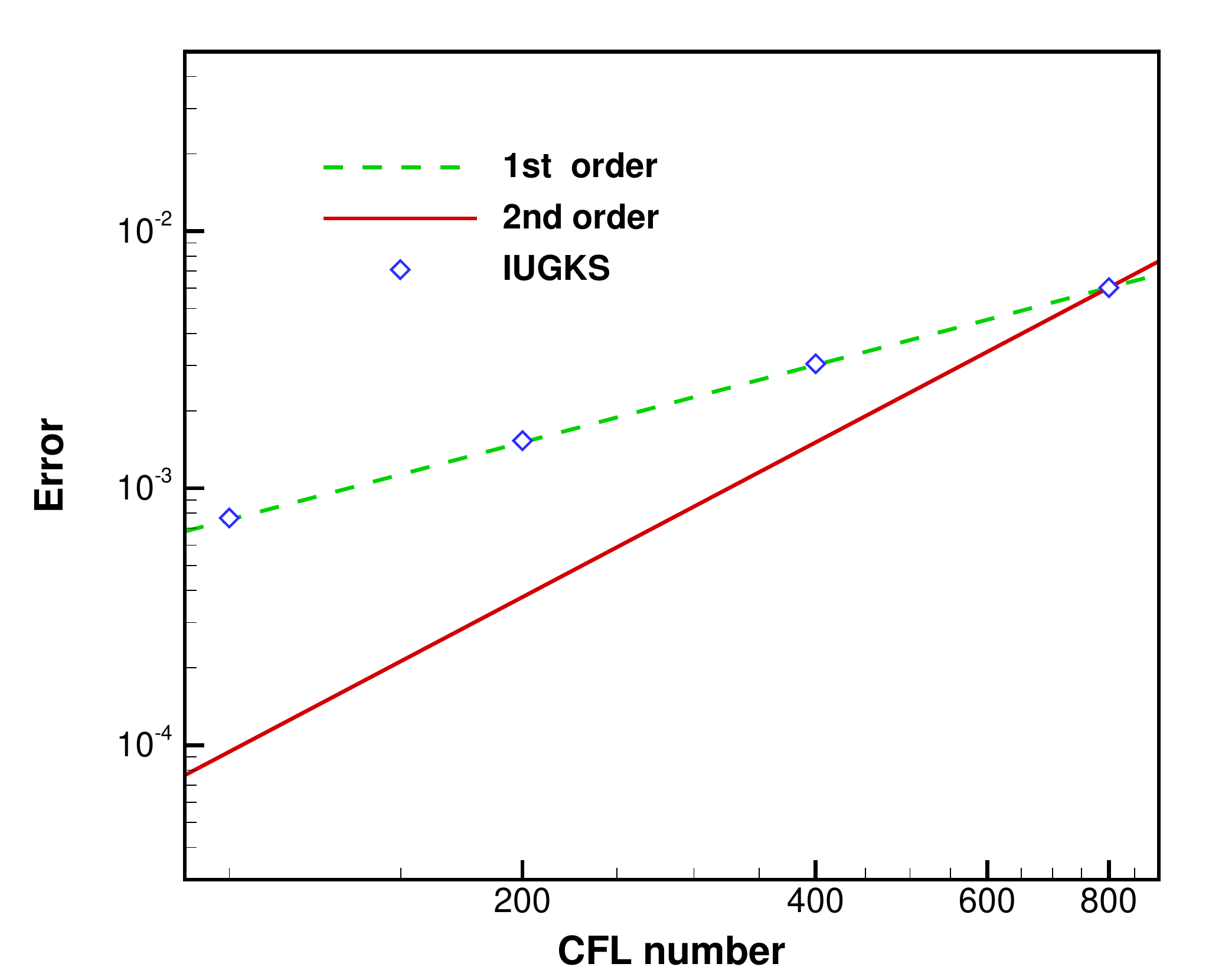}}
\caption{\label{fig:sineNorm}Temporal accuracy of the IUGKS with (a) $\epsilon = 0.5$ and (b) $\epsilon = 0.75$ \cite{zhu2019iugks}.}
\end{figure}

\begin{figure}[htbp]
\centering
\subfloat[\label{fig:sineResA}]
{\includegraphics[width=0.4\textwidth]{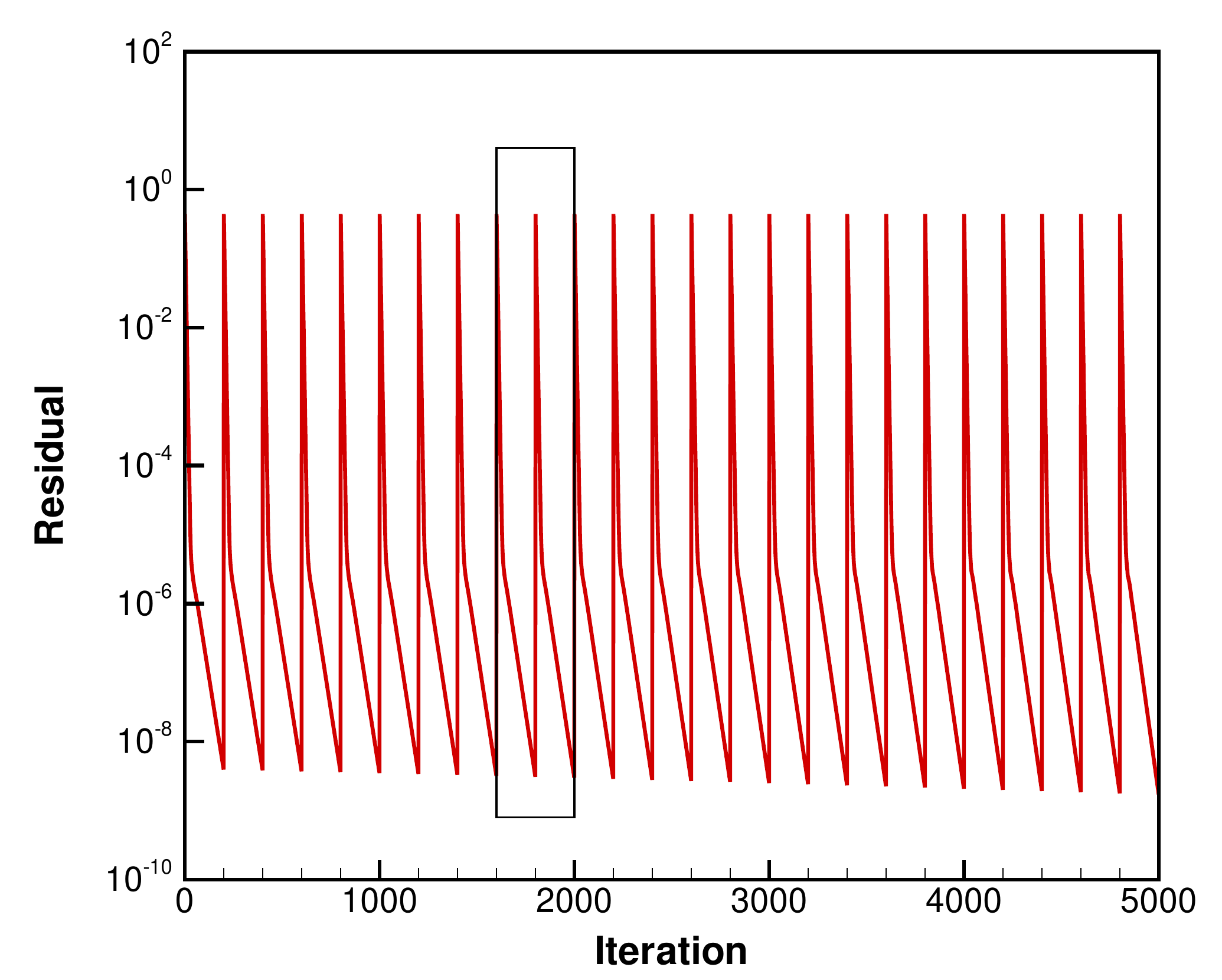}}
\subfloat[\label{fig:sineResB}]
{\includegraphics[width=0.4\textwidth]{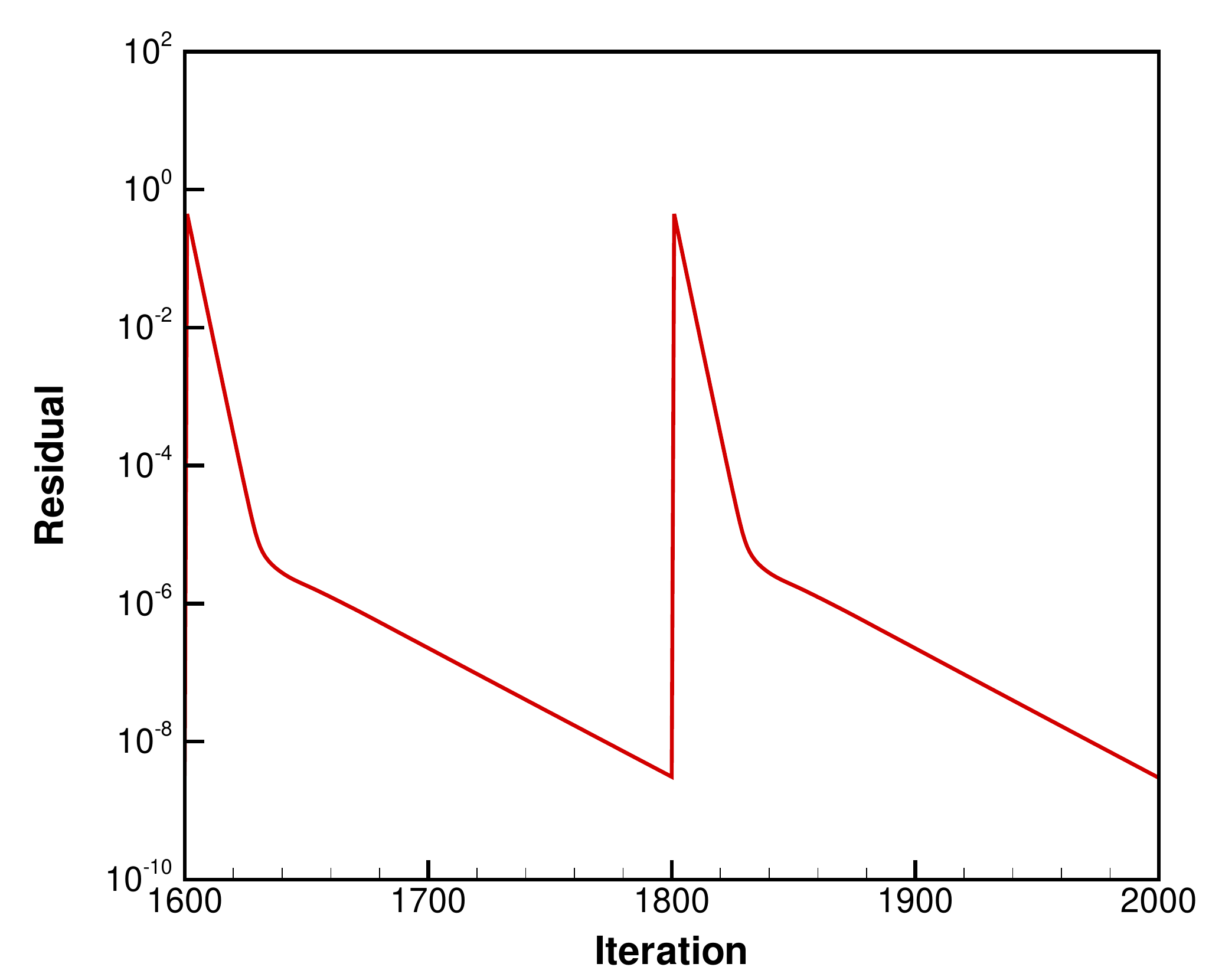}}
\caption{\label{fig:sineRes}Convergence history of the unsteady flow of sine wave propagation for the case at ${\text{CFL}}=400$ with $\epsilon=0.5$ \cite{zhu2019iugks}. (a) Residual variation with respect to total iteration steps, (b) local enlargement for two evolution steps. }
\end{figure}

Figure \ref{fig:sineDensity} shows the density distribution at time $t=2$ obtained by the IUGKS with different time steps.
The errors for different time step cases are evaluated to measure the temporal accuracy of the IUGKS.
From Fig.~\subref*{fig:sineNormSecond}, it shows that the C-N scheme with $\epsilon=0.5$ achieves second order temporal accuracy.
The accuracy of IUGKS with a temporal discretization of $\epsilon =0.75$ has also been tested, see in Fig.~\subref*{fig:sineNormFirst}.
As expected, the IUGKS for the case with $\epsilon > 0.5$ has a first-order accuracy in time.

\begin{figure}[htbp]
\centering
\subfloat[]
{\includegraphics[width=0.32\textwidth]{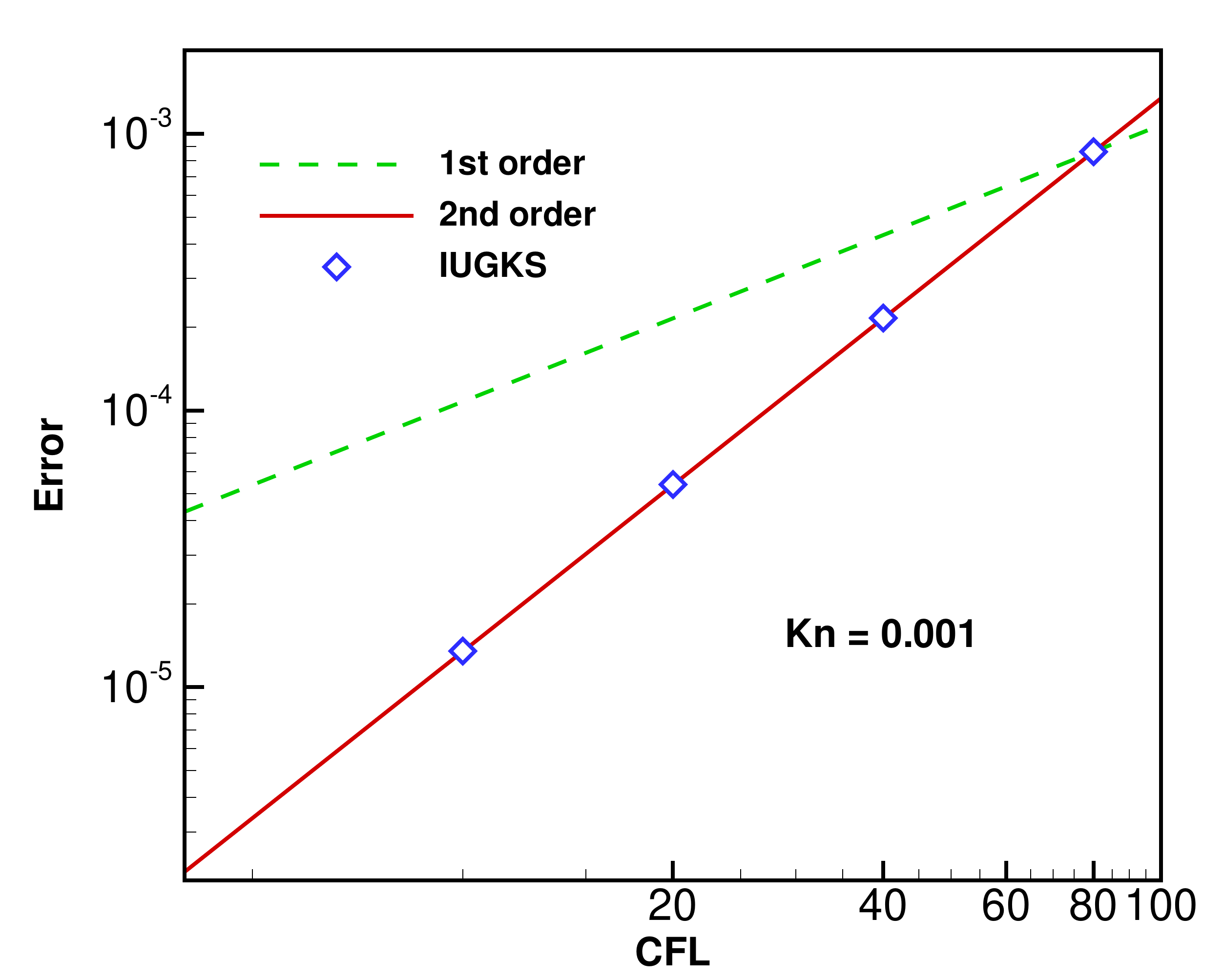}}
\subfloat[]
{\includegraphics[width=0.32\textwidth]{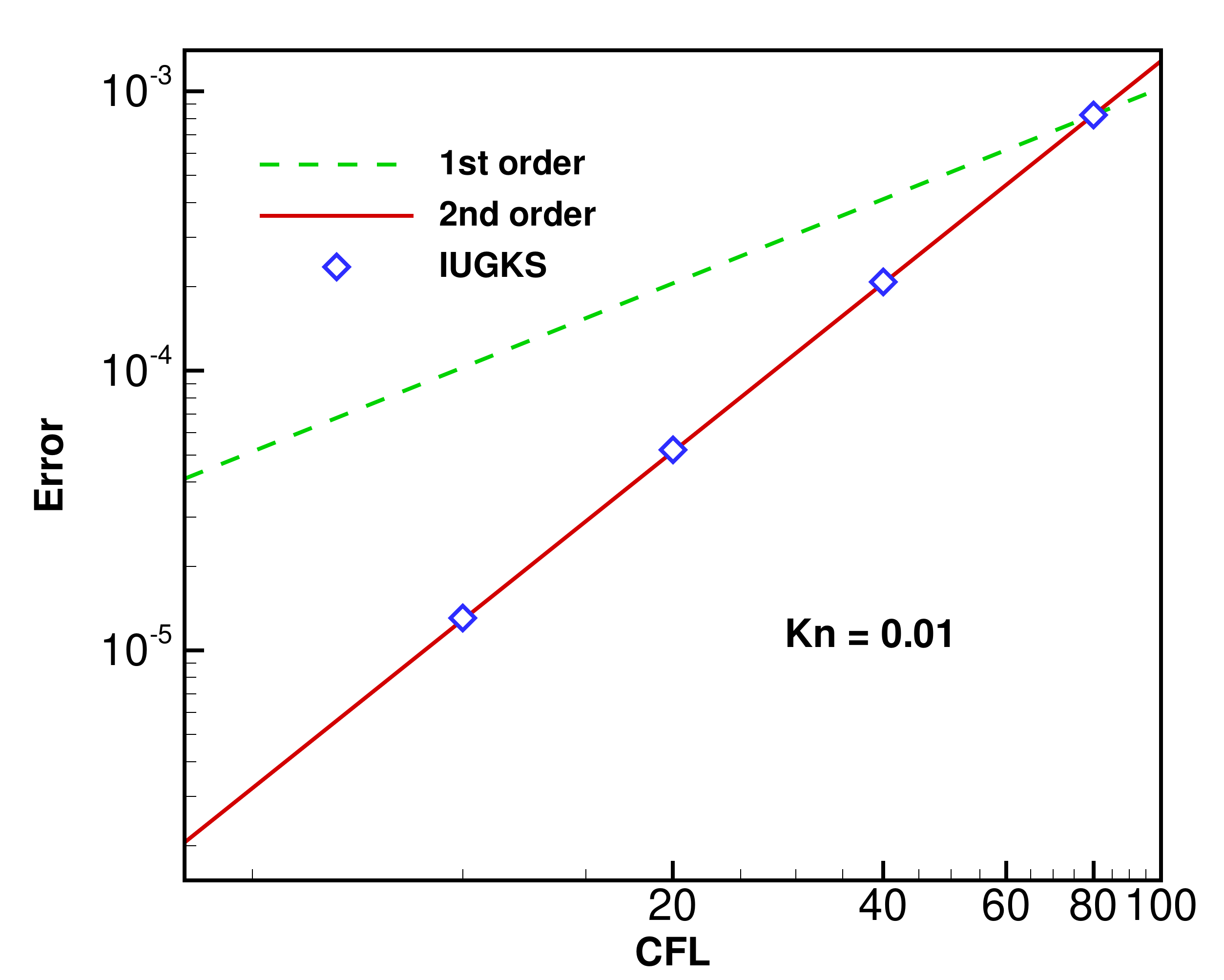}}\\
\subfloat[]
{\includegraphics[width=0.32\textwidth]{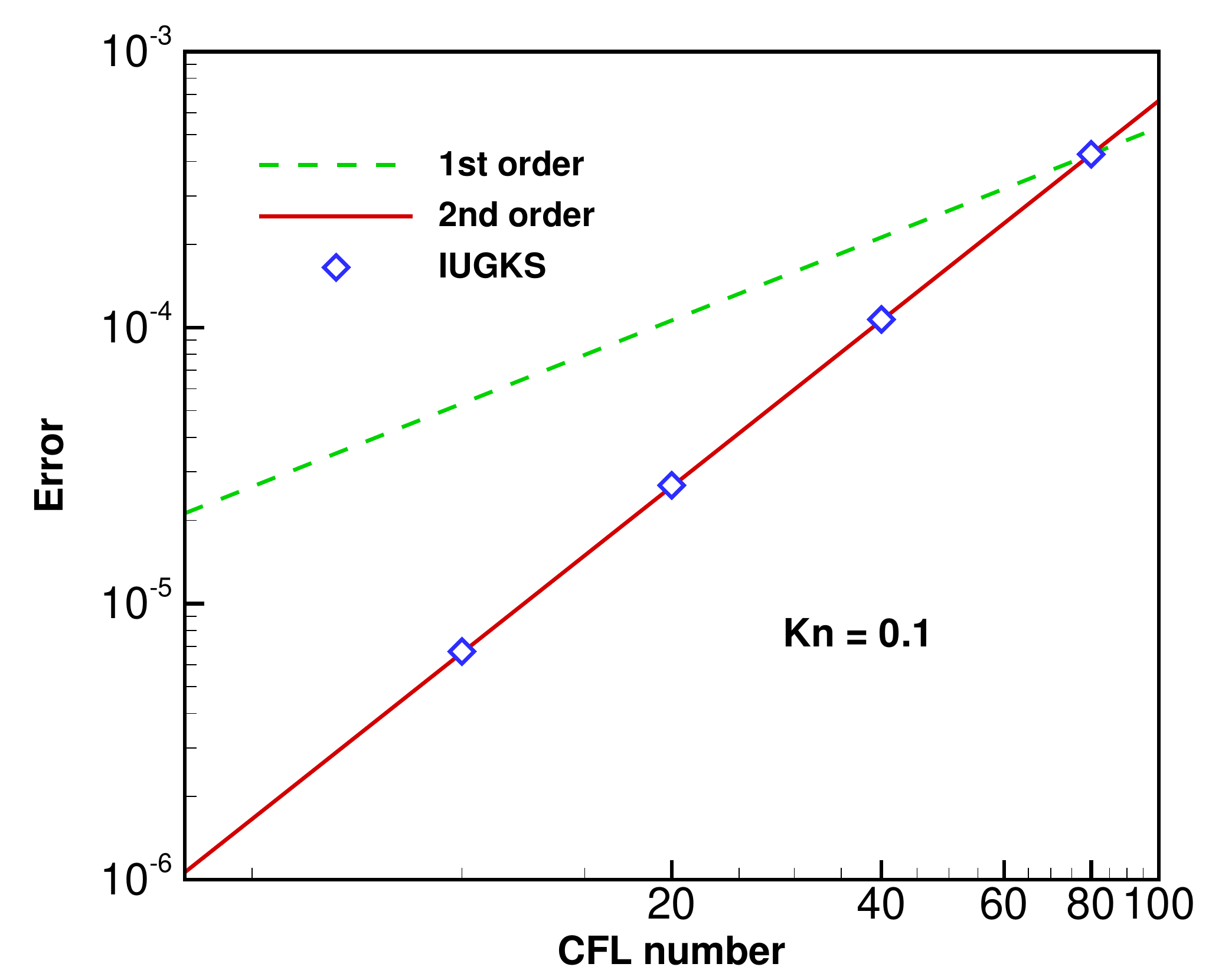}}
\subfloat[]
{\includegraphics[width=0.32\textwidth]{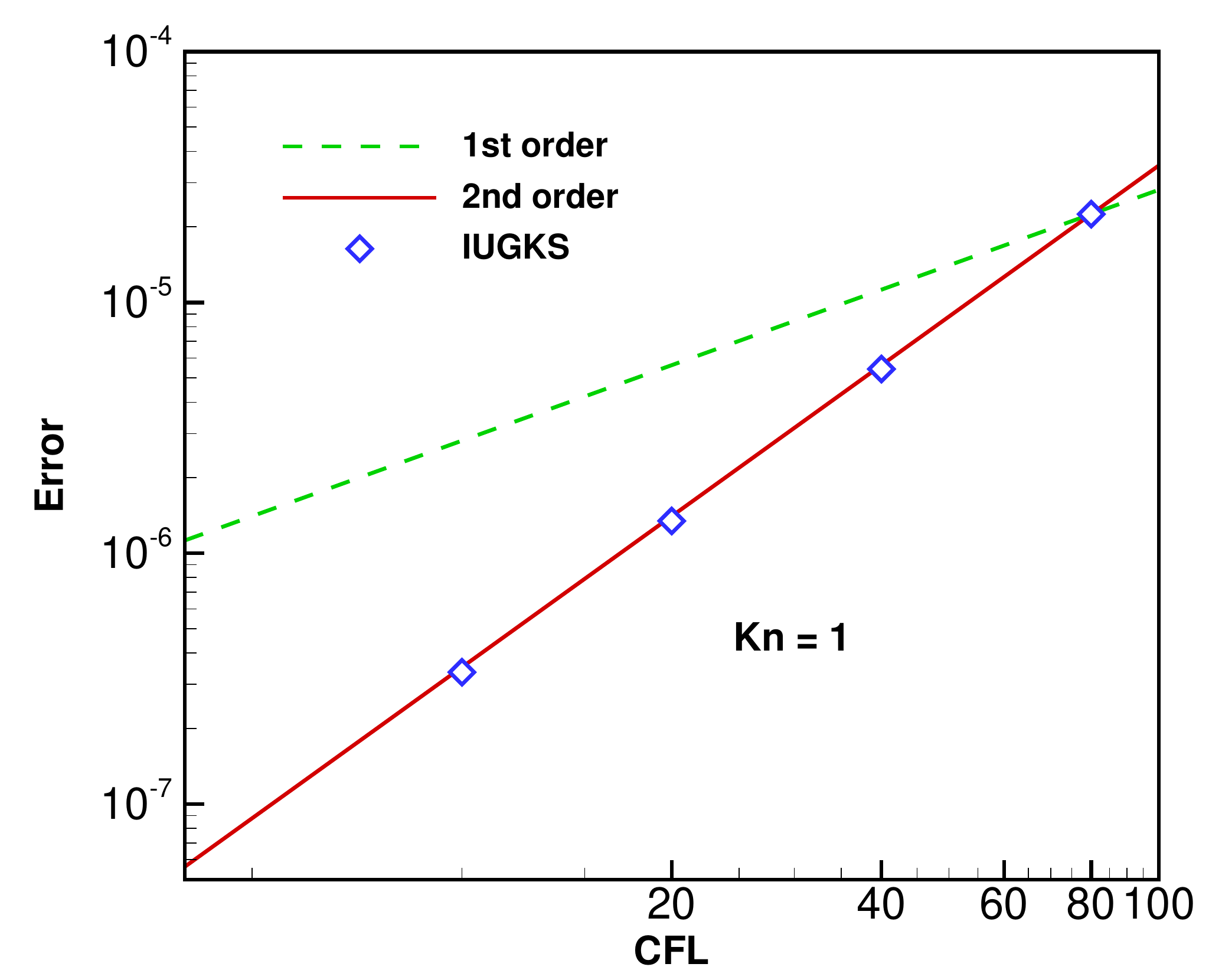}}
\subfloat[]
{\includegraphics[width=0.32\textwidth]{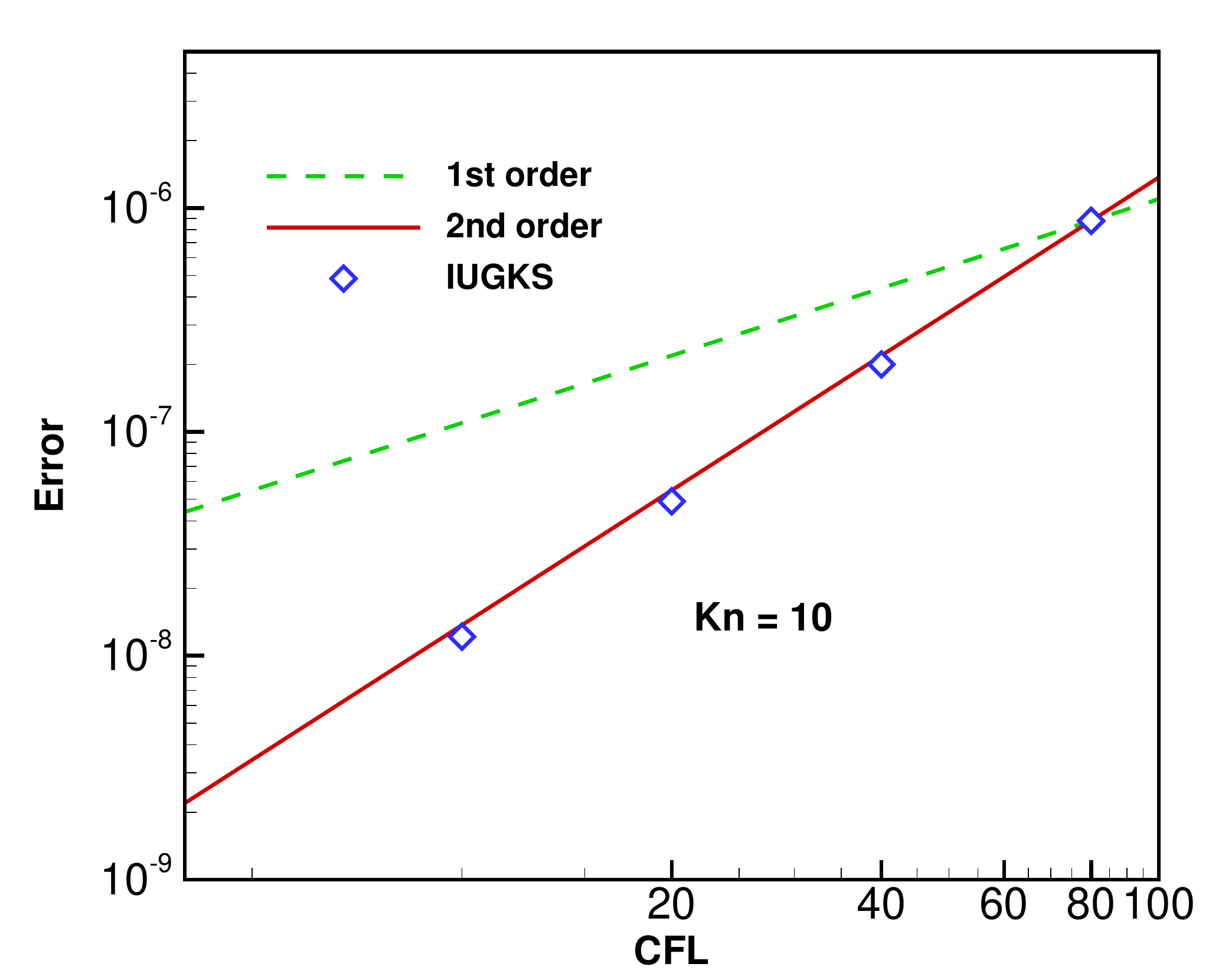}}
\caption{\label{fig:sineAccuracyRarefied}Accuracy test of the IUGKS in different flow regimes \cite{zhu2019iugks}.
(a) ${ Kn} = 0.001$, (b) ${ Kn} = 0.01$, (c) ${ Kn} = 0.1$,(d) ${ Kn} = 1$, and (e) ${ Kn} = 10$.}
\end{figure}

The convergence history of the unsteady flow of sine wave propagation is shown in Fig.~\ref{fig:sineRes}.
For this case, the maximum inner iteration is set as $200$.
As shown in Fig.~\subref*{fig:sineResA}, for each evolution step the residual of the unsteady governing equation can be reduced to the order of $10^{-8}$, and the error of flow variable change for each evolution step is about $10^{-12}$.
Usually, there is no need to constrain the residual of each evolution step to such a small value.
Considering the temporal accuracy of numerical scheme, a residual of $\mathcal{O}(\Delta t^m)$ for each step is sufficient, where $m$ is the order of temporal accuracy.
In practice, two or three orders of residual reduction is acceptable \cite{tan2017time}.
The details of the converge history in two time-marching steps are enlarged in the Fig.~\subref*{fig:sineResB}, where it gives a typical convergence curve.
Specifically, the high-frequency error can be more efficiently eliminated by the iterations than the low-frequency error.
After the high-frequency error is mainly eliminated, the convergence rate will decrease and more iterations are required for low-frequency error.
The multigrid method \cite{zhu2017multigrid} could be adopted to further improve the convergence property for a higher efficiency.

In addition, based on this periodic case, the accuracy of the IUGKS has been investigated in different flow regimes.
The time accuracy test is conducted for the cases at ${ Kn} = 0.001$, $0.01$, $0.1$, $1$ and $10$ on a uniform mesh with $400$ cells.
The residuals are plotted in Fig.~\ref{fig:sineAccuracyRarefied} with respect to the time step. The IUGKS achieves second order accuracy for all cases from continuum to free molecular flows.

\noindent
\textbf{(b) Rayleigh flow}

The Rayleigh flow is an unsteady gas flow around a vertical plate with infinite length.
Initially, the argon gas with molecular mass $m_0 = 6.63 \times 10^{-26} { kg}$ is stationary and has a temperature of $273{ K}$, and suddenly the plate obtains a constant vertical velocity of $10{ m/s}$ with a higher temperature of $373{ K}$.
The computational domain is $1 { m}$ long, which is the characteristic length to define the Knudsen number by the VHS model.

\begin{figure}[htbp]
\centering
\subfloat[]{\includegraphics[width=0.4\textwidth]{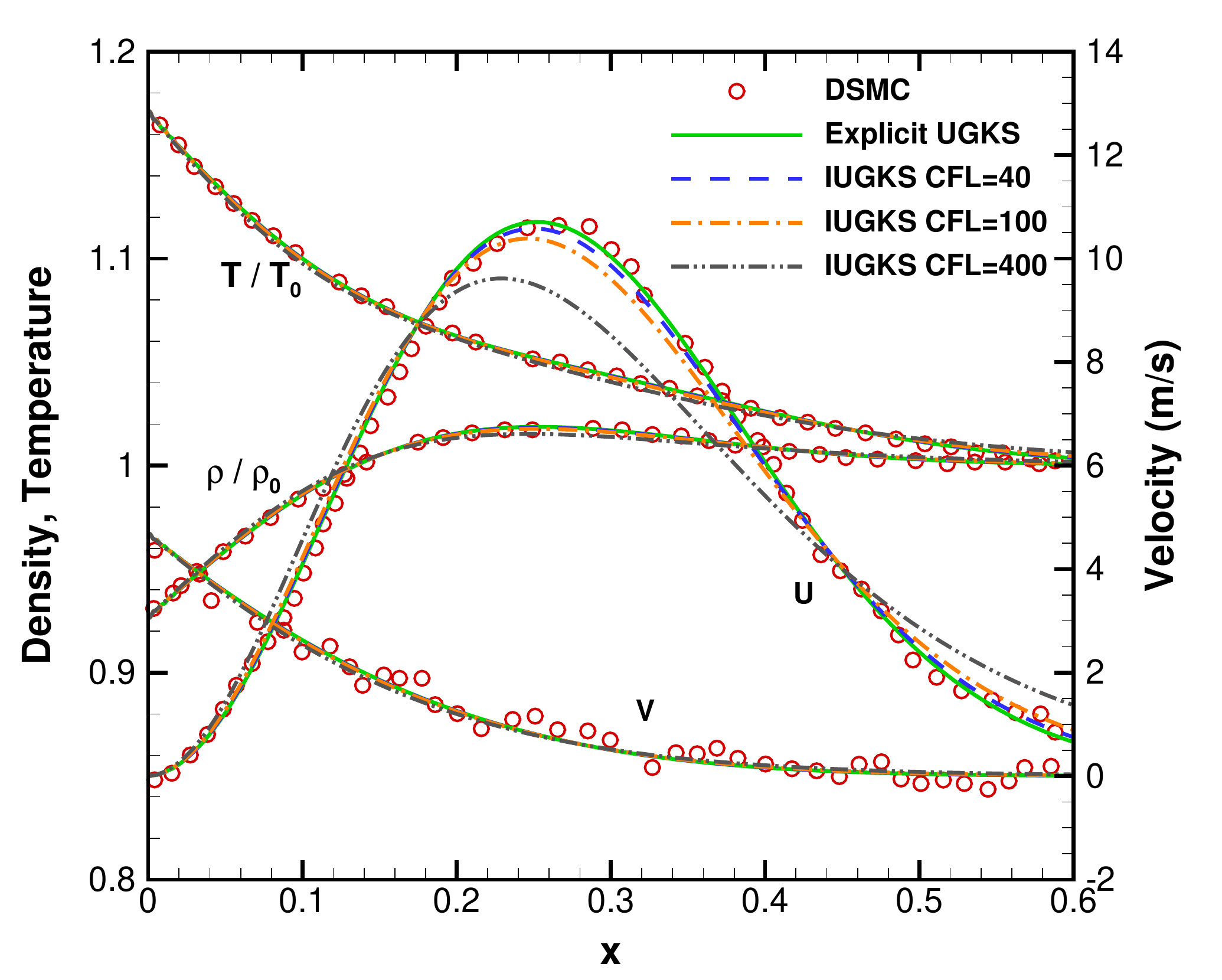}} \subfloat[]{\includegraphics[width=0.4\textwidth]{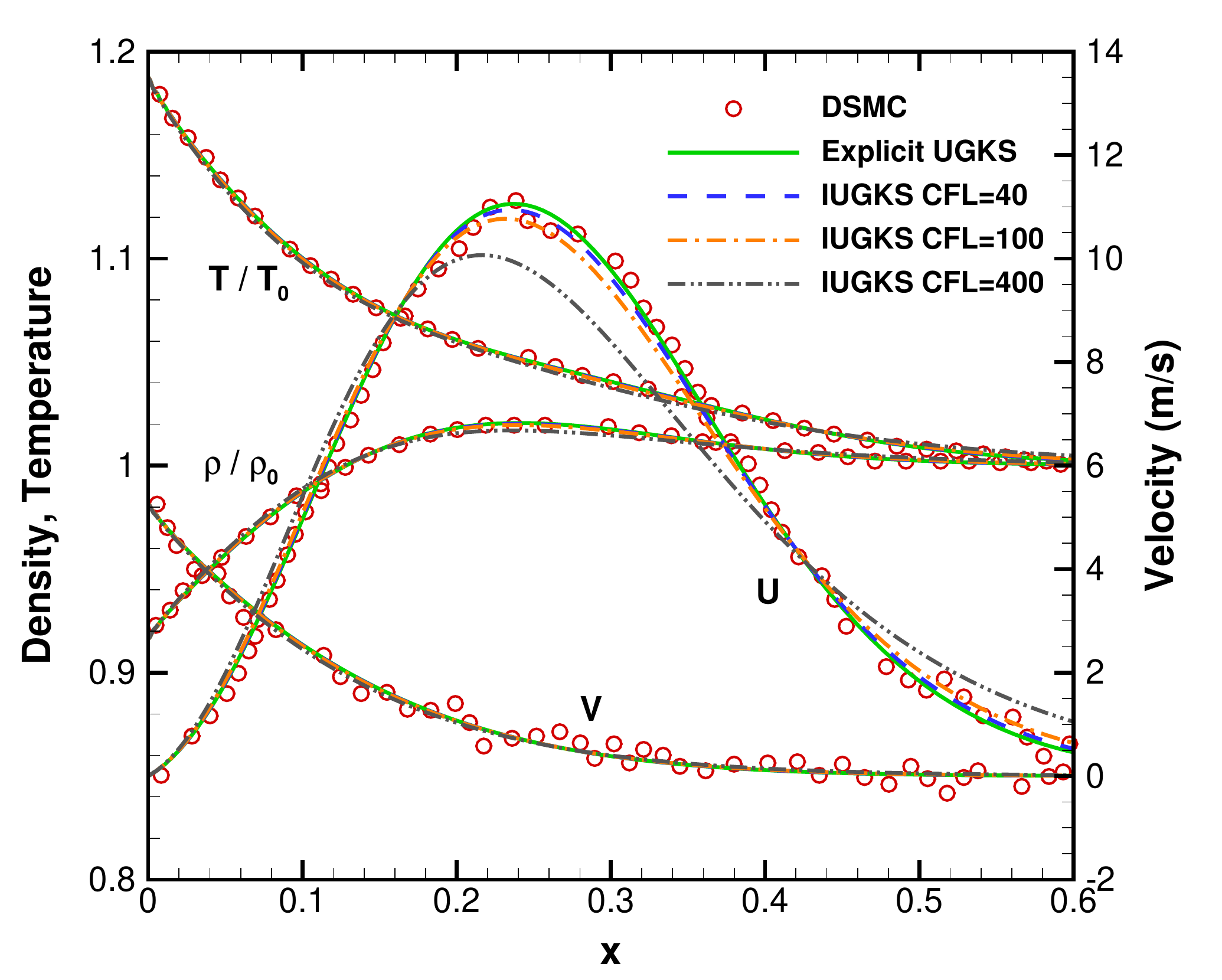}} \\
\subfloat[]{\includegraphics[width=0.4\textwidth]{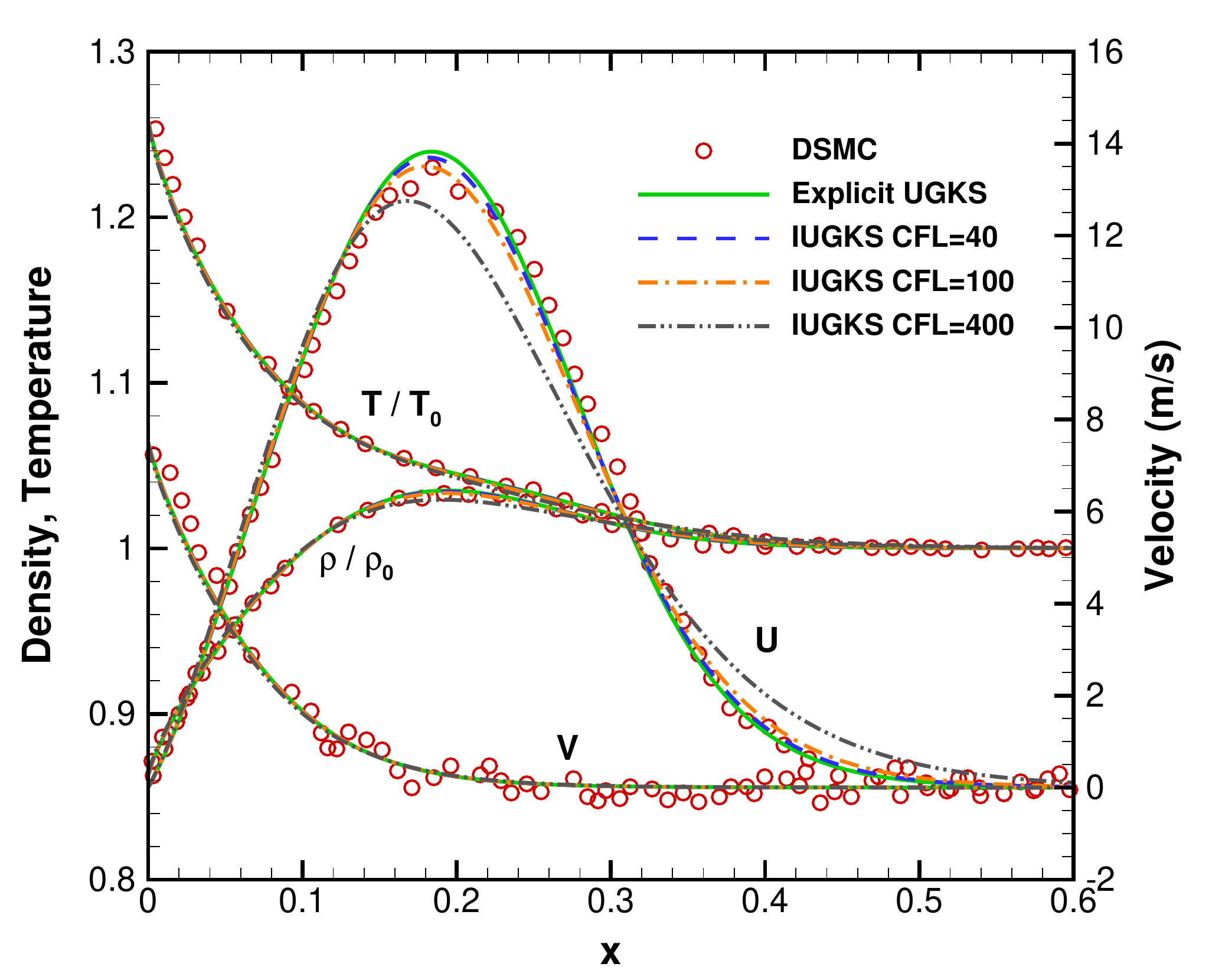}}
\subfloat[]{\includegraphics[width=0.4\textwidth]{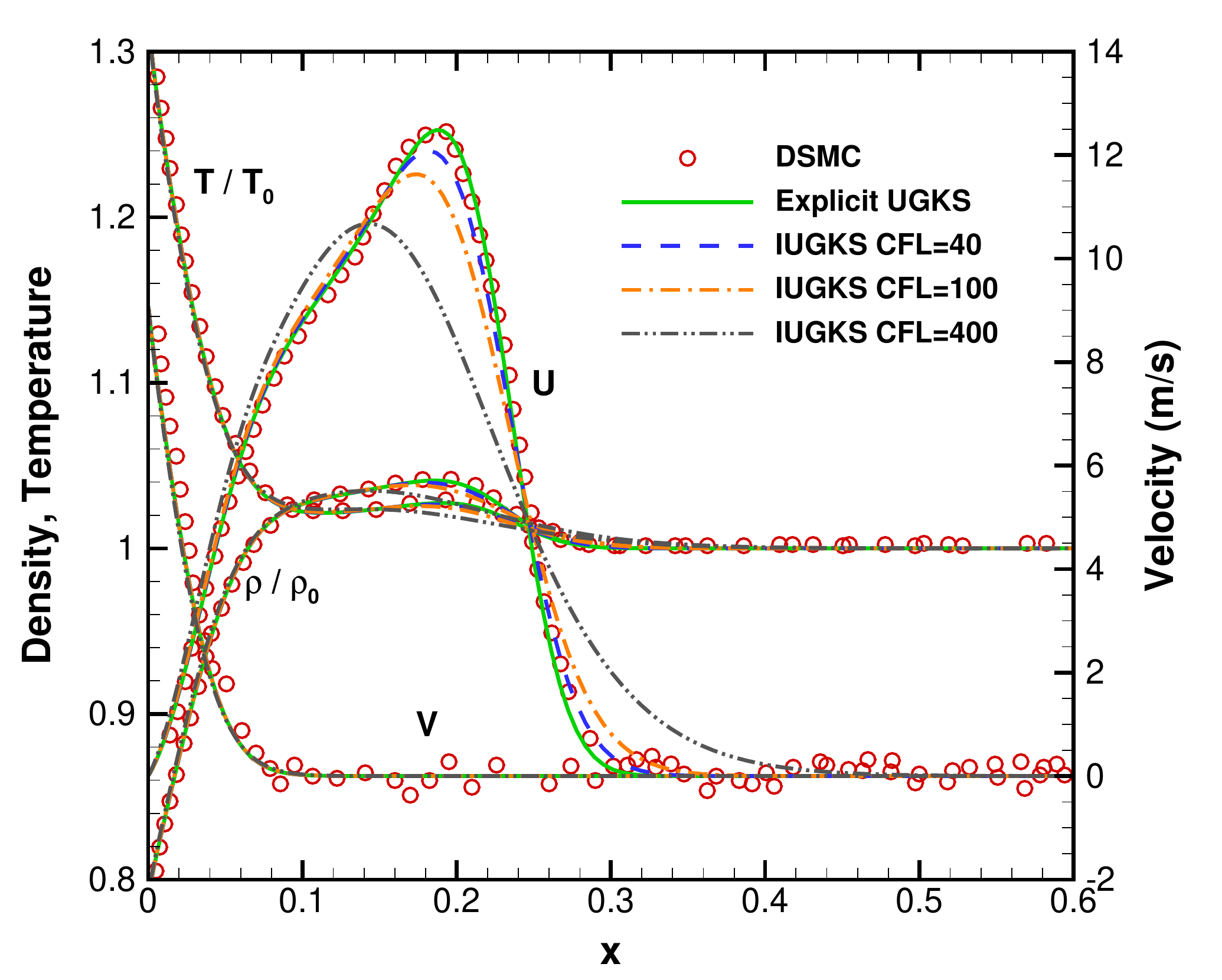}}
\caption{\label{fig:RayleighResults}Rayleigh problems at different Knudsen numbers \cite{zhu2019iugks}.
(a) ${ Kn}=2.66$, (b) ${ Kn}=0.266$, (c) ${ Kn}=0.0266$, and (d) ${ Kn} = 0.00266$.
 }
\end{figure}

The target of the IUGKS is to release the time step restriction from small-size cells on non-uniform mesh and thus to accelerate overall computational efficiency.
For this test, non-uniform mesh with the minimum cell size of $0.0005{ m}$ near the plate is used.
The results at time $t=0.7{ ms}$ for different Knudsen numbers are plotted in Fig.~\ref{fig:RayleighResults}, where the density and temperature are normalized by $\rho_0$ and $T_0$.
In comparison with the DSMC results obtained from the reference paper \cite{huang2013micro} and those from the explicit UGKS simulation, the IUGKS can give satisfactory results with a relatively large time step.
The collisionless limit requires the interval of discrete velocities to satisfy $\Delta u \sim \Delta x / t$ due to the ray effect \cite{zhu2020ray}.
In order to get smooth solutions, $600$ velocity points uniformly in $u$-direction and 100 points in $v$-direction are used to cover a range from $-2023{ m/s}$ to $2023{ m/s}$. Extra more discrete points have been employed for discretization of the physical and velocity space in this case so that it gives more reliable results in the efficiency testing due to long program running time.
The computational cost is provided in Table.~\ref{tab:RayleighEfficiency} at different Knudsen numbers with various time steps.
Generally more inner iterations are required in the small Knudsen number case.
The increase of computational efficiency for near continuum flows is not as much as that in rarefied cases, but the IUGKS with CFL=40  is still about ten times faster than the explicit scheme.

\begin{table}[htbp]
\centering
\begin{threeparttable}[c]
\caption{\label{tab:RayleighEfficiency}Computational cost for Rayleigh problem with different time steps.}
\begin{tabular*}{0.98\textwidth}{@{\extracolsep{\fill}}llcrrrr}
\toprule
& \multirow{2}{*}{Kn} & Explicit UGKS & \multicolumn{4}{c}{IUGKS with $\epsilon = 0.75$}\\\cline{4-7}
& ~    & CFL=0.5   & 40    & 100   & 400   & 1200\\
\midrule
\multirow{4}{*}{\begin{tabular}{c} CPU time \\(min)\end{tabular}}
 &2.66      & 437.8  & 22.5 & 10.3  & 4.0  & 1.9\\
~&0.266     & 435.0  & 25.9 & 11.5  & 4.8  & 1.8\\
~&0.0266    & 436.7  & 31.0 & 14.3  & 7.1  & 3.3\\
~&0.00266   & 439.8  & 48.2 & 28.2  & 13.7  & 11.1\\
\midrule
\multirow{4}{*}{Speedup}
 &2.66      & 1.0     & 19.5  & 42.3 & 108.7 & 236.6\\
~&0.266     & 1.0     & 16.8  & 37.7 & 91.0 & 236.8\\
~&0.0266    & 1.0     & 14.1  & 30.6 & 61.1 & 132.4\\
~&0.00266   & 1.0     & 9.1  & 15.6 & 32.1 & 39.5\\
\bottomrule
\end{tabular*}
\end{threeparttable}
\end{table}

\subsection{Implicit UGKS for steady flow}
The implicit UGKS for steady flow can be regarded as a simplified scheme for the unsteady flow, where temporal accuracy can be ignored and the time discretization uses $\epsilon = 1$ for a backward Euler scheme.

The governing equations for implicit iteration become
\begin{equation}\label{eq:delta_macro_siugks}
\frac{1}{\Delta t}\Delta \vec{w}_i^{n}
+ \frac{1}{V_i} \sum_{j \in N(i)}{\Delta \vec{F}_{ij}^{n} S_{ij}}
= \vec{R}_i^{n},
\end{equation}
and
\begin{equation}\label{eq:delta_micro_siugks}
\left(\frac{1}{\tau_i^{n+1}} + \frac{1}{\Delta t}\right) \Delta  f_i^{n}
+ \frac{1}{V_i} \sum_{j \in N(i)} {\Delta \mathcal{F}_{ij}^{n} S_{ij}}
= r_i^{n},
\end{equation}
where the quantities in the $\Delta$ form denotes $\Delta Q^{n} = Q^{n+1} - Q^{n}$, which is the correction of a specific variable $Q$.
The residuals in the right hand sides of Eqs.~\eqref{eq:delta_macro_siugks} and \eqref{eq:delta_micro_siugks} are
\begin{equation}\label{eq:residual_macro_siugks}
\vec{R}_i^{n} = - \frac{1}{V_i}\sum_{j \in N(i)}{\vec{F}_{ij}^{n}S_{ij}},
\end{equation}
and
\begin{equation}\label{eq:residual_micro_siugks}
r_i^{n} = \frac{\tilde{g}_i^{n+1} - f_i^{n}}{\tau_i^{n}} - \frac{1}{V_i} \sum_{j \in N(i)} {\mathcal{F}_{ij}^{n} S_{ij}}.
\end{equation}
$\Delta t$ is a pseudo time step only for stability consideration.

In the study of Zhu et al. \cite{zhu2016iugks}, the Couette flow, lid-driven cavity flow and high-speed flow around a circular cylinder are calculated to validate the computational efficiency of the IUGK for steady state solutions.
It can be found that the IUGKS is about one or two order of magnitudes faster than the explicit UGKS.

In order to overcome the stiffness problem in the continuum flow, the macroscopic equations are solved with the implicit Euler flux to drive the convergence of the IUGKS, see Eqs.~\eqref{eq:residual_macro} and \eqref{eq:implicit_flux_macro}.
It makes the IUGKS efficient in the low Knudsen number cases.
In order to further improve the performance of the IUGKS for viscous flows, Yuan et al. \cite{yuan2021multi} take into account the NS terms during the macroscopic iterations.
Since the NS equations are invalid in the rarefied regimes, a limiting factor is employed to constrain the viscous terms.
In the continuum flows, due to that an NS solution is predicted by the macroscopic equations, the multiscale implicit scheme could be more efficient for low Knudsen number cases.
Similar idea to drive the convergence in the continuum regimes can be found in the development of the general synthetic iterative scheme (GSIS) \cite{su2020can,su2020fast}.
From the study of Yuan et al. \cite{yuan2021multi}, it can be found that the efficiency can be further improved by one order of magnitude on the basis of the original IUGKS for NS solutions in the continuum flows.

\subsection{Multigrid method}
Multigrid method is also a commonly used acceleration algorithm in the CFD simulations.
The multigrid method may originate from 1960s \cite{fedorenko1962relaxation,fedorenko1964speed}, and gets a fast development in the engineering applications \cite{brandt2011guide,trottenberg2000multigrid,stuben2001review,wesseling1991introduction,mavriplis1995multigrid} since Brandtl's studies \cite{brandt1977multi}.
It has been widely used in CFD simulation to solve the Euler and NS equations \cite{blazek2015cfd,jameson1983solution,yoon1986multigrid} as well as the GKS \cite{xu1994fvs,jiang2012implicit}.
The basic idea behind all multigrid strategies is to accelerate the solution at fine grid by computing corrections on a coarse grid to eliminate low-frequency errors efficiently \cite{mavriplis1995multigrid}.
In general, an iterative algorithm can reduce the high-frequency errors faster than the low-frequency ones.
The multigrid method reduces the low-frequency errors on a fine mesh more efficiently by transitioning them to a coarser mesh, where the errors become high-frequency ones with respect to the coarse mesh size, and then can be eliminated faster.
Zhu et al. \cite{zhu2017multigrid} developed the UGKS with multigrid acceleration which further improves the convergence of the IUGKS for steady flow simulations.

A two-grid cycle method is the basis for any multigrid algorithm.
Usually, it consists of pre-smoothing, coarse grid correction, and post-smoothing processes.
The pre-smoothing and post-smoothing can be regarded as taking numerical iterations to solve the governing equations on the fine mesh, while the coarse grid correction is the process to obtain the correction of flow variables by taking numerical iterations on the coarse mesh.
Therefore, the basic procedures in the multigrid method include restriction, smoothing, and prolongation.

\begin{figure}[htbp]
	\centering
	\includegraphics[width=1.0\textwidth]{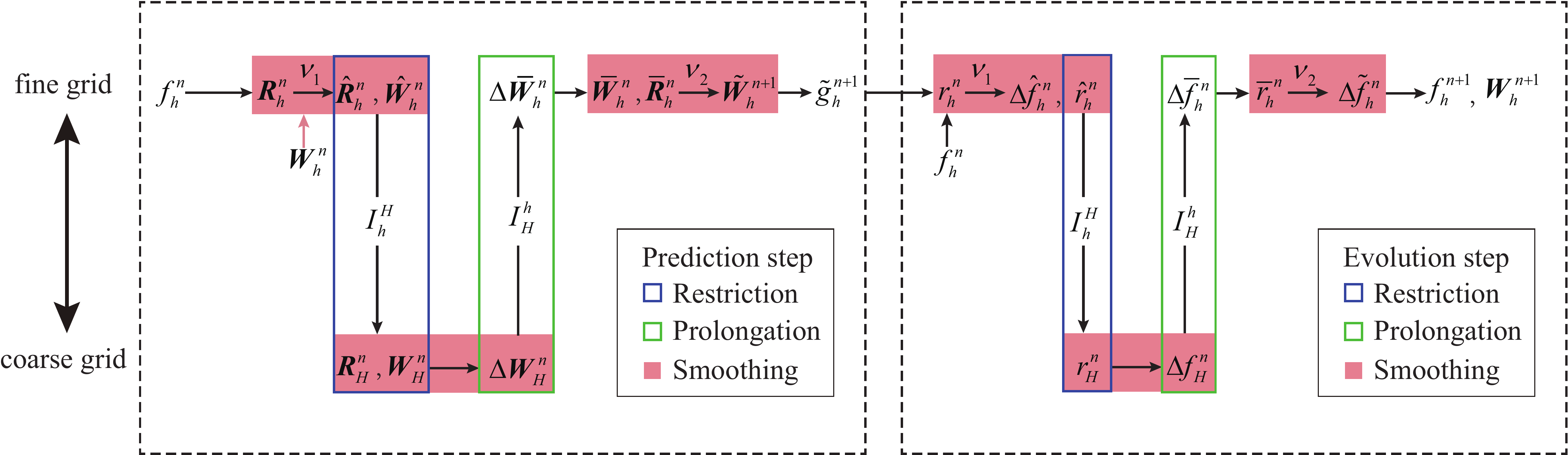}
	\caption{\label{fig:two_grid_cycle} Two-grid cycle for IUGKS \cite{zhu2017multigrid}   .}
\end{figure}
For a fine mesh denoted by $\Omega_h$ and a coarse mesh referred to as $\Omega_H$, the two-grid cycle for the IUGKS is illustrated in Fig.~\ref{fig:two_grid_cycle}.
Here, the prediction and evolution steps denote the processes of solving the macroscopic and microscopic equations, respectively.
$I_h^H$ and $I_H^h$ are the transfer operators between the fine and coarse meshes.
$\nu_1$ and $\nu_2$ are the number of pre-smoothings and post-smoothings, respectively.

\subsubsection{Restriction}
In order to obtain the correction of macroscopic flow variables and microscopic distribution function on the coarse mesh, the residual and flow variables need to be interpolated from the fine mesh to the coarse one.
From Eqs.~\eqref{eq:implicit_macro} and \eqref{eq:implicit_micro}, we can find that the governing equations for implicit iteration are nonlinear and linear, respectively for the macroscopic flow variables and the microscopic distribution function.
Therefore, the full approximation storage (FAS) scheme \cite{brandt2011guide} is used in the prediction step for the conservative flow variables, and the correction scheme (CS) \cite{brandt2011guide,trottenberg2000multigrid} for solving linear equation is utilized in the evolution step of distribution function.

For nonlinear equations, both the macroscopic flow variables and the residuals are required to be restricted from $\Omega_h$ to $\Omega_H$; while only the residuals are needed for linear equation of distribution function.
Specifically, for the cell $I$ on the coarse mesh, we have
\begin{equation}\label{eq:macro_restriction}
\begin{aligned}
(\vec{w}_H^n)_I=(I_h^H \hat{\vec{w}}_h^n)_I &=  \frac{1}{V_I} \sum_{j \in S(I)}{(\hat{\vec{w}}_h^n)_j V_j},\\
(\vec{R}_H^n)_I = (I_h^H \hat{\vec{R}}_h^n)_I &= \frac{1}{V_I} \sum_{j \in S(I)}{(\hat{\vec{R}}_h^n)_j V_j },
\end{aligned}
\end{equation}
and
\begin{equation}\label{eq:micro_restriction}
(r_H^n)_I=(I_h^H \hat{r}_h^n)_I = \frac{1}{V_I} \sum_{j \in S(I)} {(\hat{r}_h^n)_j V_j},
\end{equation}
where $S(I)$ denotes the children cells on the fine mesh of the cell $I$ on the coarse mesh.

\subsubsection{Smoothing}
The smoothing process to obtain the correction of macroscopic flow variables on the coarse mesh can be implemented by solving
\begin{equation}\label{eq:macro_smoothing}
\begin{aligned}
\frac{1}{V_I} \sum_{J \in N(I)}{S_{IJ}  \vec{F}_{IJ} (\vec{w}_H^{(m+1)}) }
-\frac{1}{V_I} \sum_{J \in N(I)}{S_{IJ}  \vec{F}_{IJ}(\vec{w}_H^{(m)}) }
= \\
(\vec{P}_H^n )_I
-\frac{1}{V_I} \sum_{J \in N(I)}{S_{IJ}  \vec{F}_{IJ}(\vec{w}_H^{(m)}) },
\end{aligned}
\end{equation}
where $\vec{w}_H^{(m)}$ is the intermediate solution for $m$-th smoothing process, $\vec{P}_H^n$ is the forcing function \cite{jameson1987lower} defined as the difference between the residuals directly transferred from the fine grid and the residuals determined from the macroscopic evolution equations which are re-computed on the coarse mesh, i.e.,
\begin{equation}\label{eq:forcingFunction}
(\vec{P}_H^n)_I = (I_h^H \vec{R}_h^n )_I+ \frac{1}{V_I} \sum_{J \in N(I)}{S_{IJ} \vec{F}_{IJ}(\vec{w}_H^{n})}.
\end{equation}
After $m$ times of smoothing, the macroscopic residuals can be renewed by
\begin{equation}\label{eq:renewed_macro_residual}
(\vec{R}_H^{(m)})_I = (\vec{P}_H^n )_I
-\frac{1}{V_I} \sum_{J \in N(I)}{S_{IJ}  \vec{F}_{IJ}(\vec{w}_H^{(m)}) }.
\end{equation}

The smoothing process to obtain the correction of microscopic distribution function on the coarse mesh can be carried out by solving
\begin{equation}\label{eq:micro_defect}
D_{I} \Delta{f}_{I}^{(m+1)}
+\sum_{J\in N(I)}{D_{J} \Delta{f}_{J}^{(m+1)}}
=(I_h^H r_{h}^n)_I - \left(D_{I} \Delta{f}_{I}^{(m)}
+\sum_{J\in N(I)}{D_{J} \Delta{f}_{J}^{(m)}}\right),
\end{equation}
where $\Delta{f}_I^{(m)} = f_I^{(m)}-f_I^{(0)}$ represents the correction of distribution function.
After $m$ times of smoothing, the microscopic residual can be renewed by
\begin{equation}\label{eq:renewed_micro_residual}
(r_H^{(m)})_I = (I_h^H r_{h}^n)_I - \left(D_{I} \Delta{f}_{I}^{(m)}
+\sum_{J\in N(I)}{D_{J} \Delta{f}_{J}^{(m)}}\right).
\end{equation}

\subsubsection{Prolongation}
The correction of the flow variables solved on the coarse mesh should be interpolated onto the fine mesh to eliminate the low-frequency errors.
The prolongation operator is implemented by spatial interpolation.
Specifically, for an arbitrary variable $Q$, the interpolated correction on the fine mesh is
\begin{equation}\label{eq:prolongation}
(I_H^h Q_H)_i = \frac{\sum_{J\in S(i)}{w_J Q_J}}{\sum_{J \in S(i)}{w_J}},
\end{equation}
where $S(i)$ is the stencil of cells on coarse mesh for interpolating the values of cell $i$ on the fine mesh, $\omega_J$ is the interpolation weight of coarse cell $J$.
By interpolation, the correction of conservative variables $\Delta \vec{w}_H^{n} = \vec{w}_H^{n+1} - \vec{w}_H^{n}$ and the correction of distribution function $\Delta f_H^n$ can be transferred on the fine mesh, which are used to update the flow variables and residuals on the fine mesh.

With the basic two-grid cycle, the multigrid algorithms on multiple levels of grids, such as W-cycle and V-cycle, can be constructed by recursive combination of two-grid cycles.

\subsubsection{Lid-driven cavity flow}

Simulations of lid-driven cavity flows have been carried out at different Knudsen numbers \cite{zhu2017multigrid} to validate the efficiency of the multigrid IUGKS.
The gas in the cavity is argon with molecular mass $m_0 = 6.63\times10^{-26} { kg}$ and with an initial temperature $T_0 = 273{ K}$.
The cavity has a fixed wall temperature $T_w=273{ K}$ and a moving lid at a constant velocity $U_w =50 { m/s}$.
The Knudsen number is defined as the ratio of mean free path to the length of cavity side wall.
Cases at three different Knudsen numbers, i.e., $Kn=10, 1, 0.075$ have been tested.
For the cases of ${ Kn}=10$ and $1$, the mean free path is defined using VHS model, while for ${ Kn}=0.075$, the HS model is employed.
The dynamic viscosity is evaluated by Eq.~\eqref{eq:dynamic_viscosity} with $\omega = 0.81$ for all the three cases.
In physical space, the computational domain is discretized with a mesh of $64\times64$ cells. 
In velocity space, $120\times120$, $100\times 100$ and $80\times 80$ discrete velocity points are used respectively for cases at $Kn=10, 1.0$ and $0.075$.
The steady state is thought to be reached when the mean squared residuals of the conservative variables are reduced to a level being less than $1.0\times10^{-6}$, where the residuals are computed by
\begin{equation}\label{eq:Residual_Def}
\vec{R}^n = \sqrt{\sum_{i=1}^{N_c}{\vec{R}_i^2} / N_c},
\end{equation}
which denotes the variation rate of the conservative variables.
Here $N_c$ is the total number of discrete cells in the computational domain.

\begin{figure}[htbp]
\centering
\subfloat[]{\includegraphics[width=0.32\textwidth]{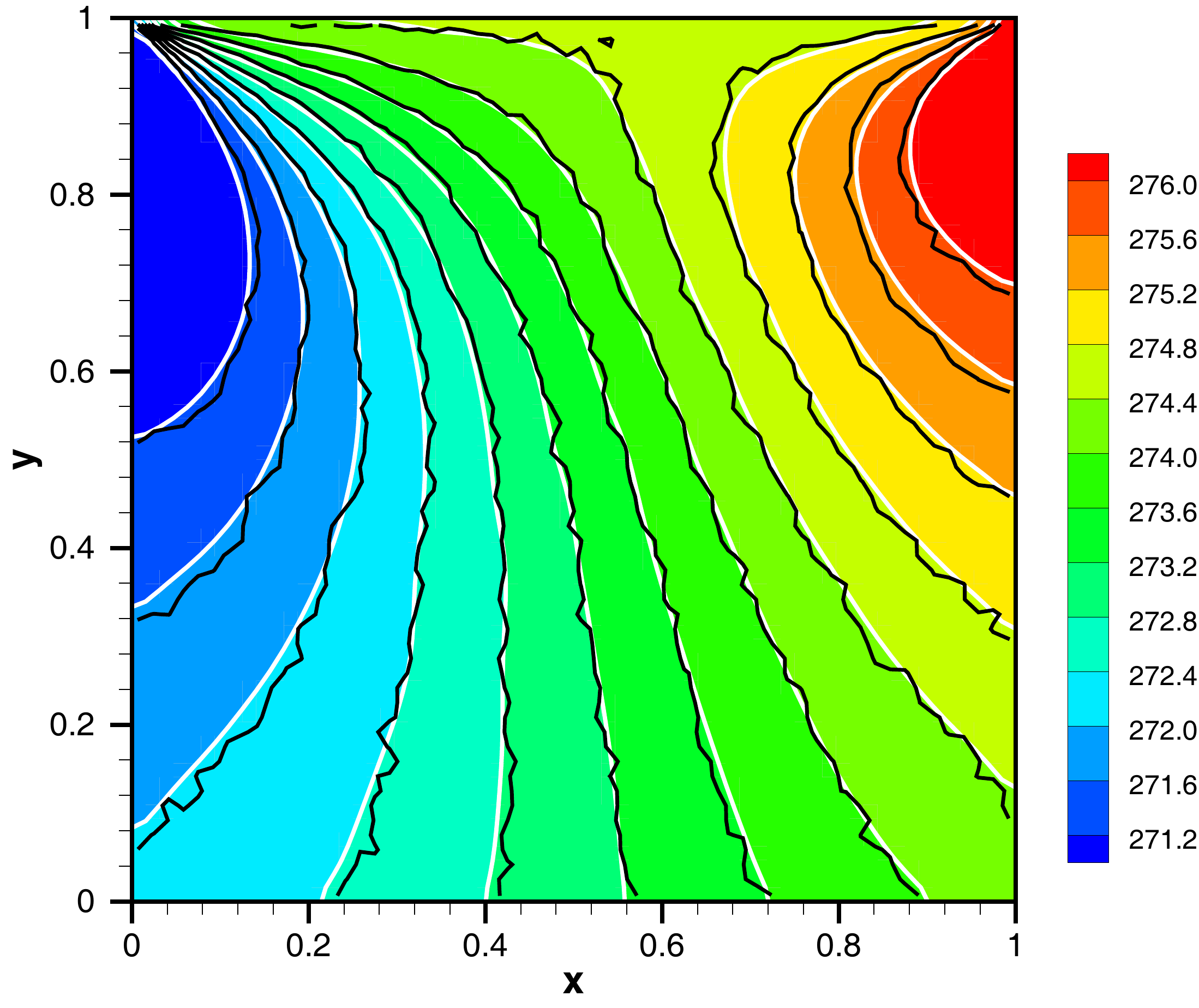}}
\subfloat[]{\includegraphics[width=0.32\textwidth]{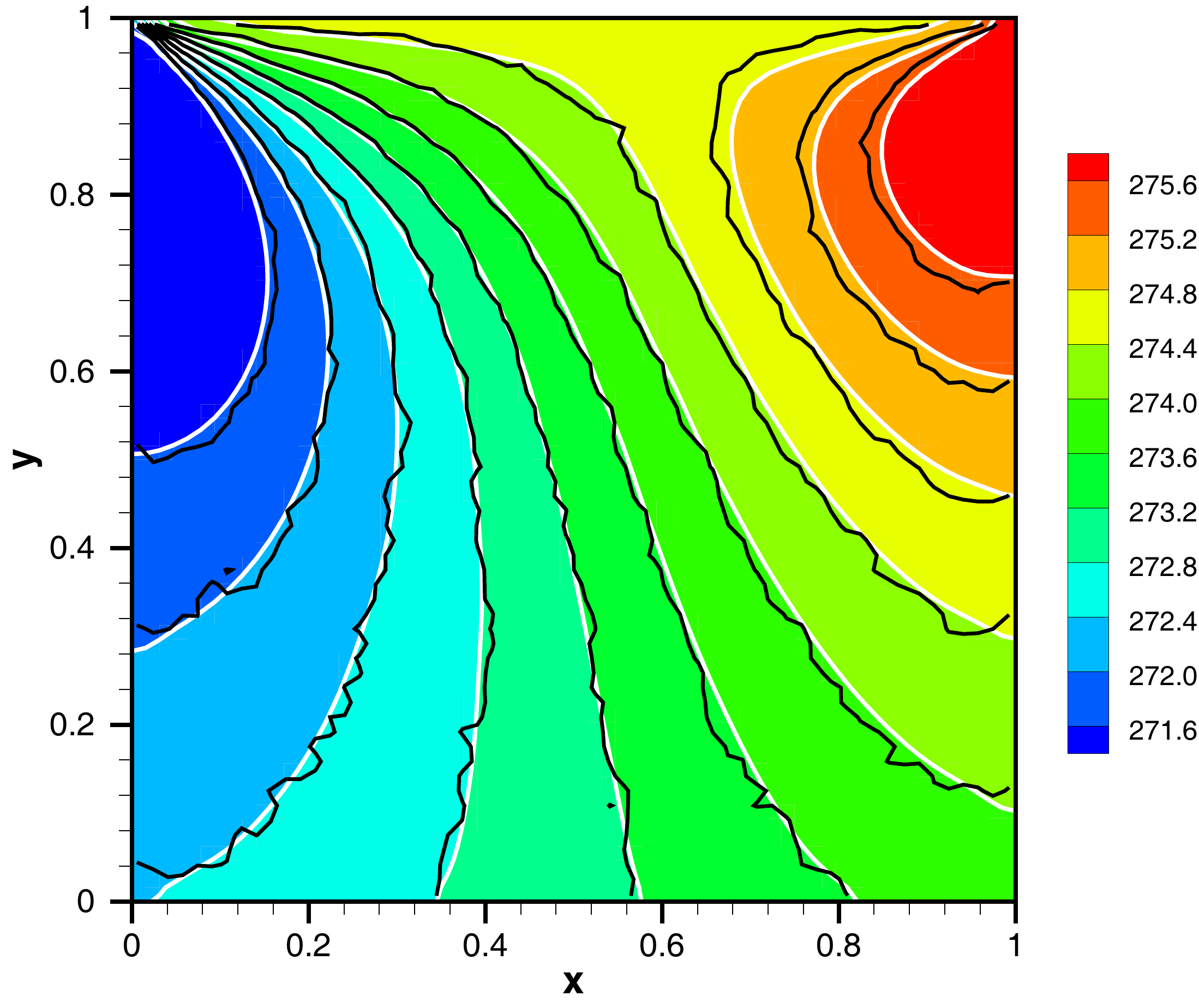}}
\subfloat[]{\includegraphics[width=0.32\textwidth]{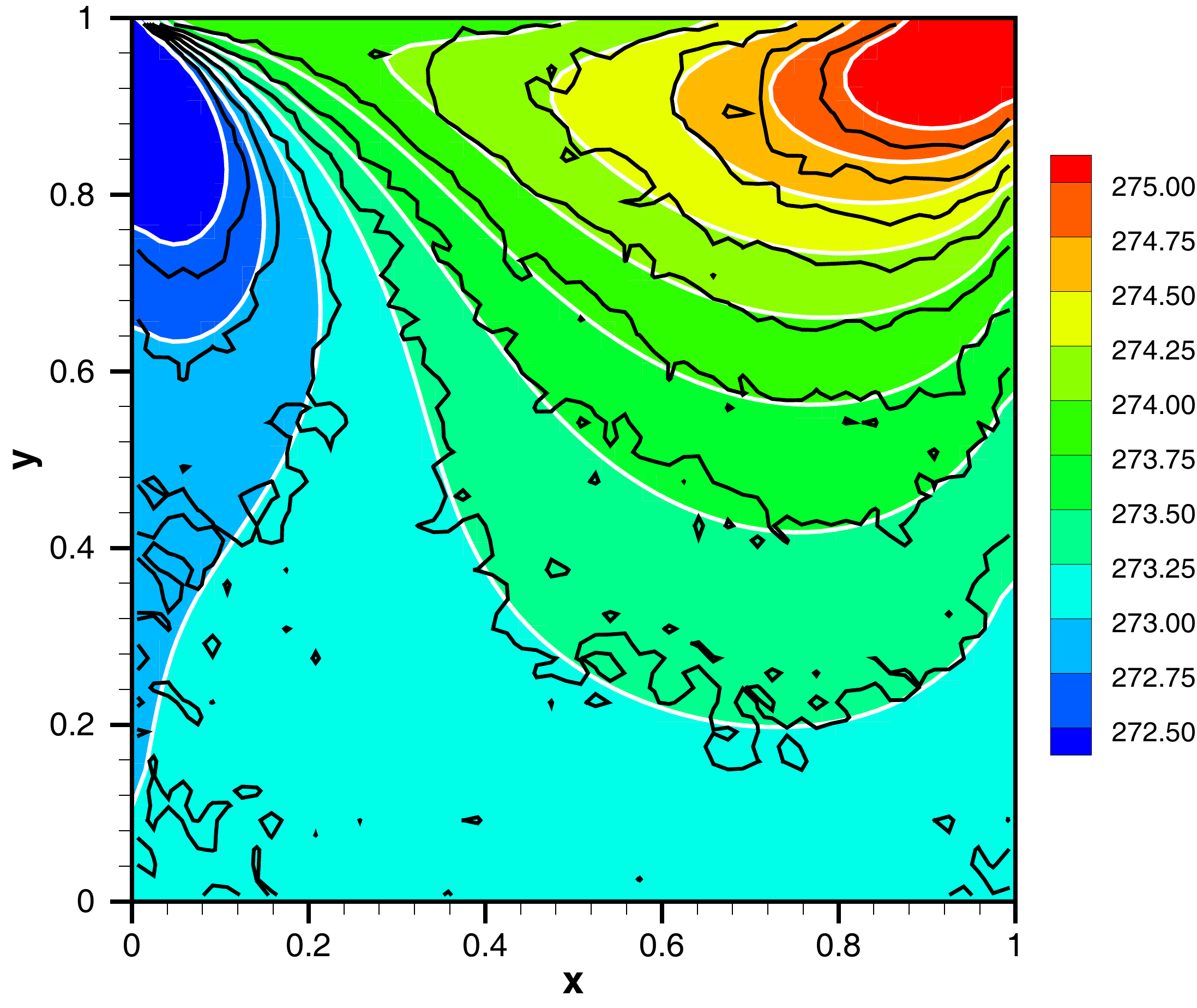}}
\caption{\label{fig:cavityTemperature}Temperature distribution in the cavity at different Knudsen numbers \cite{zhu2017multigrid}. (a) ${ Kn}=10$, (b) ${ Kn}=1$ and (c) ${ Kn}=0.075$. Background: the multigrid IUGKS; white solid line: the IUGKS; black solid line: the DSMC results \cite{john2011effects}.}
\end{figure}

\begin{figure}[htbp]
\centering
\subfloat[]{\includegraphics[width=0.32\textwidth]{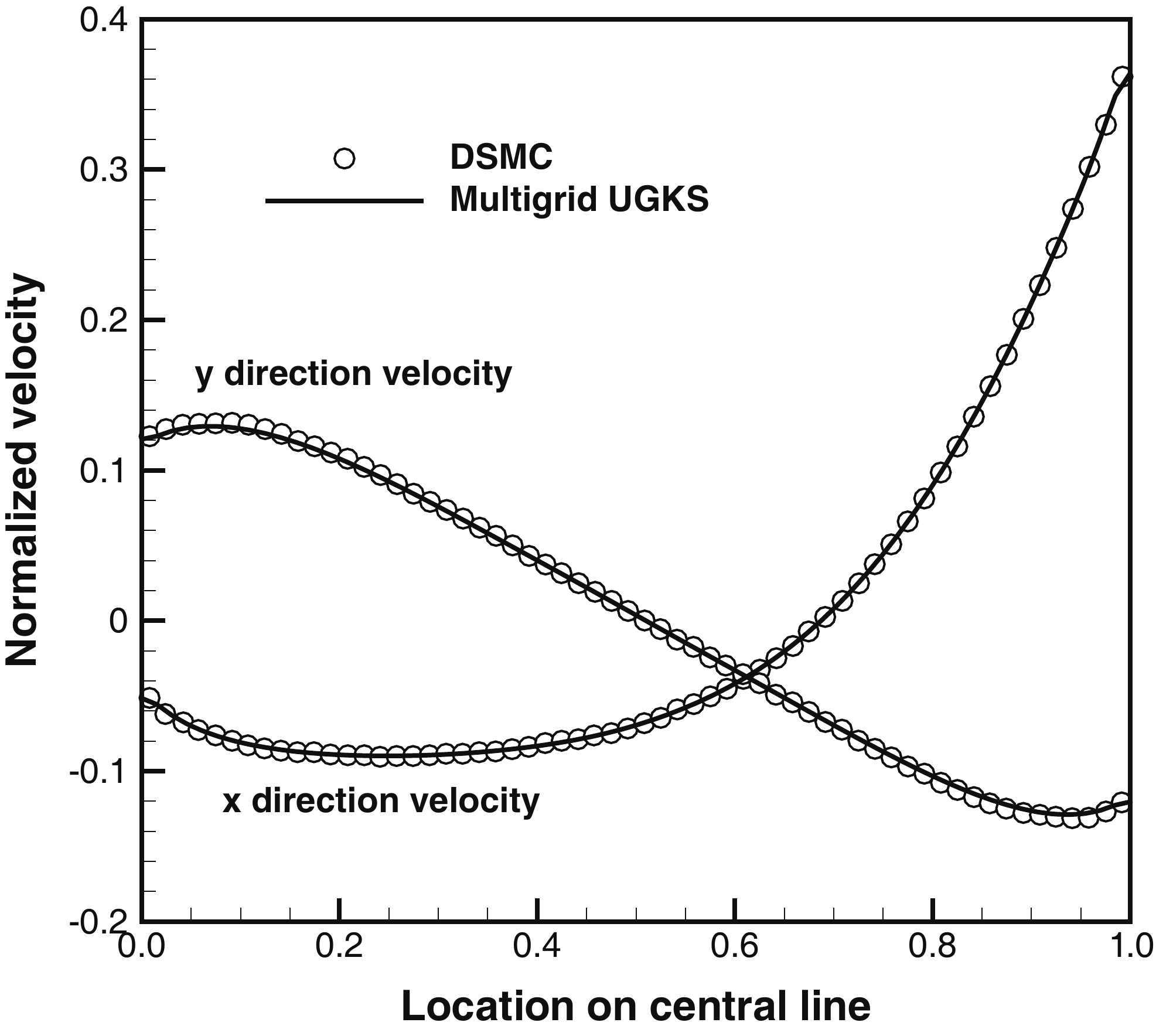}}
\subfloat[]{\includegraphics[width=0.32\textwidth]{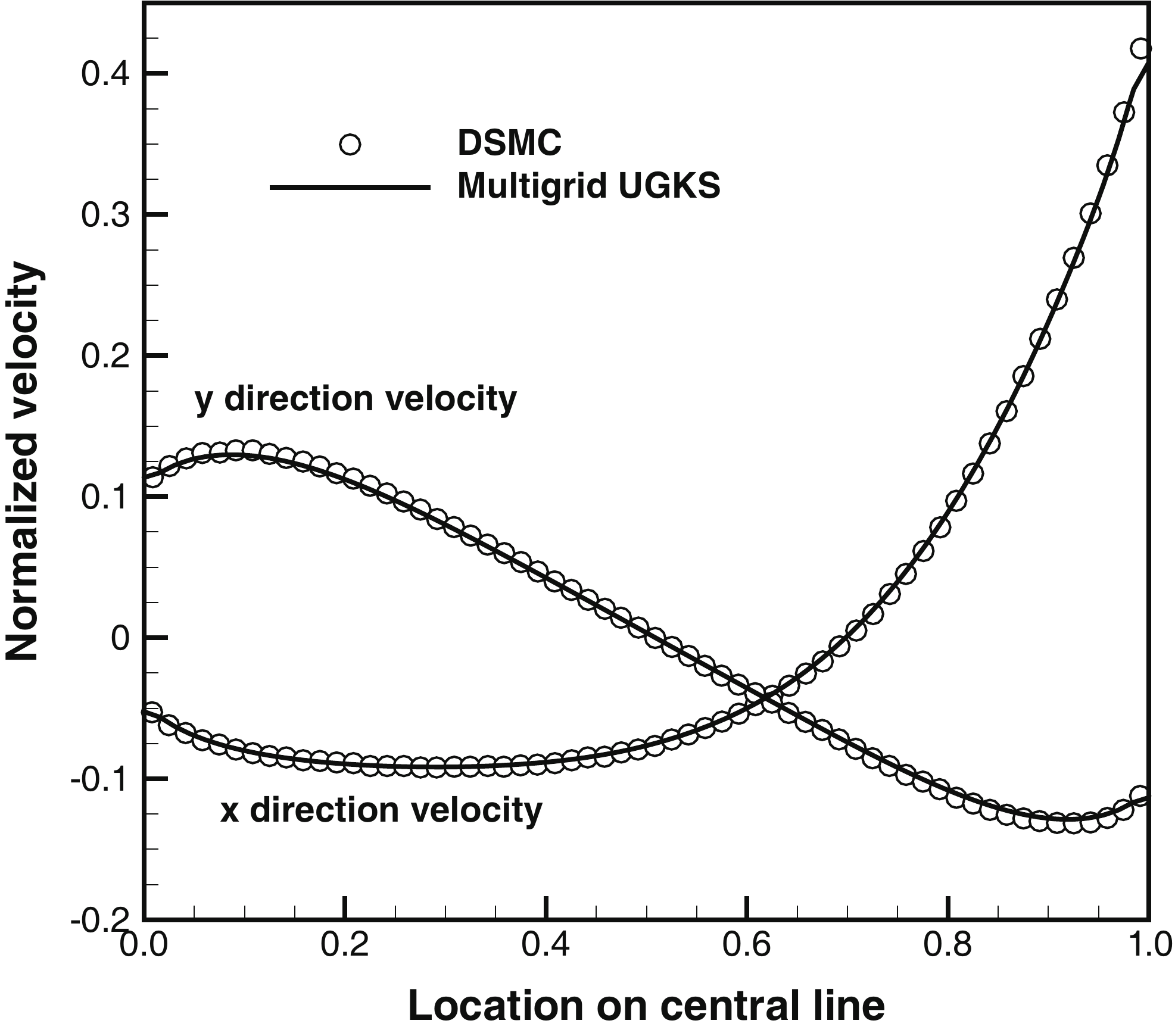}}
\subfloat[]{\includegraphics[width=0.32\textwidth]{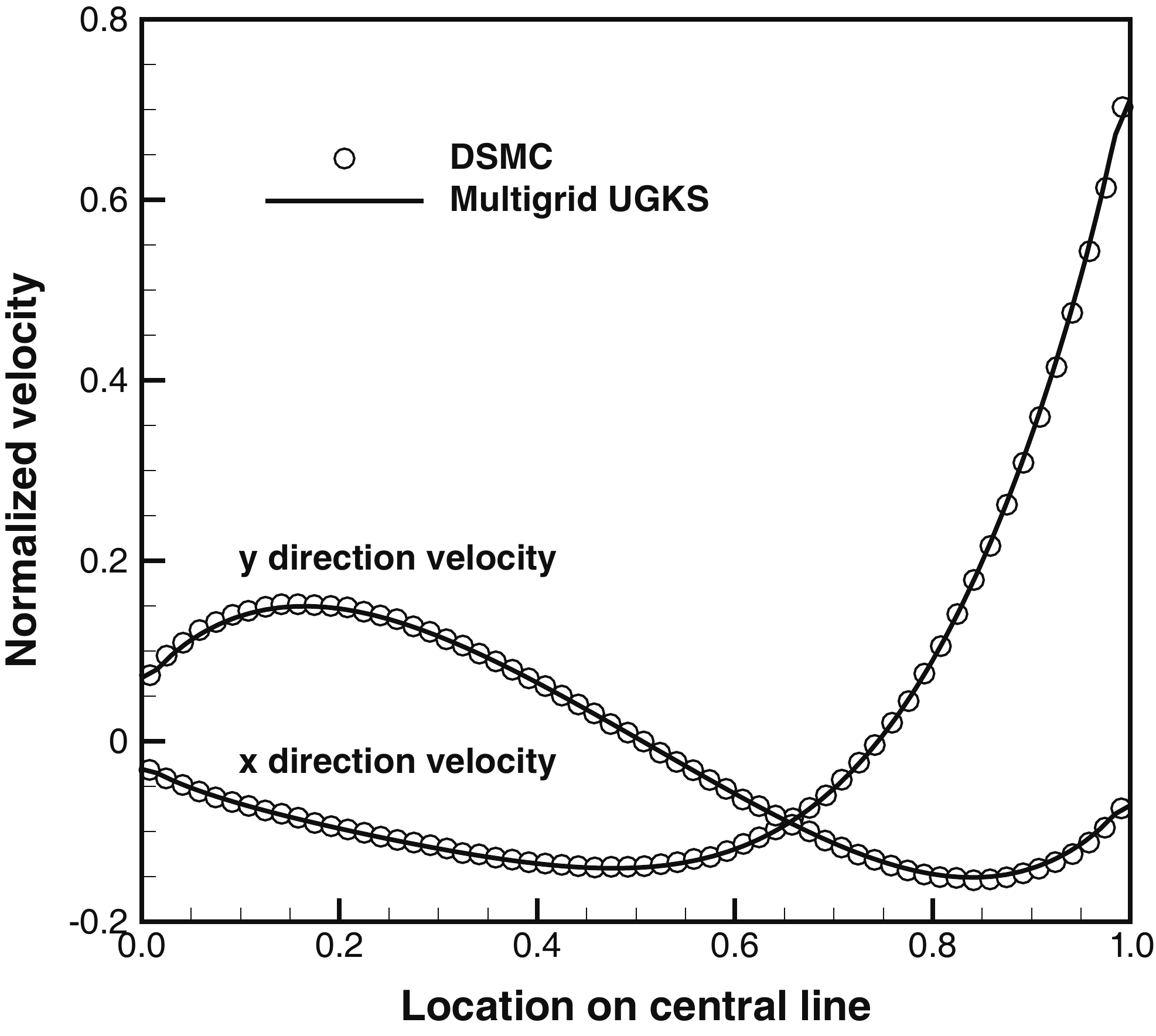}}
\caption{\label{fig:cavityVelocity}Normalized $x$- and $y$-component velocity along central lines at different Knudsen numbers \cite{zhu2017multigrid}. (a) ${ Kn}=10$, (b) ${ Kn}=1$, and (c) ${ Kn}=0.075$.}
\end{figure}

The results of the temperature distribution in the cavity at different Knudsen numbers have been plotted in Fig.~\ref{fig:cavityTemperature}.
The results of the multigrid method are consistent with those of the original IUGKS with a single level of grid, and agree well with the DSMC results.
The distribution of the normalized velocities along the vertical and horizontal central lines are plotted and compared with the DSMC data in Fig.~\ref{fig:cavityVelocity}, which also shows good agreement between the multigrid UGKS and the DSMC method.

\begin{figure}[htbp]
\centering
\subfloat[]{\includegraphics[width=0.32\textwidth]{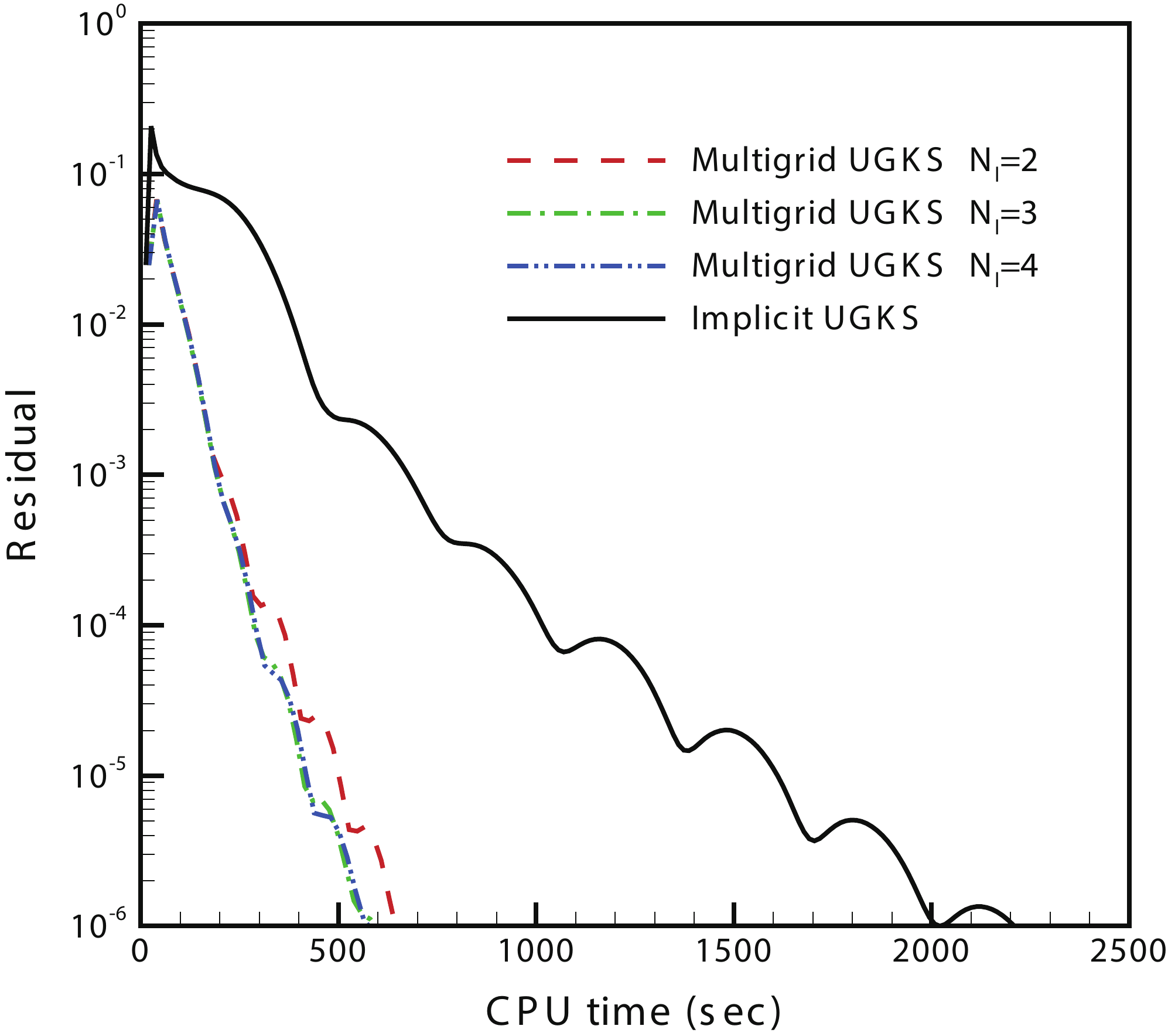}}
\subfloat[]{\includegraphics[width=0.32\textwidth]{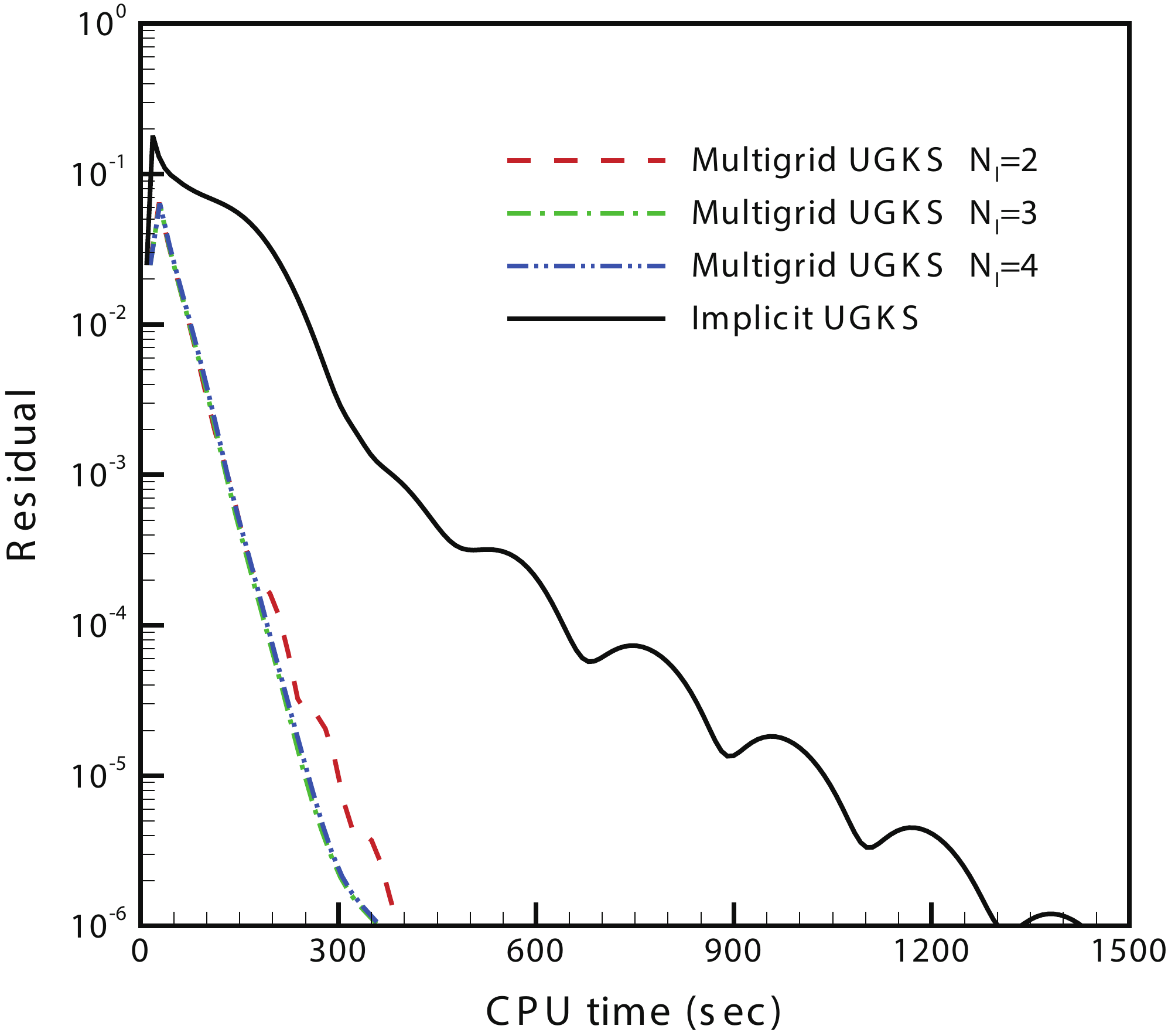}}
\subfloat[]{\includegraphics[width=0.32\textwidth]{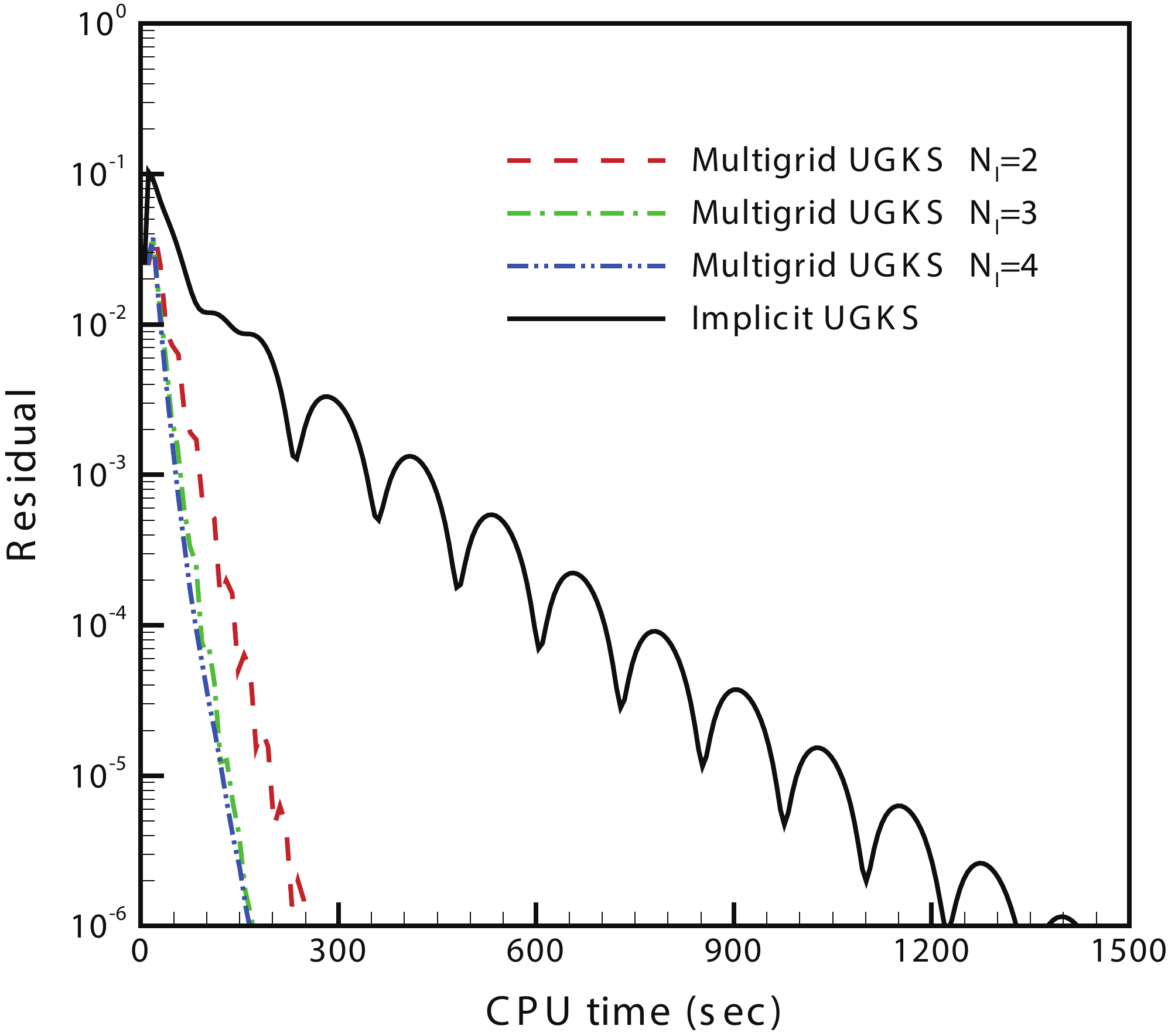}}
\caption{\label{fig:cavityHistory}Convergence history of the cavity flow indicated by the residual of energy density at different Knudsen numbers \cite{zhu2017multigrid}. (a) ${ Kn}=10$, (b) ${ Kn}=1$, and (c) ${ Kn}=0.075$.}
\end{figure}

Figure \ref{fig:cavityHistory} plots the convergence histories of the energy density with respect to CPU time to show the acceleration of the multigrid method.
Obvious accelerating effects of the multigrid method on the IUGKS can be observed.
In the high Knudsen number cases at ${ Kn}=10$ and $1$, the multigrid method is about $3$ times faster than the original implicit scheme and in the case at ${ Kn}=0.075$ the acceleration rate can be increased up to $8$ times.
With a single machine (Intel(R) Core(TM) i5-4570 CPU@3.2GHz), the multigrid IUGKS can get convergent solution at ${ Kn} = 0.075$ with the CPU time being less than $3$ minutes, while the DSMC solution needs parallel supercomputers for that \cite{john2011effects}.

\subsection{Parallel strategy}
For the CFD algorithms to solve the Navier--Stokes equations, the domain decomposition method is commonly used in the parallel computations.
Since the kinetic solvers take numerical discretization on both the physical space and the velocity space, the parallel strategy is more flexible for large scale simulations.

\begin{figure}[htbp]
\centering
\subfloat[]{\includegraphics[width=0.44\textwidth]{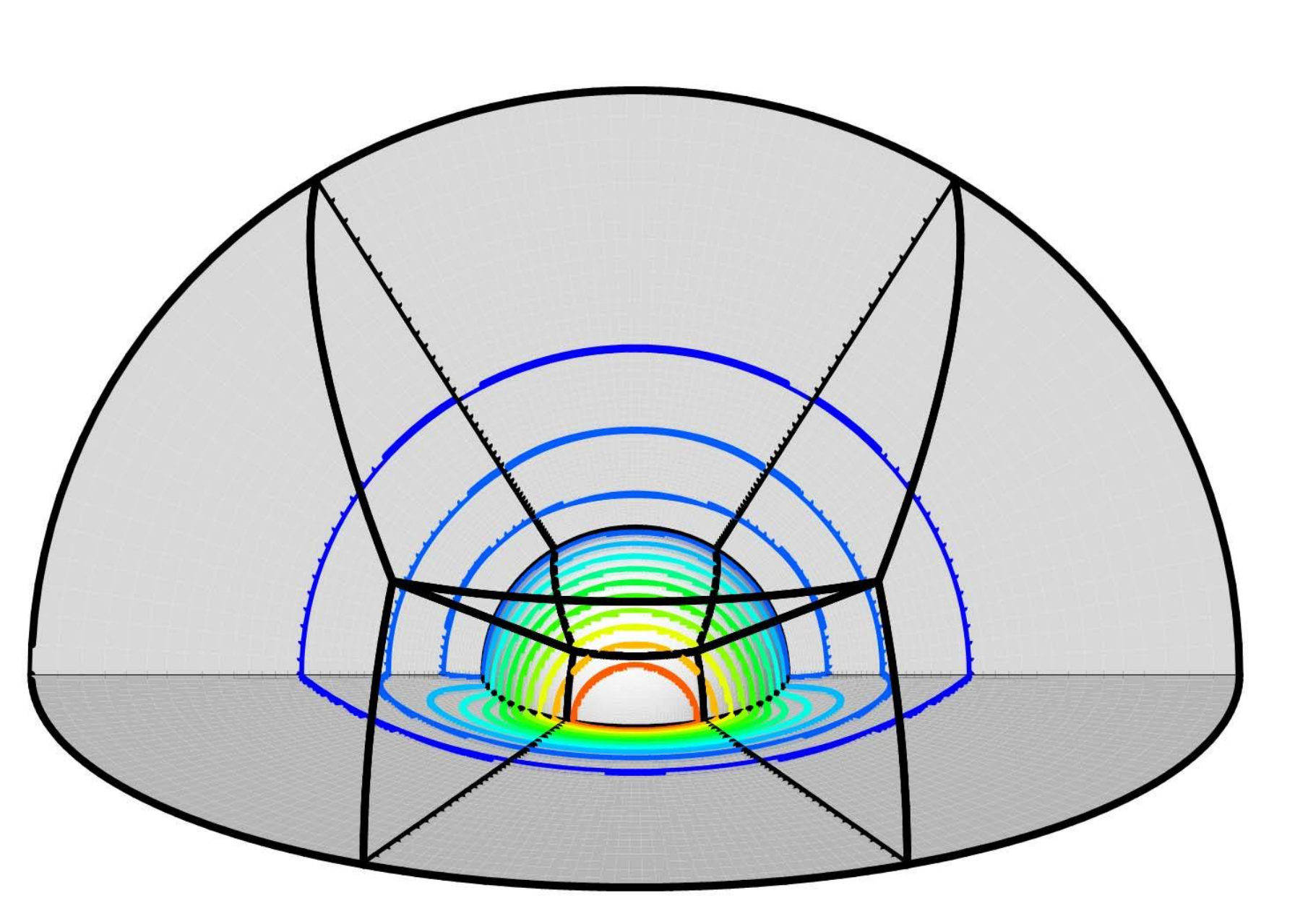}} \hspace{1em}
\subfloat[]{\includegraphics[width=0.44\textwidth]{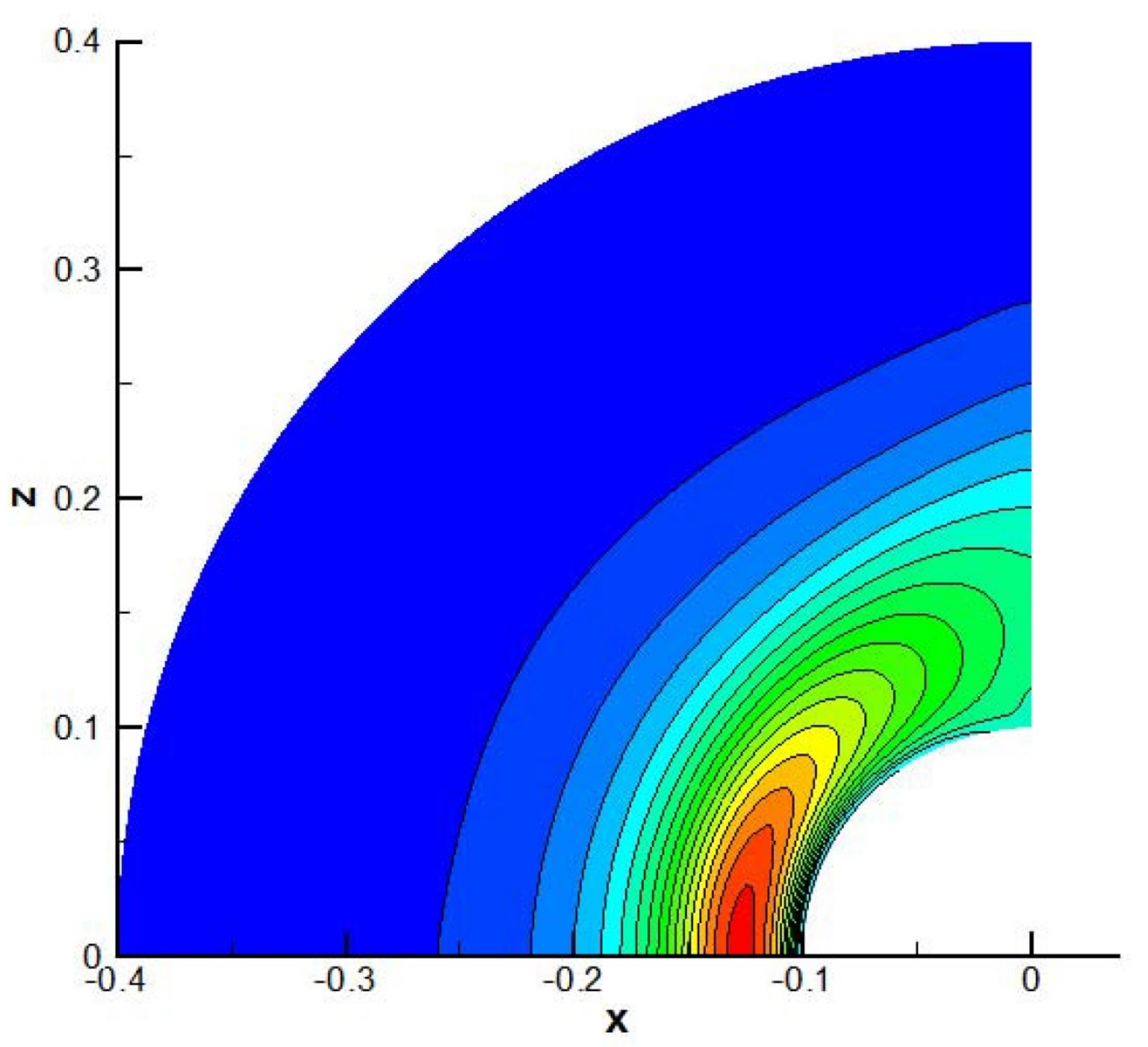}}
\caption{\label{fig:sphere_flow} Supersonic flow over a sphere at ${ Ma}=5$ and ${ Kn}=0.236$ \cite{li2016parallel}. (a) Blocks in physical space with pressure contours, (b) temperature distribution on the symmetric plane.}
\end{figure}

\begin{figure}[htbp]
\centering
\subfloat[]{\includegraphics[width=0.44\textwidth]{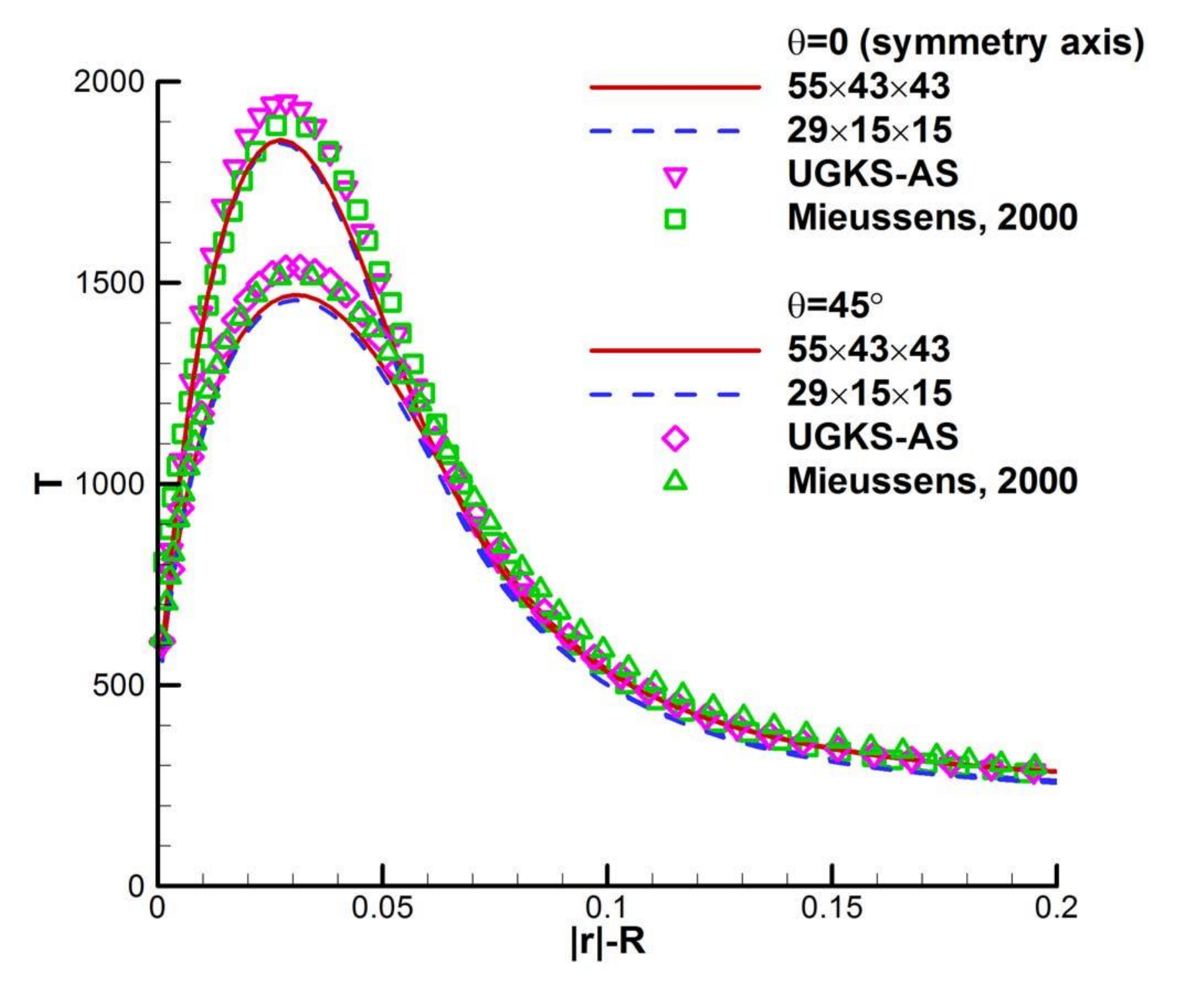}} \hspace{1em}
\subfloat[]{\includegraphics[width=0.44\textwidth]{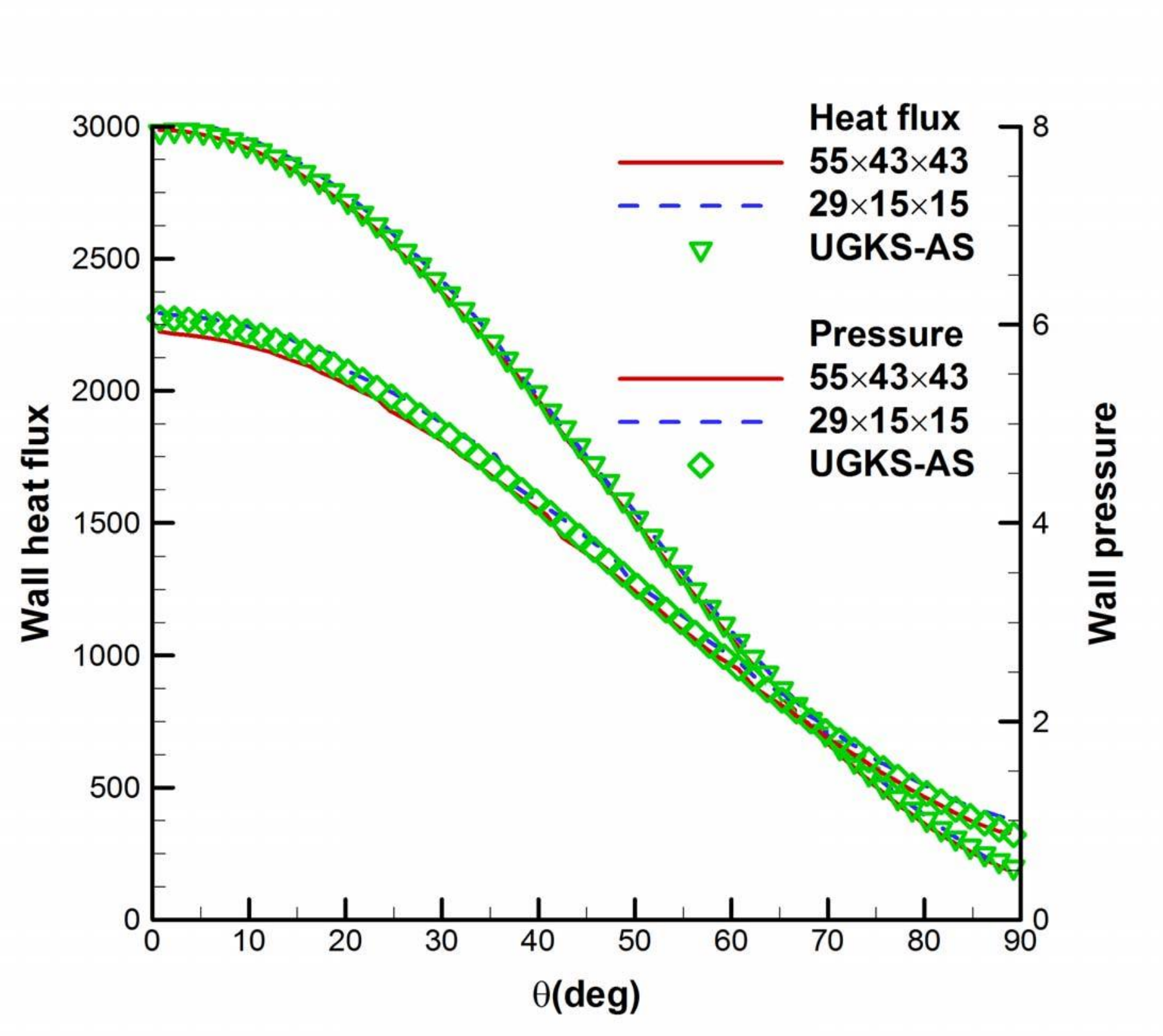}}
\caption{\label{fig:sphere_surface} Supersonic flow over a sphere at ${ Ma}=5$ and ${ Kn}=0.236$ \cite{li2016parallel}. (a) Temperature profile along the stagnation line, (b) surface flux and pressure distribution, which are compared with the reference data \cite{mieussens2000discrete}.}
\end{figure}

\begin{figure}[htbp]
\centering
\subfloat[]{\includegraphics[width=0.44\textwidth]{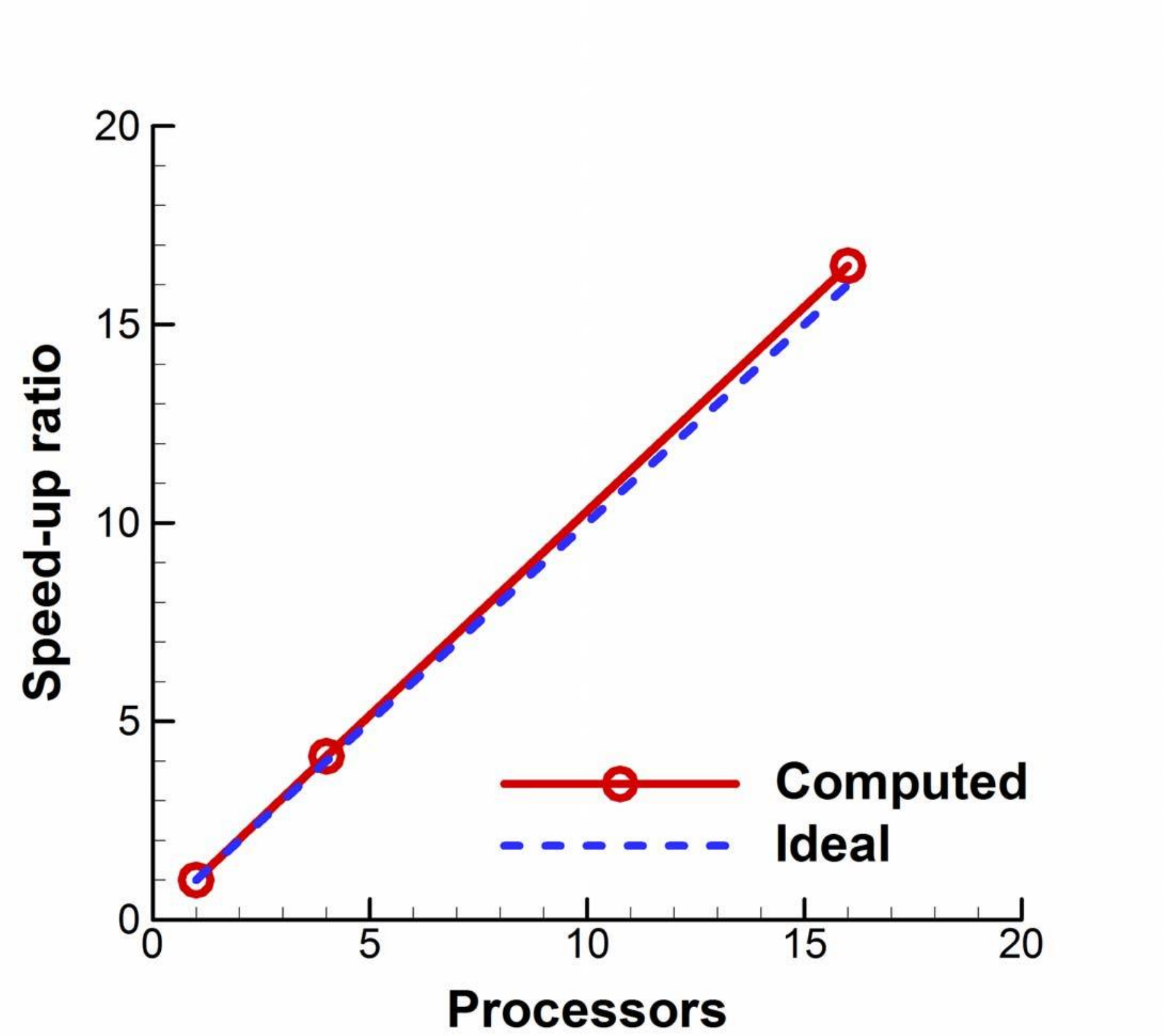}} \hspace{1em}
\subfloat[]{\includegraphics[width=0.44\textwidth]{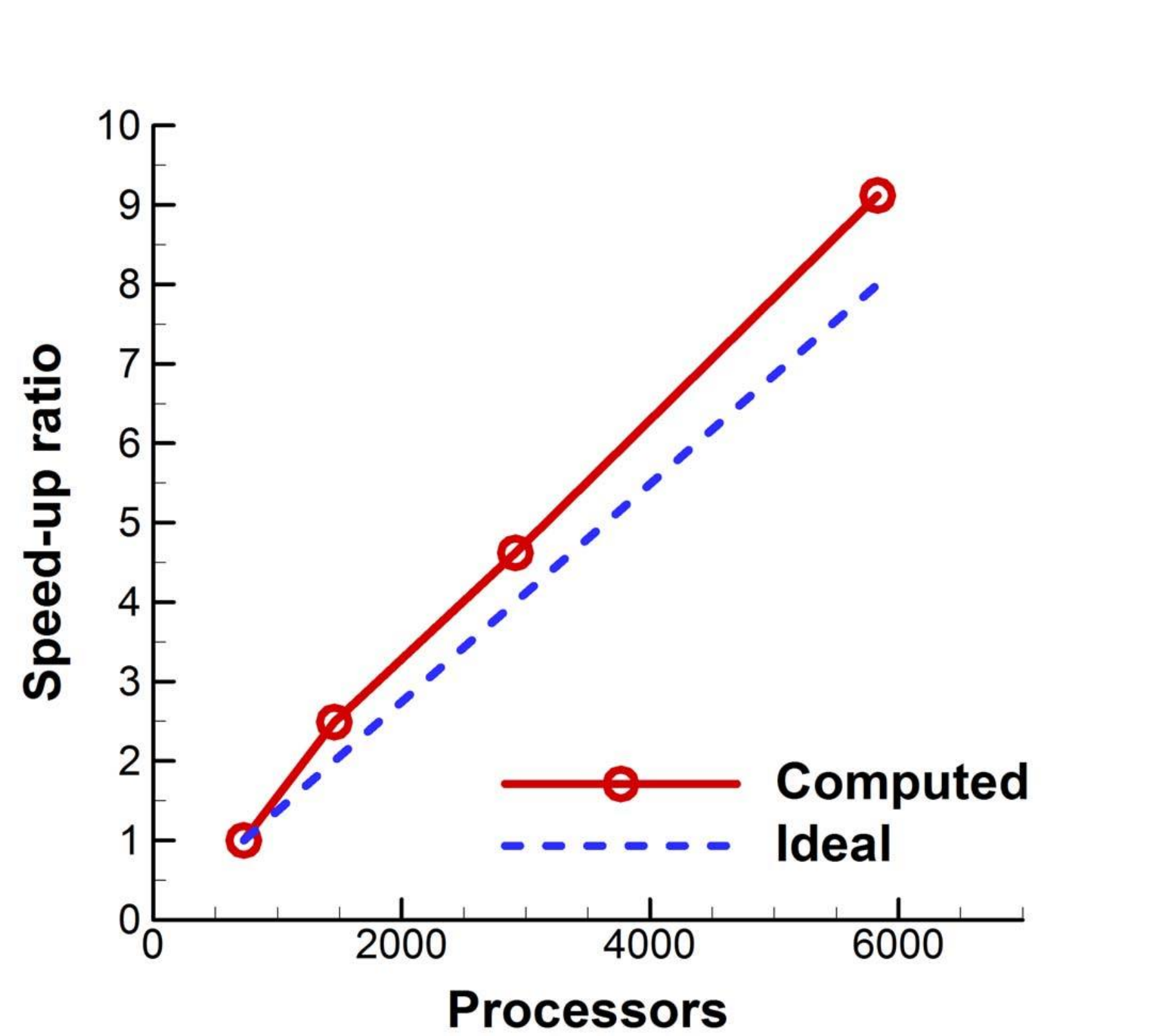}}
\caption{\label{fig:sphere_parallel} Parallel speedup ratio of the UGKS \cite{li2016parallel}. (a) Small-scale problem, and (b) large-scale problem.}
\end{figure}

For the UGKS on Cartesian grids, Ragta et al. \cite{ragta2017efficient} adopt the parallel strategy based on the domain decomposition, and investigate the parallelization performance up to thousands of cores.
Chen et al. \cite{chen2017memory} developed an implicit kinetic method with memory reduction techniques, which significantly reduces the memory consumption, and the parallel computation based on the discrete velocity distribution function is employed.
Li et al. \cite{li2016parallel} and Tan et al. \cite{shuang2019parallel} implement a hybrid MPI strategy for parallel algorithms in both physical space and velocity space, where a two-dimensional Cartesian topology is used to arrange the physical and velocity blocks.
Similarly, Jiang et al. \cite{jiang2019aia} take parallel computing with MPI for the decomposed physical mesh, and use several threads with OpenMP in every MPI process for parallel computation of discrete distribution function.
Parallel algorithm is also implemented on GPU devices for two dimensional UGKS \cite{liu2020gpu}.

In the study of Li et al. \cite{li2016parallel}, the parallel speedup ratio has been tested based on a supersonic flow over a sphere.
Figure \ref{fig:sphere_flow} shows the mesh blocks in the physical space and the temperature distribution on the symmetric plane.
Detailed results, such as the temperature profile along the stagnation line and the surface flux and pressure, are given in Fig.~\ref{fig:sphere_surface}, which are compared with the reference data \cite{mieussens2000discrete}.
The speed up ratios with respect to the number of processors are illustrated in Fig.~\ref{fig:sphere_parallel}, where the parallel efficiency is around $1$ up to $5832$ processors, which reveals the good scalability of the parallel UGKS.

\subsection{Adaptive mesh}

In order to reduce the computational cost, adaptive mesh refinement (AMR) is commonly used for a better spatial discretization with fewer grid points.
For the kinetic solvers with discrete phase space, the AMR technique can be applied in the velocity space as well \cite{arslanbekov2013kinetic}.
In the study of Chen et al. \cite{chen2012mesh}, an adaptive UGKS was proposed through introduction of an adaptive quadtree structure in the velocity space. Together with moving mesh in the physical space, the UGKS is able to simulate moving solid-gas interaction in the high-speed flows with a better efficiency.
Qin et al. \cite{qin2017adaptive} presented a simple local discrete velocity space in the UGKS, adopting a uniform background velocity mesh, which avoids the interpolation between different levels of velocity grids.
Xiao et al. \cite{xiao2020velocity} replace the discrete velocity points in the UGKS by a continuous velocity space in the region where the NS solutions provided by the GKS are valid.
So, a breakdown parameter is employed to handle the transition between continuous and discrete velocity spaces.
A multiscale flow with both continuum and rarefied regions, the computational cost can be reduced.

\begin{figure}[htbp]
\centering
\subfloat{\includegraphics[width=0.44\textwidth]{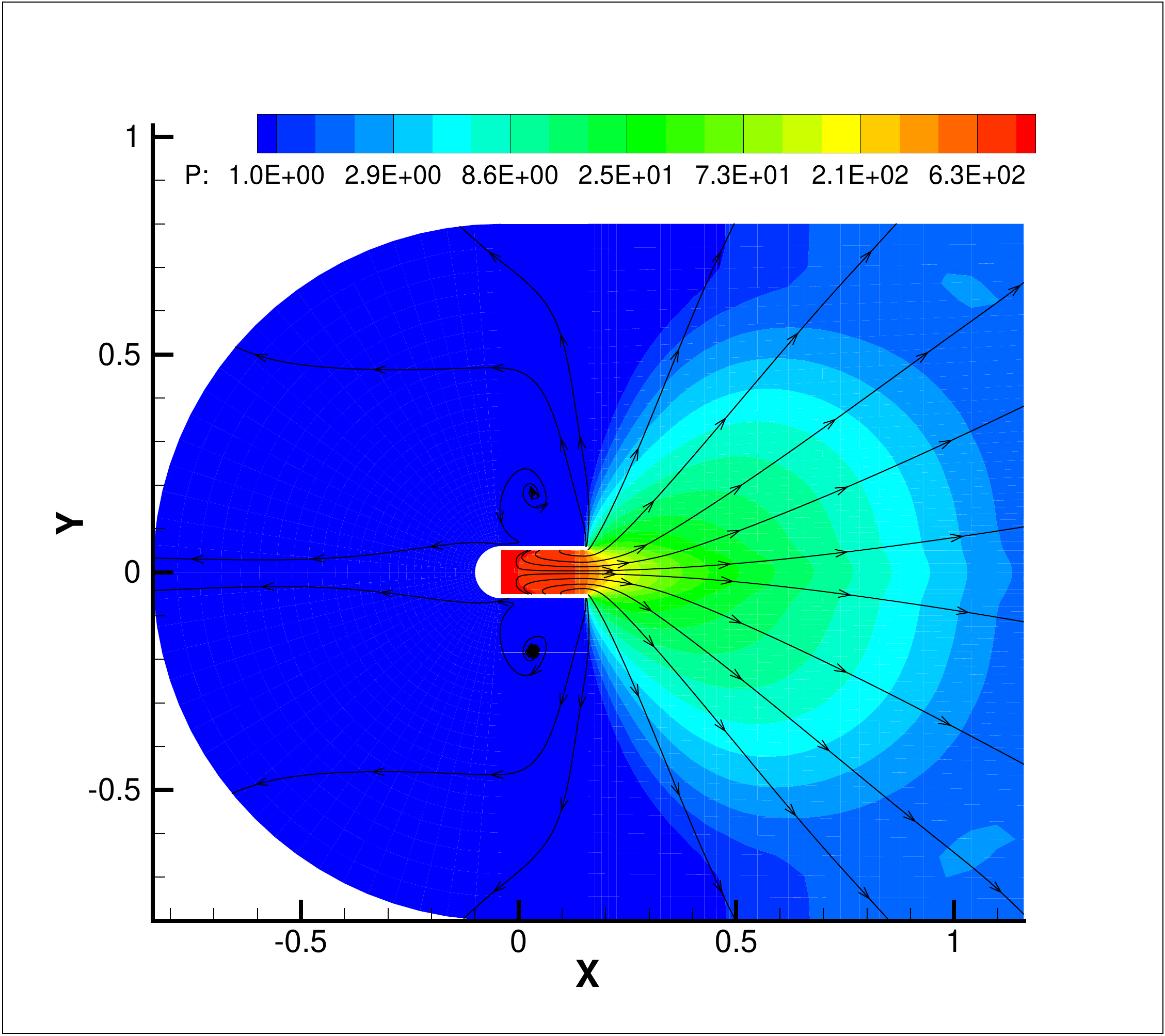}} \hspace{1em}
\subfloat{\includegraphics[width=0.44\textwidth]{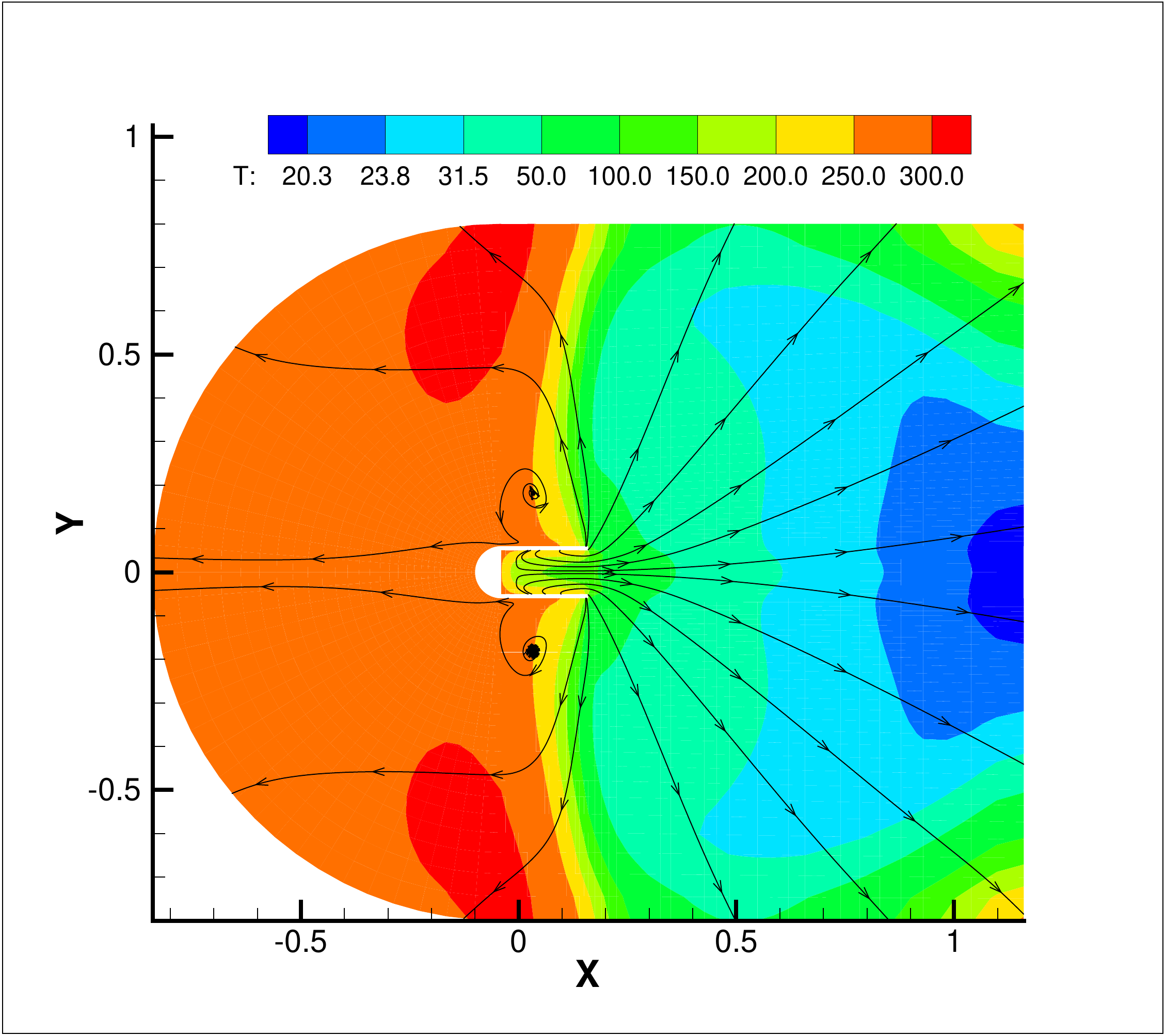}}
\caption{\label{fig:nozzle_expansion}Flow field during the expansion stage \cite{chen2012mesh}. (a) Pressure, and (b) temperature.}
\end{figure}

\begin{figure}[htbp]
\centering
\subfloat{\includegraphics[width=0.44\textwidth]{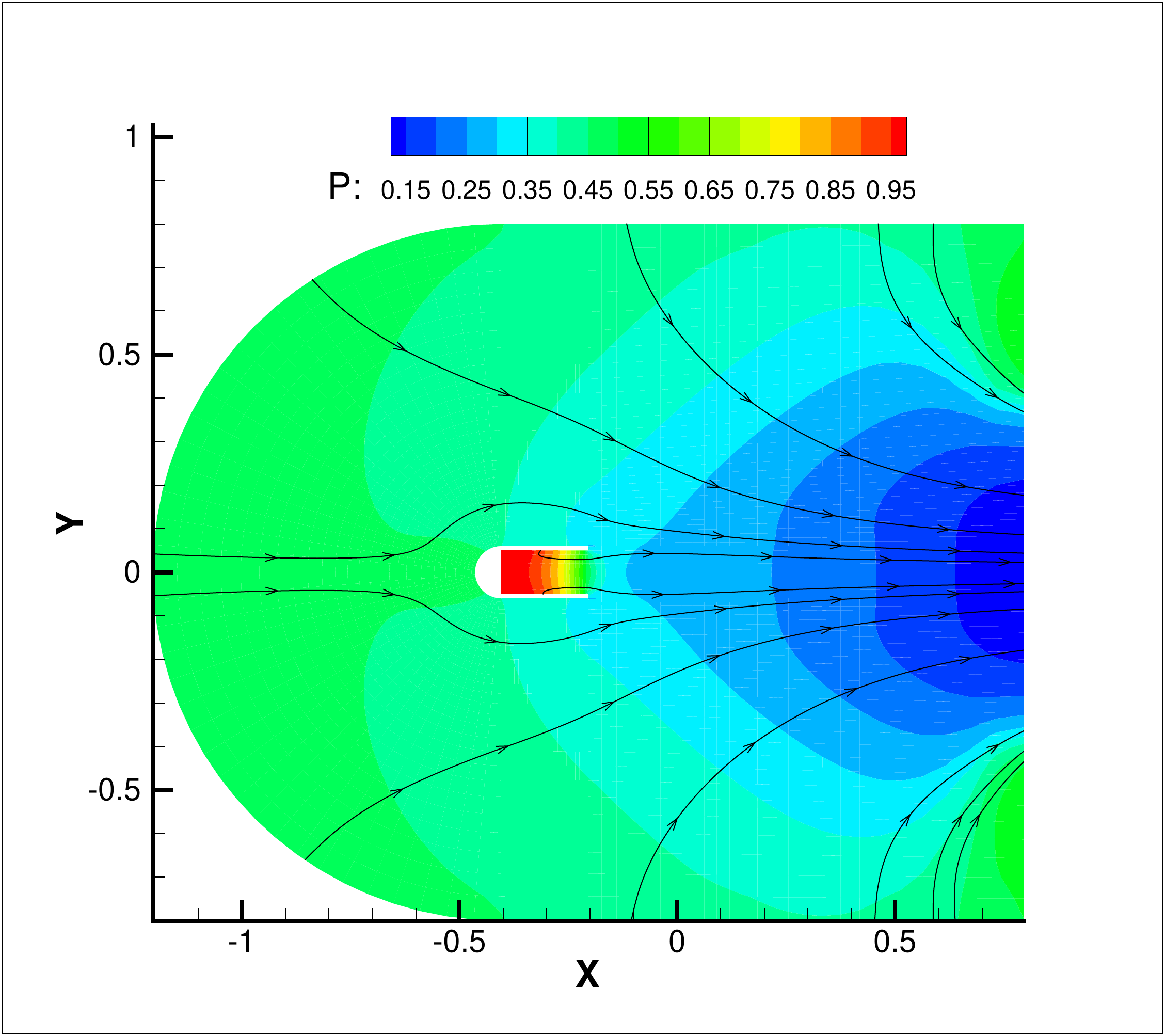}} \hspace{1em}
\subfloat{\includegraphics[width=0.44\textwidth]{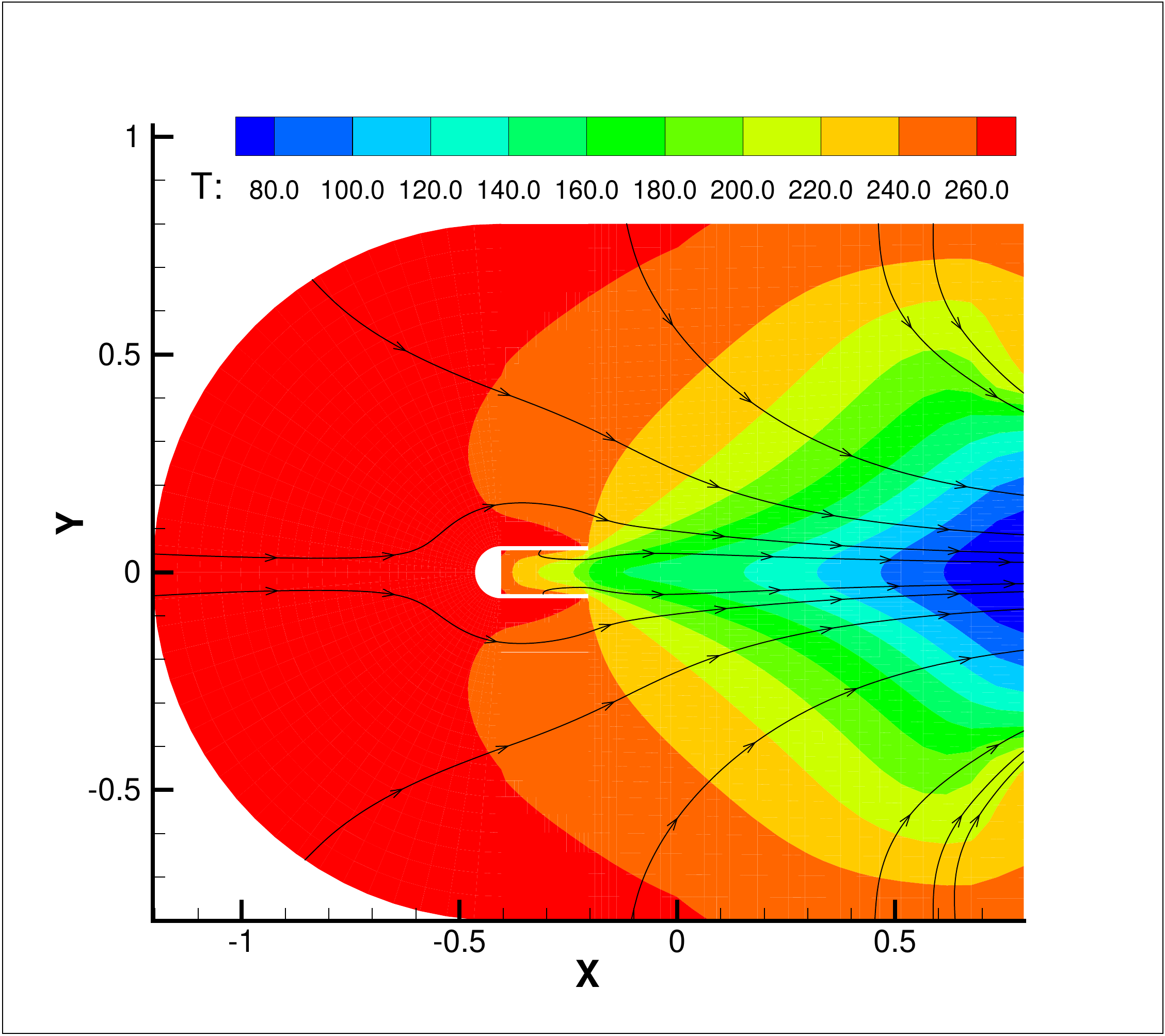}}
\caption{\label{fig:nozzle_jet} Flow field during the intermediate stage \cite{chen2012mesh}. (a) Pressure, and (b) temperature.}
\end{figure}

\begin{figure}[htbp]
\centering
\subfloat{\includegraphics[width=0.44\textwidth]{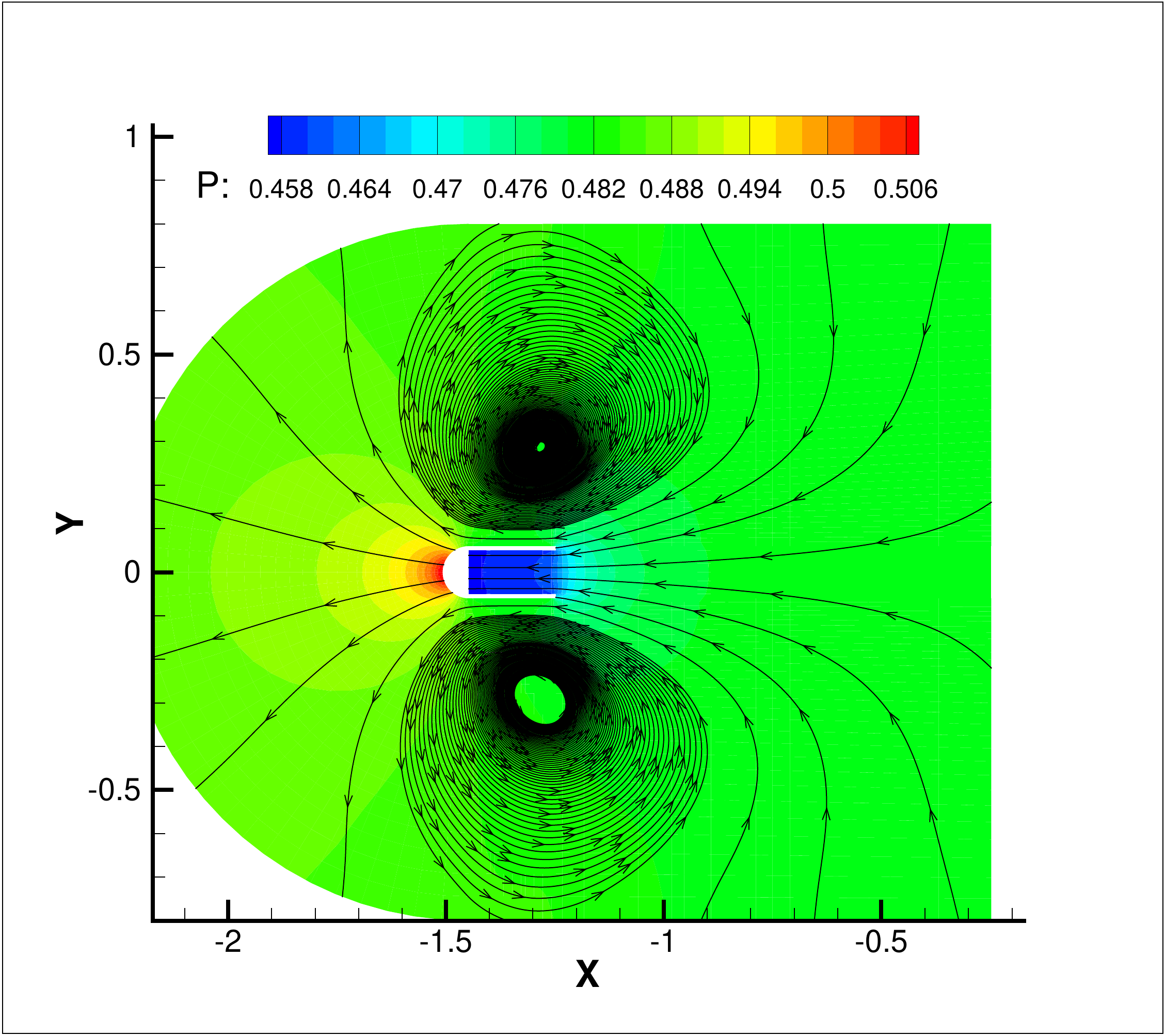}} \hspace{1em}
\subfloat{\includegraphics[width=0.44\textwidth]{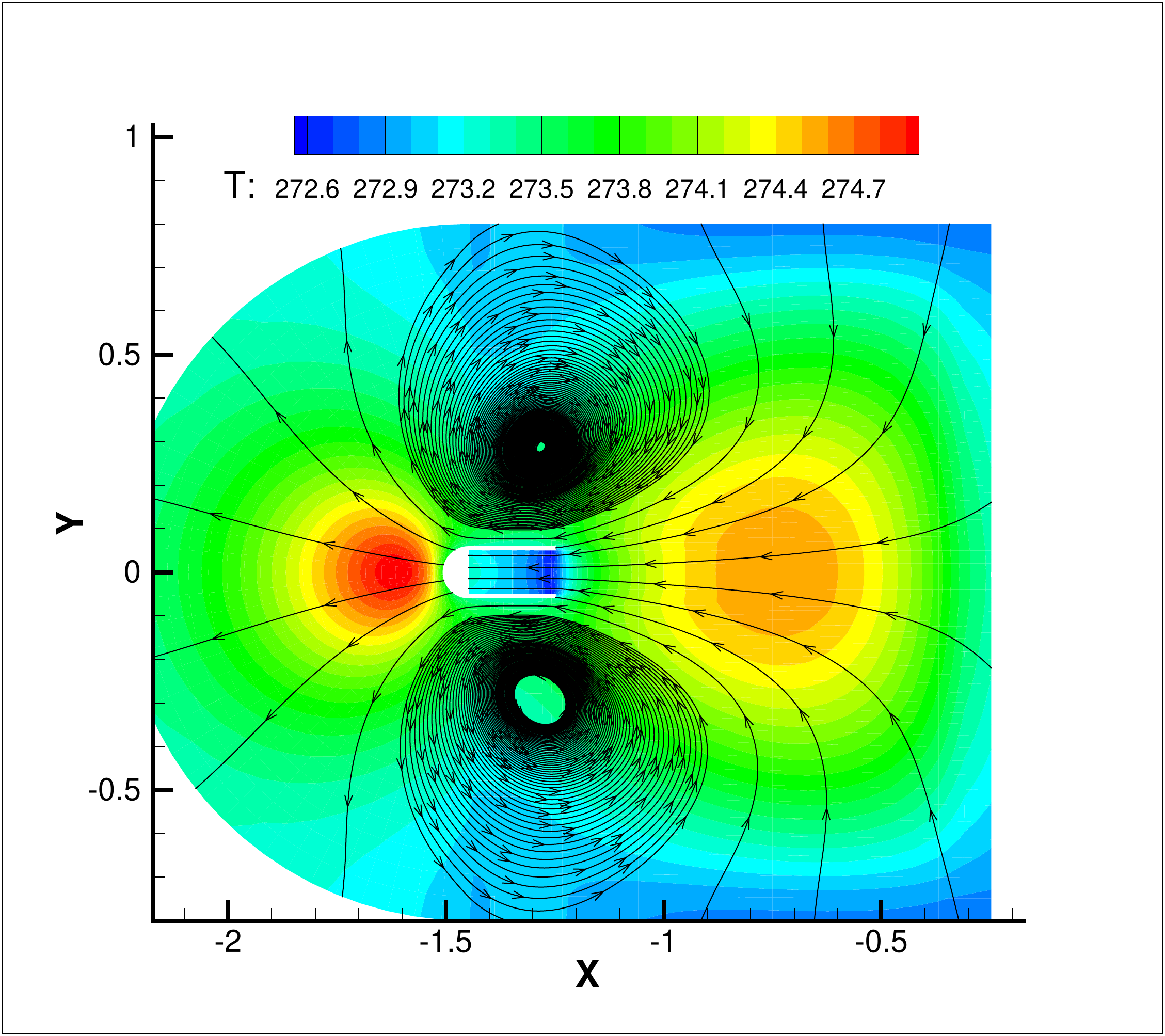}}
\caption{\label{fig:nozzle_deceleration}Flow field during the deceleration stage \cite{chen2012mesh}. (a) Pressure, and (b) temperature.}
\end{figure}

With adaptive mesh in velocity space, Chen et al. \cite{chen2012mesh} computed a moving nozzle flow with the co-existence of both continuum and rarefied regimes in a single computation.
The nozzle is designed to have a rectangle shape with a round head.
The mass of the nozzle is $1.57\times10^{-2} { kg}$.
Both the dense gas inside the nozzle and the dilute gas outside the nozzle are stationary initially.
The temperature is $273 { K}$ for all gas and solid walls.
Density inside the nozzle is $\rho_{in} = 8.582\times 10^{-2} {kg/m^3}$, and outside $\rho_{out} = 8.582\times 10^{-6} { kg/m^3}$, with a viscosity coefficient $2.117\times 10^{-5} {kg/(m \cdot s)}$.
Even with the density ratio of $10^4$, the same cell size is used inside and outside the nozzle in this calculation.
The Knudsen number of the external gas is $0.05$ with the diameter of nozzle head as a reference length.
The whole expansion consists of three stages.
Initially, the gas behaves as a free expansion, and the stream lines become radial due to the rarefied environment outside, see Fig.~\ref{fig:nozzle_expansion}.
Associated with the intensive expansion, the temperature of gas drops quickly to almost $10 { K}$.
The distribution function becomes a narrow-kernel shape in the velocity space.
After free expansion, the follow-up gas out of the nozzle is confined in a narrow region, owing to the reduced density difference outside.
Figure \ref{fig:nozzle_jet} shows the flow field.
At the same time, the nozzle gets accelerated and then moves by its own inertia, due to the re-action force from the expansion.
Eventually, the expansion stage terminates.
Figure \ref{fig:nozzle_deceleration} shows that the pressure inside the nozzle is lower than the surroundings.
The streamline around the nozzle presents a typical characteristic flow at low speed.
Because the environment density is very low, the deceleration is very weak. This stage will sustain a long time until the nozzle gets stopped.

\begin{figure}[htbp]
\centering
\subfloat[]{\includegraphics[width=0.3\textwidth]{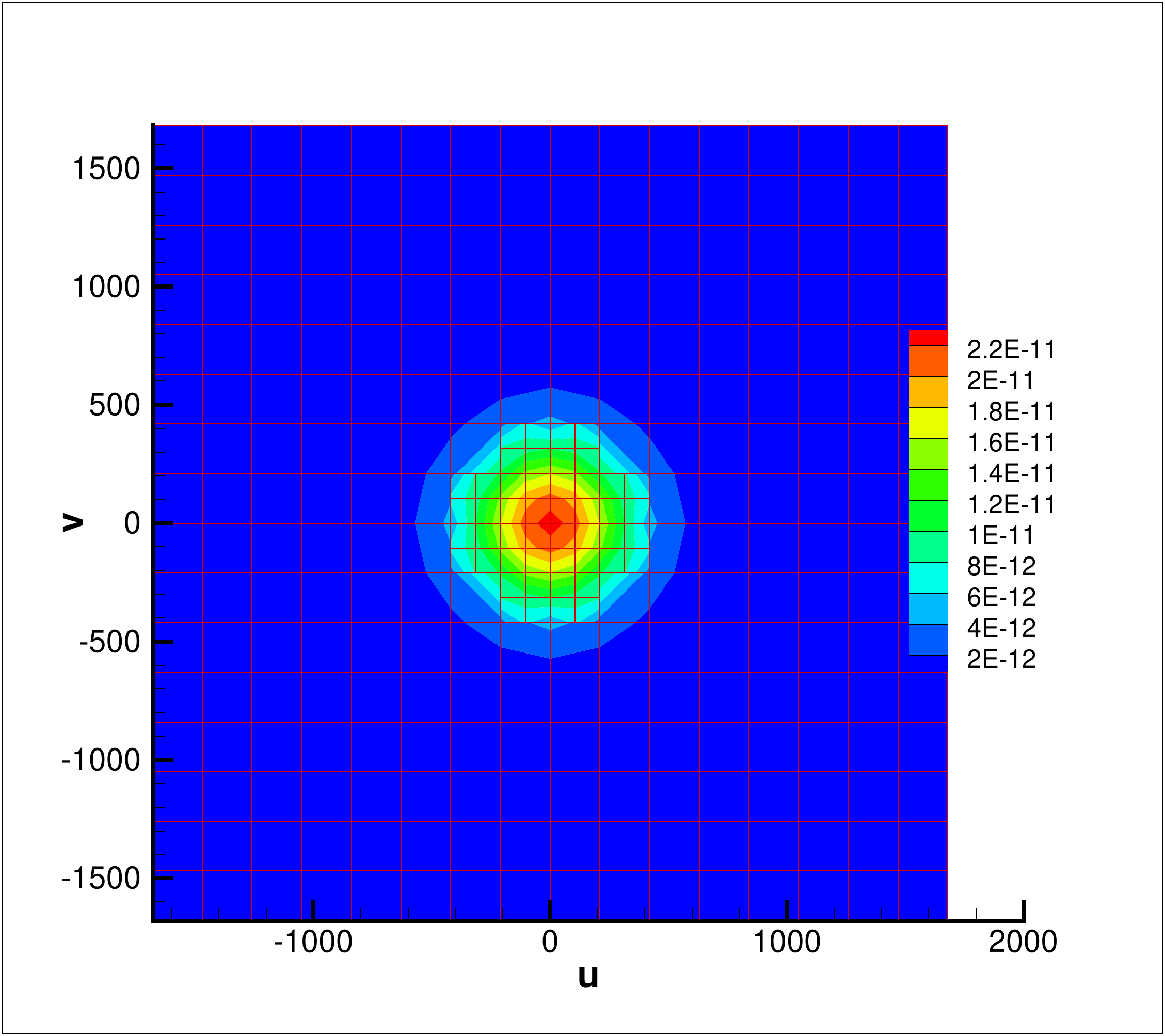}} \hspace{0.5em}
\subfloat[]{\includegraphics[width=0.3\textwidth]{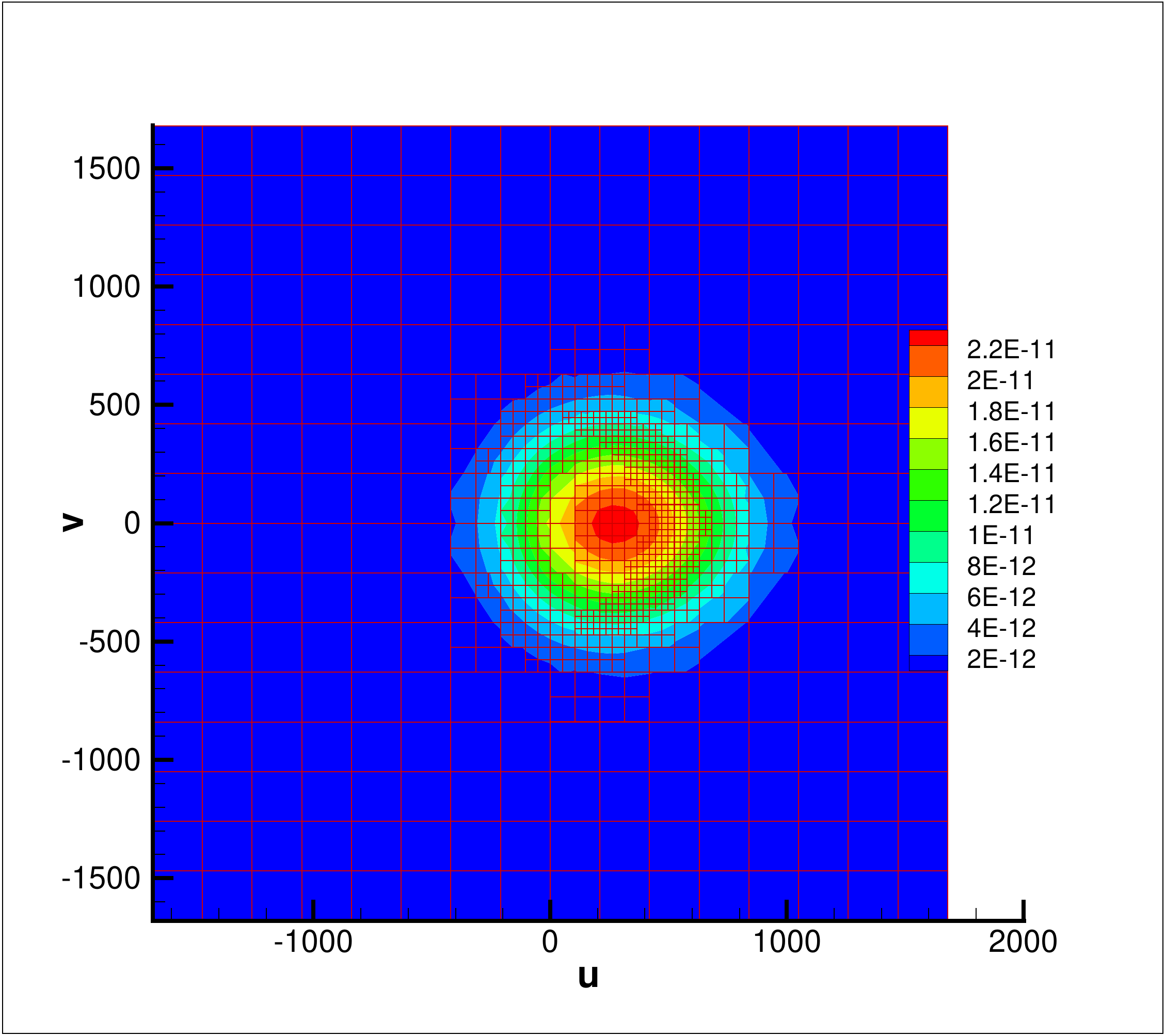}} \hspace{0.5em}
\subfloat[]{\includegraphics[width=0.3\textwidth]{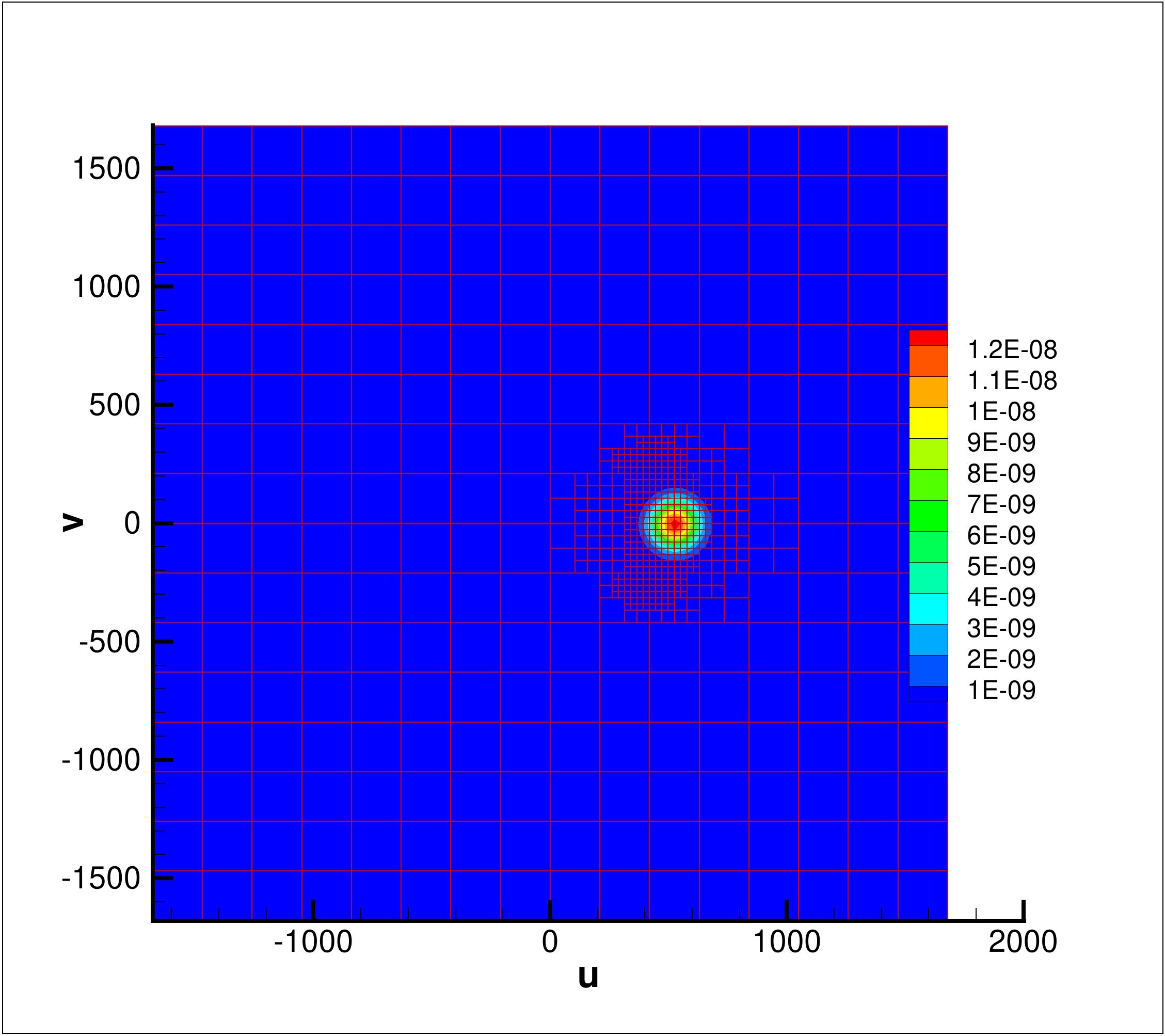}}
\caption{\label{fig:nozzle_pdf}Distribution function in the adaptive velocity space \cite{chen2012mesh}.}
\end{figure}

Figure \ref{fig:nozzle_pdf} shows the quadtree mesh and contour of particle distribution function at a location $(1.2,0)$ for the evolution process.
At first the distribution function keeps a Maxwellian distribution, then the fast particles arrive, and distort the distribution function a little bit.
The corresponding macroscopic picture shows that a shock wave sweeps this location.
The followed expansion makes the temperature drop significantly to a very low value, and the kernel of distribution function becomes much more concentrated than the initial one.

This is a multiscale and challenging problem for any single scale-based numerical scheme.
However, this is exactly the place where the UGKS can be faithfully applied to get accurate solution in different regimes.
For such a multiscale problem, a hybrid approach with domain decomposition may usually be adopted \cite{sun2004hybrid,hash1996assessment,carlson2004hybrid}.
While, for the unsteady case, the hybrid approach can be hardly applied due to the absence of a valid interface.

\subsection{Memory reduction}
For rarefied flow study, direct Boltzmann solver based on the discrete velocity distribution function is a common approach to simulate the non-equilibrium flow.
However, it requires a huge memory to follow the evolution of discrete distribution function in a computational space with seven dimensions.
It would become more severe for numerical computation of high-speed rarefied flows, where very large region in velocity space has to be covered.
For steady state simulation, a memory reduction technique was proposed by Chen et al. \cite{chen2017memory,xu2017overcome} for solving the Boltzmann kinetic model equations.
Yang et al. \cite{yang2018memory} applied the memory reduction method in the implicit UGKS.
Due to the fact that only the spatial connection and variation take effect in a steady state solution, the non-equilibrium distribution function can be obtained from spatial iterations at each velocity point, therefore only the memory for distribution function at one discrete velocity point in the physical domain is required, which reduces the memory cost into the same level as that in the hydrodynamic solvers for Euler and Navier--Stokes solutions.

\begin{figure}[htbp]
\centering
\subfloat[]{\includegraphics[width=0.6\textwidth]{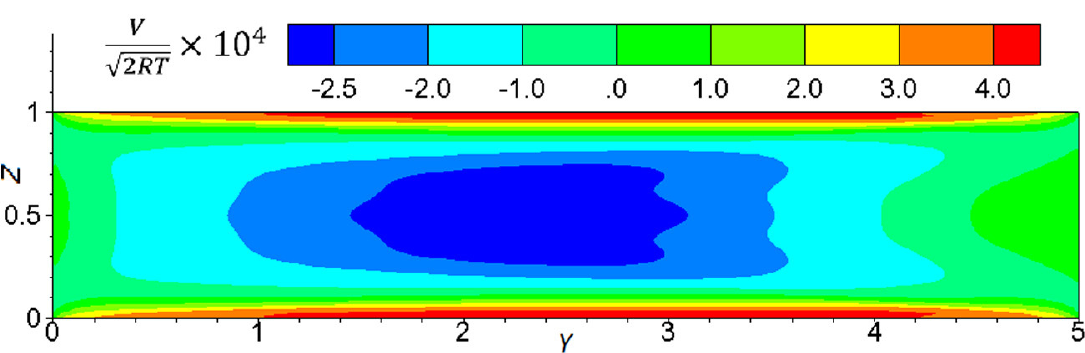}} \\
\subfloat[]{\includegraphics[width=0.31\textwidth]{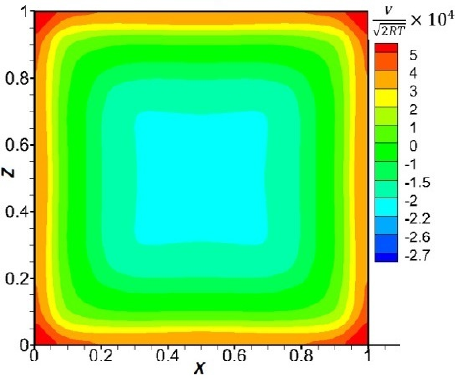}}
\subfloat[]{\includegraphics[width=0.31\textwidth]{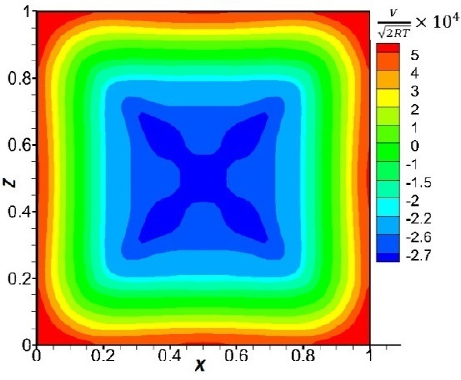}}
\subfloat[]{\includegraphics[width=0.31\textwidth]{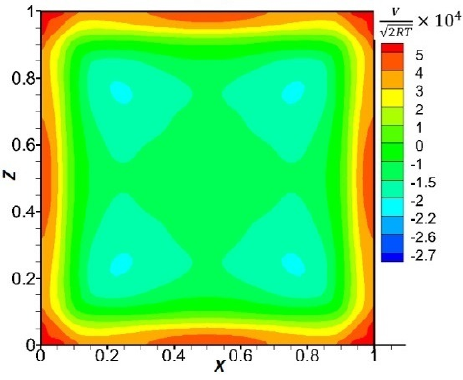}}
\caption{\label{fig:longtube} $y$-component velocity distribution on different slices of flow field \cite{chen2017memory}. (a) $x=0.5$, (b) $y=1$, (c) $y=2.5$, and (d) $y=4$. }
\end{figure}

Chen et al. \cite{chen2017memory} computed a thermal transpiration flow in a three-dimensional square cube with closed walls.
The temperature of the solid walls varies spatially from a high temperature on the left to a lower temperature on the right.
Figure \ref{fig:longtube} shows the $y$-component velocity distribution on different slices of flow fields.
For this computation, the spatial domain is discretized by a structured mesh with $60\times60\times120$ cells, and the velocity space employs $101\times101\times101$ discrete velocity points.
If the distribution function at all the discrete velocity space is stored, the memory requirement would be approximately 13 terabytes (TB) with double precision calculation. Here, only one discrete distribution function value is stored in each cell center and cell face for each discrete velocity.
With the memory reduction technique, the observed data storage for a parallel code is only 181 megabytes (MB) for each process in a total of 128 MPI processes.
The memory technique makes it possible to calculate the three dimensional problems using a discrete velocities based kinetic method on a small cluster.

\subsection{Wave-particle adaptation}

For rarefied gas simulations, the DSMC method is more preferable in the engineering applications than the discrete velocity method, due to the fact that the stochastic particle method requires much less computational cost than the discrete velocity method, especially for high speed calculations \cite{bird1990application,blanchard1997aerodynamic,fang2020dsmc}.
Different from the discrete velocity method where the discrete velocities are predetermined, the stochastic particle method  can be regarded as the optimal velocity space adaptation method, because the particles collide and change their velocity, and appear in the phase space where there should be a non-zero distribution function.

\begin{figure}[htbp]
\centering
\includegraphics[width=0.5\textwidth]{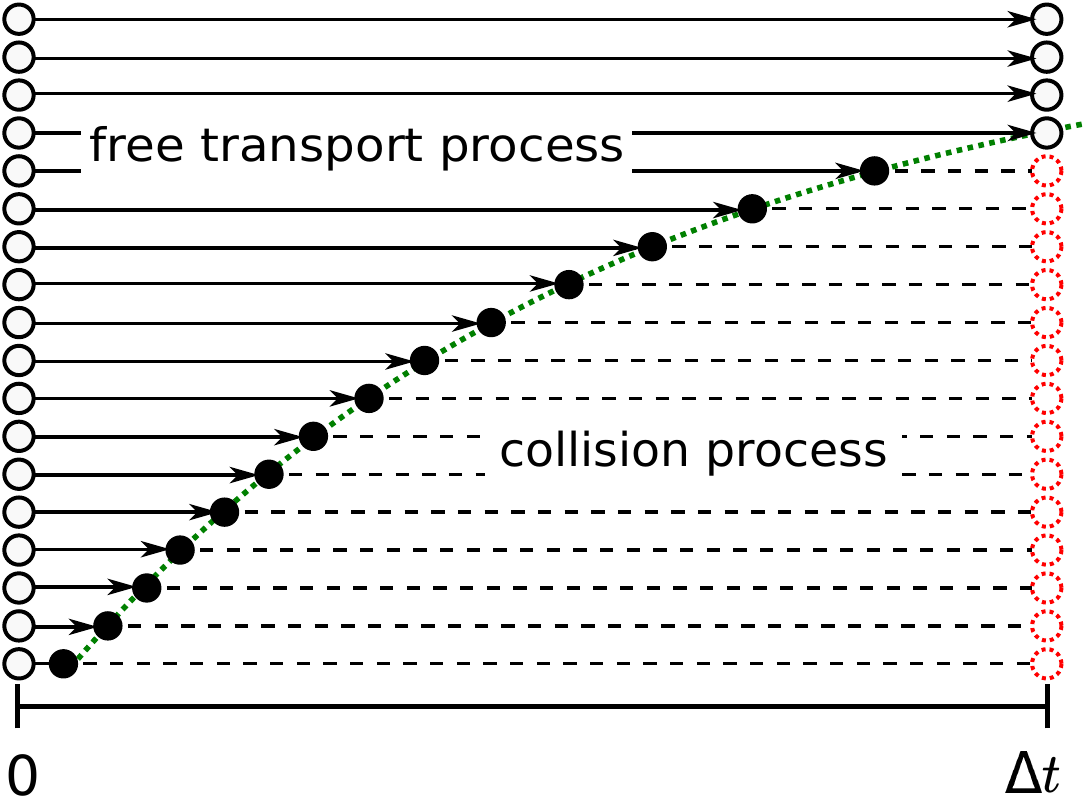}
\caption{\label{fig:particle_modeling}Free transport and collision processes of stochastic particles \cite{zhu2020thesis}.}
\end{figure}

In order to incorporate the adaptive property of the stochastic methods, the unified gas-kinetic particle (UGKP) method has been developed \cite{li2018ugkp,shi2020ugkp,shi2020improved,wang2020ugkp} according to the direct modeling methodology, where stochastic particles are employed to record the non-equilibrium distribution, and the evolution processes are carried out on both the macroscopic and microscopic levels.
The UGKP method recovers the UGKS solutions by following the local evolution solution of the kinetic models, which is the key to construct a multiscale method for all Knudsen number flows.
Specifically, the particles' evolution described by the gas distribution function follows
\begin{equation}
f(\vec{x}, t) = (1-e^{-t/\tau}) g(\vec{x}^\prime, t^\prime) + e^{-t/\tau} f_0(\vec{x}-\vec{u} t),
\end{equation}
where
\begin{equation}\label{eq:particle_evolution_equation}
\vec{x}^\prime = \vec{u} \left(\frac{t e^{-t/\tau}}{1-e^{-t/\tau}} - \tau\right), \quad
t^\prime = \frac{t}{1-e^{-t/\tau}} - \tau.
\end{equation}
The evolution process of the microscopic particles is illustrated in Fig.~\ref{fig:particle_modeling}.
According to Eq.~\eqref{eq:particle_evolution_equation}, only a portion of particles can move freely in a time step, and the others will encounter collisions with other particles during the transport process.
Define the process before the particles taking their first collision as the free transport process, and the following process of particles' collision and streaming as the collision process, the evolution can be described as that in the free transport process particles keeps free streaming for a specific time carrying the initial information, which recovers the term $f_0$ in Eq.~\eqref{eq:particle_evolution_equation}, and in the collision process, particles get collided and continue streaming and collision, which makes the collision particles' distribution approaching to the local equilibrium.
The free transport process can be fully resolved by tracking all the particles, and the contribution of the particles' free motion to the macroscopic fluxes across cell interface can be recorded as well.
Different from the particles that could move freely within a whole time step, the collided particles will encounter collision during a time step evolution.
Since a portion of particles will suffer many times of collision during the collision process, the colliding particles will not be followed by using particle tracking methods to evolve to the end of time step.
Fortunately, the collective behavior of the colliding particles in the collision process can be well modeled on the macroscopic level using the calculation of the equilibrium state $\mathcal{F}_{ij}^{eq}$ in Eq.~\eqref{eq:fij_flux}, which is only related to the macroscopic flow variables and gradients.
Similar to the UGKS, the conservative variables can be updated first with the fluxes contributed in the free transport and collision processes, and then the colliding particles will be re-sampled from a local equilibrium state according to their corresponding conservative variables.
By this way, the UGKP method is constructed based on the conservation laws in the framework of finite volume method, where multiscale flow physics is recovered by driving the particles' evolution according to the integral solution of kinetic model.

\begin{figure}[htbp]
\centering
\includegraphics[width=0.8\textwidth]{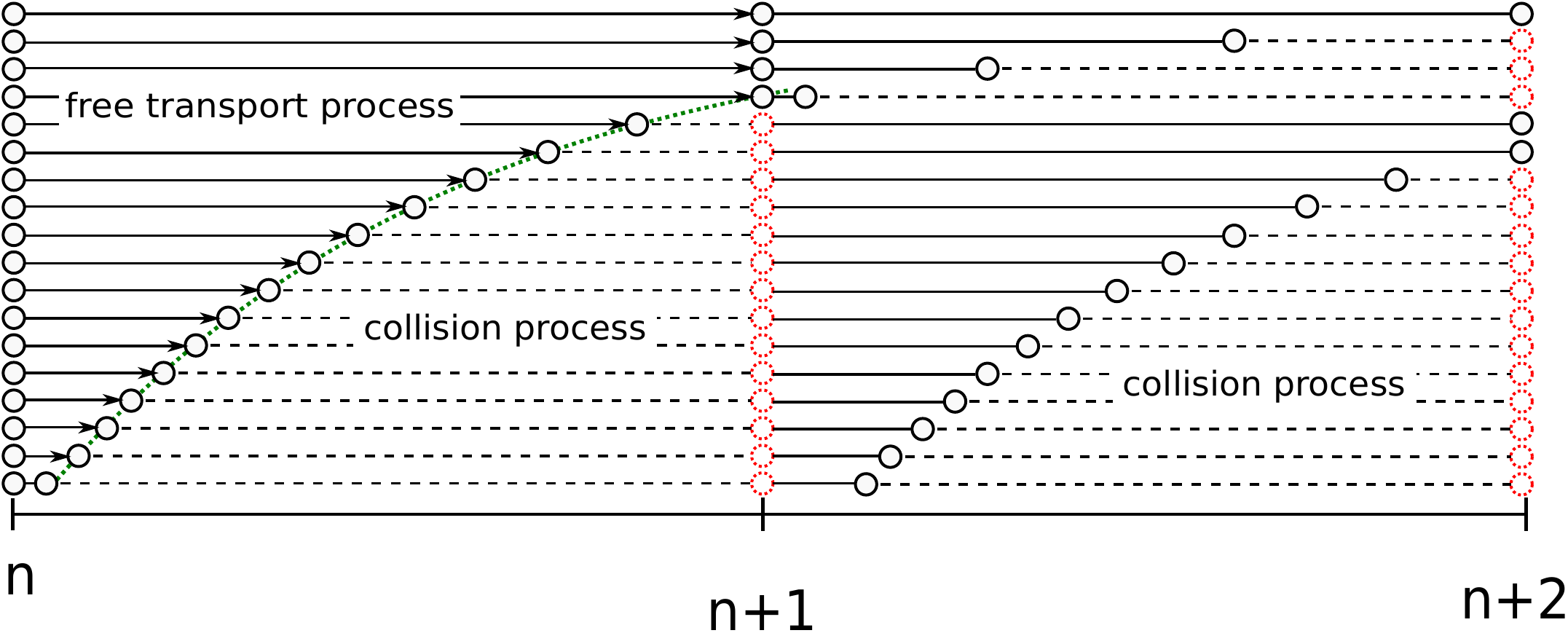}
\caption{\label{fig:wp_modeling}Free transport and collision processes of wave-particle formulation \cite{zhu2020thesis}.}
\end{figure}

Considering the fact that the colliding particles are re-sampled from a known equilibrium state, there is no need to describe the equilibrium state by discrete particles, but using the analytic expression itself with high efficient and low memory consumption.
Therefore, a novel adaptive wave-particle formulation is proposed to describe the multiscale flow physics, resulting in the unified gas-kinetic wave-particle (UGKWP) method \cite{zhu2019ugkwp,liu2020ugkwp,li2020ugkwp,chen2020ugkwp,xu2021tau}.
The idea of wave-particle evolution can be illustrated by Fig.~\ref{fig:wp_modeling}.
It can be found that if the colliding particles (denoted by red dashed circle) are not re-sampled at the time $t^{n+1}$ and instead expressed by analytic formulation, its contribution to the free transport and collision can be analytically computed.
It only needs to sample a portion of particles that becomes free transport in the whole time step from $t^n$ to $t^{n+1}$.
There is a competitive balance between the analytic waves and stochastic particles according to local physics.
Particles become waves due to collisions, and waves transform into particles in order to capture the partially free transport mechanism.

With the adaptive wave-particle formulation, the UGKWP method can reduce to a particle method in the highly rarefied flow regimes, and becomes a hydrodynamic NS solver in the continuum regimes automatically.
It is a novel technique which combines and optimizes the adaptive velocity space method \cite{chen2012mesh} and GKS/UGKS hybrid algorithm \cite{xiao2020velocity}.
Different from the hybrid methods \cite{wadsworth1990one,sun2004hybrid,degond2010hybrid} which are based on the domain decomposition and solver hybridization, the UGKWP method describes the physical state by an adaptive wave-particle decomposition in each cell with a unified treatment in the whole computational domain.

\subsubsection{Laminar boundary layer}
The laminar boundary layer over a flat plate has been computed to validate the UGKWP method for viscous NS solutions in the continuum limit \cite{zhu2019ugkwp,liu2020ugkwp}.
The computational domain is $[-50, 120]\times[0, 50]$, where a non-uniform mesh with $120\times50$ cells is employed.
The free stream is monatomic gas flow at Reynolds number ${ Re} = 10^{5}$ and Mach number ${ Ma}=0.3$ with constant viscosity.
The Reynolds number and Mach number is defined with respect to the length of the flat plate $L = 120 L_0$.
The reference variables $U_0$ and $t_0$ are used to non-dimensionalize the velocity and time by $U_0 = \sqrt{2 k_B T_0 / m_0}$ and $t_0 = {L_0}/{U_0}$, where $T_0$ is the temperature in the free stream.

\begin{figure}[htbp]
\centering
\subfloat[]{\includegraphics[width=0.48\textwidth]{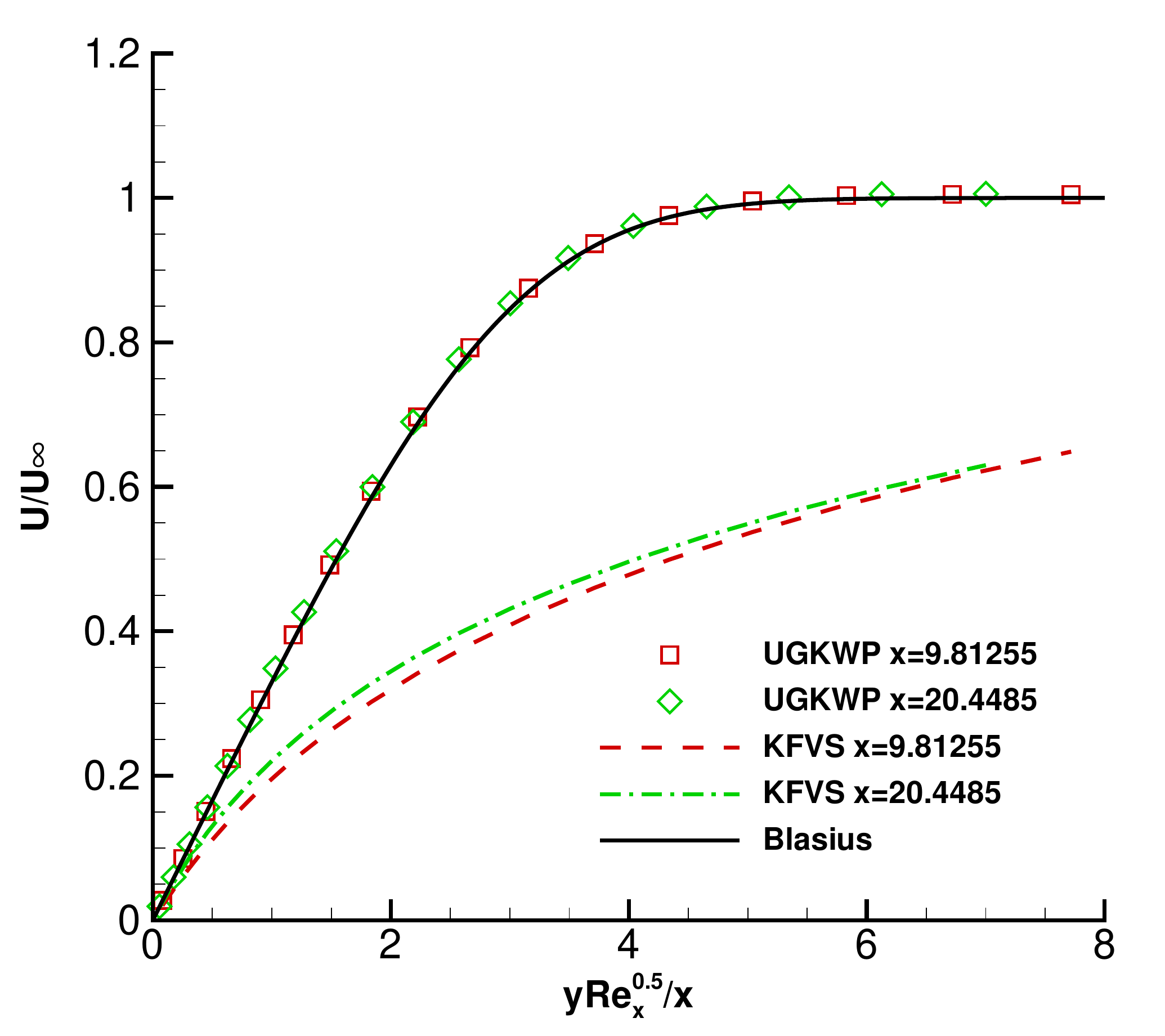}}
\subfloat[]{\includegraphics[width=0.48\textwidth]{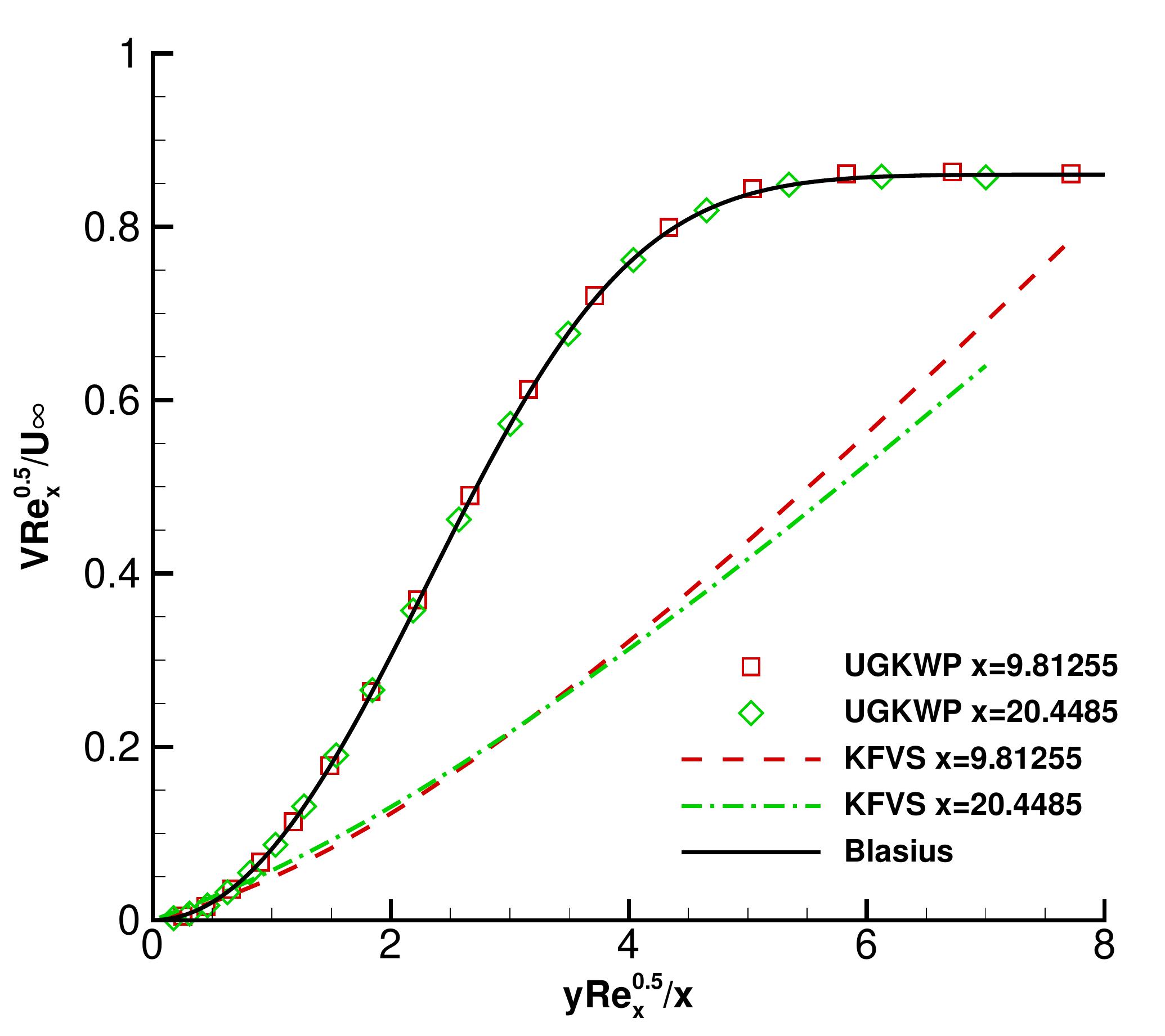}}
\caption{\label{fig:bl_line}Velocity distribution in the laminar boundary layer obtained by the UGKWP method\cite{zhu2020thesis}. (a) Normalized $x$-component velocity and (b) normalized $y$-component velocity.}	
\end{figure}

Numerical solutions at time $t=1000$ is obtained as the convergent steady state solution.
Fig.~\ref{fig:bl_line} shows the comparison among the UGKWP results, the Blasius solutions, and KFVS results.
Since the flux evaluation of the kinetic flux vector scheme (KFVS) is based on the free transport mechanism, the numerical dissipation overwhelms the physical viscosity.
In the computation, the time step $\Delta t$ and particle collision time $\tau$ are $0.02$ and $6.57\times10^{-4}$, respectively.
Since the ratio $e^{-\Delta t/\tau}$ has a very small value of $6\times10^{-14}$, the hydrodynamic wave is dominant and the particle contribution can be neglected.
The computational time for $50000$ step simulations is $15$ minutes and the memory cost is $24$ MB.
Under such condition, the UGKWP method automatically becomes a hydrodynamic fluid solver, such as the GKS \cite{xu2001gks,xu1994numerical}.
Due to the multiscale transport, the UGKWP method can recover NS solutions without the requirement of the mesh size and the time step being less than the mean free path and the particle collision time.
Moreover, the computational cost is comparable to the hydrodynamic fluid solver in the continuum regime, which is much more efficient than the UGKS with a discretized particle velocity space.

\subsubsection{Flow around a circular cylinder}
Hypersonic flow past a circular cylinder have been simulated to show the capability of the UGKWP method for high-speed rarefied flow simulations.
The free stream is initialized with the monatomic gas flow of argon with an initial temperature $T_{\infty} = 273 { K}$.
The diameter $D$ of the cylinder is $1{ m}$ long.
The solid boundaries are isothermal walls with a constant temperature $T_{w} = 273 { K}$.
The Knudsen number is defined with respect to the diameter of the cylinder.

\begin{figure}[htbp]
	\centering
	\subfloat[]{\includegraphics[width=0.32\textwidth]{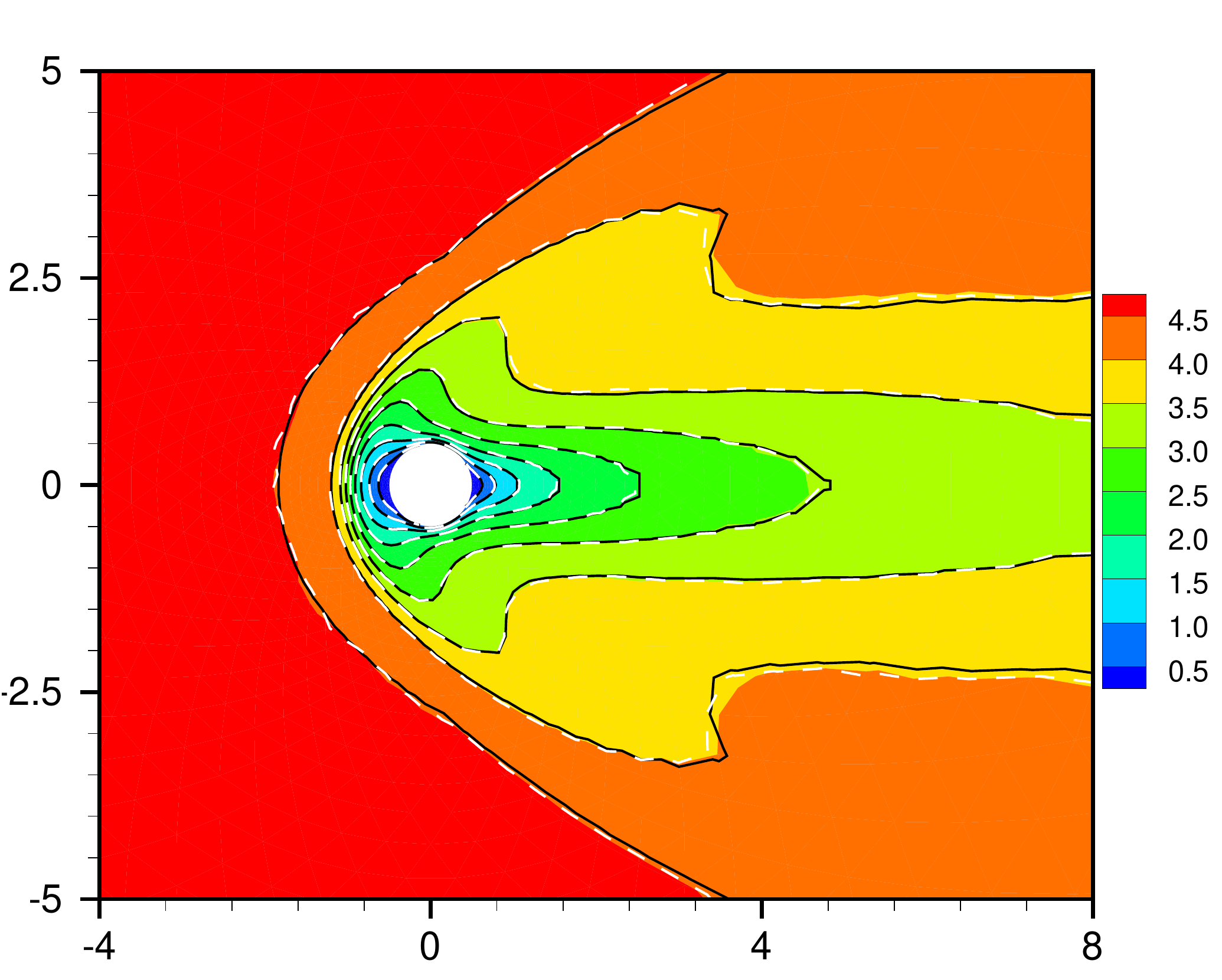}}
	\subfloat[]{\includegraphics[width=0.32\textwidth]{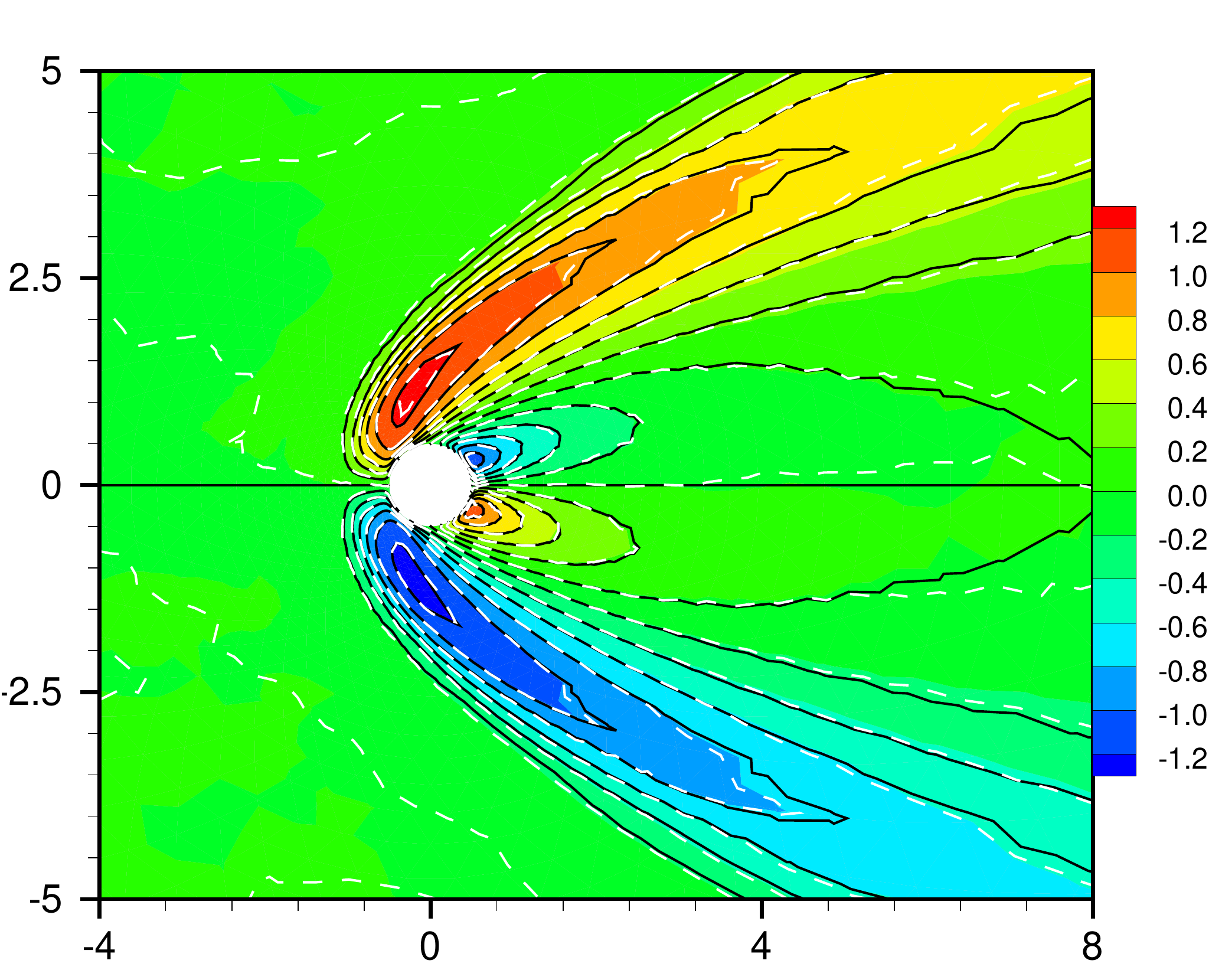}}
	\subfloat[]{\includegraphics[width=0.32\textwidth]{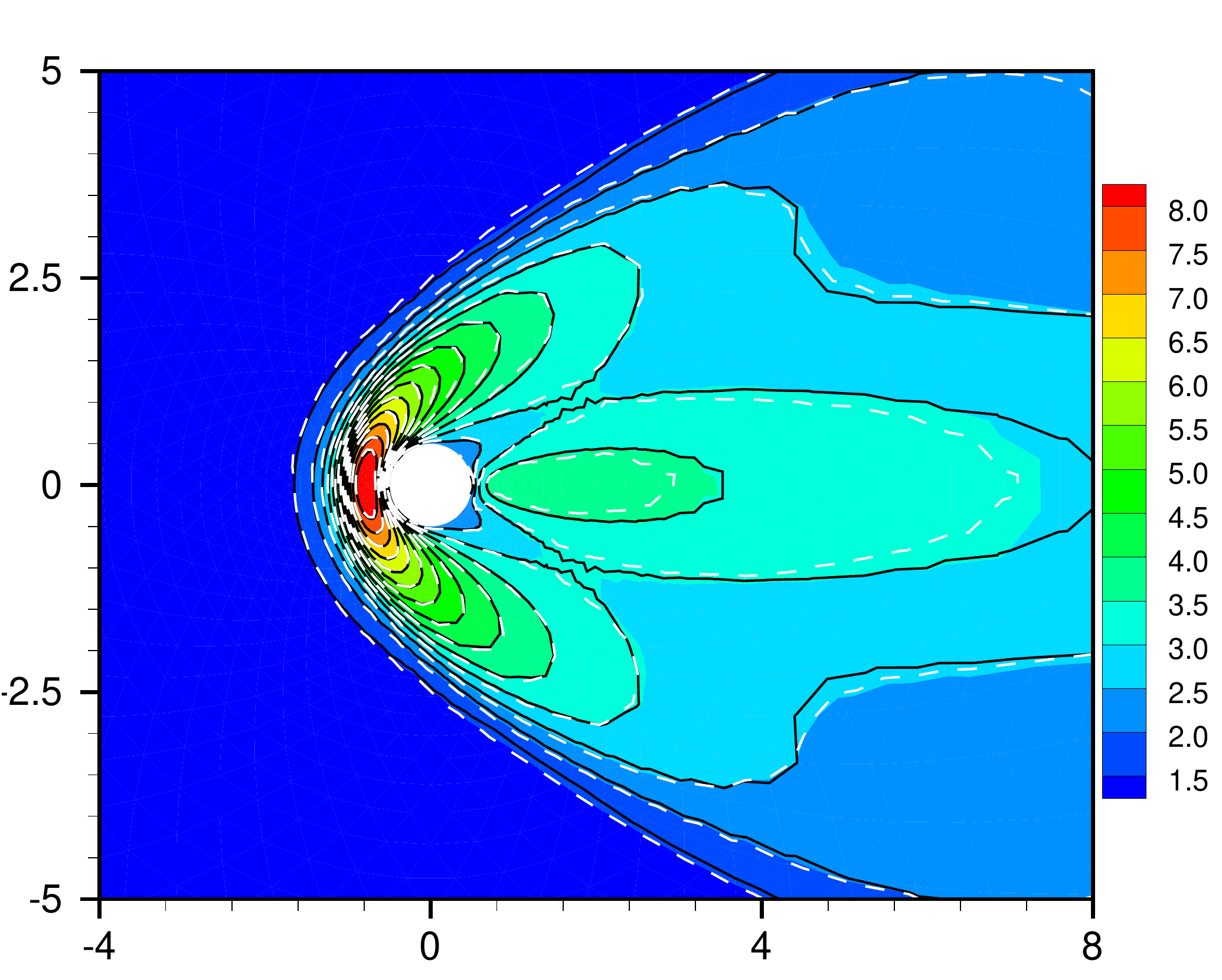}}
	\caption{\label{fig:cylinder_kn01} Hypersonic flow at ${ Ma}=5$ around a circular cylinder at ${ Kn} = 0.1$ \cite{zhu2020thesis}. (a) $x$-component velocity, (b) $y$-component velocity, and (c) temperature.}
\end{figure}

\begin{figure}[htbp]
	\centering
	\subfloat[]{\includegraphics[width=0.32\textwidth]{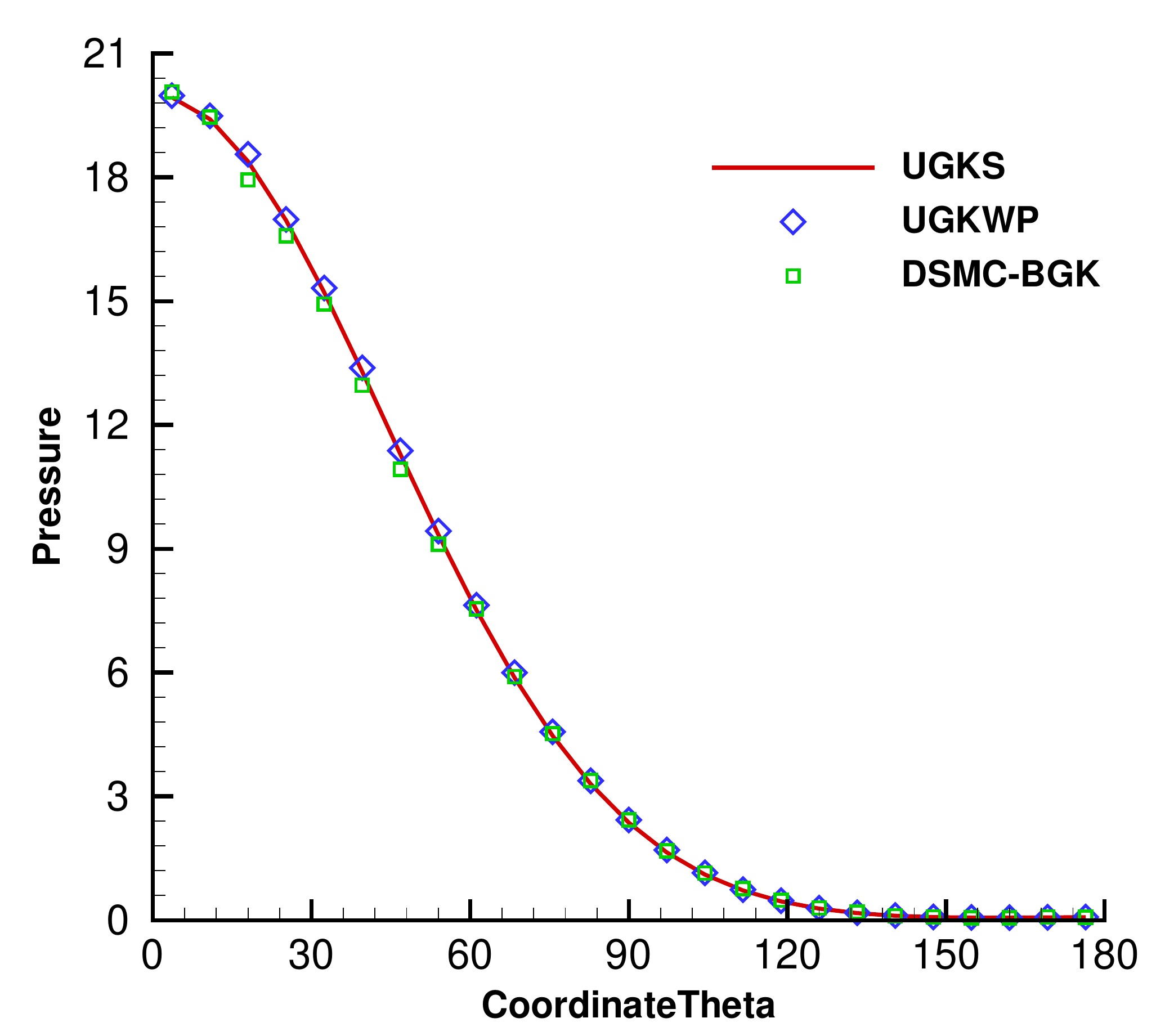}}
	\subfloat[]{\includegraphics[width=0.32\textwidth]{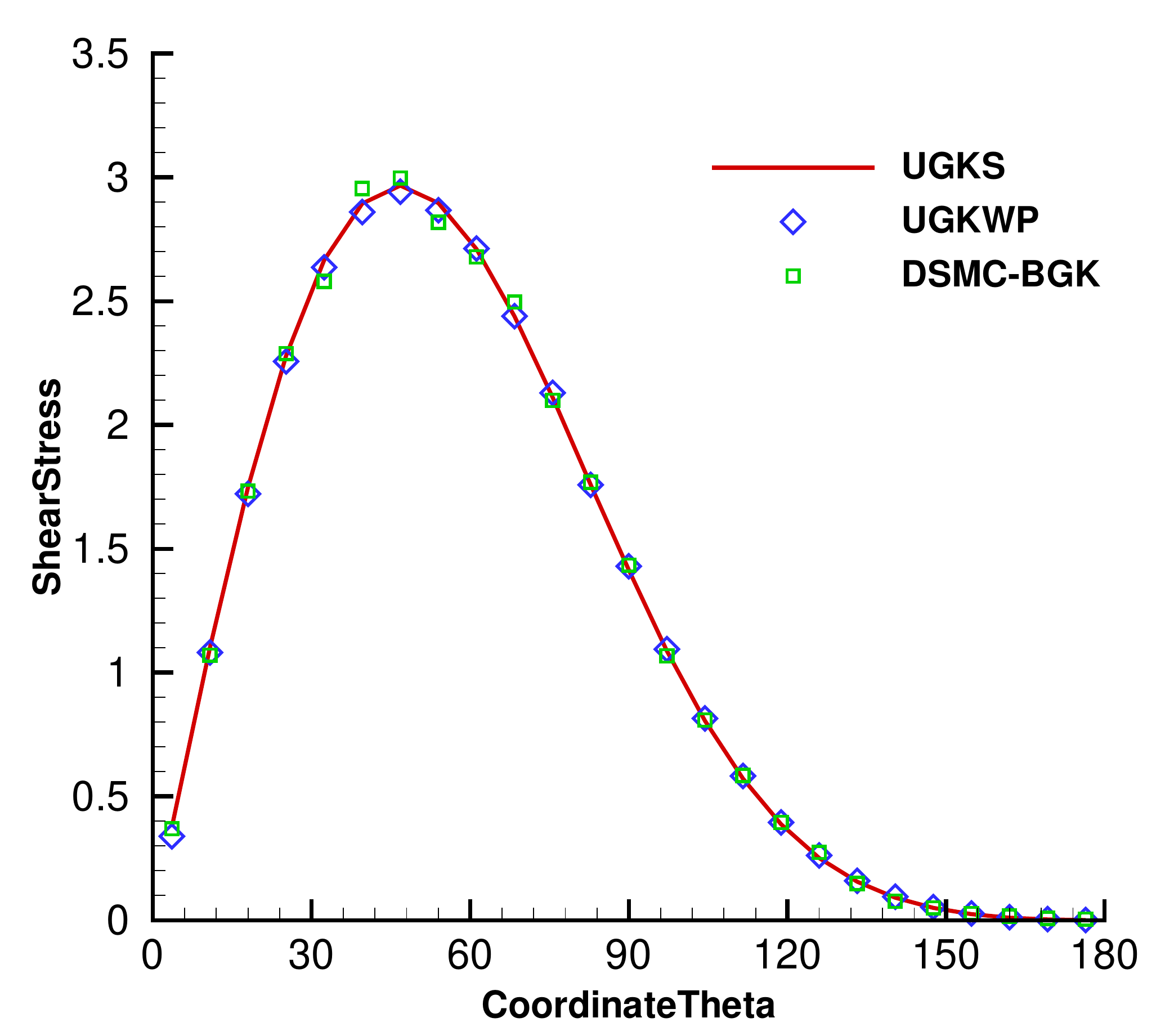}}
	\subfloat[]{\includegraphics[width=0.32\textwidth]{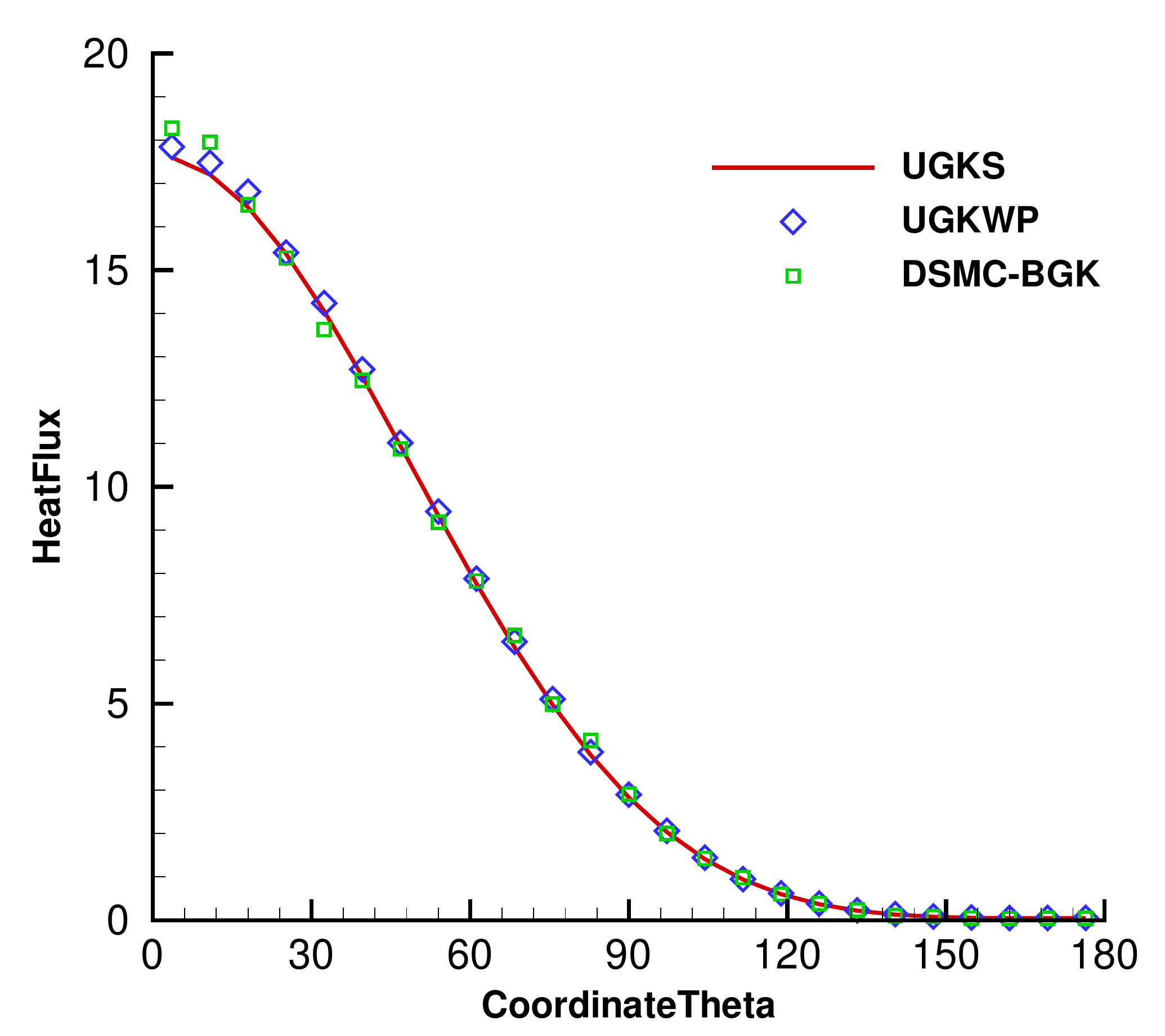}}
	\caption{\label{fig:cylinder_surface_kn01}Surface quantities around a circular cylinder at ${ Ma}=5$ and ${ Kn}=0.1$ \cite{zhu2020thesis}. (a) Pressure, (b) shear stress, and (c) heat flux.}
\end{figure}

\begin{figure}[htbp]
	\centering
	\subfloat[]{\includegraphics[width=0.32\textwidth]{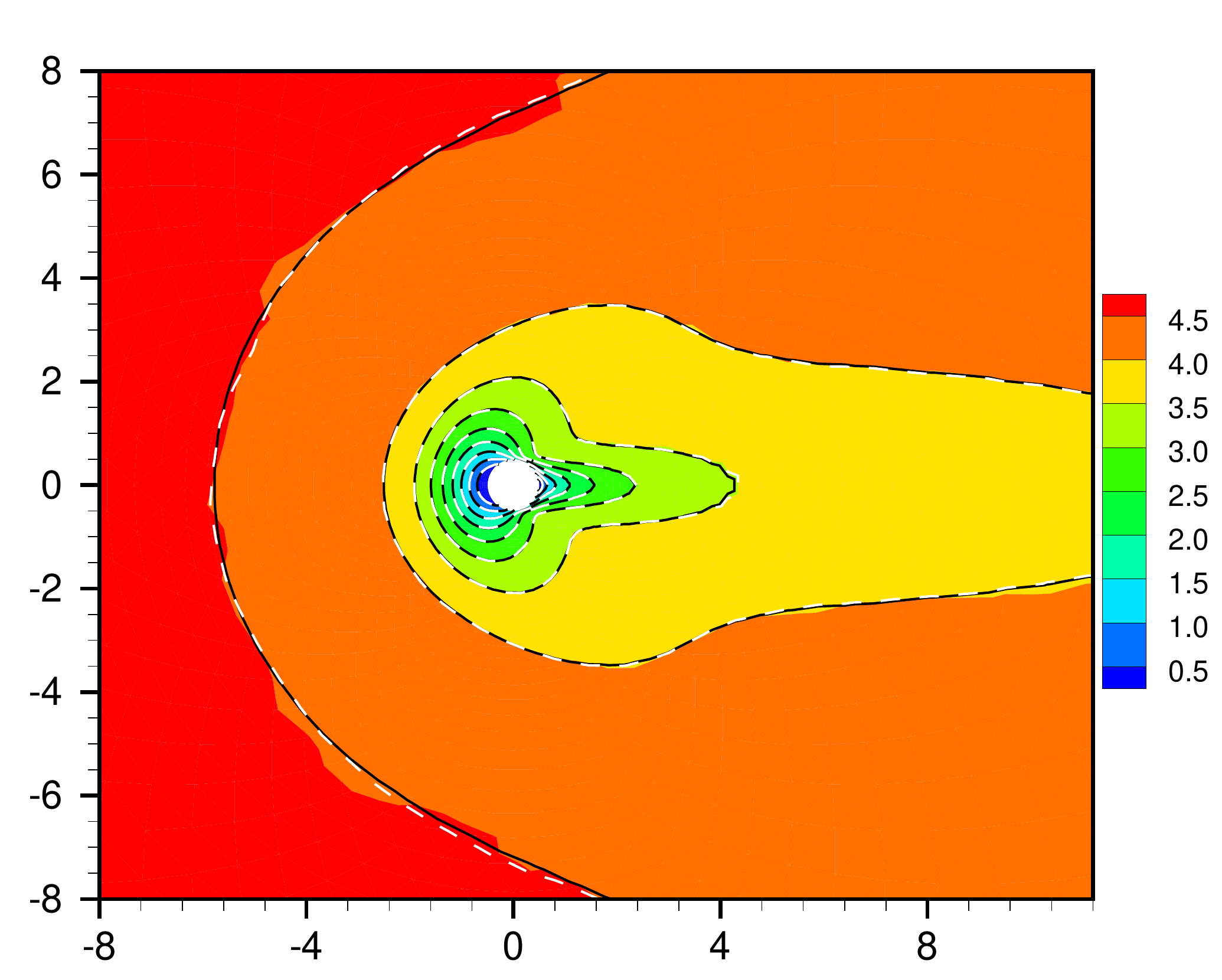}}
	\subfloat[]{\includegraphics[width=0.32\textwidth]{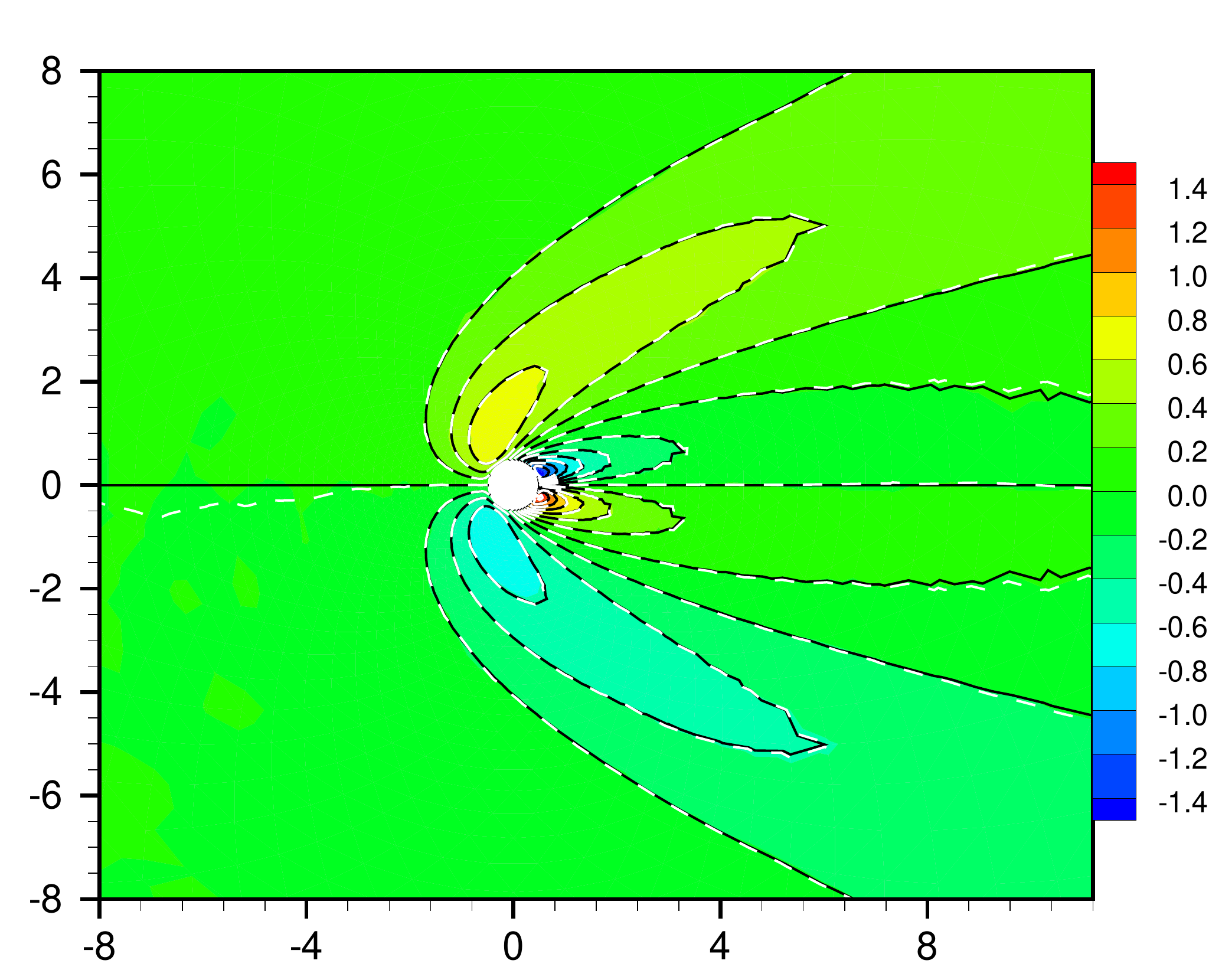}}
	\subfloat[]{\includegraphics[width=0.32\textwidth]{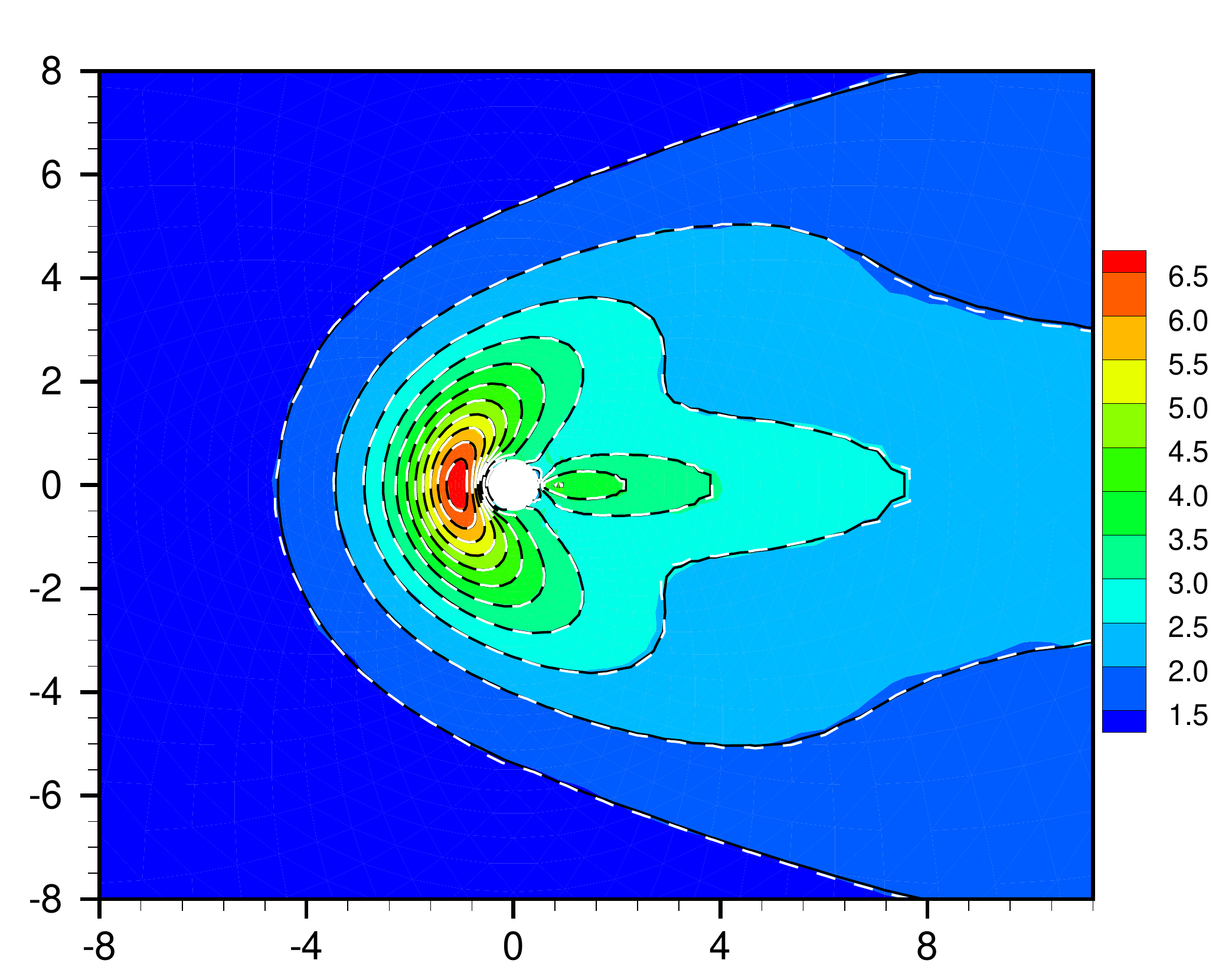}}
	\caption{\label{fig:cylinder_kn1}Hypersonic flow at ${ Ma}=5$  around a circular cylinder for ${ Kn} = 1$ \cite{zhu2020thesis}. (a) $x$-component velocity, (b) $y$-component velocity, and (c) temperature.}
\end{figure}

\begin{figure}[htbp]
	\centering
	\subfloat[]{\includegraphics[width=0.32\textwidth]{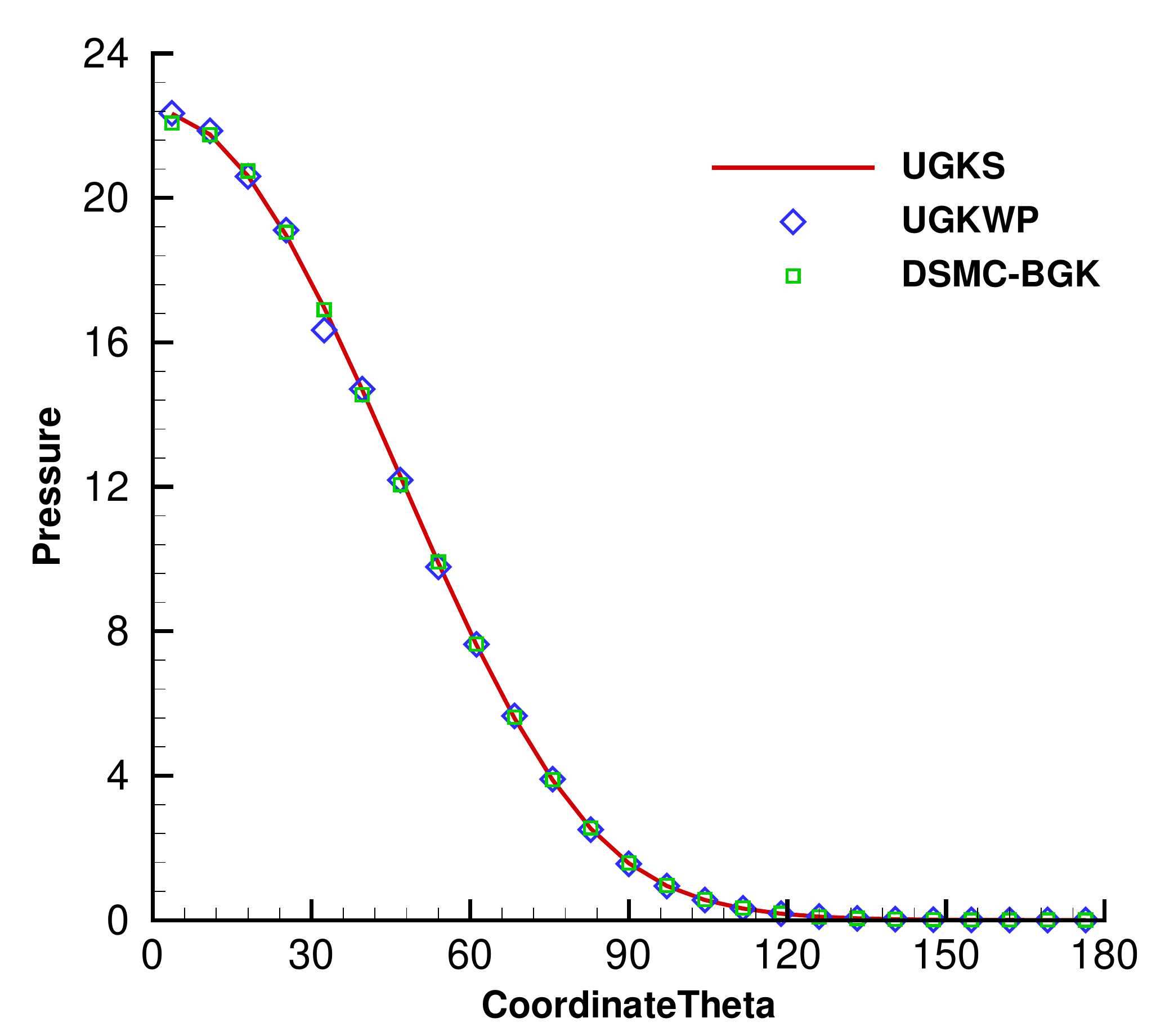}}
	\subfloat[]{\includegraphics[width=0.32\textwidth]{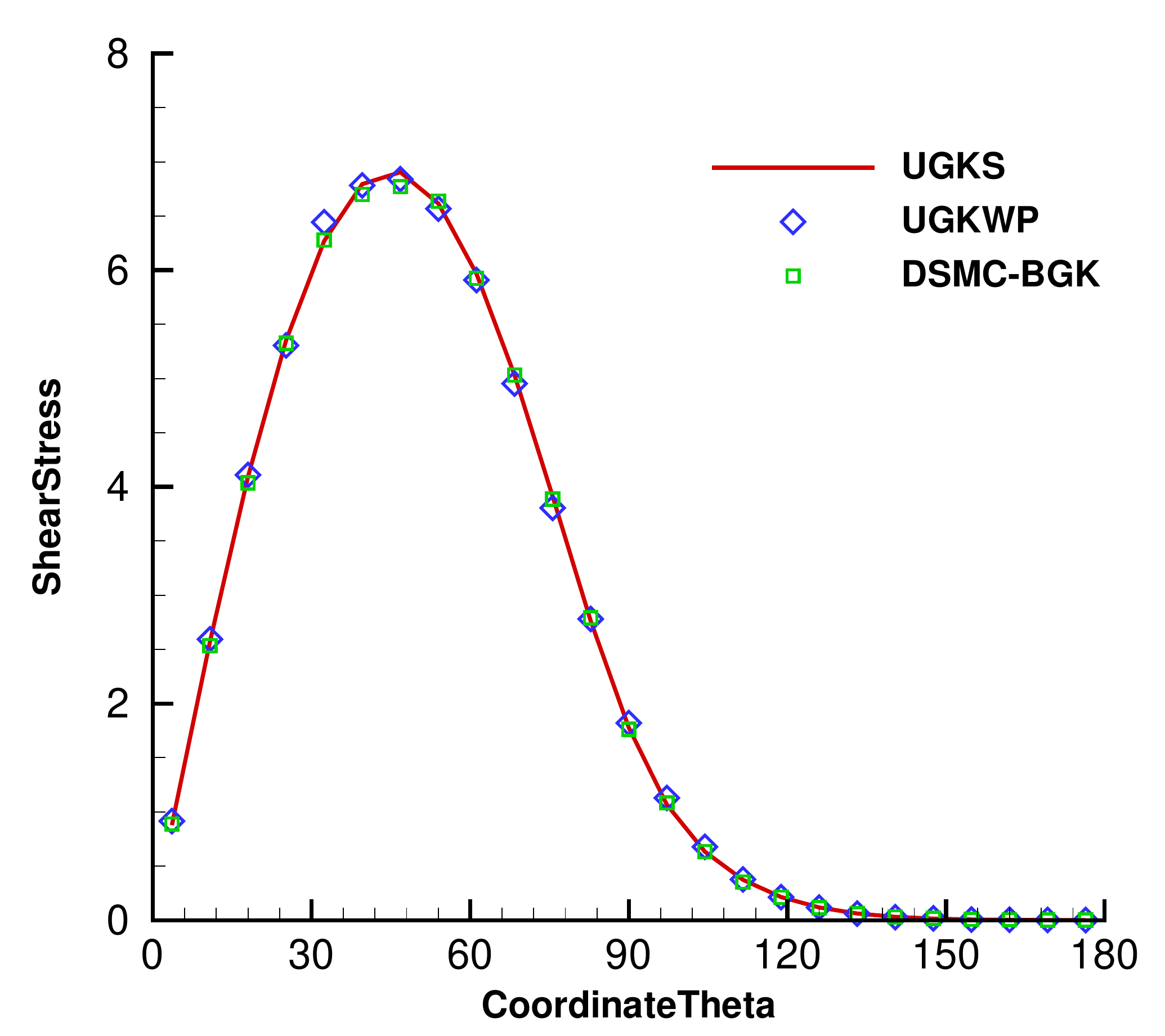}}
	\subfloat[]{\includegraphics[width=0.32\textwidth]{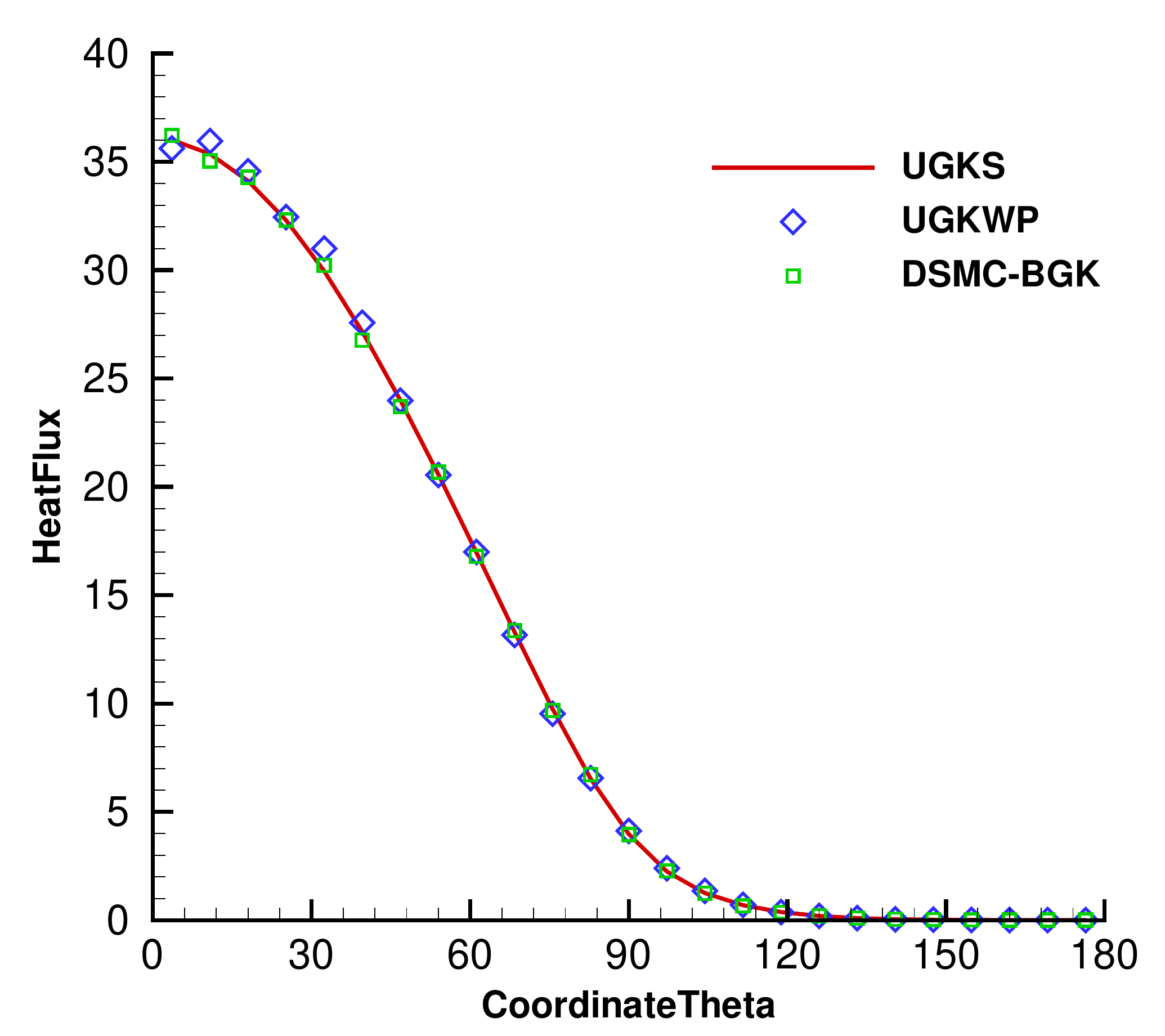}}
	\caption{\label{fig:cylinder_surface_kn1}Surface quantities around the circle cylinder at ${ Ma}=5$ and ${ Kn}=1$ \cite{zhu2020thesis}.  (a) Pressure, (b) shear stress, and (c) heat flux.}
\end{figure}

For the free stream with a relatively low Mach number ${ Ma}=5$, the cases at the Knudsen numbers $0.1$ and $1$ are computed.
The computational domain is discretized by $50 \times 50 \times2$ triangular cells, which covers a region of $\pi (15D)^2$.
Along the radial direction, the minimum heights of the triangles near the boundaries are $0.01{ m}$ and $0.03{ m}$ for ${ Kn} = 0.1$ and $1$, respectively.
The UGKS employs $100\times100$ velocity points in the velocity space, and the initial reference number of particles $N_r$ for the UGKWP method is set as $400$.
In comparison with the UGKS solutions, the flow fields computed by the UGKWP method are shown in Figs.~\ref{fig:cylinder_kn01} and \ref{fig:cylinder_kn1}.
It can be seen that the UGKWP results agree well with those obtained from the UGKS computations.
Detailed comparisons of the surface quantities, such as the pressure, shear stress, and heat flux, are given in Figs.~\ref{fig:cylinder_surface_kn01} and  \ref{fig:cylinder_surface_kn1}.
The UGKS solutions are fully recovered by the UGKWP method on the unstructured meshes, but with one-order-of-magnitude lower in computational cost and memory consumption in the UGKWP computation.

\begin{figure}[htbp]
	\centering
	\subfloat[]{\includegraphics[width=0.32\textwidth]{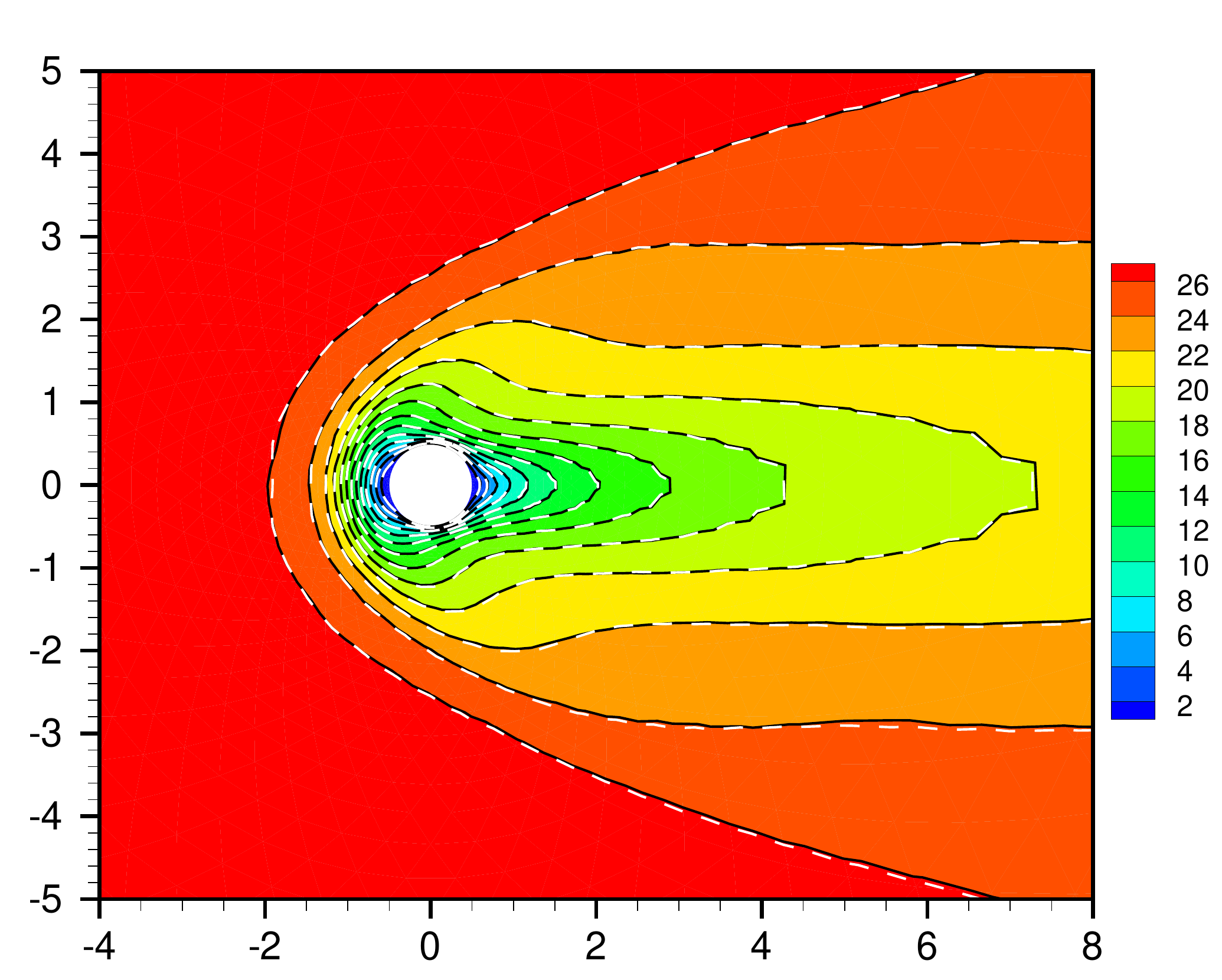}}
	\subfloat[]{\includegraphics[width=0.32\textwidth]{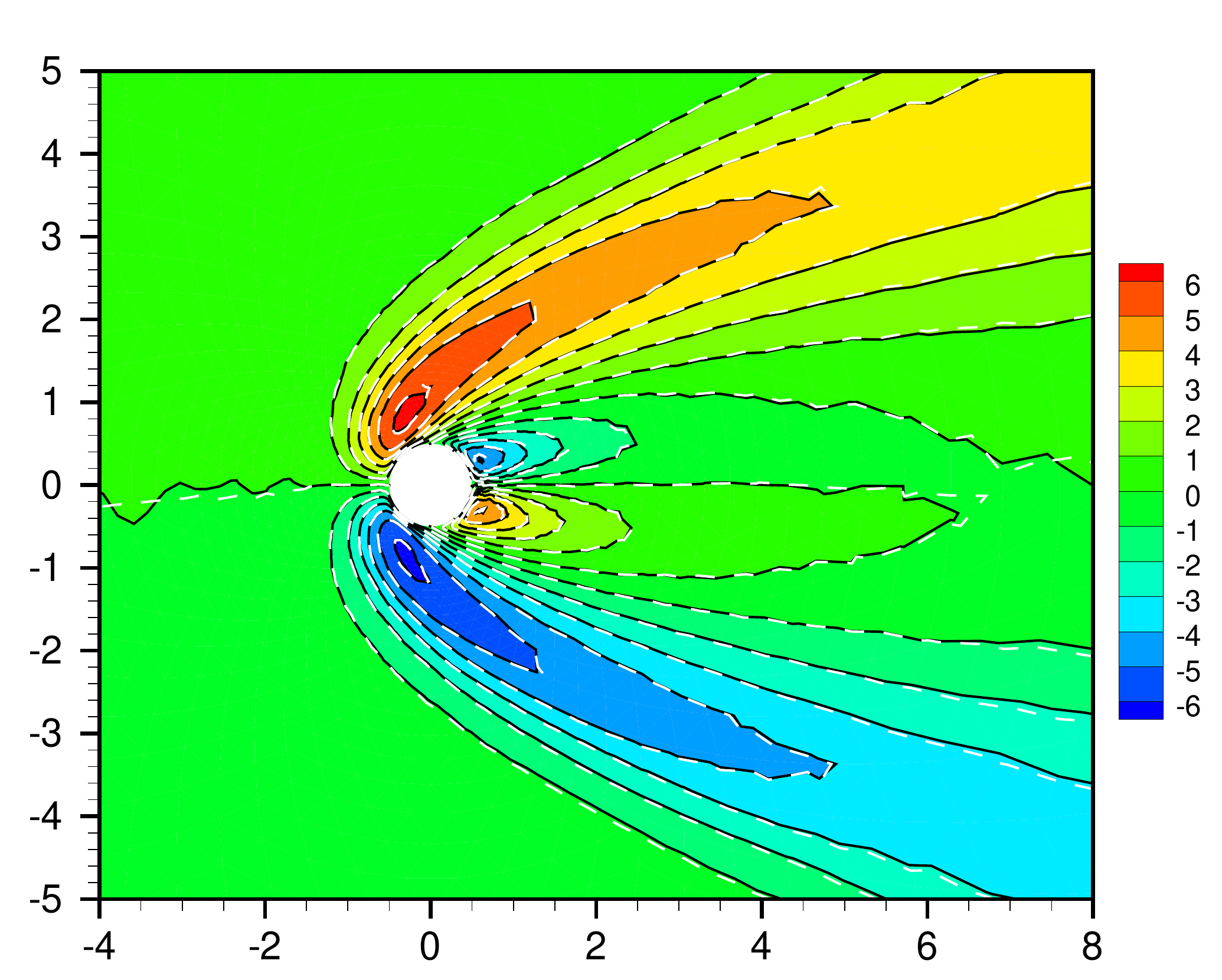}}
	\subfloat[]{\includegraphics[width=0.32\textwidth]{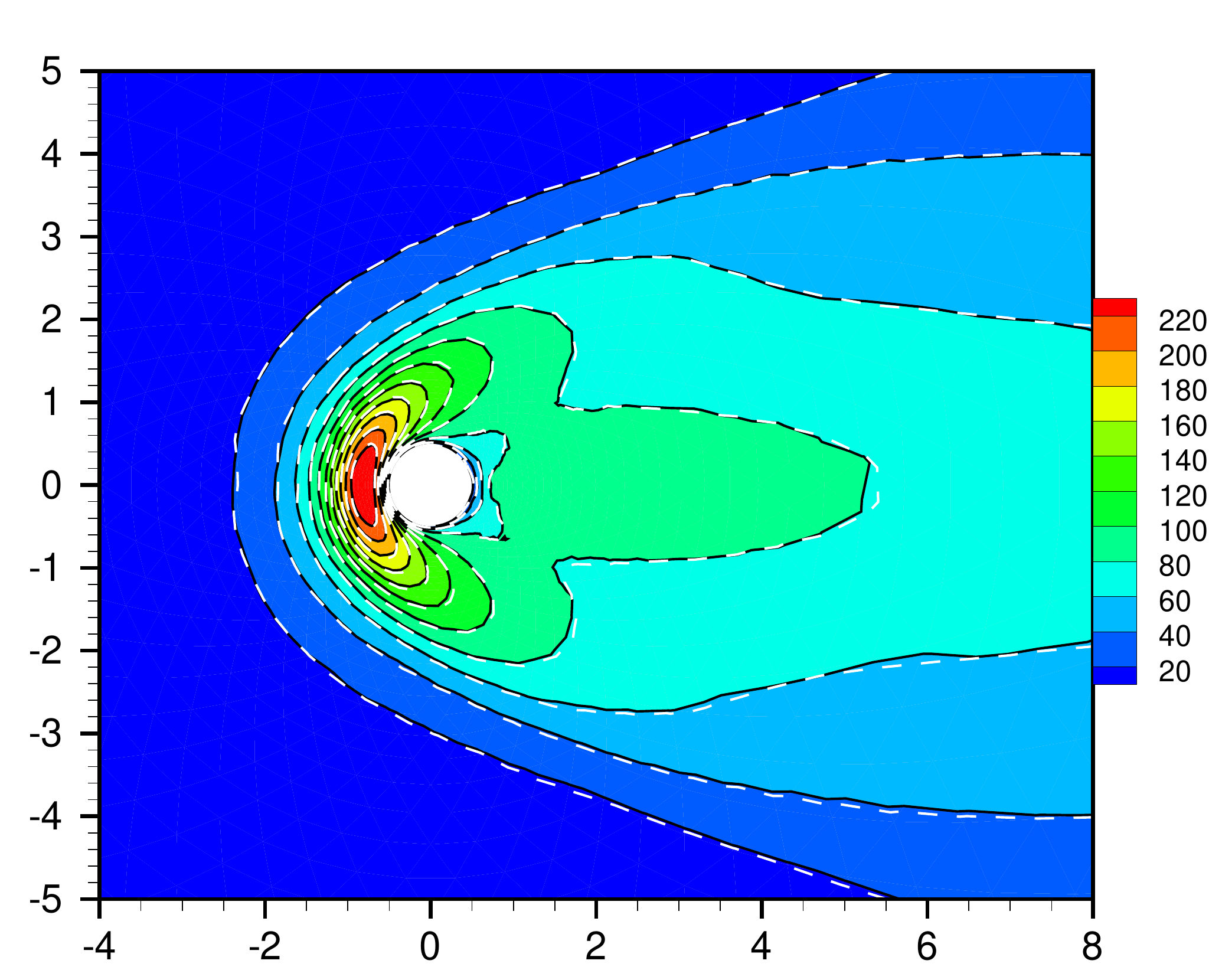}}
	\caption{\label{fig:cylinder_ma30}Hypersonic flow at ${ Ma}=30$ around a circle cylinder for ${ Kn} = 0.1$ obtained by the UGKWP method \cite{zhu2020thesis}. (a) $x$-component velocity, (b) $y$-component velocity, and (c) temperature.}
\end{figure}

\begin{figure}[htbp]
	\centering
	\subfloat[]{\includegraphics[width=0.32\textwidth]{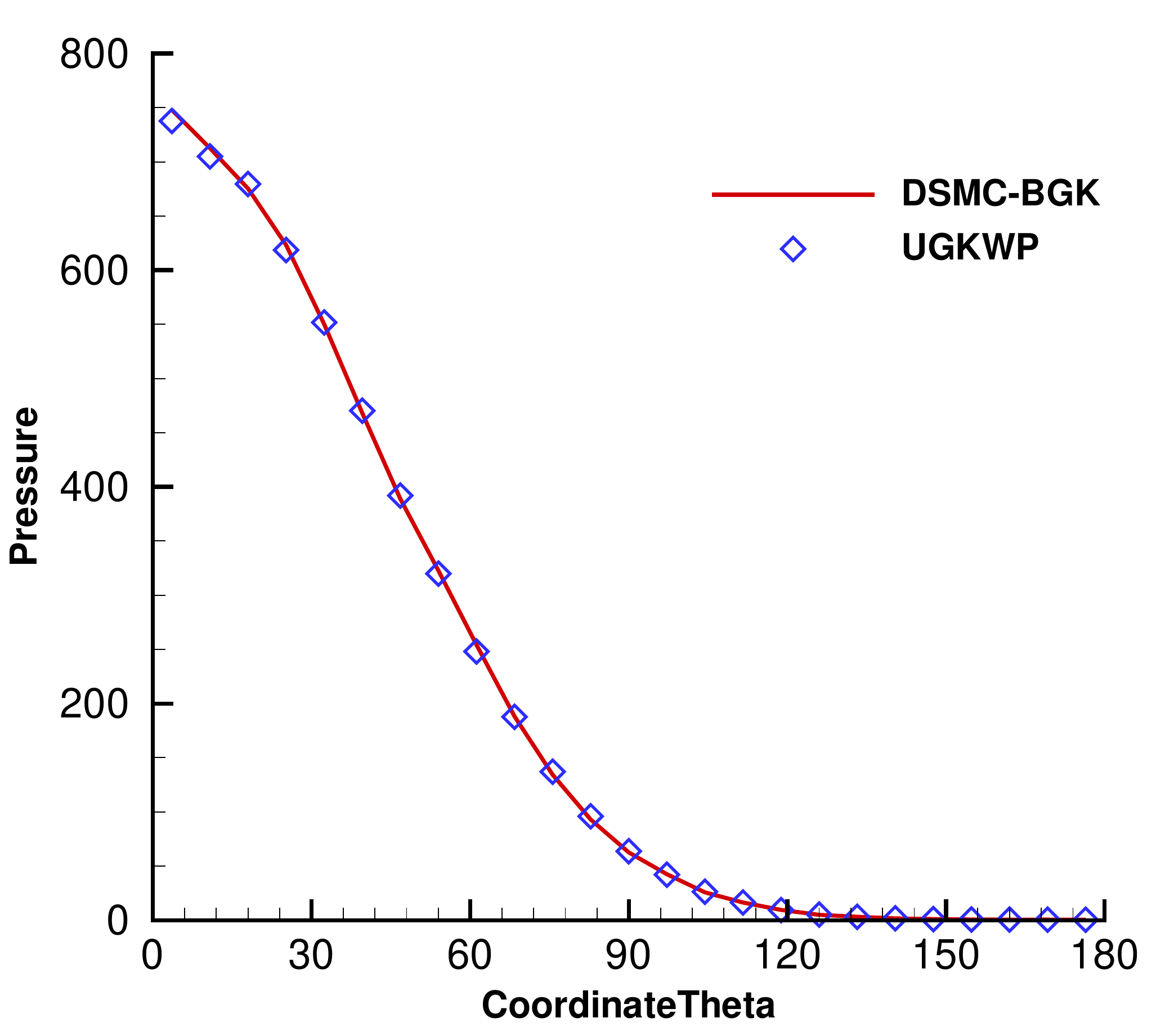}}
	\subfloat[]{\includegraphics[width=0.32\textwidth]{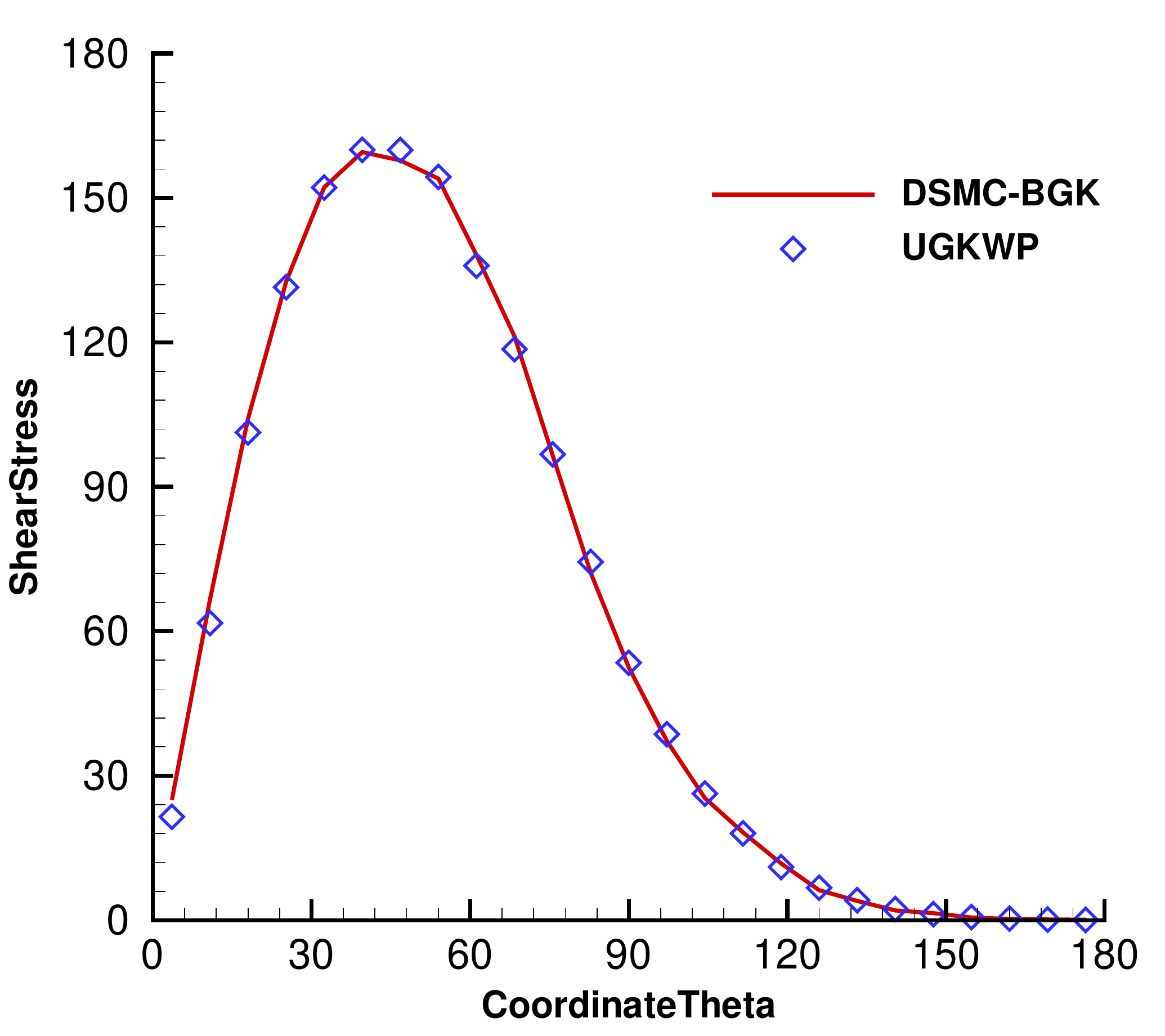}}
	\subfloat[]{\includegraphics[width=0.32\textwidth]{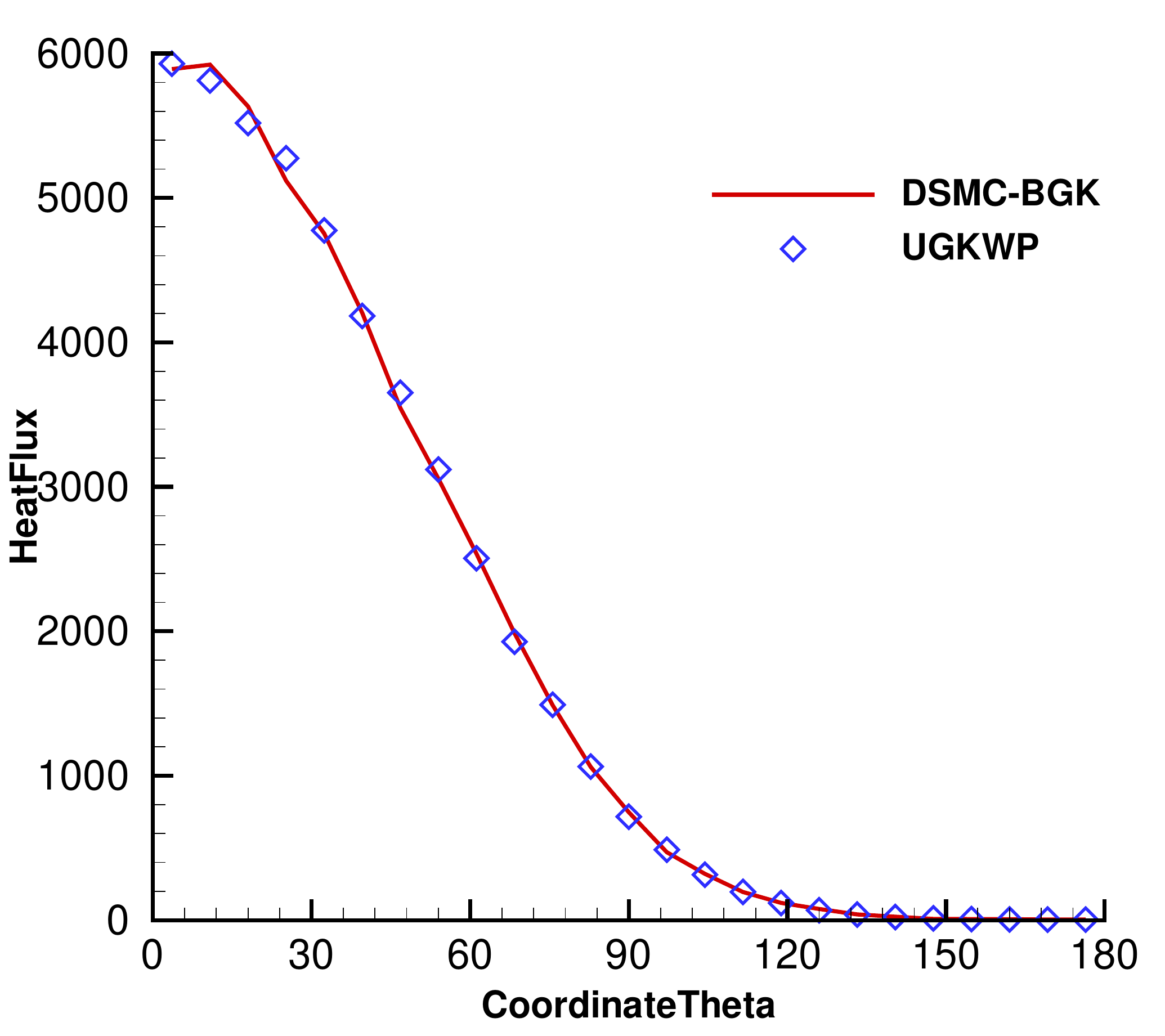}}
	\caption{\label{fig:cylinder_surface_ma30}Surface quantities around the circle cylinder at ${ Ma}=30$ and ${ Kn}=0.1$ \cite{zhu2020thesis}.  (a) Pressure, (b) shear stress, and (c) heat flux.}
\end{figure}

Furthermore, a very high-speed flow at ${ Ma}=30$ is computed for the case ${ Kn} = 0.1$ on the same unstructured mesh.
Since the memory requirement of the discrete velocity points for the UGKS is unaffordable for such high Mach number computation, we only show the results of the UGKWP method and the particle method in Figs.~\ref{fig:cylinder_ma30} and \ref{fig:cylinder_surface_ma30}.
In the computation, the memory cost of the UGKWP method is only $375$ MB.
The advantage of the particle method with a nature adaptivity in the phase space through particles is well inherited by the UGKWP method for high speed rarefied flow computations.

\subsubsection{High speed flow around X38-like space vehicle}

The high speed flow at ${ Ma}=10$ over a space vehicle in the transition regime at ${ Kn}=10^{-3}$ has been calculated to show the efficiency and capability of the UGKWP method for simulating three dimensional hypersonic flow over complex geometry configuration \cite{chen2020ugkwp}.
The angle of attack is $20$ degree.
The computational domain is discretized by an unstructured mesh with  $560593$  cells.
The reference length to define the Knudsen number is $L=0.28{ m}$.
Diffusive boundary condition is implemented for the solid walls, on which the temperature maintains at $T_w = 300 { K}$.
Due to the symmetry, only half of the vehicle is simulated.
The inflow is monatomic argon gas at $T_{\infty} = 300 { K}$.
The dynamic viscosity is evaluated by the power formula (\ref{eq:dynamic_viscosity}) with temperature index $\omega = 0.81$.

\begin{figure}[htbp]
	\centering
	\subfloat[]{\includegraphics[width=0.44\textwidth]{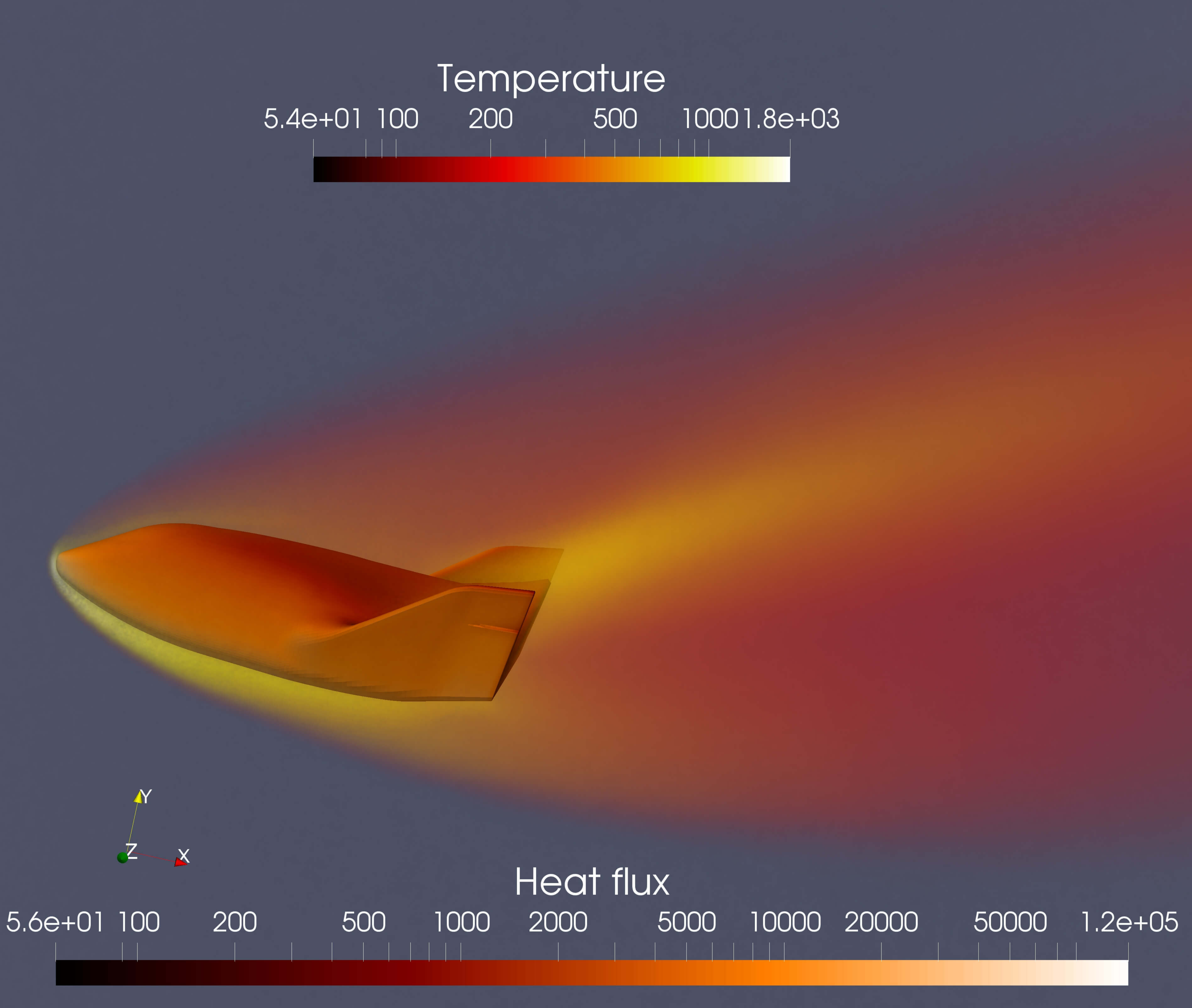}} \hspace{1em}
	\subfloat[]{\includegraphics[width=0.44\textwidth]{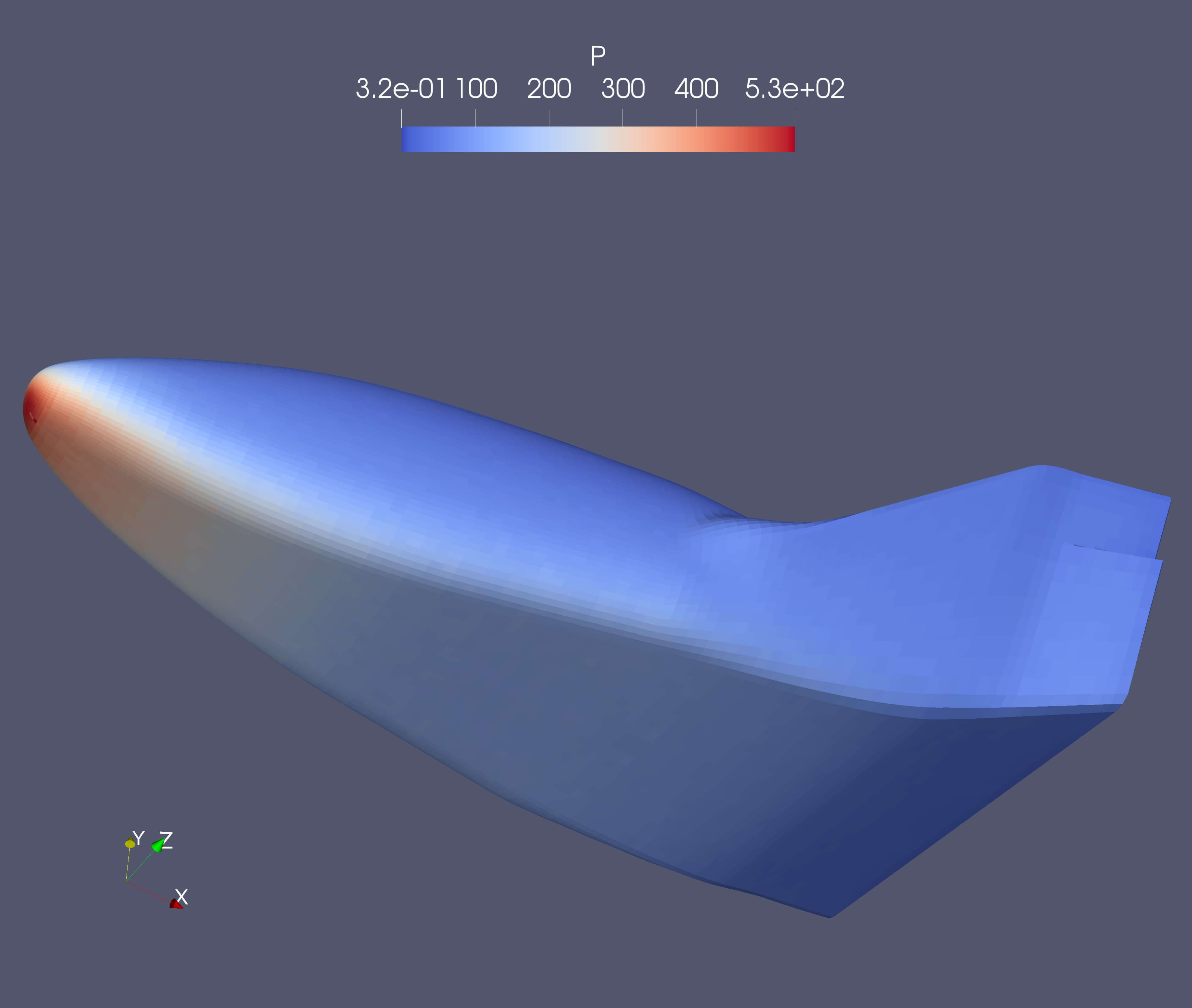}} \\
	\subfloat[\label{fig:ugkwp_x38_kn}]{\includegraphics[width=0.44\textwidth]{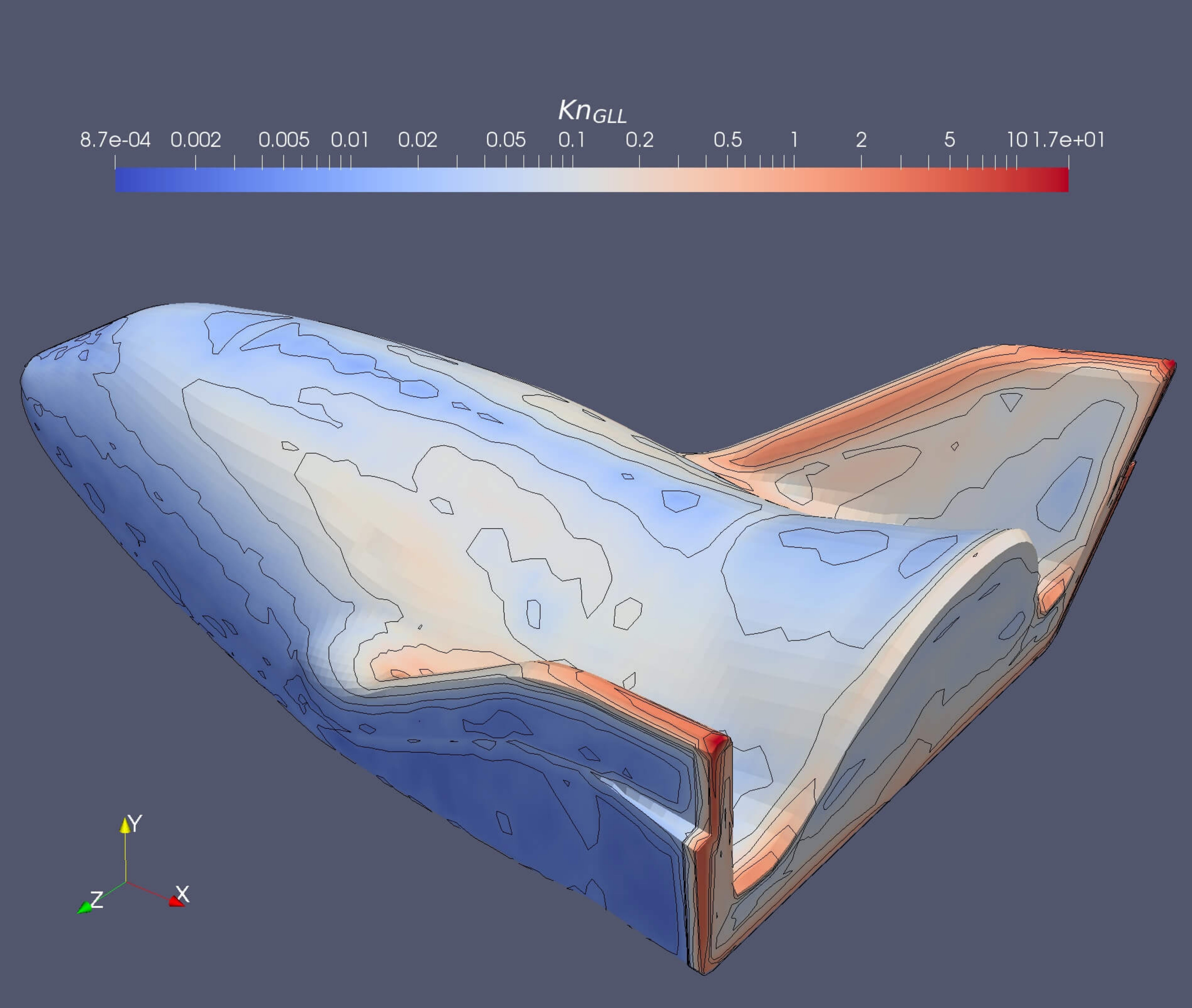}} \hspace{1em}
	\subfloat[]{\includegraphics[width=0.44\textwidth]{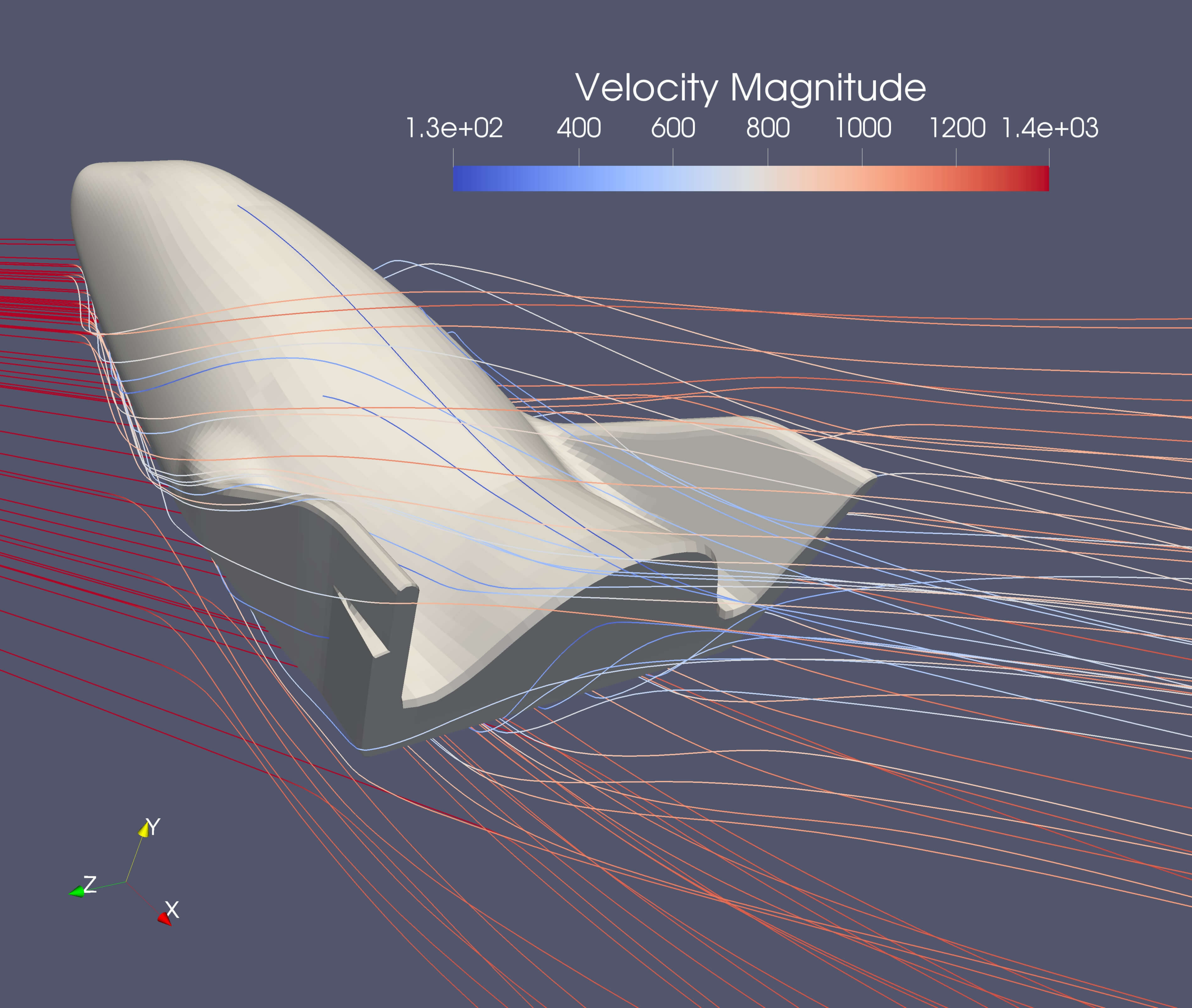}}
	\caption{\label{fig:ugkwp_x38_ma10}Flow around space vehicle at ${ Ma}=10$ and ${ Kn}=10^{-3}$ \cite{chen2020ugkwp}. (a) Temperature field and surface heat flux, (b) pressure, (c) Knudsen number distribution and (d) streamlines colored by the magnitude of velocity.}
\end{figure}

Figure \ref{fig:ugkwp_x38_ma10} presents the distribution of temperature, heat flux, pressure, local Knudsen number, and streamlines around the vehicle at ${ Ma}=10$.
From Fig.~\subref*{fig:ugkwp_x38_kn},  the density gradient based local Knudsen number can cover a wide range of values with five orders of magnitude difference.
Therefore, a multiscale method, like the UGKWP, is necessary to capture flow physics in different regimes.
Despite the low intensity of heat exchange upon the vehicle surface, a high temperature region is observed at the leeward side, which is mainly caused by the particle collisions in the strong recompression region with a relatively low freestream Knudsen number.
The simulation takes $22.8$ hours on Tianhe-2 with 8 nodes or 192 cores.
The UGKWP method shows great power to solve the engineering problems involving high speed non-equilibrium flows, such as flow around near space or re-entry vehicles.

\section{Multiscale transports and modelings}
The methodology of the UGKS that taking direct modeling on the discretization scale can be applied in the construction of multiscale method for many other transport processes, such as radiative transport \cite{sun2015radiative,sun2015frequency,sun2017unstructured,sun2017iugks,sun2018iugks,xu2020fe}, plasma \cite{liu2017plasma}, granular flow \cite{liu2019granular,wang2018granular,wang2019monodisperse}, neutron transport \cite{shuang2019parallel,tan2020neutron}.
The key ingredient is that in the framework of finite volume method, a multiscale flux function coupling microscopic particles' transport and collisions is constructed from the integral solution of a BGK-type kinetic model equation.
The integration over a time step implies a scale transition process from a kinetic scale to a macroscopic scale through accumulation of particles' collision.

\subsection{Radiative transfer}
The physics of radiation transport is very important in many areas, such as astrophysics, inertial confinement fusion, and high temperature flow systems.
Due to the variation of photons' mean free path and interaction with the material, multiscale transport in different regimes is involved from optically thin and optically thick ones.
The physics of radiation energy transport and the energy exchanges with the background material is modeled by the radiative transfer equations
\begin{equation}\label{eq:radiative_equation}
	\begin{dcases}
	\frac{\epsilon^2}{c} \frac{\partial I}{\partial t} + \epsilon \vec{\Omega} \cdot \nabla I = \sigma \left(\frac{1}{4\pi} a c T^4 - I\right), \\
	\epsilon^2 C_v \frac{\partial T}{\partial t} \equiv \epsilon^2 \frac{\partial U}{\partial t} = \sigma \left(\int{I d\vec{\Omega}} - a c T^4\right),
	\end{dcases}
\end{equation}
where $I(\vec{x}, \vec{\Omega}, t)$ is the radiation intensity, $\vec{\Omega}$ represents the angular variable.
$T$ is the material temperature, and $U$ means the material energy density.
$\sigma$ denotes the opacity, $a$ is the radiation constant, $c$ denotes the speed of light, and $\epsilon > 0$ is the Knudsen number.
For the gray radiative transfer equations, the optical opacity is only a function of local material temperature.

In the study of Mieussens \cite{mieussens2013asymptotic}, a UGKS to solve a linear kinetic model of radiative transfer equation was constructed, and the asymptotic preserving property for the diffusion limit is analyzed.
Sun et al. \cite{sun2015radiative} developed a multiscale method for solving the above gray radiative transfer equation by applying the integral solution in the construction of flux function, where the free transport of phonon and the interactions with background material are coupled in a dynamic evolution process.
Similar to the UGKS for gas dynamics, the time dependent solution at the cell interface can be written in the form of
\begin{equation}\label{eq:radiation_flux}
	I(t) = \tilde{I}^n + \tilde{\phi}^{n+1},
\end{equation}
where $\tilde{I}^n$ denotes the terms related to the initial distribution of radiation intensity at time level $n$, and $\tilde{\phi}^{n+1}$ represents the emission terms implicitly discretized at time level $n+1$, such as
\begin{equation}
	\phi_t \approx \frac{\phi^{n+1} - \phi^n}{\Delta t}, \quad
	\phi_x(x_0) \approx \frac{\phi^{n+1}(x_0 + \Delta x) - \phi^{n+1}(x_0)}{\Delta x} ,
\end{equation}
where $\phi = a c T^4$.
The difference between the UGKS for radiative transfer and that for rarefied gas dynamics is that the equilibrium state adopts an implicit discretized form at time level $n+1$.

Denoting $\rho = \int{I d\vec{\Omega}}$, the macroscopic system can be obtained by taking integration of the kinetic equation (\ref{eq:radiative_equation}), i.e.,
\begin{equation}
\begin{dcases}
	\frac{\epsilon^2}{c} \frac{\partial \rho}{\partial t} + \epsilon \nabla \cdot \vec{\Phi} = \sigma (\phi - \rho), \\
	{\epsilon^2} \frac{\partial \phi}{\partial t} = \beta \sigma (\rho - \phi),
\end{dcases}
\end{equation}
where $\vec{\Phi} = \smallint{\vec{\Omega} I d\vec{\Omega}}$ denotes the flux vector, and $\beta$ is defined as
\begin{equation}
	\beta = \frac{\partial \phi}{\partial U} = \frac{4 a c T^3}{C_v},
\end{equation}
representing the relation between the material energy density $U$ and the radiation energy density $\phi$.
Substituting the integral solution \eqref{eq:radiation_flux} into the discretization of the macroscopic system
\begin{equation}\label{eq:radiation_macro}
	\begin{cases}
	\begin{aligned}
	\rho_i^{n+1} = \rho_i^{n}
		&- \frac{\Delta t}{V_i} \sum_{j \in N(i)}{ \tilde{\Phi}_{ij}(\tilde{I}^n) S_{ij}}
		- \frac{\Delta t}{V_i} \sum_{j \in N(i)}{ \tilde{\Phi}_{ij}(\tilde{\phi}^{n+1}) S_{ij}} \\
		&+ \frac{\sigma_i^{n+1} c \Delta t}{\epsilon^2} ({\phi}_i^{n+1}-\rho_i^{n+1}),
	\end{aligned}	\\
	\phi_i^{n+1} = \phi_i^n + \frac{(\beta \sigma)_i^{n+1} \Delta t}{\epsilon^2} (\rho_i^{n+1} - \phi_i^{n+1}),
	\end{cases}
\end{equation}
$\rho^{n+1}$ and $\phi^{n+1}$ can be solved by implicit iteration, which could overcome the stiffness in the optical thick regimes.
Consequently, the radiation intensity can be evolved by
\begin{equation}
\begin{aligned}
	I_i^{n+1} = I_i^n
	&- \frac{\Delta t}{V_i} \sum_{j \in N(i)}{\Omega_{n,ij} \tilde{I}^n S_{ij}}
	- \frac{\Delta t}{V_i} \sum_{j \in N(i)}{\Omega_{n,ij} \tilde{\phi}^{n+1} S_{ij}} \\
	&+ \frac{c \Delta t \sigma_i^{n+1}}{\epsilon^2} \left(\frac{1}{2\pi}\phi_i^{n+1} - I_i^{n+1}\right).
\end{aligned}
\end{equation}
From the updated $I^{n+1}$, $\phi^{n+1}$ can be renewed by the second equation in \eqref{eq:radiation_macro}, and the material temperature can be calculated by $T^{n+1} = (\phi^{n+1} / (a c))^{1/4}$.
Due to the coupled treatment of photon transport and collision in the flux function, the UGKS has asymptotic preserving property in the capturing of diffusion solution without using a mesh size being smaller than the photon mean free path.

\begin{figure}[htbp]
\centering
\subfloat[]{\includegraphics[width=0.46\textwidth]{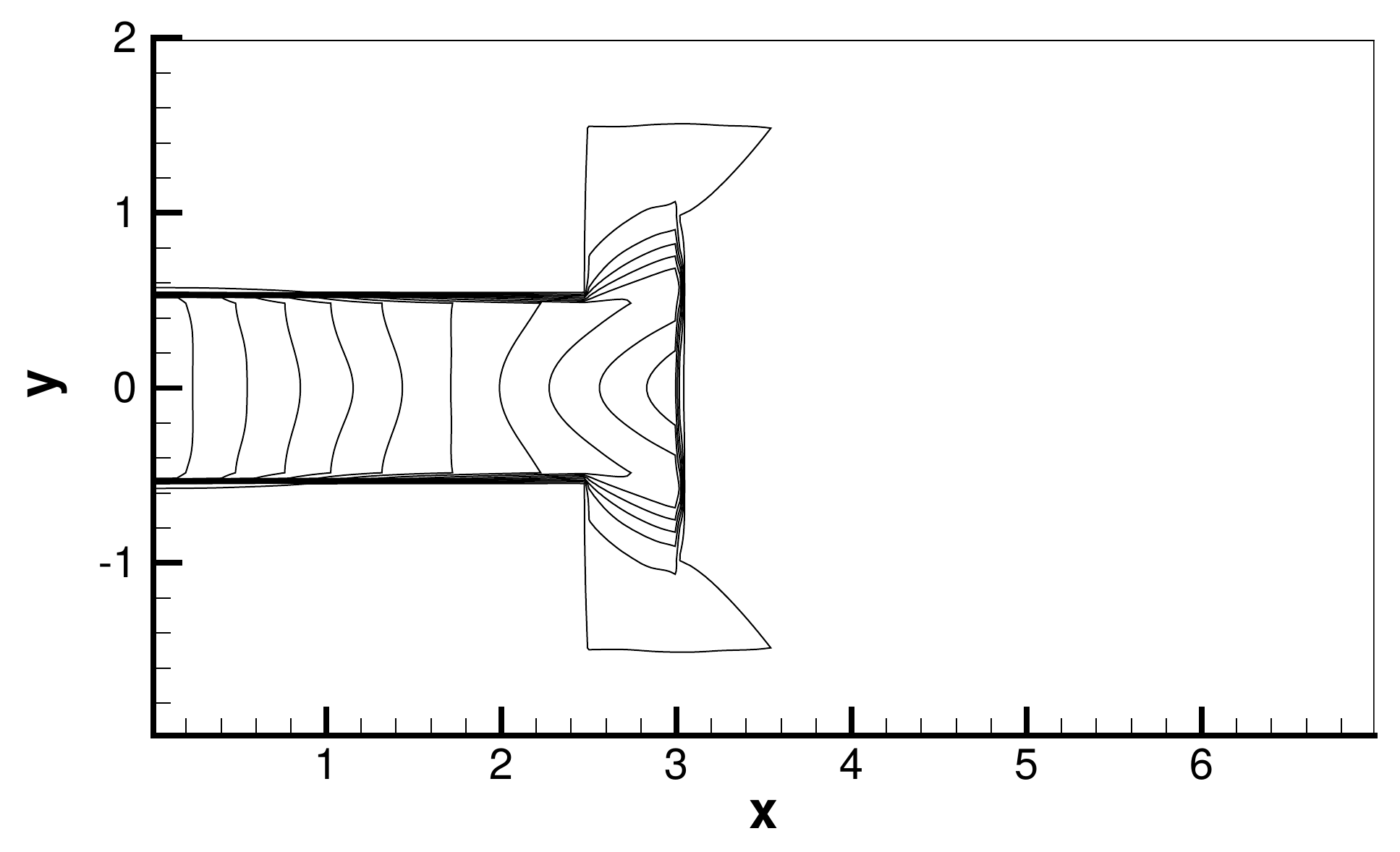}} \hspace{1em}
\subfloat[]{\includegraphics[width=0.46\textwidth]{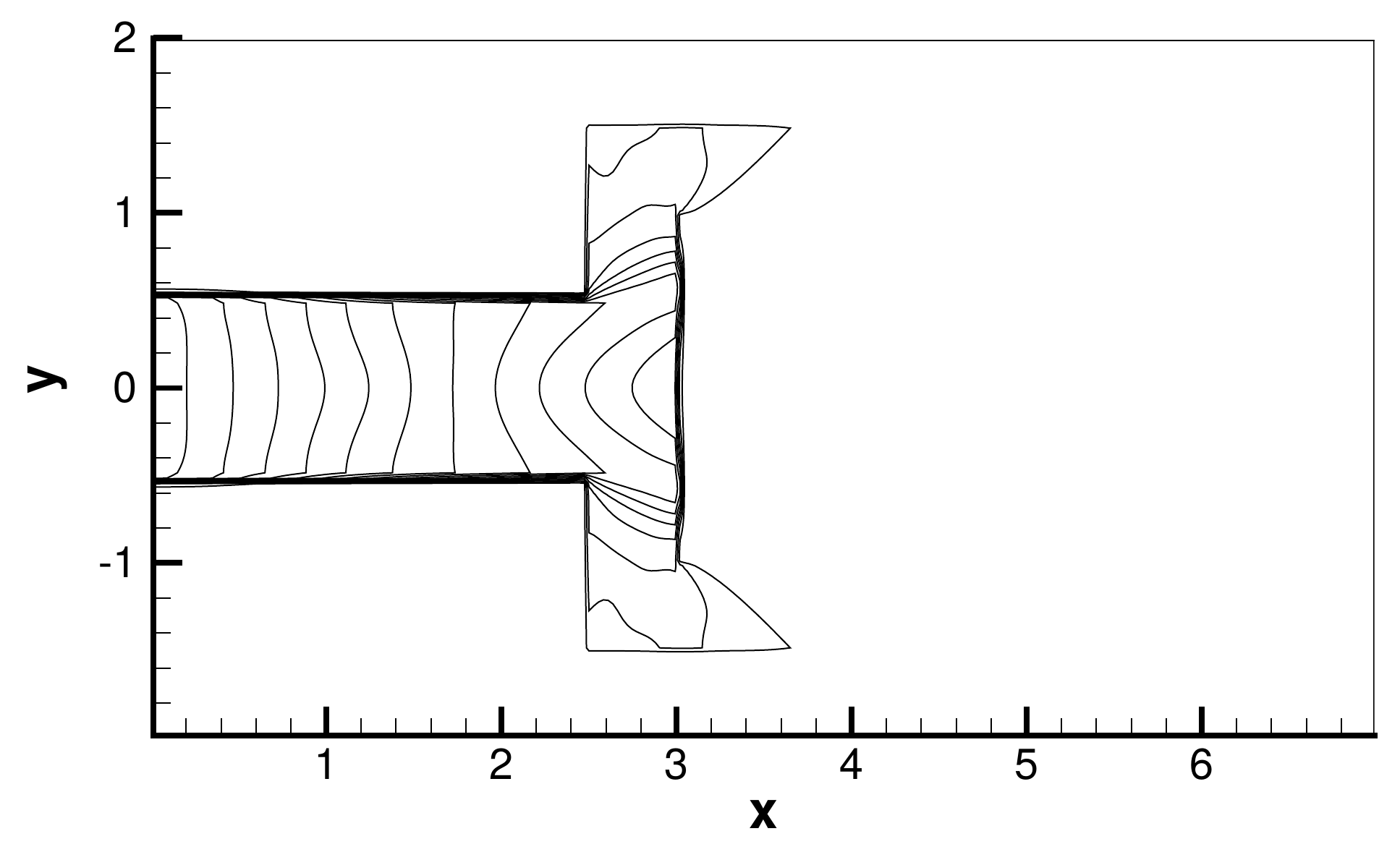}} \\
\subfloat[]{\includegraphics[width=0.46\textwidth]{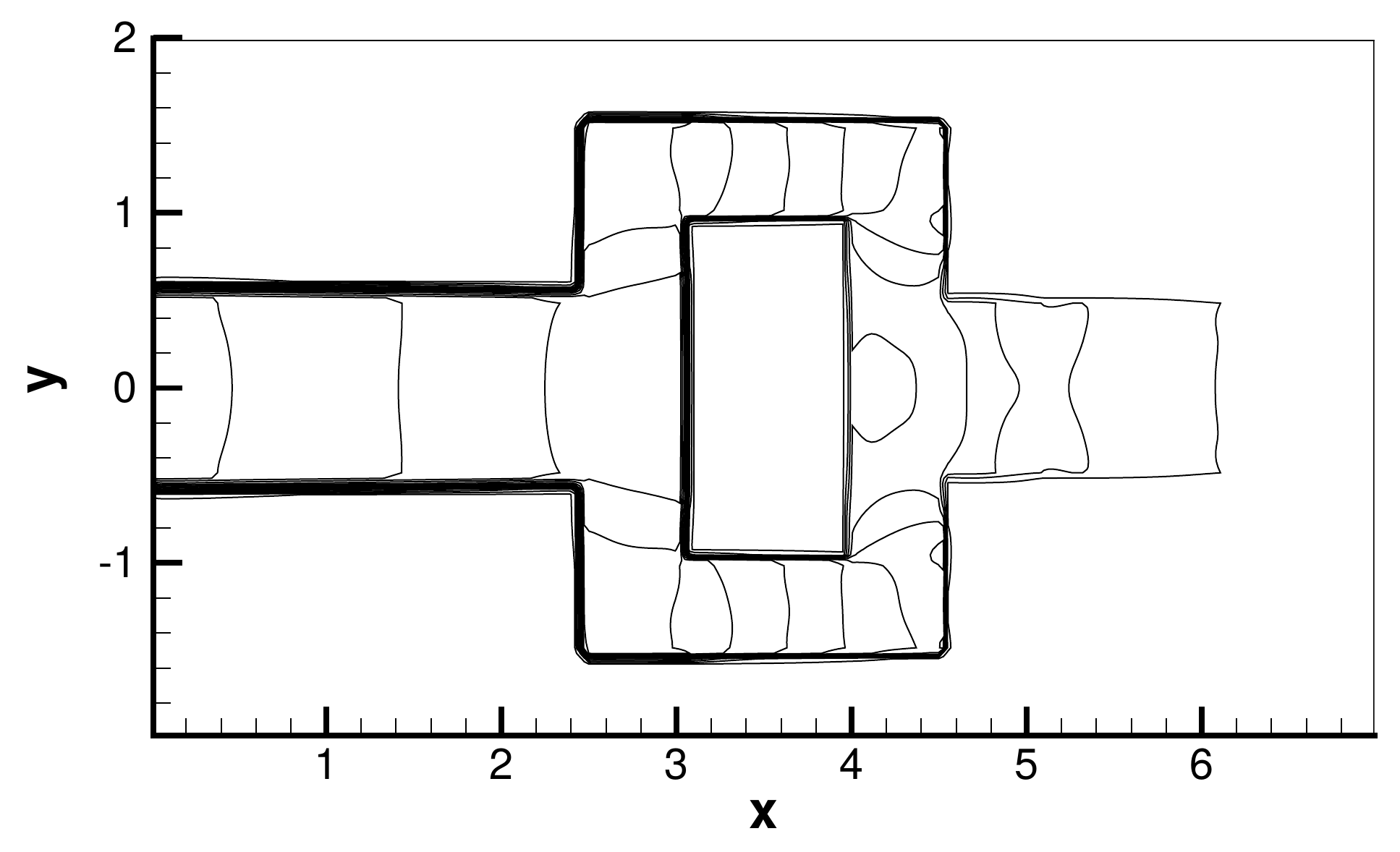}} \hspace{1em}
\subfloat[]{\includegraphics[width=0.46\textwidth]{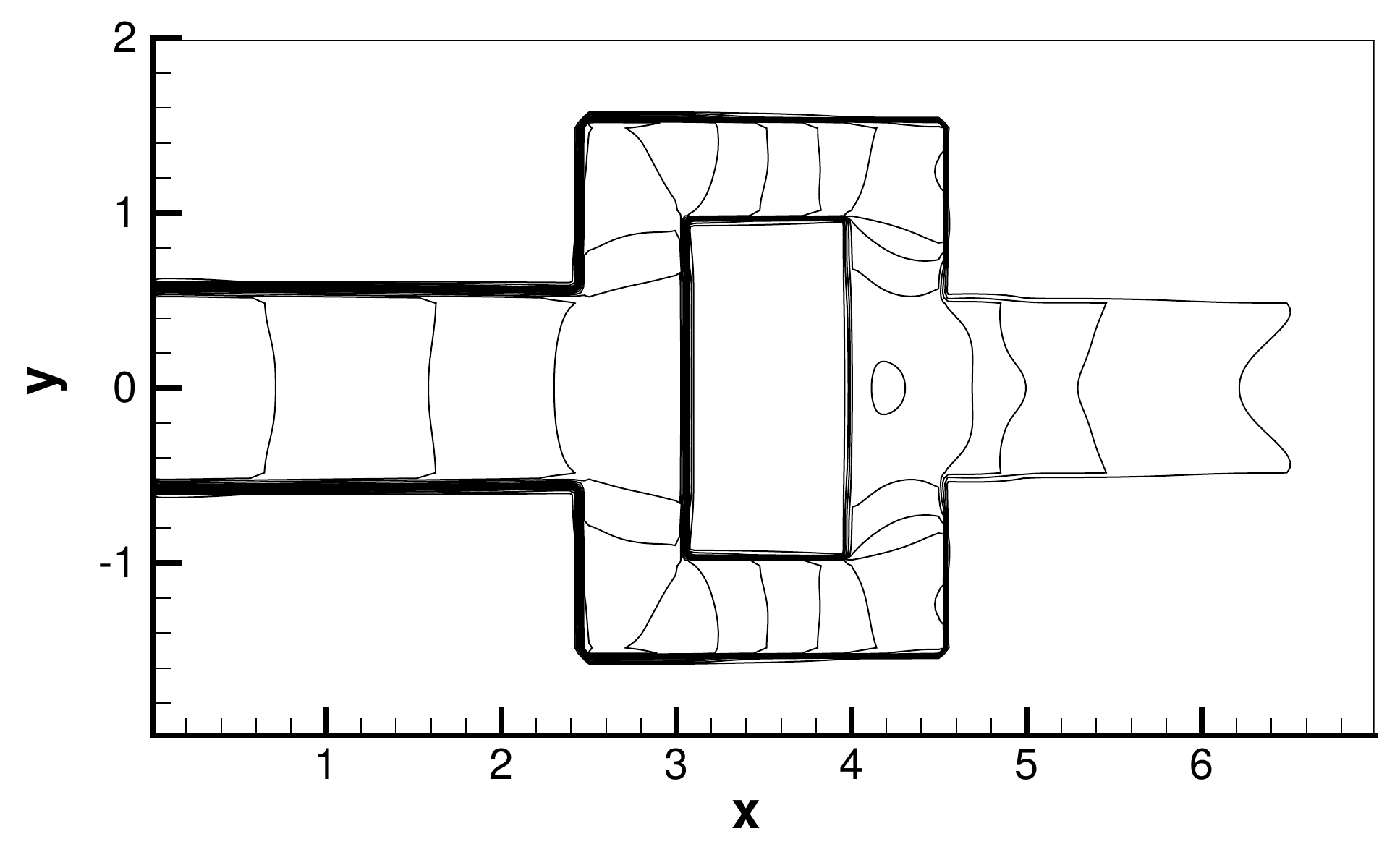}}
\caption{\label{fig:radiation_tophat}The contours of the material and radiation temperature of Tophat problem \cite{sun2015radiative}.
(a) material temperature and (b) radiation temperature at $8{ ns}$, (c) material temperature and (d) radiation temperature at $94{ ns}$.}
\end{figure}

\begin{figure}[htbp]
\centering
\includegraphics[width=0.44\textwidth]{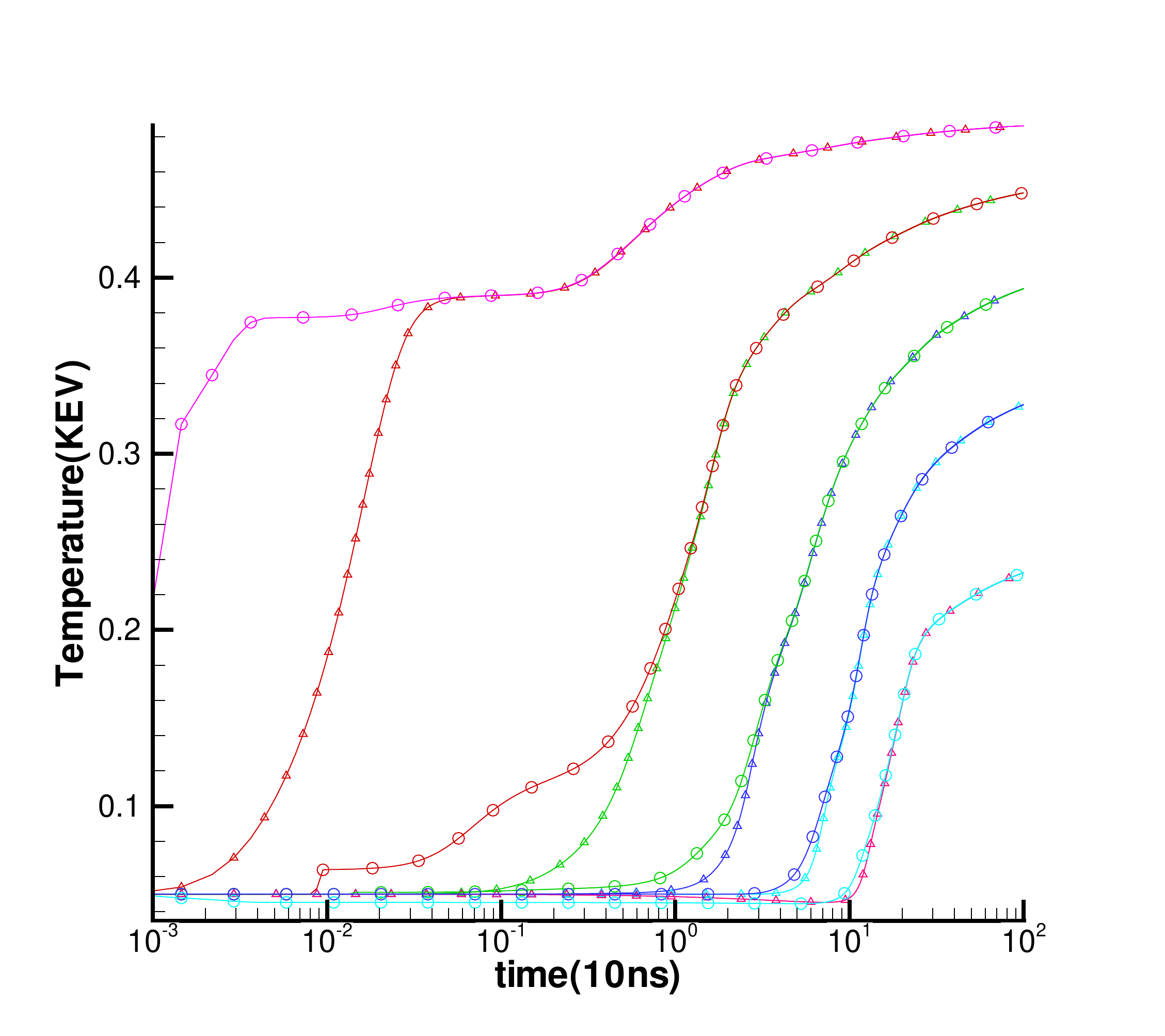}
\caption{\label{fig:radiation_tophat_temperature}The time evolution of material and radiation temperature at five probe points \cite{sun2015radiative}. Circle: the radiation temperature. Triangle: the material temperature.}
\end{figure}

In the study of Sun et al. \cite{sun2015radiative}, the Tophat problem is solved in 2D Cartesian coordinate.
The size of the simulation domain is $[0,7]\times[-2,2]$.
Dense, opaque material with density $10 { g/{{cm}^3}}$ and opacity $\sigma=2000 { {cm}^{-1}}$ is located in the following regions: $(3,4)\times(-1,1)$, $(0,2.5)\times(-2,-0.5)$, $(0,2.5)\times(0.5,2)$, $(4.5,7)\times(-2,-0.5)$, $(4.5,7)\times(0.5,2)$,  $(2.5,4.5)\times(-2,-1.5)$, and $(2.5,4.5)\times(1.5,2.0)$.
The pipe, with density $0.01{  g/{{cm}^3}}$ and opacity $\sigma=0.2 { {cm}^{-1}}$, occupies all other regions.
The heat capacity is $0.1 { GJ/g/{Kev}}$.
Initially, the material has a temperature $0.05 { Kev}$ everywhere, and the radiation and material temperature are in equilibrium.
A heating source with a fixed temperature $0.5 { Kev}$ is located on the left boundary for $-0.5 < y < 0.5$.
All the other  boundary conditions are outflow.
Five probes are placed at $(0.25,0)$,  $(2.75,0)$,  $(3.5,1.25)$, $(4.25,0)$,  and $(6.75,0)$ to monitor the change of the temperature in the  thin opacity material.
Figure \ref{fig:radiation_tophat} shows the distributions of material and radiation temperatures from the UGKS simulations on the mesh with $256\times 128$ cells.
In comparison with the results of \cite{gentile2001imc}, the interface  between the opacity thick and thin materials is captured sharply by the UGKS.
The time evolution of the temperatures at five probe points are given in Fig.~\ref{fig:radiation_tophat_temperature}.
At the fifth probe point, it shows that the temperature cools off slightly before being heated up by the radiation wave, which is consistent with the observation in \cite{gentile2001imc}.

The UGKS has also been extended to the frequency-dependent radiative transfer system \cite{sun2015frequency}, and to the numerical simulations on the unstructured mesh \cite{sun2017unstructured} and cylindrical coordinate \cite{sun2019cylindrical}.
The implicit schemes \cite{sun2017iugks,sun2018iugks} are developed in order to accelerate the numerical computation.
For mitigation of ray effect using discrete ordinate method, the UGKS with an angular discretization by the linear finite element method is constructed \cite{xu2020fe}.
Moreover, the UGKS for non-equilibrium radiative transfer is coupled with the hydrodynamic evolution, where the compressible inviscid Euler flow is solved by the gas kinetic scheme \cite{sun2020hydrodynamics}.

\subsection{Multi-component and plasma}
The direct modeling methodology of the UGKS has also been applied in the plasma simulation \cite{liu2017plasma,pan2018plasma}.
The dynamics of a fully ionized plasma are modeled by the Fokker-Planck-Landau (FPL) equations on the kinetic level
\begin{equation}\label{eq:fpl}
  \frac{\partial f_\alpha}{\partial t}
  +\vec{u} \cdot \nabla_{\vec{x}} f_{\alpha}
  +\frac{\vec{F}_\alpha}{m_\alpha}
  \cdot\nabla_{\vec{u}} f_{\alpha}=\sum_\beta Q_{\alpha,\beta}(f_\alpha,f_\beta),
\end{equation}
where $f_\alpha(\vec{x},\vec{u},t)$ is the velocity distribution function of species $\alpha$ ($\alpha=i$ for ion and $\alpha=e$ for electron).
$\vec{F}_\alpha=e(\vec{E}+\vec{u}_\alpha\times \vec{B})$ is the averaged electromagnetic force.
The collision term $Q_{\alpha,\beta}(f_\alpha,f_\beta)$ describes the binary collisions between charged particles with long-range Coulomb interactions
\begin{equation}\label{eq:nfplc}
  Q_{\alpha,\beta}(f_\alpha, f_\beta) = \nabla_{\vec{u}}\cdot
  \left(\int_{\mathbb{R}^3} \Phi(\vec{u}-\vec{u}^\prime)
  (\nabla_{\vec{u}} f_\alpha f^\prime_\beta-\nabla_{\vec{u}}^\prime f_\beta^\prime f_\alpha)d\vec{u}^\prime \right),
\end{equation}
where $\Phi(\vec{u})$ is a $3\times3$ matrix
\begin{equation}
\Phi(\vec{u})=\frac{1}{|\vec{u}|^3}(|\vec{u}|^2 I_3-\vec{u} \otimes \vec{u}).
\end{equation}
In order to overcome the complexity and high computational cost of the nonlinear Landau collision term, a single BGK-type collision operator proposed by Andries, Aoki, and Perthame (AAP model) is employed to model the collision process \cite{andries2002aap}.

In AAP model, one global collision operator is used for each component to take into account both self-collision and cross-collision, and the kinetic equations read
\begin{equation}\label{eq:plasma_kinetic}
  \frac{\partial f_\alpha}{\partial t}
  +\vec{u}\cdot \nabla_{\vec{x}} f_{\alpha}
  +\frac{\vec{F}_\alpha}{m_\alpha}
  \cdot\nabla_{\vec{u}} f_{\alpha} = \frac{f_\alpha^+ - f_\alpha}{\tau_\alpha},
\end{equation}
where post collision distribution $f^+$  is a Maxwellian distribution
\begin{equation}\label{eq:f+}
  f^{+}_\alpha=\rho_\alpha\left(\frac{m_\alpha}{2\pi k_B \overline{T}_\alpha}\right)^{3/2}
  \exp\left(-\frac{m_\alpha}{2\pi k_B \overline{T}_\alpha}(\vec{u}-\overline{\vec{U}}_\alpha)^2\right).
\end{equation}
The parameters $\overline{T}_\alpha$ and $\overline{\mathbf{U}}_\alpha$ are connected to the macroscopic properties of individual components by
\begin{equation}\label{eq:W-bar}
\begin{aligned}
  \overline{\vec{U}}_\alpha &= \vec{U}_\alpha + \tau_\alpha \sum_r \frac{2 m_r}{m_\alpha+m_r} \nu_{\alpha r}(\vec{U}_r- \vec{U}_\alpha),\\
  \frac{3}{2} k_B \overline{T}_\alpha &= \frac{3}{2} k_B T_\alpha -\frac{m_\alpha}{2}(\overline{\vec{U}}_\alpha-\vec{U}_\alpha)^2\\
  &+\tau_\alpha \sum_r  \frac{4 m_\alpha m_r \nu_{\alpha r}}{(m_\alpha + m_r)^2}  \left[\frac{3}{2} k_B (T_r -T_\alpha) + \frac{m_r}{2}(\vec{U}_r-\vec{U}_{\alpha})^2\right],
\end{aligned}
\end{equation}
where $\nu_{\alpha r}$ are the interaction coefficients that measure the strength of intermolecular collision.
The relaxation time is determined by
$\tau_{\alpha}=1/\sum_r\nu_{\alpha r}$.
The parameter $\nu_{\alpha_r}$ is determined by molecular models, and the hard sphere model is used \cite{morse1963energy}.

The averaged electric field $\vec{E}$ and magnetic field $\vec{B}$ follow the Maxwell equations,
\begin{equation}\label{eq:maxwell1}
 \begin{dcases}
  \frac{\partial \vec{B}}{\partial t}=-\nabla_{\vec{x}} \times \vec{E},\\
  \frac{\partial \vec{E}}{\partial t}=c^2\nabla_{\vec{x}} \times \vec{B}-\frac{1}{\epsilon_0} \vec{j},
 \end{dcases}
\end{equation}
where $c$ is the speed of light, $\epsilon_0$ is the vacuum permittivity which is related to the vacuum permeability $\nu_0$ by $c=(\nu_0 \epsilon_0)^{-1/2}$.
The electromagnetic field satisfies the divergence constraints
\begin{equation}\label{eq:maxwell2}
\nabla_{\vec{x}} \cdot \vec{E}=\frac{e}{\epsilon_0}
(n_i-n_e),
\qquad \nabla_{\vec{x}} \cdot \vec{B}=0,
\end{equation}
where $e$ is the charge of a proton.
Theoretically, the divergence constraints will always hold if they are initially satisfied.
However numerical techniques are needed to make sure
that the divergence constraints are satisfied by numerical solutions.
The perfectly hyperbolic Maxwell equations (PHM) \cite{munz2000divergence} are used in current work
to evolve the electromagnetic field which preserve the divergence constraints,
\begin{equation}\label{eq:Maxwell3}
\begin{aligned}
  &\frac{\partial \vec{E}}{\partial t} - c^2 \nabla_{\vec{x}} \times \vec{B} + \chi c^2 \nabla_{\vec{x}} \phi = - \frac{1}{\epsilon_0} \vec{j},\\
  &\frac{\partial \vec{B}}{\partial t} + \nabla_{\vec{x}} \times \vec{E} + \gamma \nabla_{\vec{x}} \psi= \vec{0},\\
  &\frac{1}{\chi} \frac{\partial \phi}{\partial t}+\nabla_{\vec{x}} \cdot \vec{E} = \frac{e}{\epsilon_0} (n_i-n_e),\\
  &\frac{\epsilon_0\mu_0}{\gamma} \frac{\partial \psi}{\partial t}+\nabla_{\vec{x}} \cdot \vec{B}=0,
\end{aligned}
\end{equation}
where $\vec{j}$ is the total electric current density $\vec{j}=e (n_i \vec{U}_i- n_e\vec{U}_e)$,
$\phi$, $\psi$ are artificial correction potentials.
Munz proved that the propagation speed of magnetic field divergence error and electric field divergence error are $\gamma c$ and $\chi c$ \cite{munz2000divergence}.
This scheme is built on the BGK-Maxwell system
Eqs.~\eqref{eq:plasma_kinetic}, \eqref{eq:maxwell1}, \eqref{eq:maxwell2}, which are able to cover the flow regimes of plasma from the collisionless Vlasov regime to the continuum MHD regime.

In the study of Liu et al. \cite{liu2017plasma}, the asymptotic limits of the BGK-Maxwell system have been analyzed.
In the continuum regime, when the interspecies collision is strong, the gas mixture behaves like dielectric material, and the BGK-Maxwell equations become the Euler equations.
For a conductive plasma, the BGK-Maxwell equations can span the complete range from the neutral two-fluid system to resistive-MHD, Hall-MHD, and ideal MHD equations.
Then the UGKS for multi-component plasma was developed based on the direct modeling of the plasma physics on the scale of cell size and time step.
Similarly, the multiscale flux function is constructed from the integral solution of the BGK-type AAP model equation, which provides an un-splitting treatment of particle collision, acceleration and transport for flux evaluation across cell interface.
While  the solution updating process, the particle acceleration and collision process are dealt with separately.
An implicit-explicit method is employed to renew the velocity distribution function so that the constraint of small time step due to  the large electromagnetic acceleration can be removed.
Moreover, the flow physics covered by the UGKS is more general than those from either the collisionless Vlasov equation or MHD equations in the corresponding kinetic or hydrodynamic alone.
The UGKS can provide a reliable physical solution in the transitional regime as well, which has not been fully explored before from the particle-based and MHD-based numerical methods.

\subsubsection{Landau damping}

\begin{figure}[htbp]
\centering
\subfloat[]{\includegraphics[width=0.3\textwidth]{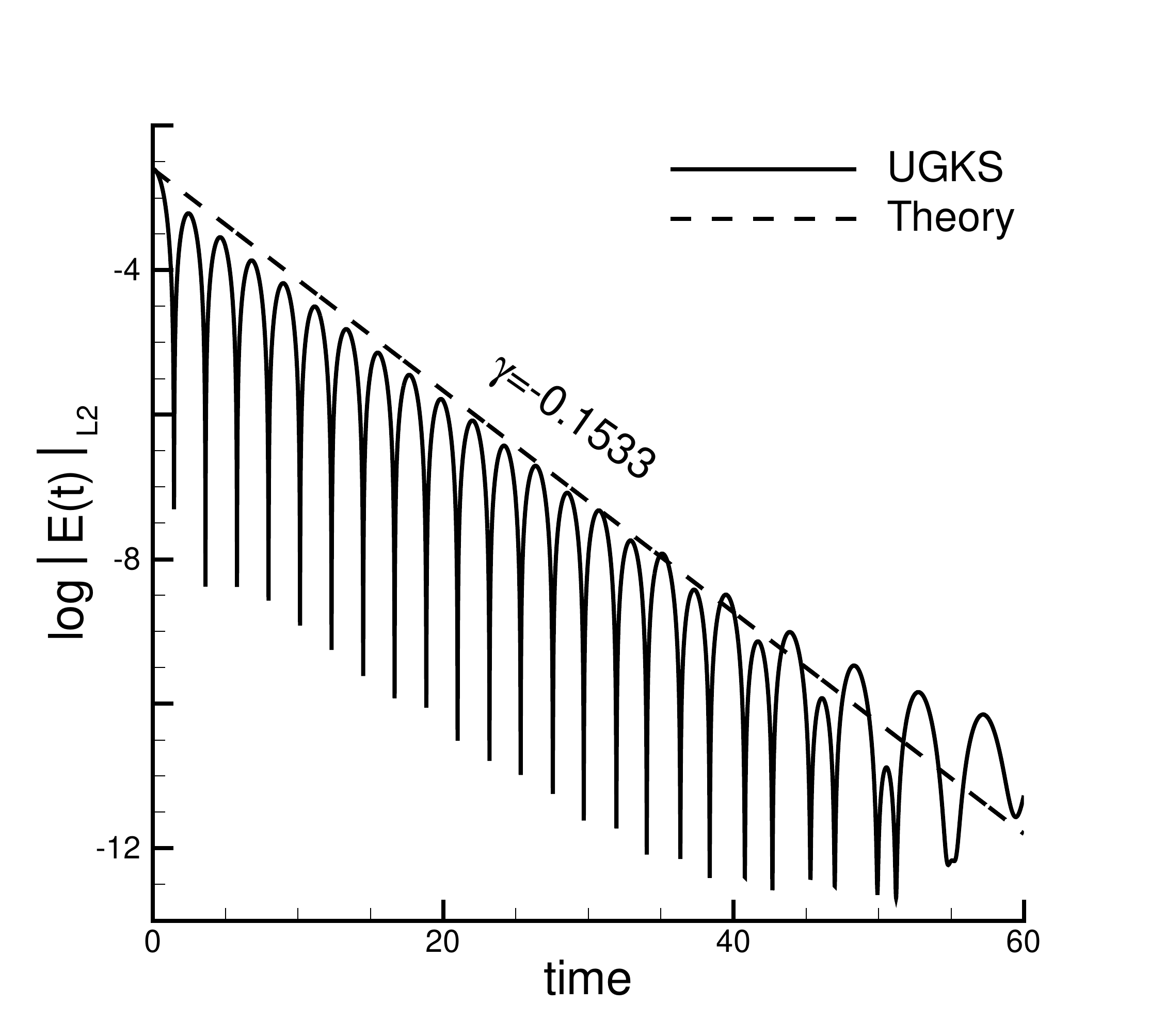}} \hspace{1em}
\subfloat[]{\includegraphics[width=0.3\textwidth]{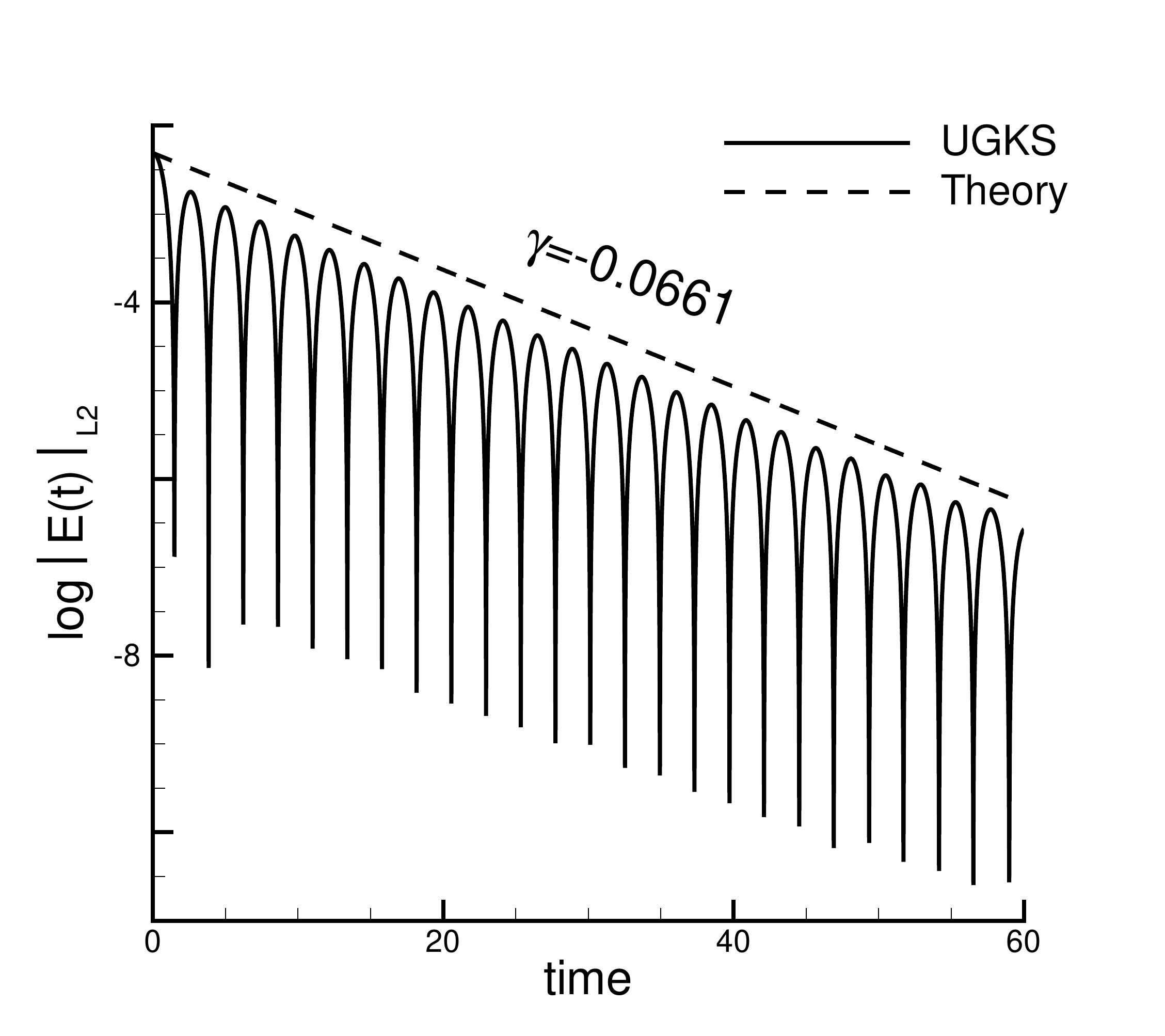}} \hspace{1em}
\subfloat[]{\includegraphics[width=0.3\textwidth]{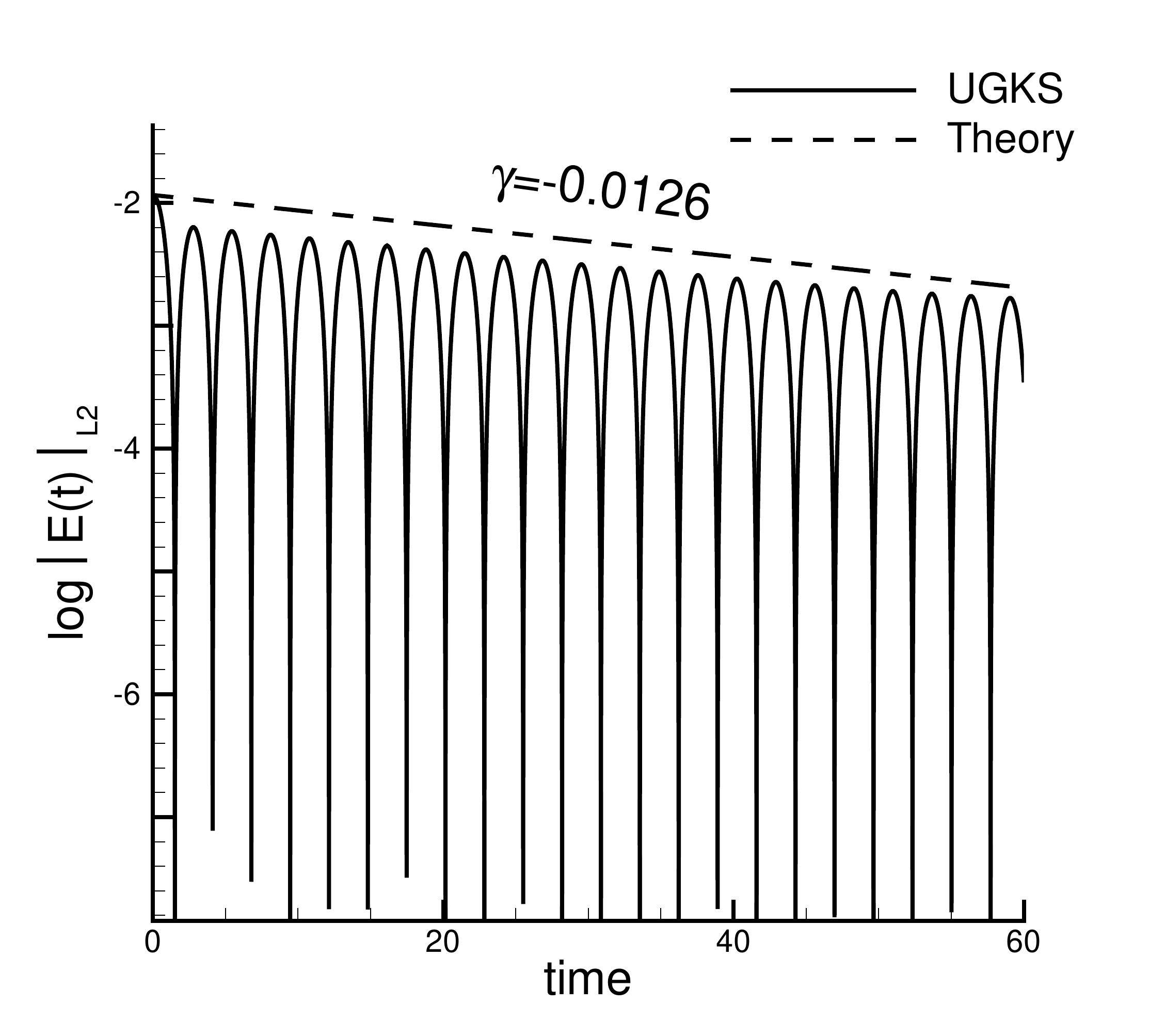}}
\caption{\label{fig:landau_linear}Weak Landau damping. Time evolution of electric field in $L^2$ norm for weak Landau damping \cite{liu2017plasma}. (a) $k=0.5$, (b) $k=0.4$, and (c) $k=0.3$.}
\end{figure}

For the Vlasov-Poisson (VP) system with a weak perturbation, the linear theory of Landau damping can be applied to predict the linear decay of electric energy with time \cite{chen2015plasma}.
The initial condition of linear Landau damping for the Vlasov Poisson system is
\begin{equation}\label{eq:landaudamping-initial}
  f_0(x,u)=\frac{1}{\sqrt{2\pi}}\left(1+\alpha\cos(kx)\right)\text{e}^{-\frac{u^2}{2}},
\end{equation}
with $\alpha=0.01$.
The length of the domain in the $x$ direction is $L=2\pi/k$.
The background ion distribution function is fixed, uniformly chosen so that the total net charge density for the system is zero.
The cases at different wave numbers are computed and compared with theoretical predictions.
The phase space is discretized with $N_x\times N_u=128 \times 128$ cells with $u_{max}=5 U_0$.
The evolution of electric field in $L^2$ norm is plotted in Fig. \ref{fig:landau_linear} for $k=0.5$, $k=0.4$, and $k=0.3$.
The correct decay rates of the electric field are observed and are matched with theoretical values.
In addition, the frequencies of oscillating solutions consist with the theoretical values of
$\omega=1.41$, $\omega=1.29$, $\omega=1.16$.

\begin{figure}[htbp]
\centering
\includegraphics[width=0.5\textwidth]{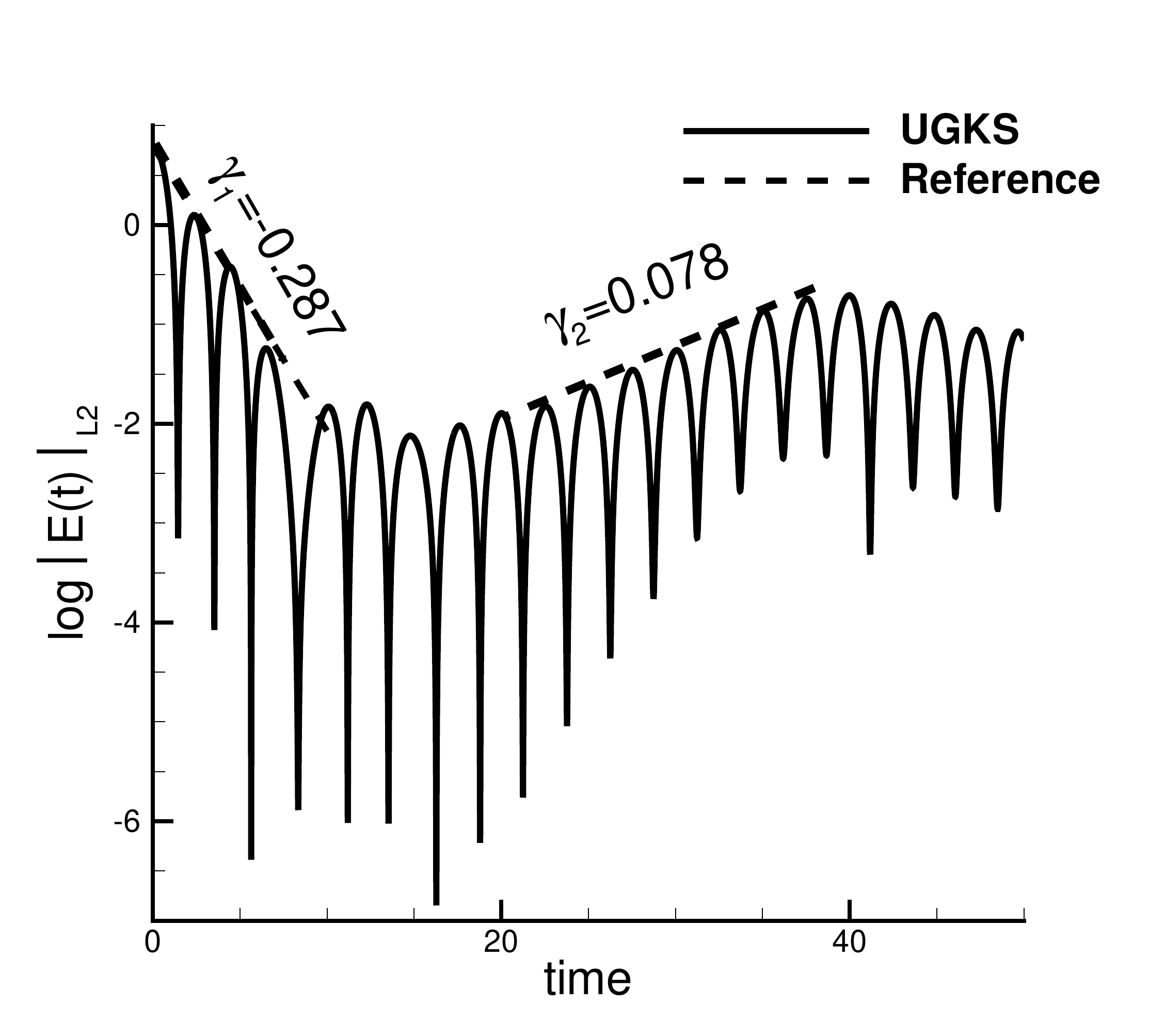}
\caption{\label{fig:landau_nonlinear}Time evolution of electric field in $L^2$ norm for strong Landau damping \cite{liu2017plasma}.}
\end{figure}

\begin{figure}[htbp]
  \centering
  \subfloat[]{\includegraphics[width=0.3\textwidth]{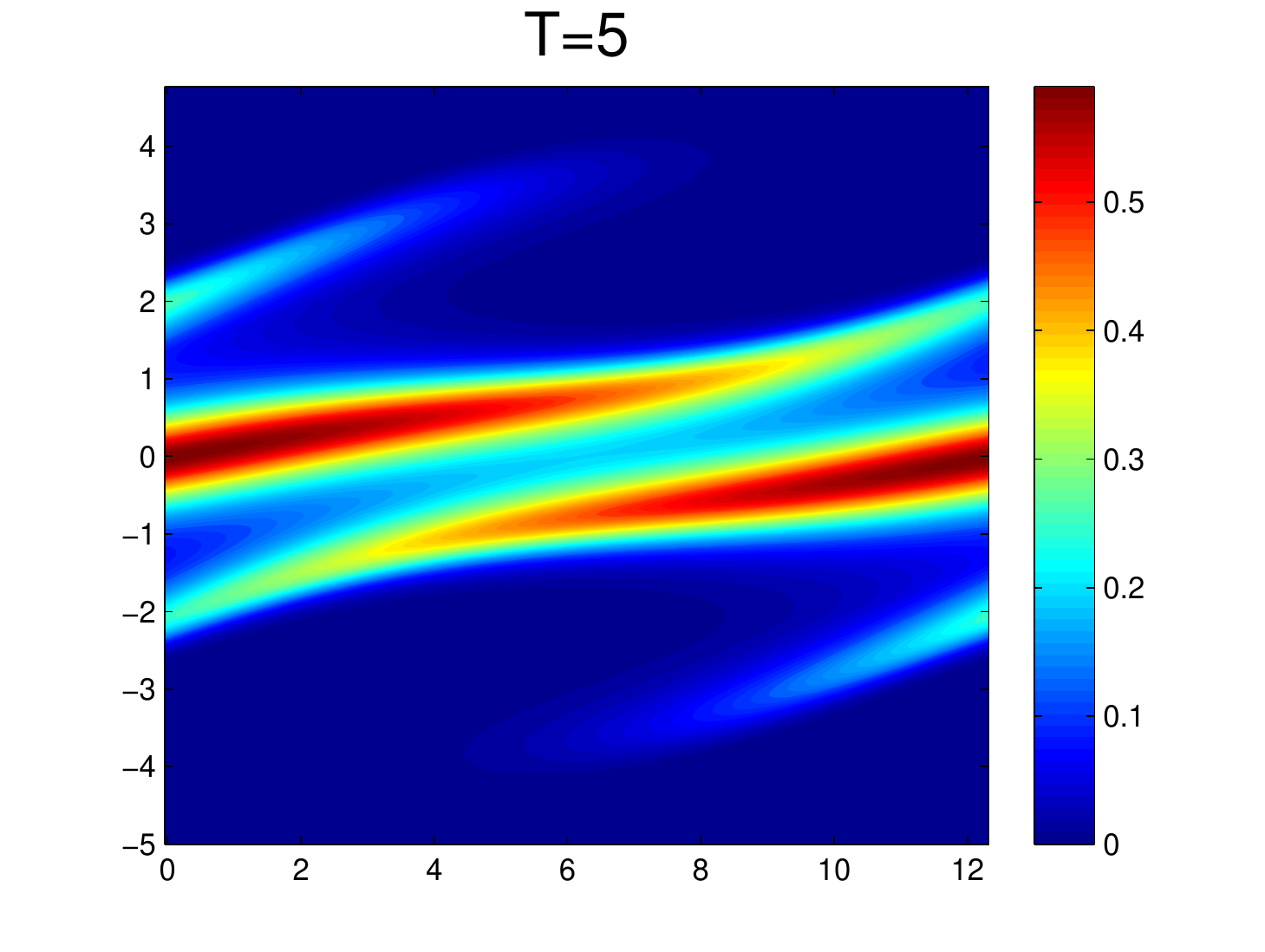}}
  \subfloat[]{\includegraphics[width=0.3\textwidth]{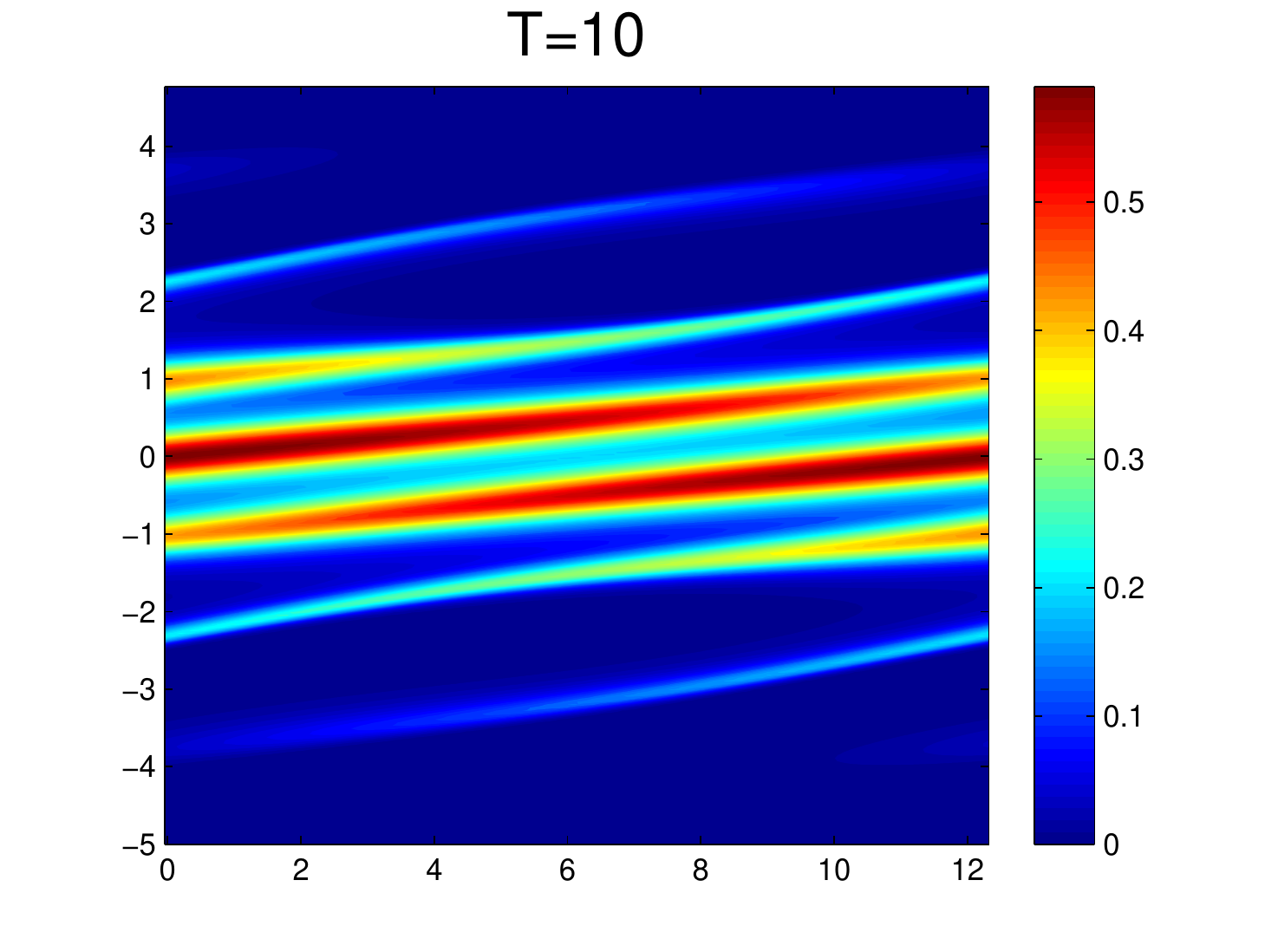}}
  \subfloat[]{\includegraphics[width=0.3\textwidth]{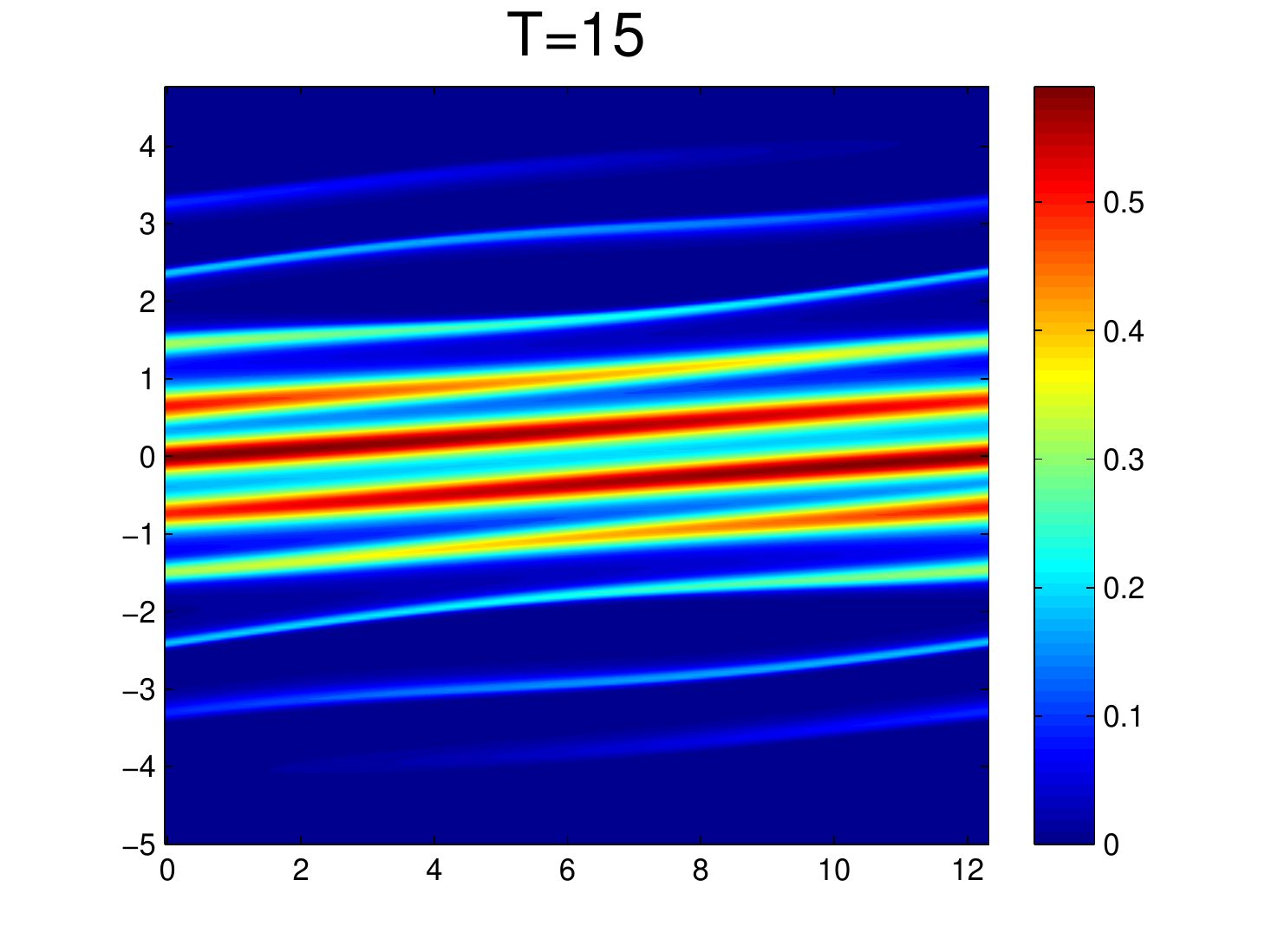}}\\
  \subfloat[]{\includegraphics[width=0.3\textwidth]{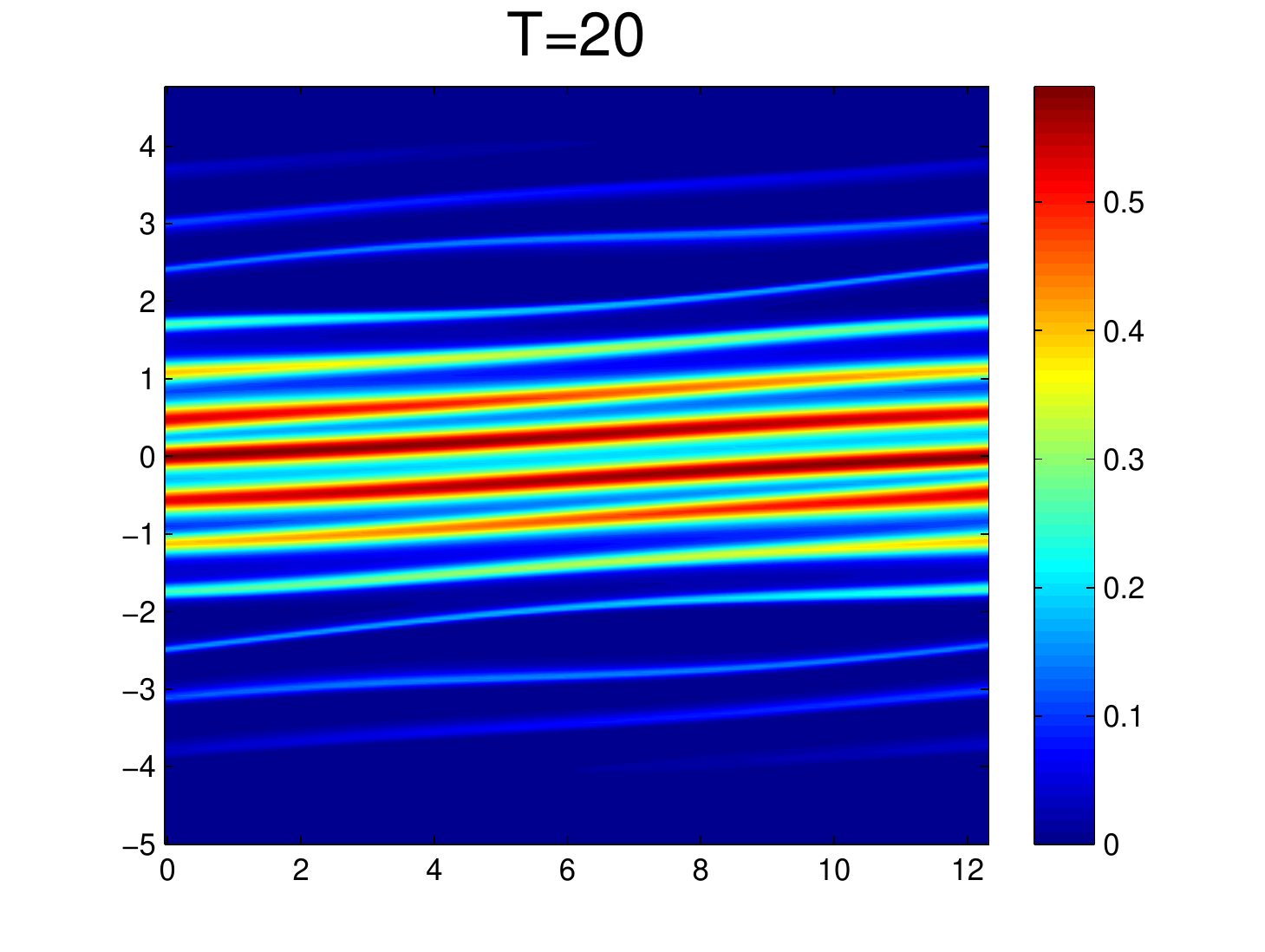}}
  \subfloat[]{\includegraphics[width=0.3\textwidth]{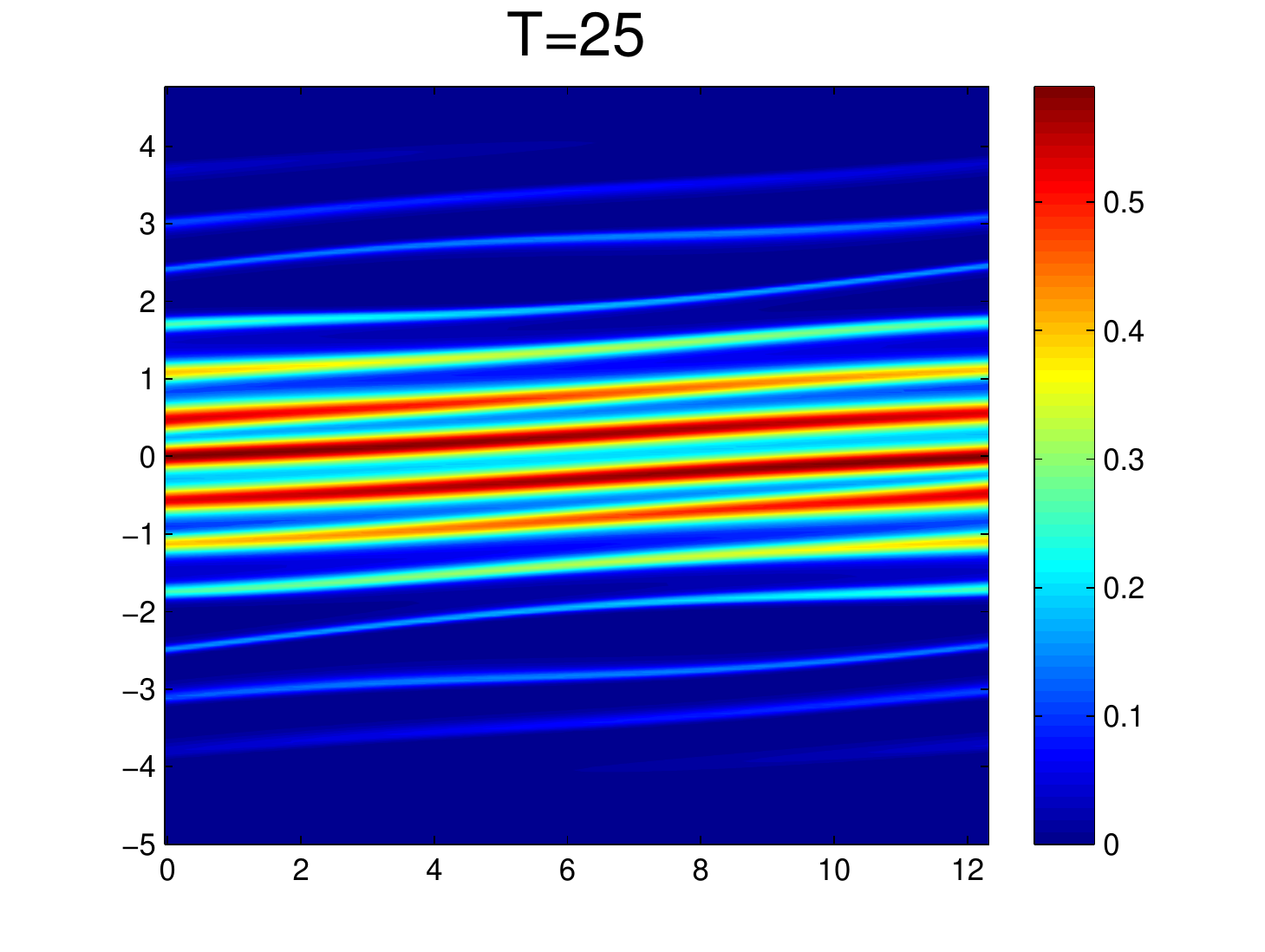}}
  \subfloat[]{\includegraphics[width=0.3\textwidth]{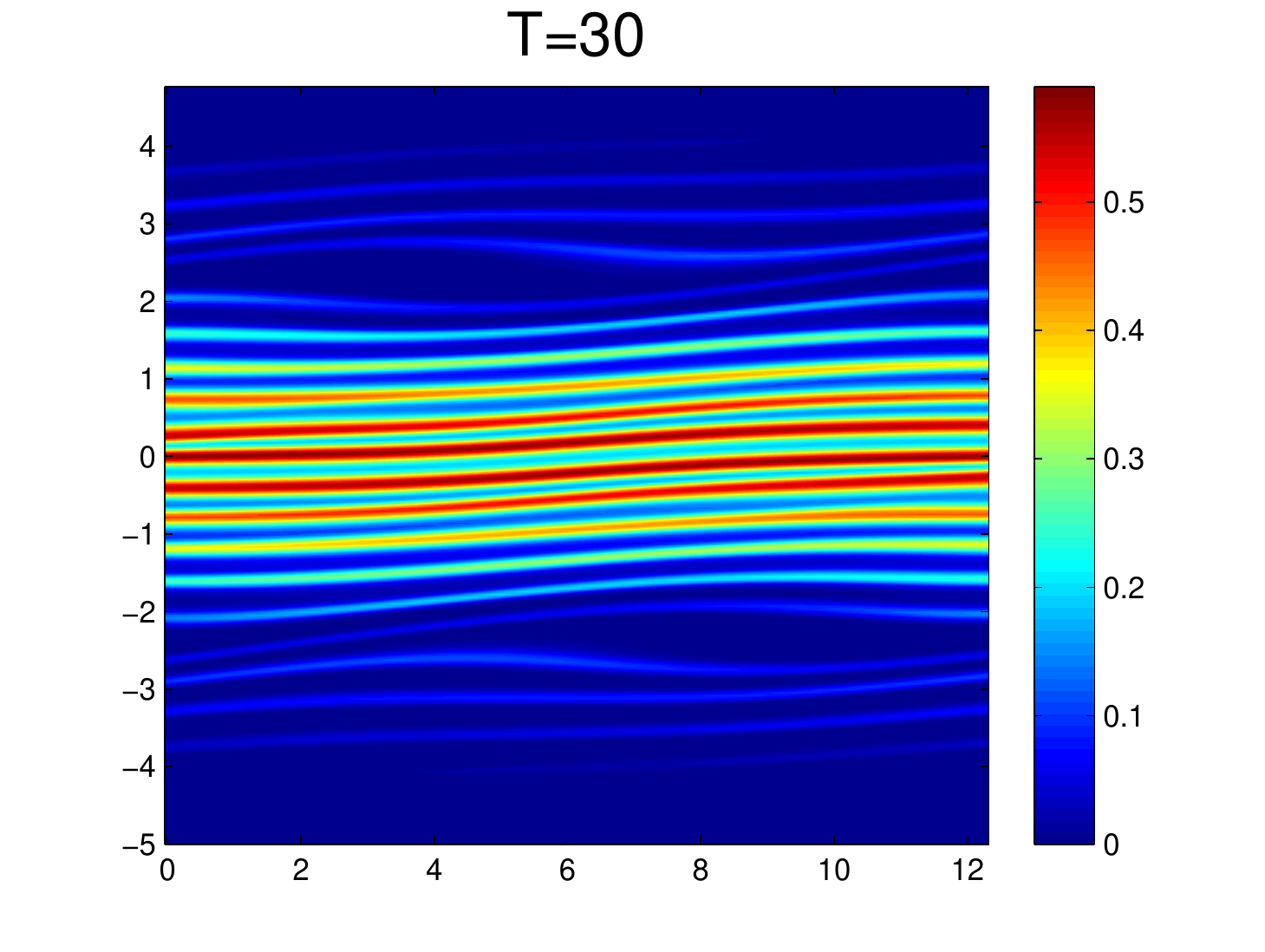}}\\
  \subfloat[]{\includegraphics[width=0.3\textwidth]{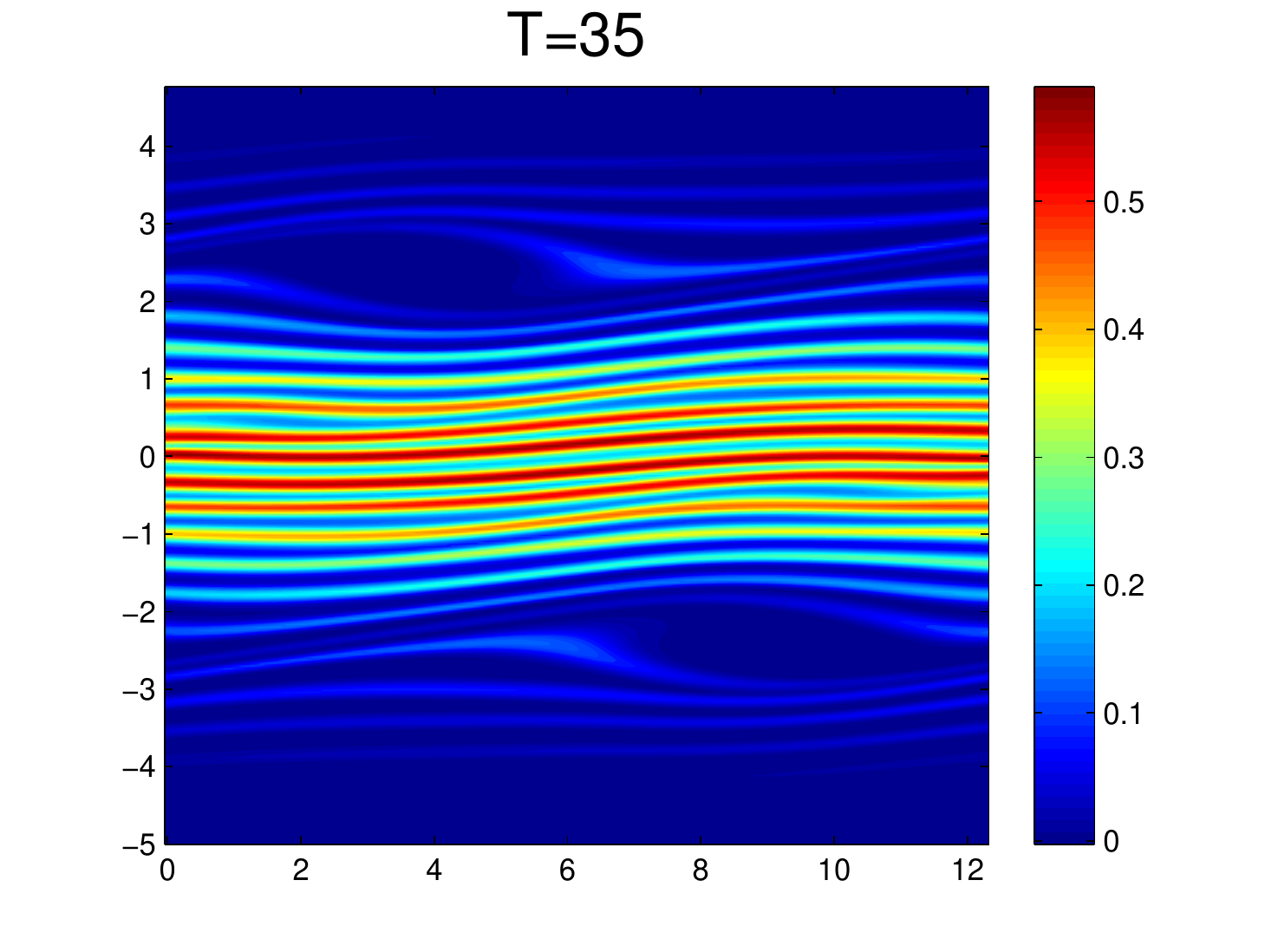}}
  \subfloat[]{\includegraphics[width=0.3\textwidth]{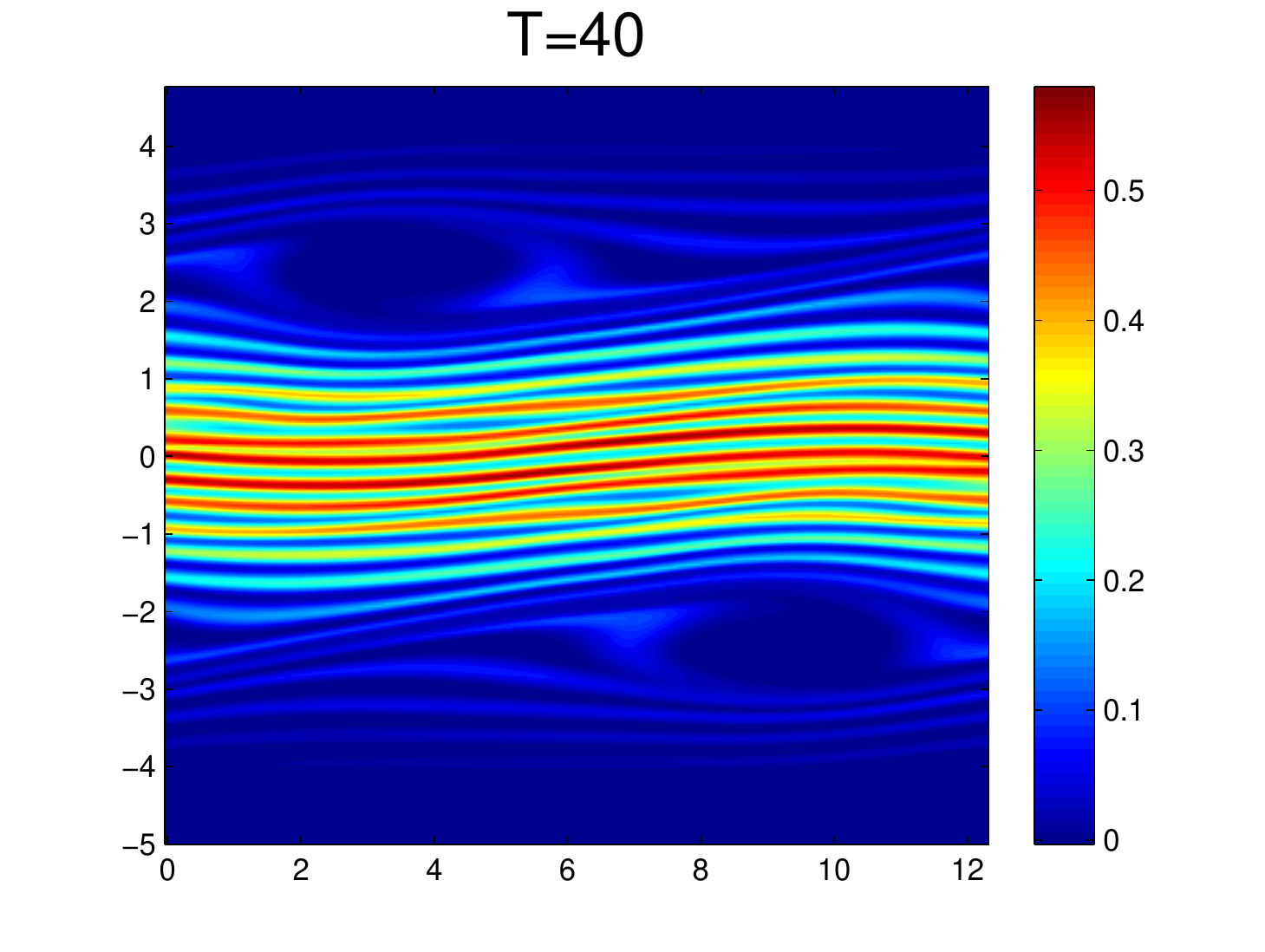}}
  \subfloat[]{\includegraphics[width=0.3\textwidth]{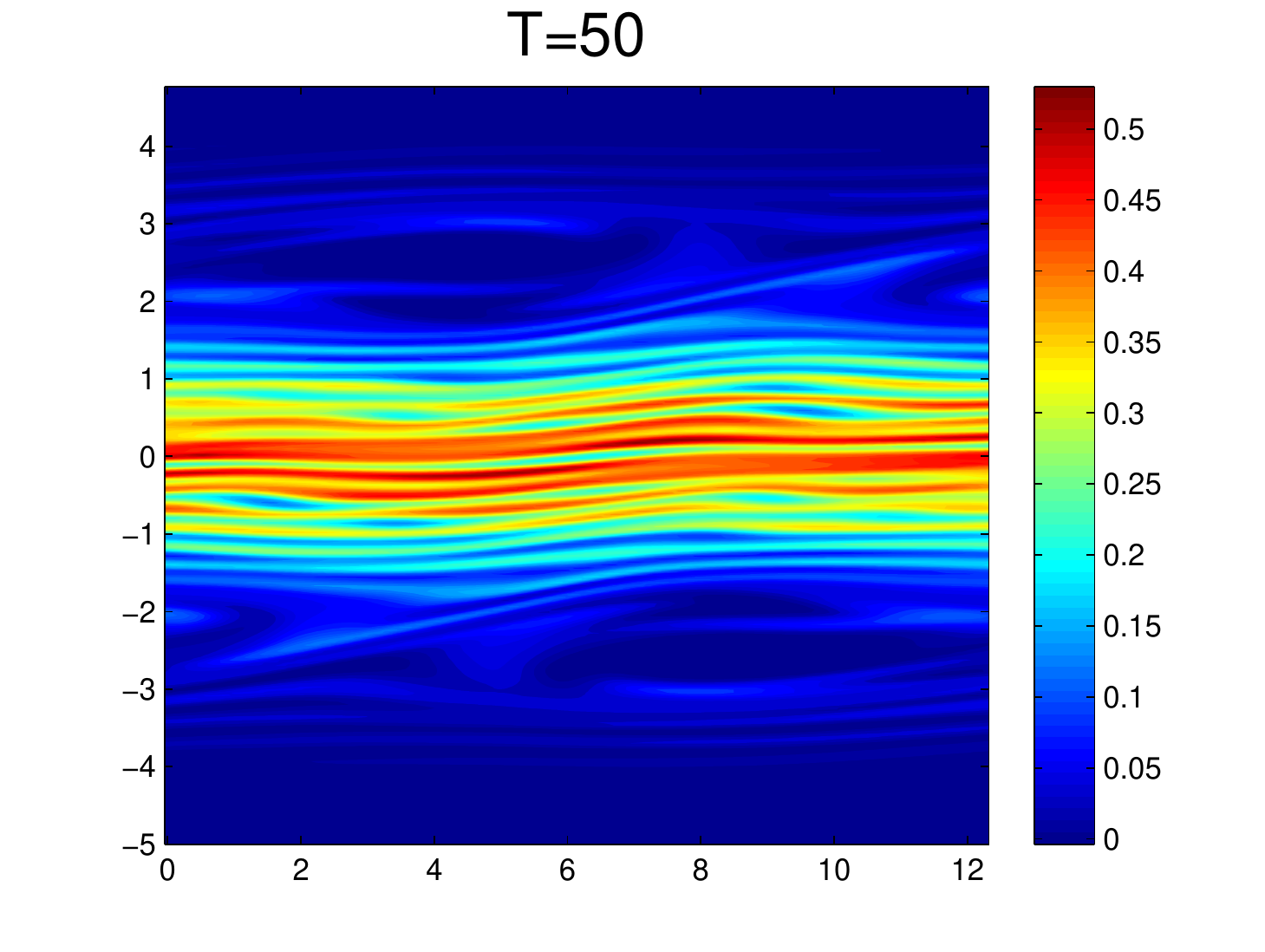}}
  \caption{\label{fig:landau_nonlinear_f} Velocity distribution contours for strong Landau damping \cite{liu2017plasma}. $N_x\times N_u=256\times 256$.}
\end{figure}

When the VP system is perturbed by a large amplitude, the nonlinear effect will appear \cite{chen2015plasma}.
For nonlinear landau damping, the initial condition is similar to the linear case, while the parameters are set as $\alpha=0.5$ and $k=0.5$.
The mesh size is set to be $N_x\times N_u=256\times 256$.
Figure \ref{fig:landau_nonlinear} shows the $L^2$ norms of electric field computed by UGKS.
The linear decay rate of electric energy is approximately equal to $\gamma_1=-0.287$, and the growth rate provided by UGKS is approximately $\gamma_2=0.078$.
The contours of velocity distribution at different times are shown in Fig. \ref{fig:landau_nonlinear_f}.

\subsubsection{Magnetic reconnection}

Magnetic reconnection is a process in which the topology of the magnetic field lines changes \cite{birn2001geospace}.
In ideal MHD, the magnetic field lines cannot be re-connected as the field lines are 'frozen' into the fluid.
Various models were used to describe the magnetic field re-connection phenomenon, for example the electron MHD \cite{hesse2001collisionless}, MHD and Hall MHD \cite{birn2001effect,ma2001hall}, full particle \cite{pritchett2001geospace}, and hybrid model \cite{kuznetsova2001collisionless}.
It was found that the reconnection initiates at a length scale on the order of the electron skin depth and the reconnection rate is governed by the ion dynamics.
The UGKS for plasma simulation is based on the Vlasov-BGK equation which can describe the physics at electron skin depth level.
Hence it can be used to describe the reconnection process.

\begin{figure}[htbp]
\centering
\includegraphics[width=0.6\textwidth]{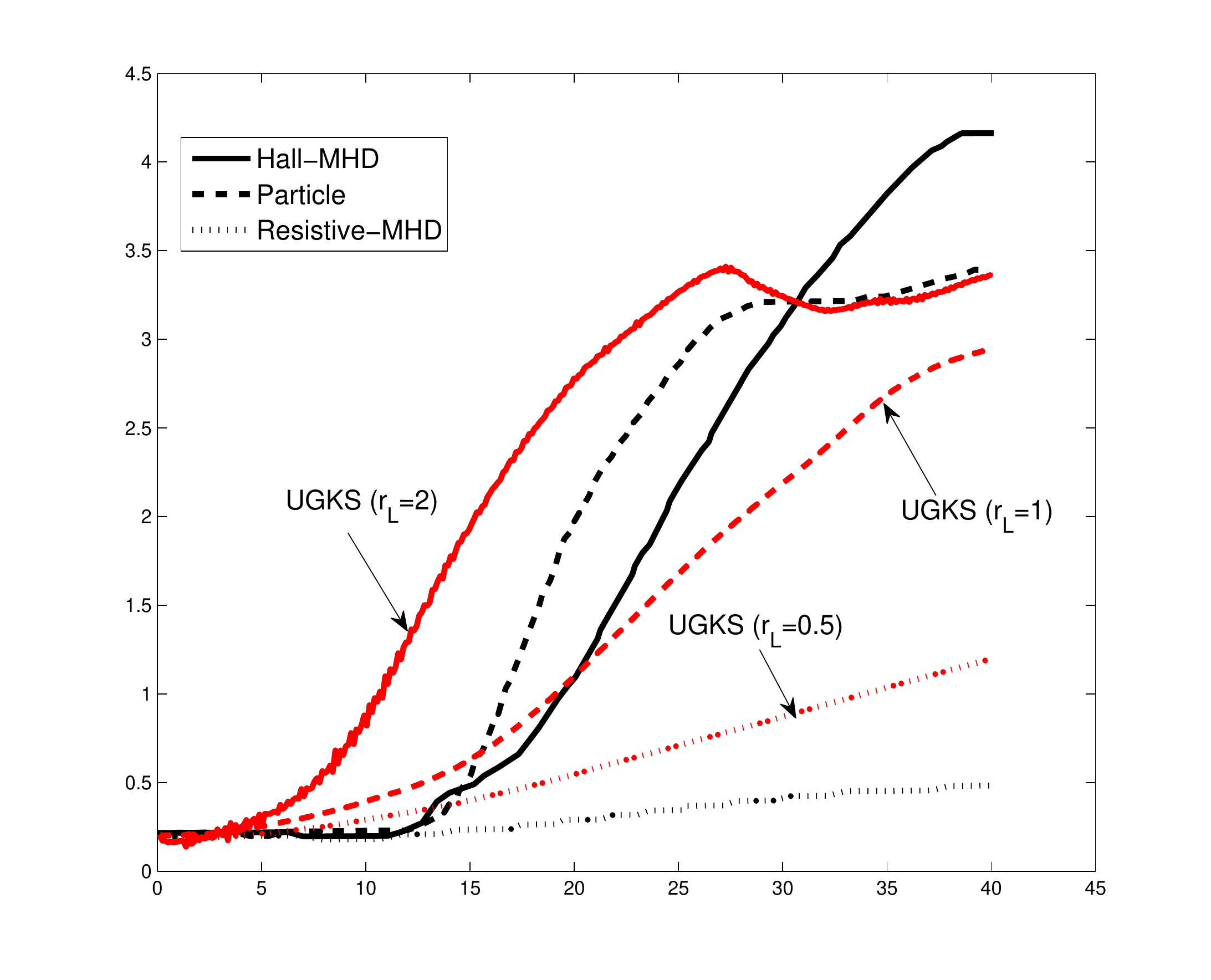}
\caption{\label{fig:reconnected_flux}The reconnected flux from the UGKS and other GEM simulations \cite{liu2017plasma}.}
\end{figure}

The simulation uses the same initial conditions as the GEM challenge problem \cite{hakim2006high}.
The initial magnetic field is given by
\begin{equation}
  \vec{B}(y)=B_0 \tanh(y/\lambda) \vec{e}_x,
\end{equation}
and a corresponding current sheet is carried by the electrons
\begin{equation}
  \vec{J}_e=-\frac{B_0}{\lambda} \text{sech} ^2(y/\lambda)\vec{e}_z.
\end{equation}
The initial number densities of electron and ion are
\begin{equation}
  n_e = n_i = 1/5 + { sech}^2 (y/\lambda).
\end{equation}
The electron and ion pressures are set to be
\begin{equation}
  p_i=5 p_e=\frac{5 B_0}{12} n(y),
\end{equation}
where $B_0=0.1$, $m_i=25m_e$ and $\lambda=0.5$.
The electromagnetic correction potentials are set $\phi=\psi=0$ initially.
The computational domain is $[-L_x/2, L_x/2]\times[-L_y/2,L_y/2]$ with $L_x=8\pi$, $L_y=4\pi$, which is divided into $200\times 100$ cells.
Periodic boundaries are applied at $x=\pm L_x/2$ and conducting wall boundaries at $y=\pm L_y/2$.
To initiate reconnection, the magnetic field is perturbed with $\delta \vec{B}=\vec{e}_z \times \nabla_{\vec{x}} \psi$, where
\begin{equation}
  \psi(x,y)=0.1 B_0 \cos(2\pi x/L_x) \cos(\pi y/L_y).
\end{equation}
The velocity space for ion is $[-3, 3]\times[-3, 3]$ and for electron is $[-25, 25]\times[-25, 25]$ with $32\times32$ velocity grids.
The computational time for UGKS is about $1342$ mins on a 3.40GHz 4-core CPU.

\begin{figure}[htbp]
\centering
\includegraphics[width=0.75\textwidth]{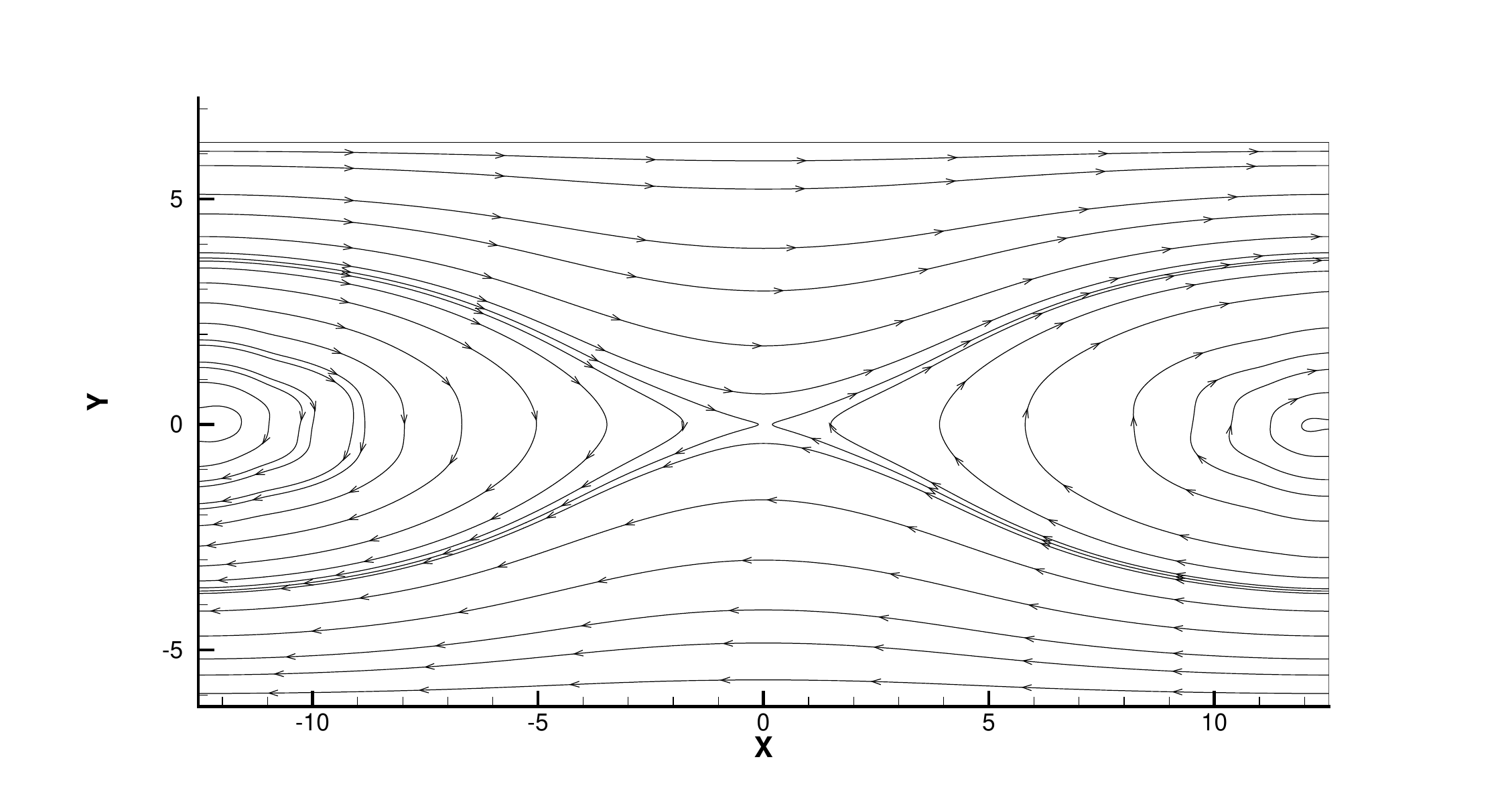}
\caption{\label{fig:reconnection_mflux}Magnetic flux at $\omega t=40$ with $r_L=1$ \cite{liu2017plasma}.}
\end{figure}

\begin{figure}[htbp]
\centering
\includegraphics[width=0.5\textwidth]{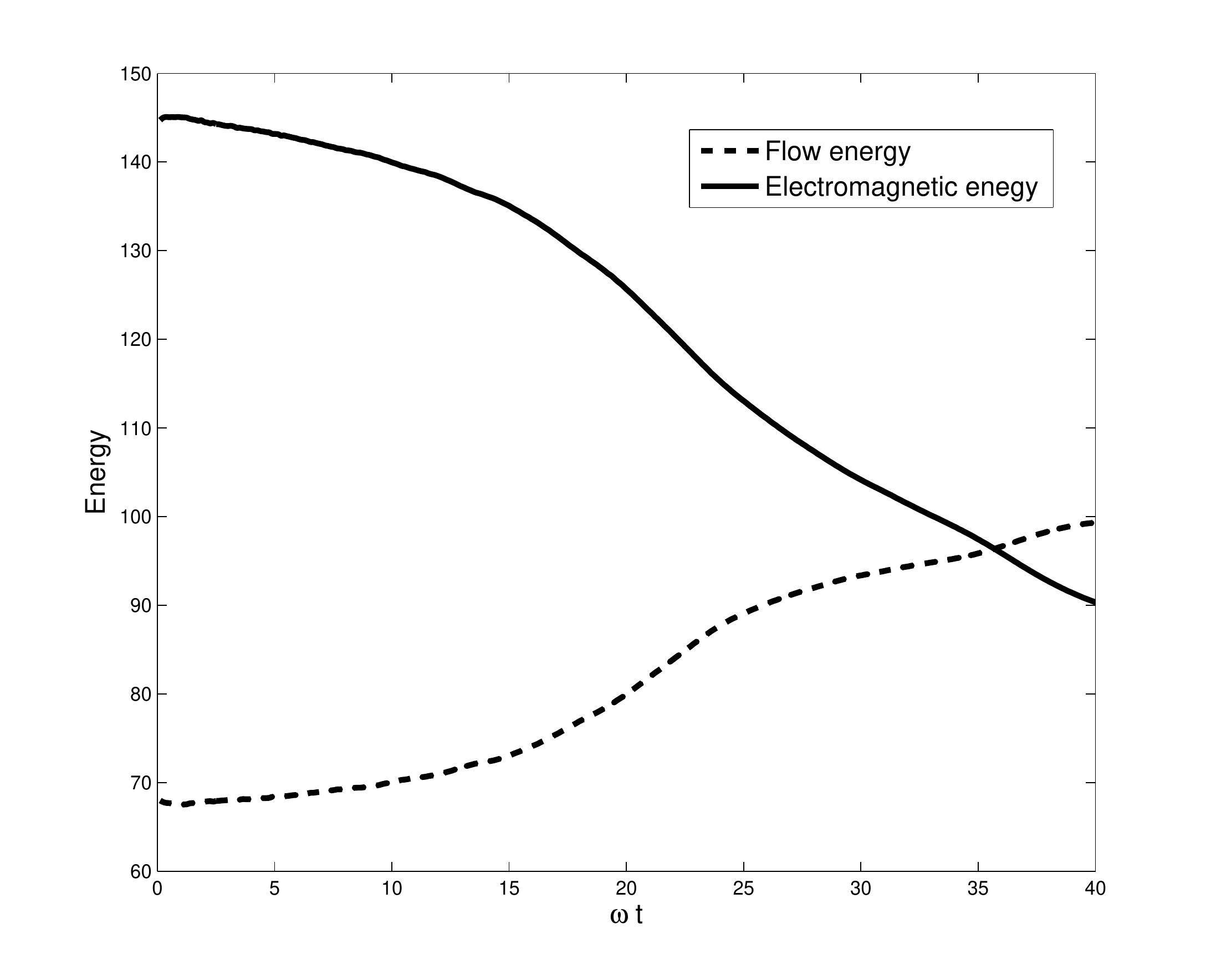}
\caption{\label{fig:reconnection_energy}Electromagnetic and flow energy evolution in the magnetic reconnection process \cite{liu2017plasma}.}
\end{figure}

\begin{figure}[htbp]
\centering
\subfloat[]{\includegraphics[width=0.44\textwidth]{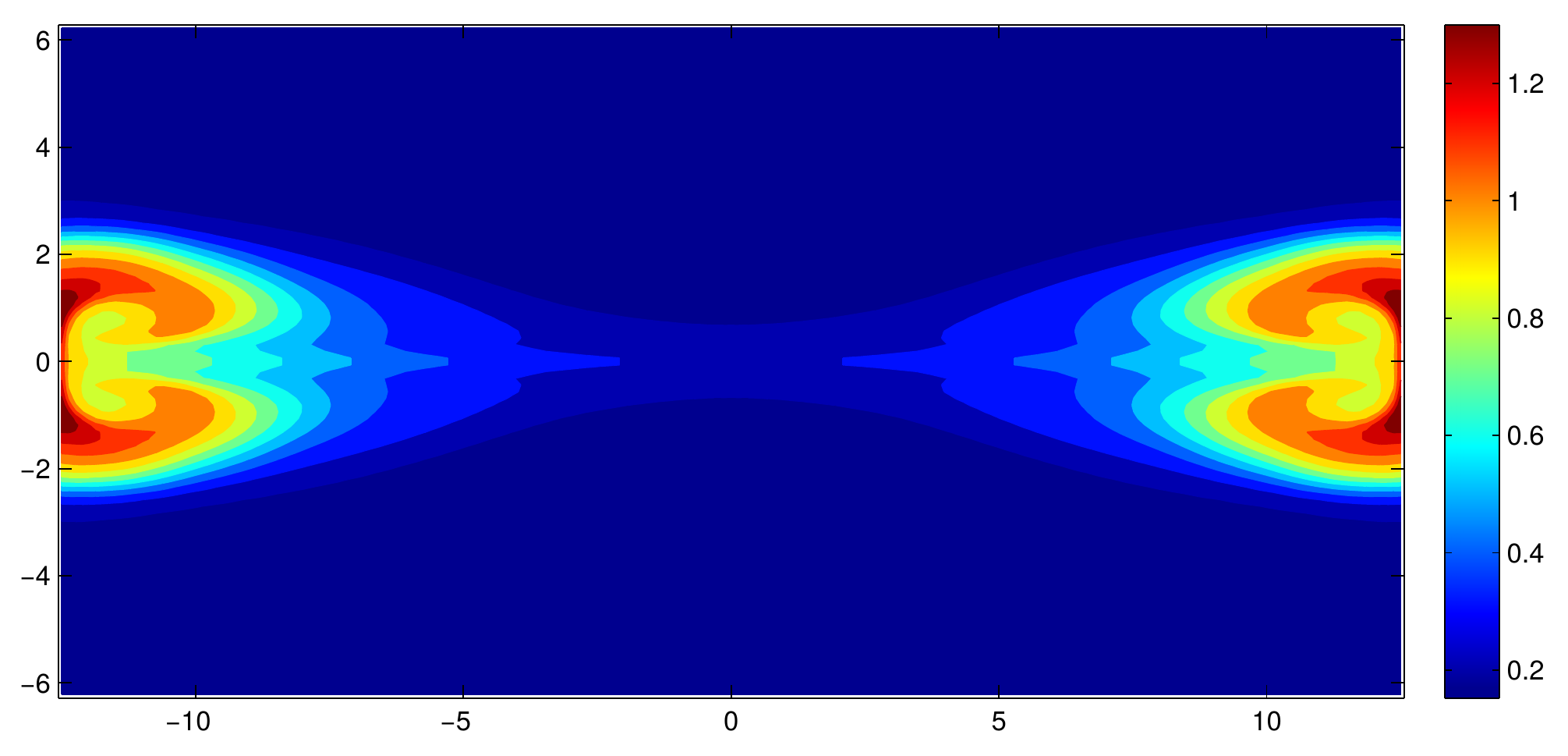}}\hspace{1em}
\subfloat[]{\includegraphics[width=0.44\textwidth]{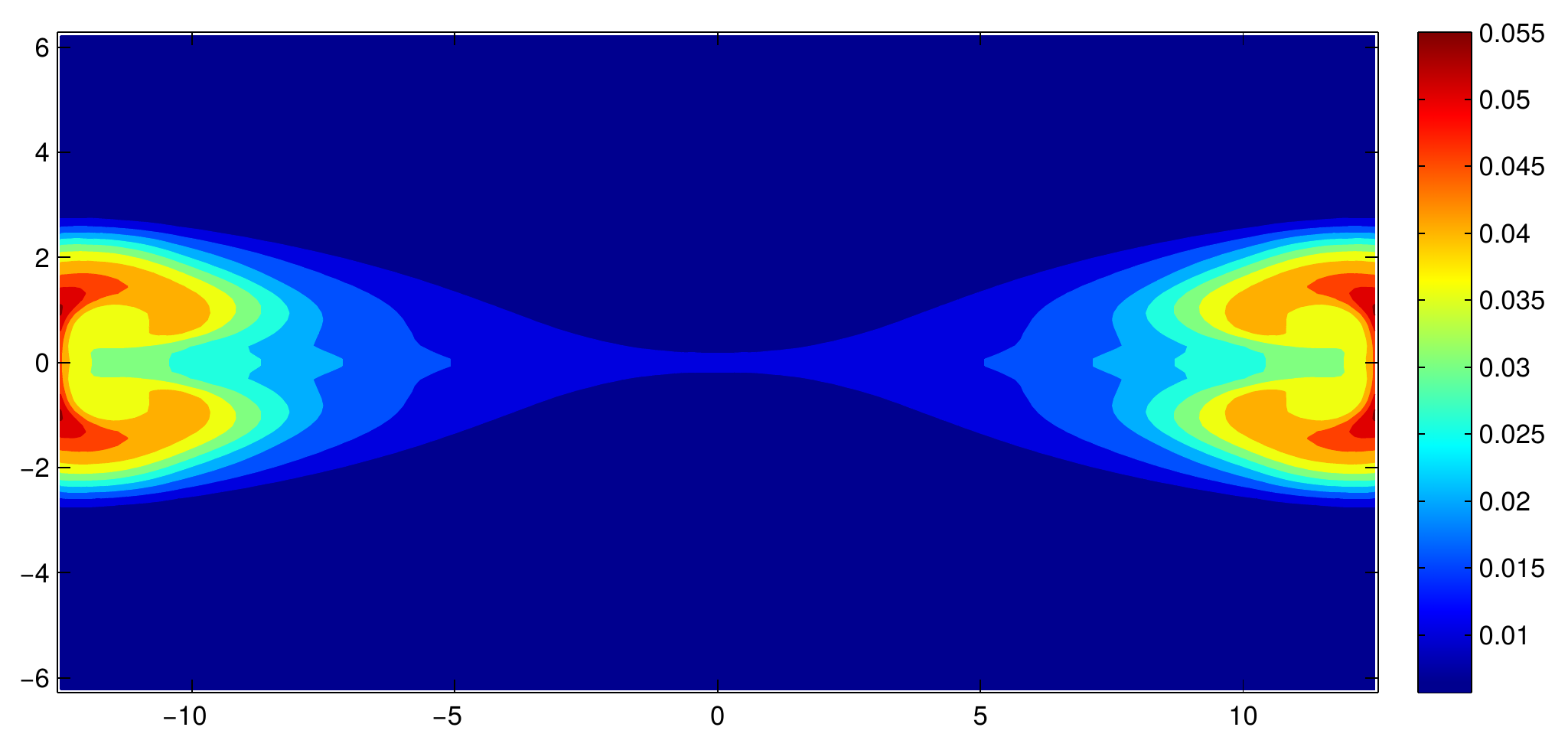}}\\
\subfloat[]{\includegraphics[width=0.44\textwidth]{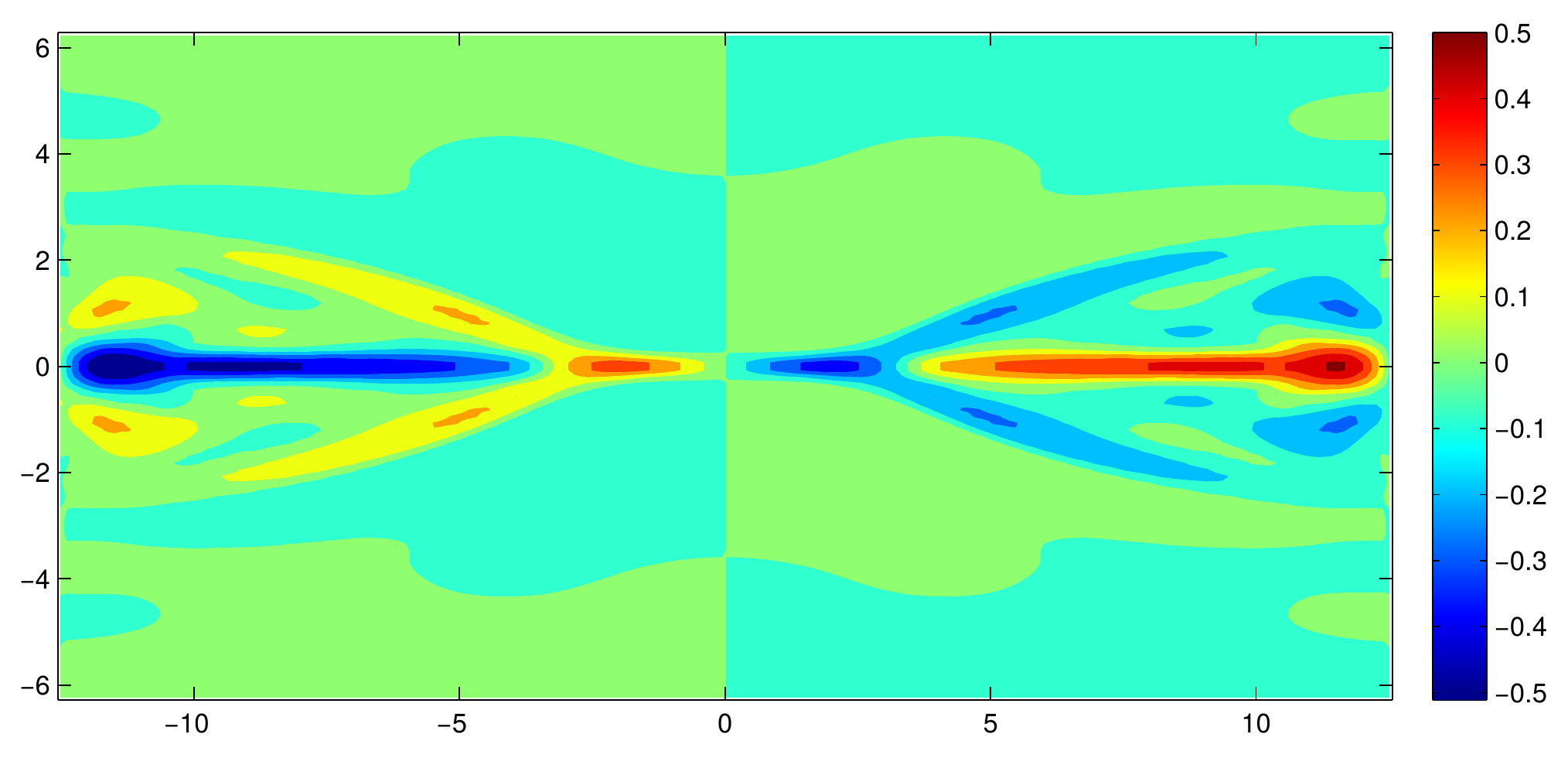}}\hspace{1em}
\subfloat[]{\includegraphics[width=0.44\textwidth]{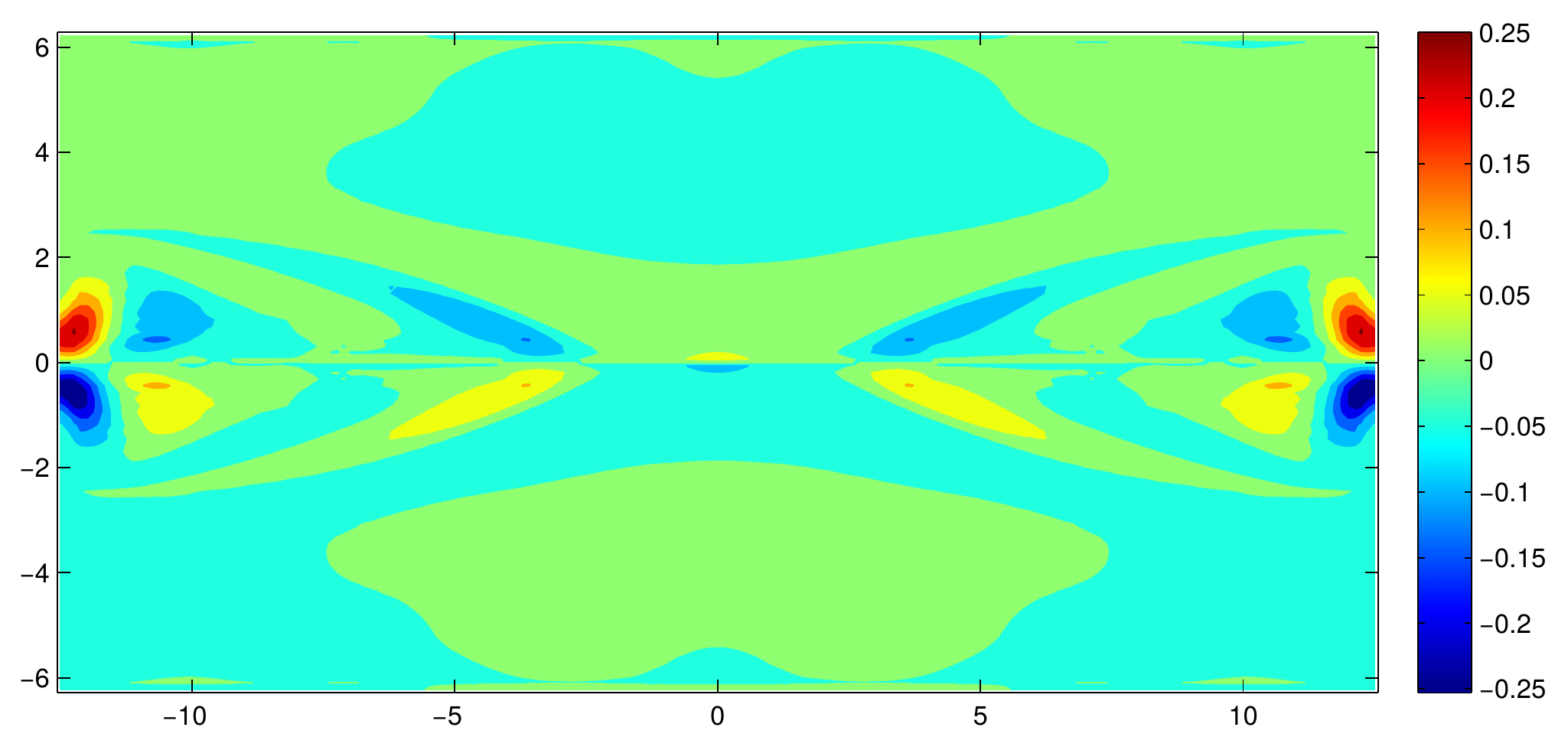}}\\
\subfloat[]{\includegraphics[width=0.44\textwidth]{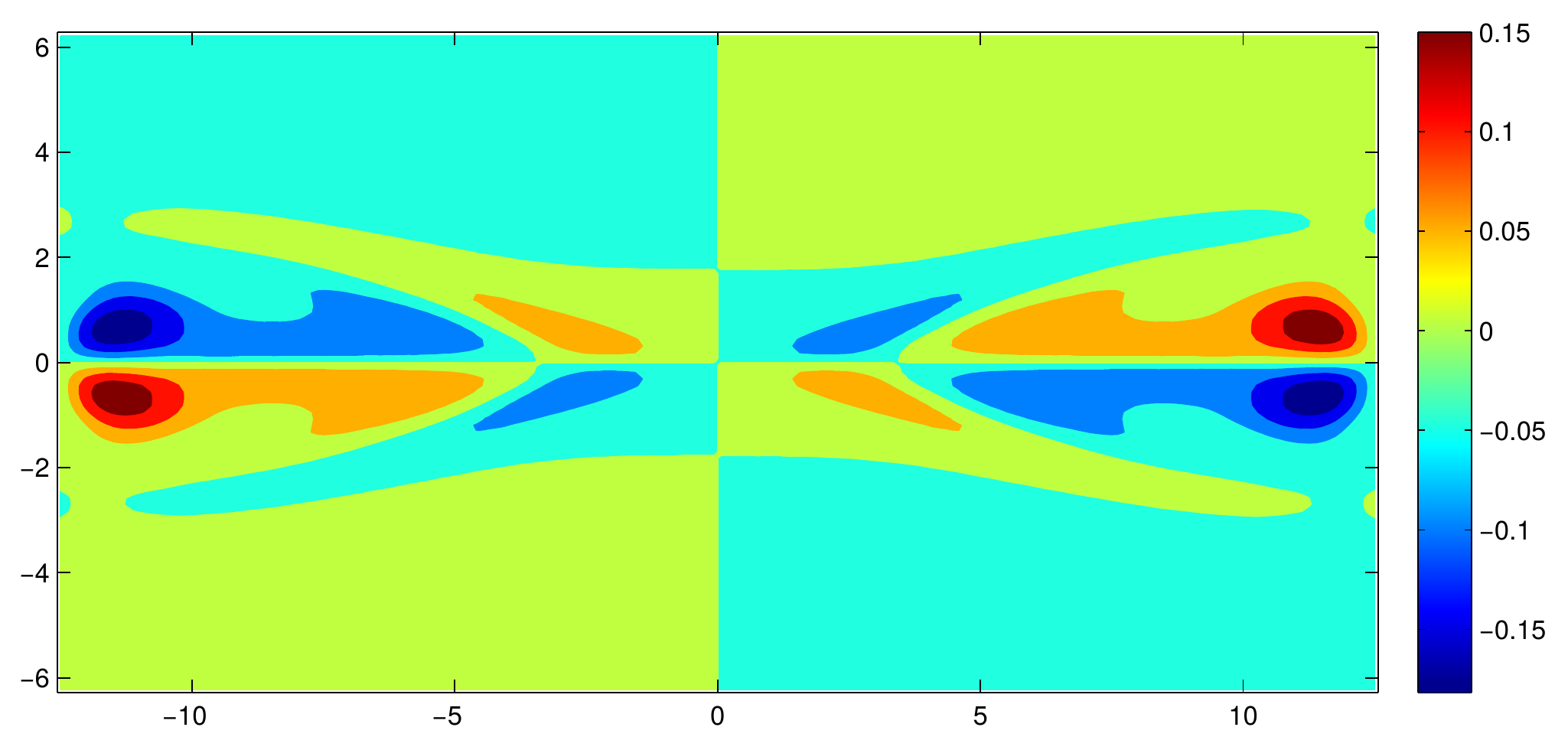}}\hspace{1em}
\subfloat[]{\includegraphics[width=0.44\textwidth]{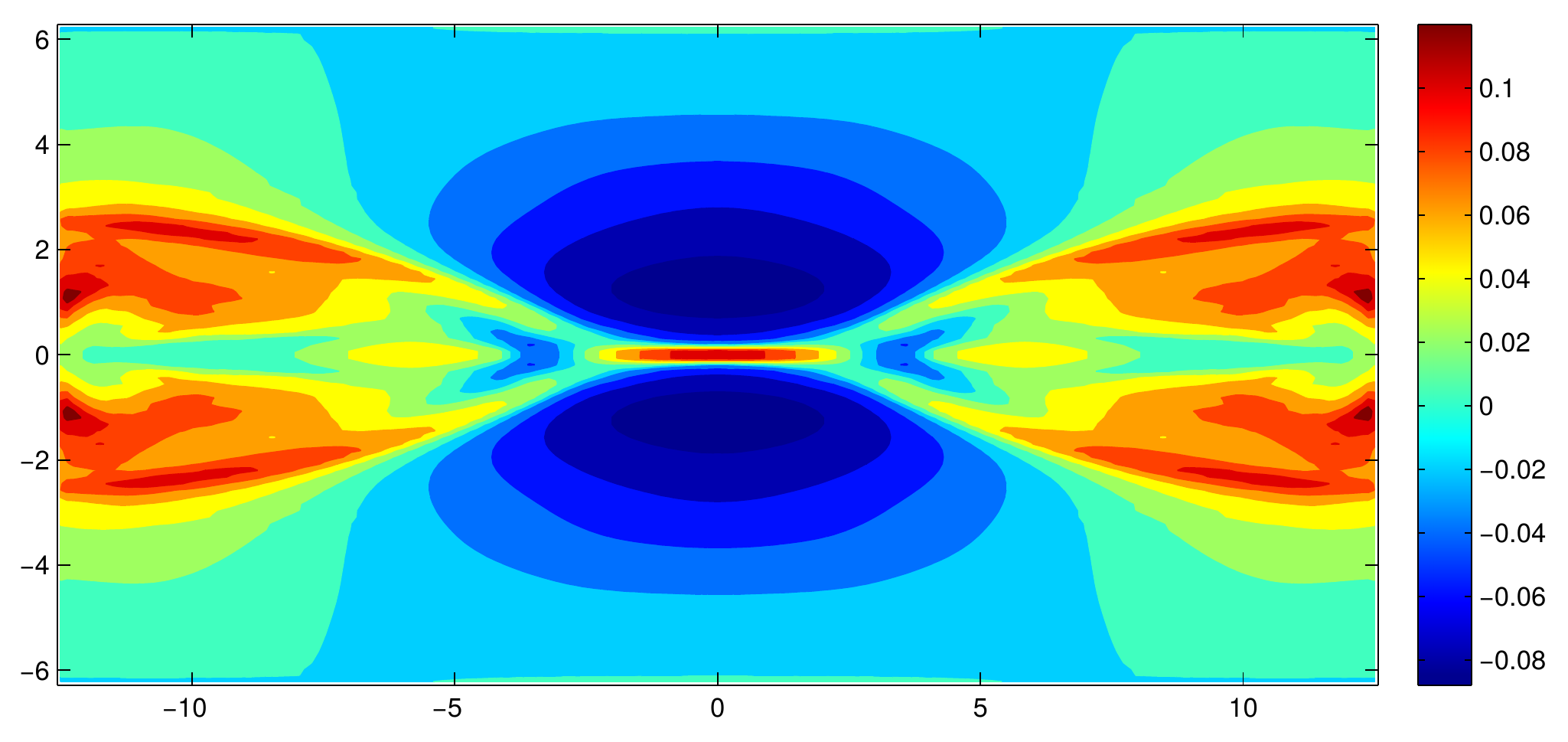}}
\caption{\label{fig:reconnection_field}The reconnection process with $r_L = 1.0$ at $t=40\omega^{-1}$ \cite{liu2017plasma}.
(a) Ion density, (b) electron density, (c) $x$-directional electric current density, (d) $y$-directional electric current density, (e) $z$-direction magnetic field, and (e) $z$-direction electric field.}
\end{figure}

Figure \ref{fig:reconnected_flux} shows the reconnected flux of UGKS defined by
\begin{equation}
\phi(t)=\frac{1}{2L_x}\int_{-L_x/2}^{L_x/2} |B_y(x,0,t)| dx,
\end{equation}
which is compared with other GEM results.
The fast reconnection rate can be predicted by UGKS, and the magnitude of reconnected flux depends on the plasma conditions.
It can be observed from the results that the reconnected flux from UGKS behaves likes resistive-MHD result when the normalized Larmor radius is small ($r_L=0.5$), and approaches to Hall-MHD solution when the normalized Larmor radius is large ($r_L=2$).
Figure \ref{fig:reconnection_mflux} shows the magnetic flux at $\omega t=40$, with $r_L=1$.
Figure \ref{fig:reconnection_energy} shows the electromagnetic and flow  energy.
The total energy of the system almost keeps a constant.
The electron and ion densities, and momentum distribution at $t=40\omega^{-1}$ are shown in Fig. \ref{fig:reconnection_field}, as well as the electromagnetic fields.

\subsection{Granular flow}
Gas-particle multiphase system represents an important class of multiphase flow, which is widely applied in many fields of engineering applications, such as the chemical process industry, aerospace engineering, and environment science.
The mechanics of a rapid granular flow granular gas is analogous to that of a classical molecular gas, and the flow regime of granular gas is characterized by the granular Knudsen number, which represents the importance of the particle collisions relative to the particle transport.
Besides the Knudsen number, the particle Reynolds number and Stokes number are important parameters to indicate the multiscale and multiphase physics of the granular flow.
A unified gas kinetic scheme for multiphase dilute gas-particle system has been proposed in \cite{liu2019granular,wang2018granular,wang2019monodisperse}, where the gas phase is assumed to be in the continuum flow regime and evolves numerically by the gas kinetic scheme for the Navier--Stokes solutions, and the particle phase is modeled by the UGKS which construct a multiscale flux taking the particles' transport, collision, and acceleration into account.

For particle phase, the evolution of the velocity distribution of solid phase apparent density $f_s(\vec{x},\vec{u},t)$ follows Williams--Boltzmann equation \cite{marchisio2013computational}, coupled with the solid temperature equation.
The equation system can be written as
\begin{equation}\label{eq:granular_kinetic}
	\frac{\partial f_s}{\partial t} + \vec{u} \cdot \nabla_{\vec{x}} f_s + \nabla_{\vec{u}} \cdot \left[\left(\vec{G} - \frac{1}{\rho_s} \nabla_{\vec{x}}p_g\right)f_s\right] + \nabla_{\vec{u}} \left(\frac{\vec{D}}{m_s} f\right) = \mathcal{J},
\end{equation}
and
\begin{equation} \label{eq:granular_temperature}
\frac{\partial C_s \rho_s \epsilon_s T_s^M}{\partial t} + \nabla_{\vec{x}} \cdot (C_s \rho_s \epsilon_s T_s^M \vec{U}_s) = r_{T_m} (1-r^2) \frac{3\epsilon_s \rho_s k_B T_s}{4 \tau_s m_s} + C_s \epsilon_s \rho_s \frac{T_g - T_s^M}{\tau_T},
\end{equation}
where $\vec{G}$ denotes the gravitational force, $\rho_s$ is the material density of the solid particle.
$p_g$ is the gas phase pressure, $m_s$ denotes the mass of one solid particle, $\vec{D}$ is the drag force from gas phase, and $\mathcal{J}$ represents the particle-particle collisions.
In Eq.~\eqref{eq:granular_temperature}, $T_s^M$ is the solid temperature, $C_s$ is the specific heat capacity of solid phase, $\epsilon_s$ is the solid volume fraction, $\vec{U}_s$ denotes the macroscopic velocity of solid phase.
$r_{T_m}$ is the proportion of the kinetic energy loss in inelastic collision which is transferred into the material thermal energy.
$r$ is the restitution coefficient of the inelastic collision, and the heat conduction between phases is modeled by a relaxation time $\tau_T$.

The collision between particles adopts a BGK-type model
\begin{equation}
	\mathcal{J} = \frac{g_s - f_s}{\tau_s},
\end{equation}
where $\tau_s$ is the relaxation time, and the equilibrium state $g_s$ is a Maxwellian distribution
\begin{equation}
	g_s = \epsilon \rho_s \left(\frac{m_s}{2 \pi k_B T_s}\right)^{\frac{3}{2}} \exp\left[-\frac{m_s}{2 k_B T_s} (\vec{u}-\vec{U}_s)^2\right].
\end{equation}
The macroscopic moments of $f_s$ and the Maxwellian distribution $g_s$ satisfy
\begin{equation}
	\int {\vec{\psi} g_s d\vec{u}} = \int {\vec{\psi}^{\prime} f_s d\vec{u}},
\end{equation}
where $\vec{\psi}^{\prime} = (1, \vec{u}, \frac{1}{2}(\vec{u}^2 + (r^2-1)(\vec{u} - \vec{U_s})^2))^T$.
The inelastic collision is considered by varying $r$ from $1$ to $0$.

In order to deal with large interspecies velocity difference and large drag coefficient, and overcome the stiffness problem, a relaxation form of drag force with implicit treatment is adopted,
\begin{equation}
	\vec{D} = \frac{m_s}{\tau_{st}} (\vec{U}_g - \vec{u}),
\end{equation}
where $\vec{U}_g$ is the macroscopic velocity of gas phase.

For the gas phase, the governing equations in the continuum regime are the Navier--Stokes equations, which correspond the kinetic equation
\begin{equation}\label{eq:gas_phase_kinetic}
	\frac{\partial \epsilon_g f_g}{\partial t} + \vec{u} \cdot \nabla_{\vec{x}} f_g + \nabla_{\vec{u}} \cdot \left[ \frac{f_s}{\tau_{st}} (\vec{U}_g - \vec{u}) - \frac{f_s}{\rho_s} \nabla p_g\right] + \nabla_{\vec{u}} \cdot (\vec{G}\epsilon_g f_g) = \frac{\epsilon_g g_g - \epsilon_g f_g}{\tau_g},
\end{equation}
and the heat conduction equation can be written as
\begin{equation}\label{eq:gas_phase_conduction}
	\frac{d T_g}{d t} = -\frac{\epsilon_s \rho_s C_s}{\epsilon_g \rho_g C_v} \frac{T_g - T_s^M}{\tau_T},
\end{equation}
where $\epsilon_g = 1 - \epsilon_s$ denotes the volume fraction of gas phase, $g_g$ is the Maxwellian equilibrium state, $C_v$ is the specific heat capacity of gas phase at constant volume.

For one step evolution, the equation system of the particle phase \eqref{eq:granular_kinetic} and \eqref{eq:granular_temperature}, and the equations of gas phase \eqref{eq:gas_phase_kinetic} and \eqref{eq:gas_phase_conduction} are solved simultaneously.
The key idea is that the acceleration terms dependent and independent on the microscopic solid particle velocities are treated separately, and a multiscale flux function is constructed the same as that in the UGKS for rarefied gas dynamics by using the integral solution of the BGK-type kinetic equation with external force, which considers the particles' transport, collision and acceleration in the evolution process.
For gas phase in the continuum regime, the gas kinetic scheme is employed to solve the kinetic equation \eqref{eq:gas_phase_kinetic}.
The interaction between the solid and gas phase is calculated by a velocity space mapping method.
The multiphase UGKS calculates the flow in the regimes from collisionless to two-fluid NS regime with different Knudsen number, and from granular flow to dusty gas dynamics with different Stokes number.
It is a true multiscale method for the gas-particle multiphase system.

\subsubsection{Particle jets impinging problem}

\begin{figure}[htbp]
\centering
\subfloat[]{\includegraphics[width=0.45\textwidth]{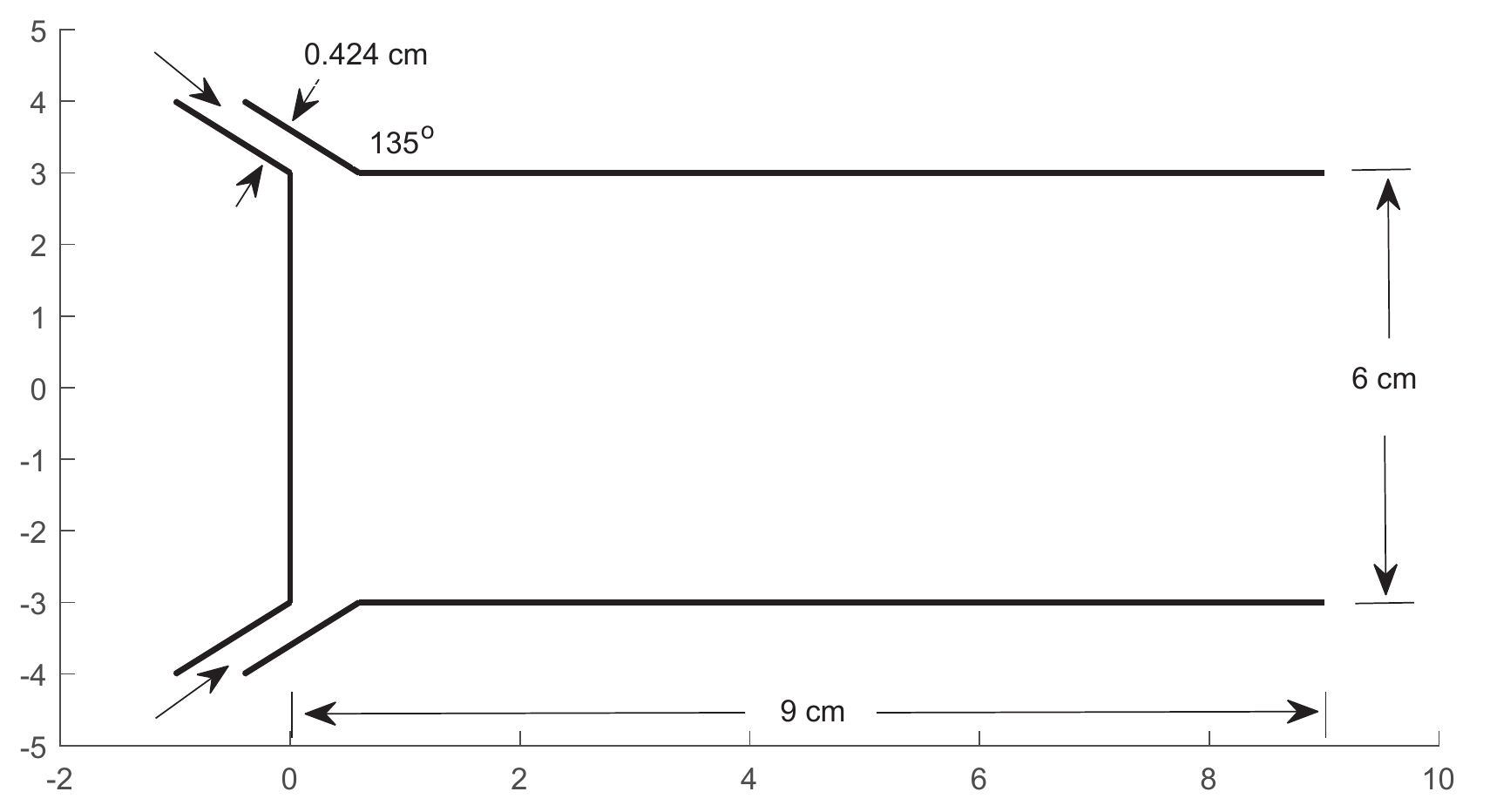}}
\subfloat[]{\includegraphics[width=0.5\textwidth]{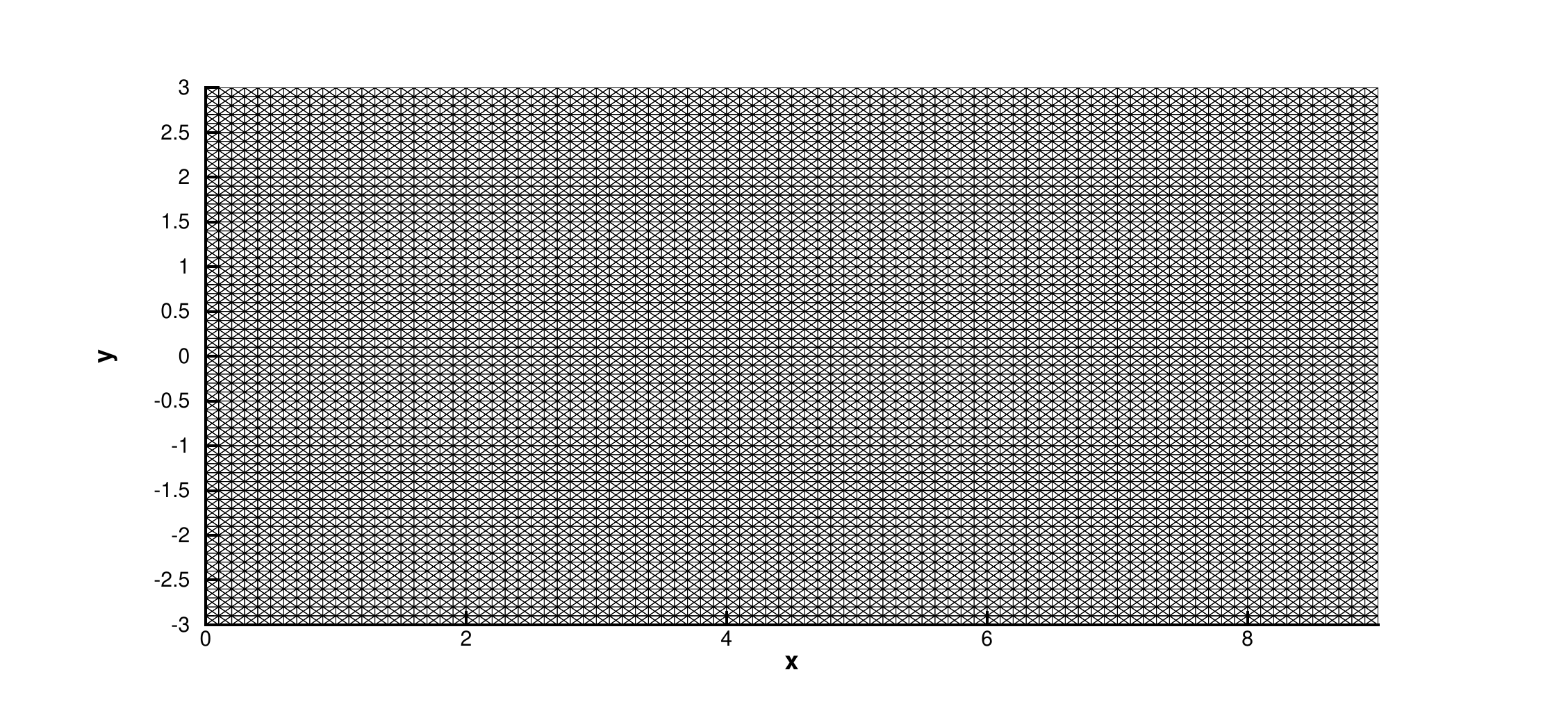}}
\caption{\label{fig:jet_initial}Setup for the calculation of two impinging particle jets \cite{liu2019granular}. (a) Channel geometry, (b) computational mesh.}
\end{figure}

\begin{figure}[htbp]
\centering
\subfloat[]{\includegraphics[width=0.44\textwidth]{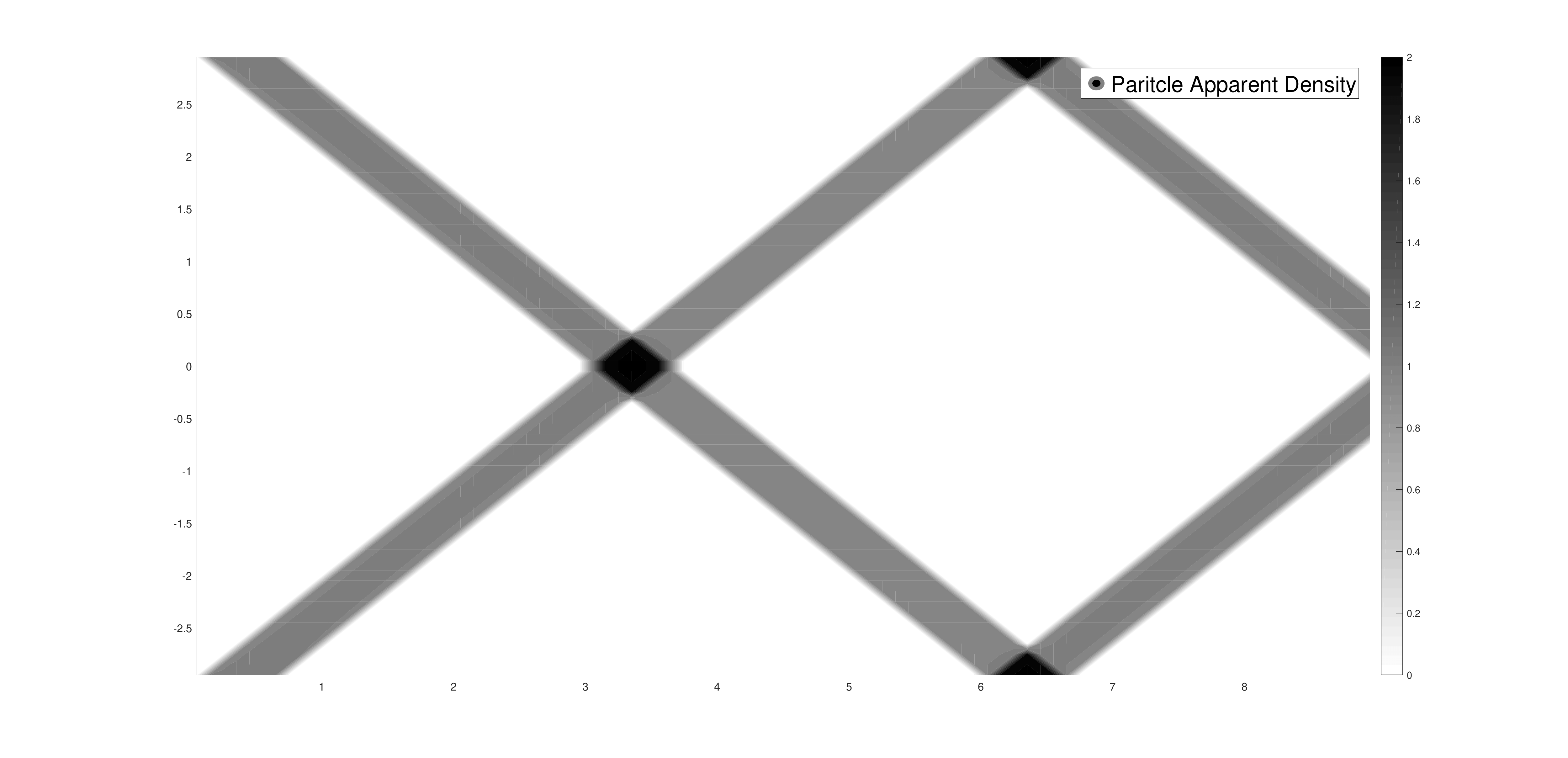}}
\subfloat[]{\includegraphics[width=0.44\textwidth]{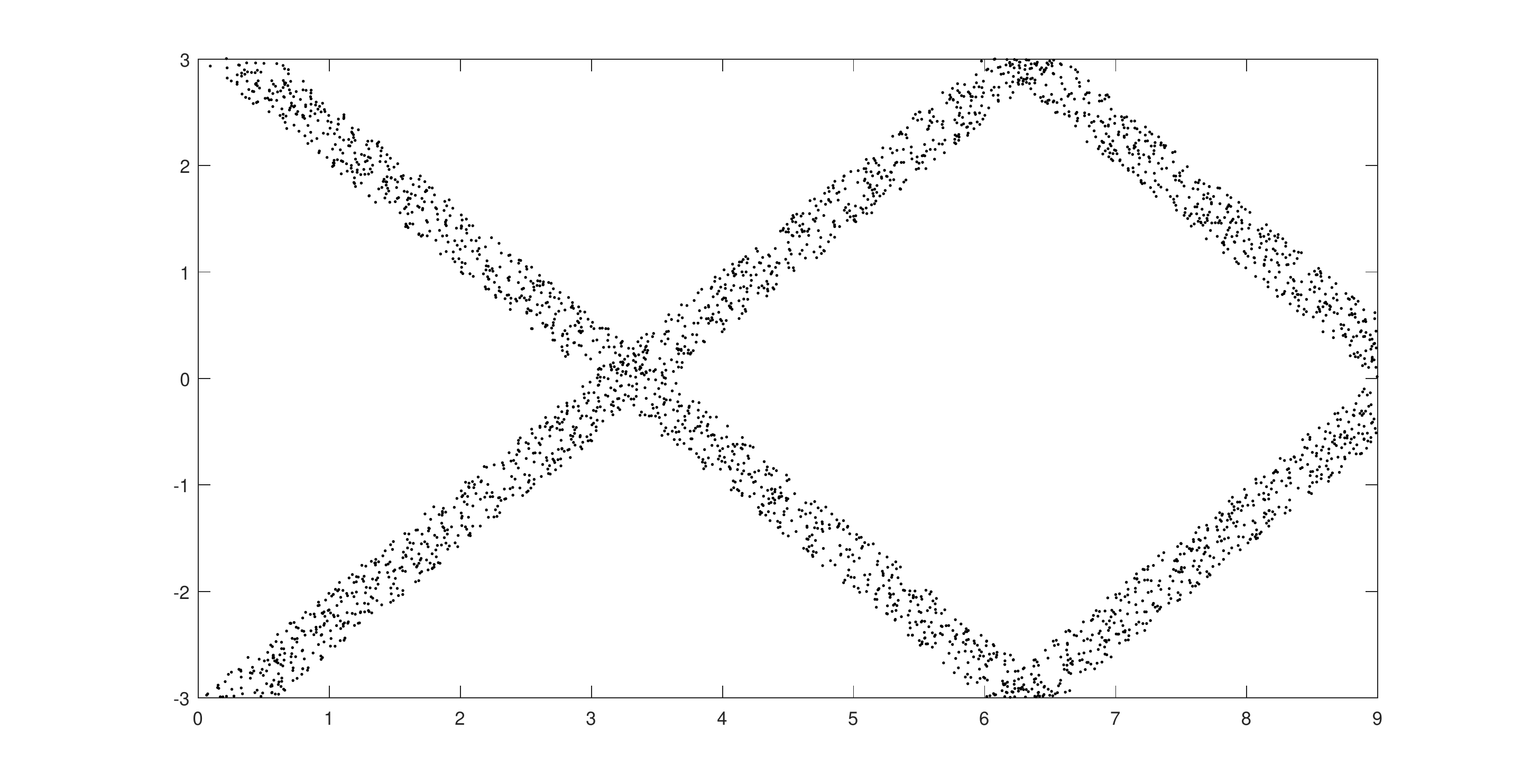}}
\caption{\label{fig:jet_collisionless}Particle number density at $t=20$ for collisionless regime \cite{liu2019granular}. (a) UGKS result, and (b) PIC result.}
\label{jets1}
\end{figure}

\begin{figure}[htbp]
\centering
\subfloat[]{\includegraphics[width=0.44\textwidth]{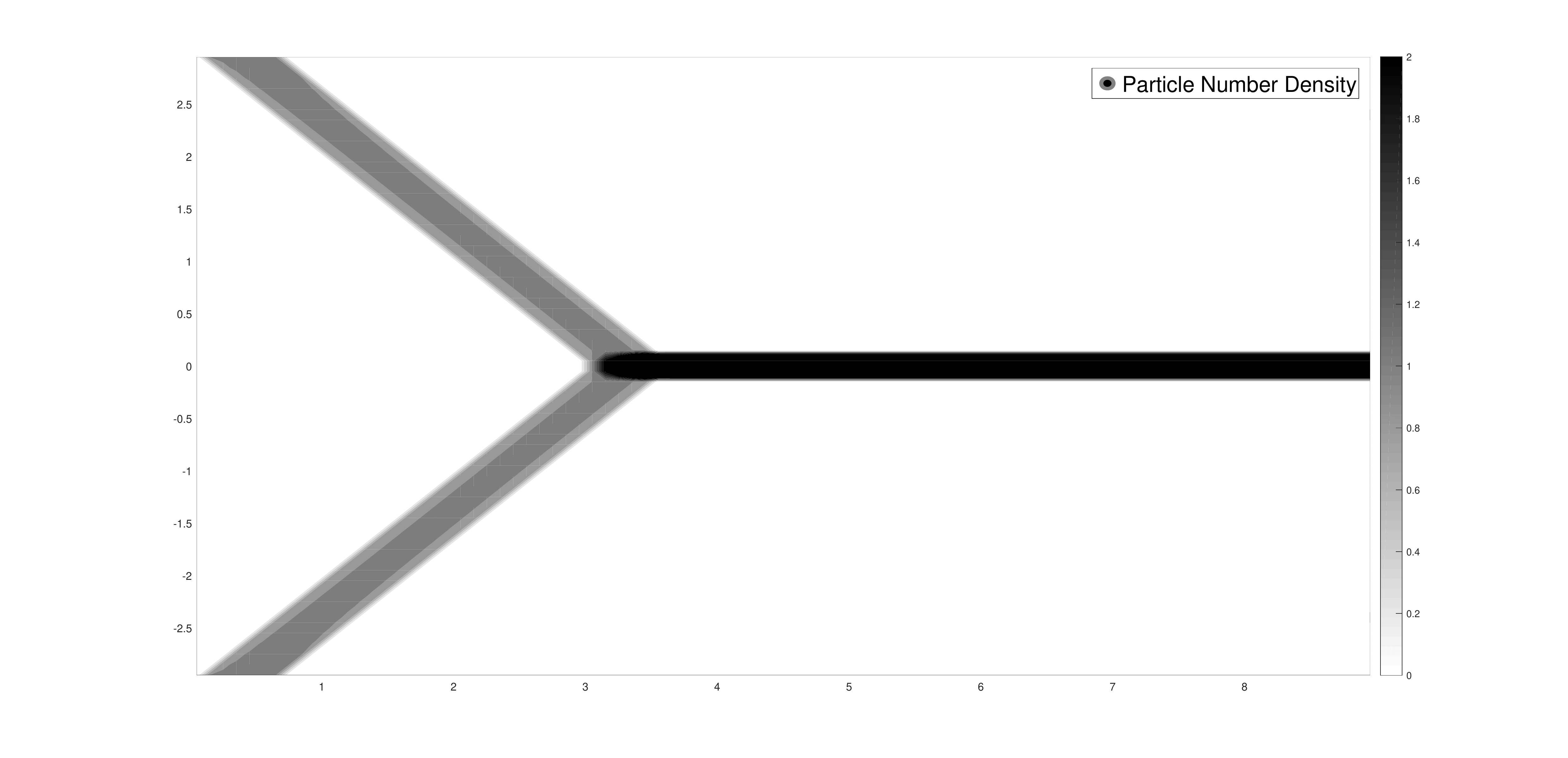}}
\subfloat[]{\includegraphics[width=0.44\textwidth]{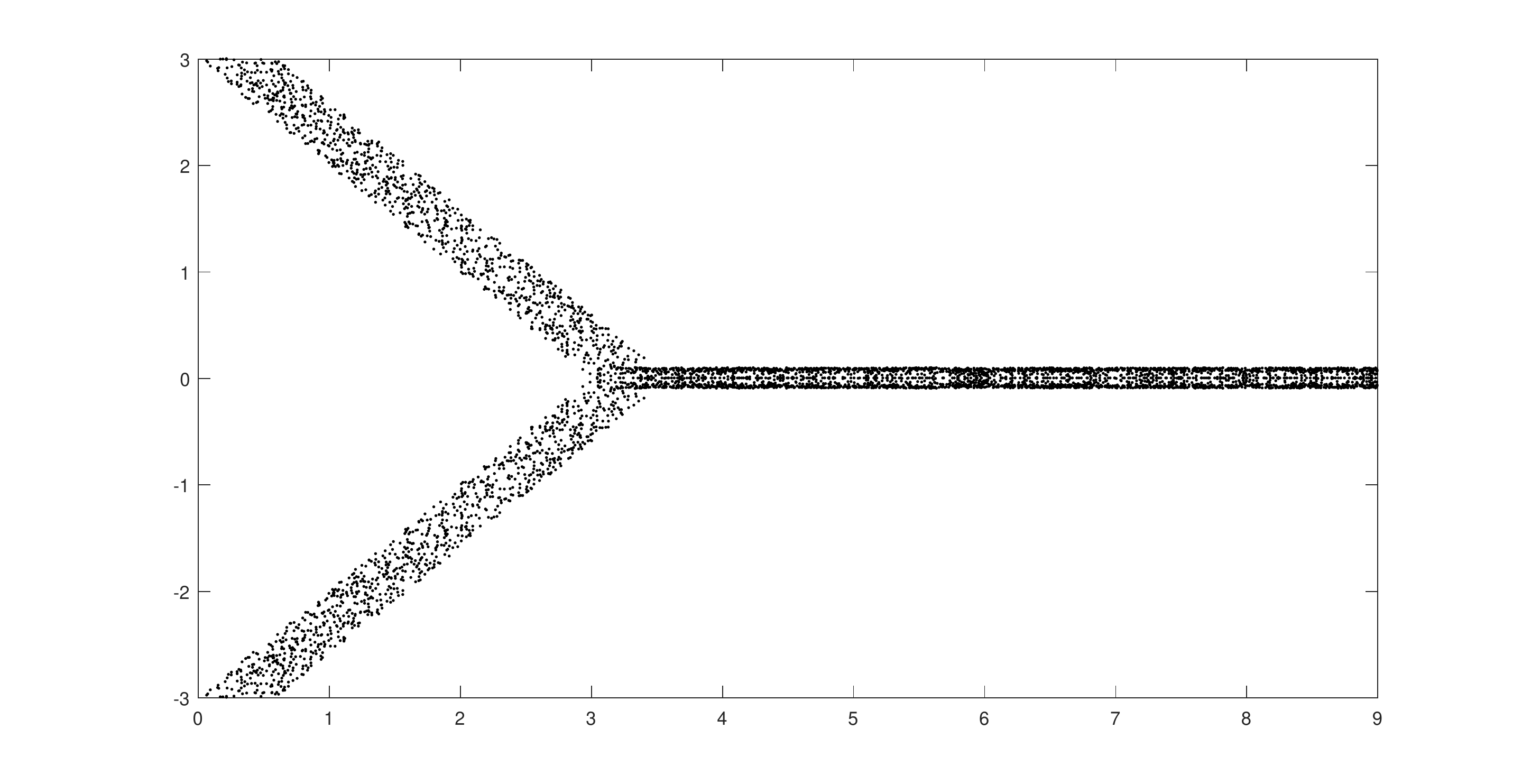}}
\caption{\label{fig:jet_collision}Particle number density at $t=20$ for highly collision regime at ${ Kn}_s=1.0\times 10^{-4}$ and $r=0$ \cite{liu2019granular}. (a) UGKS result, and (b) PIC result.}
\end{figure}

\begin{figure}[htbp]
\centering
\subfloat[]{\includegraphics[width=0.44\textwidth]{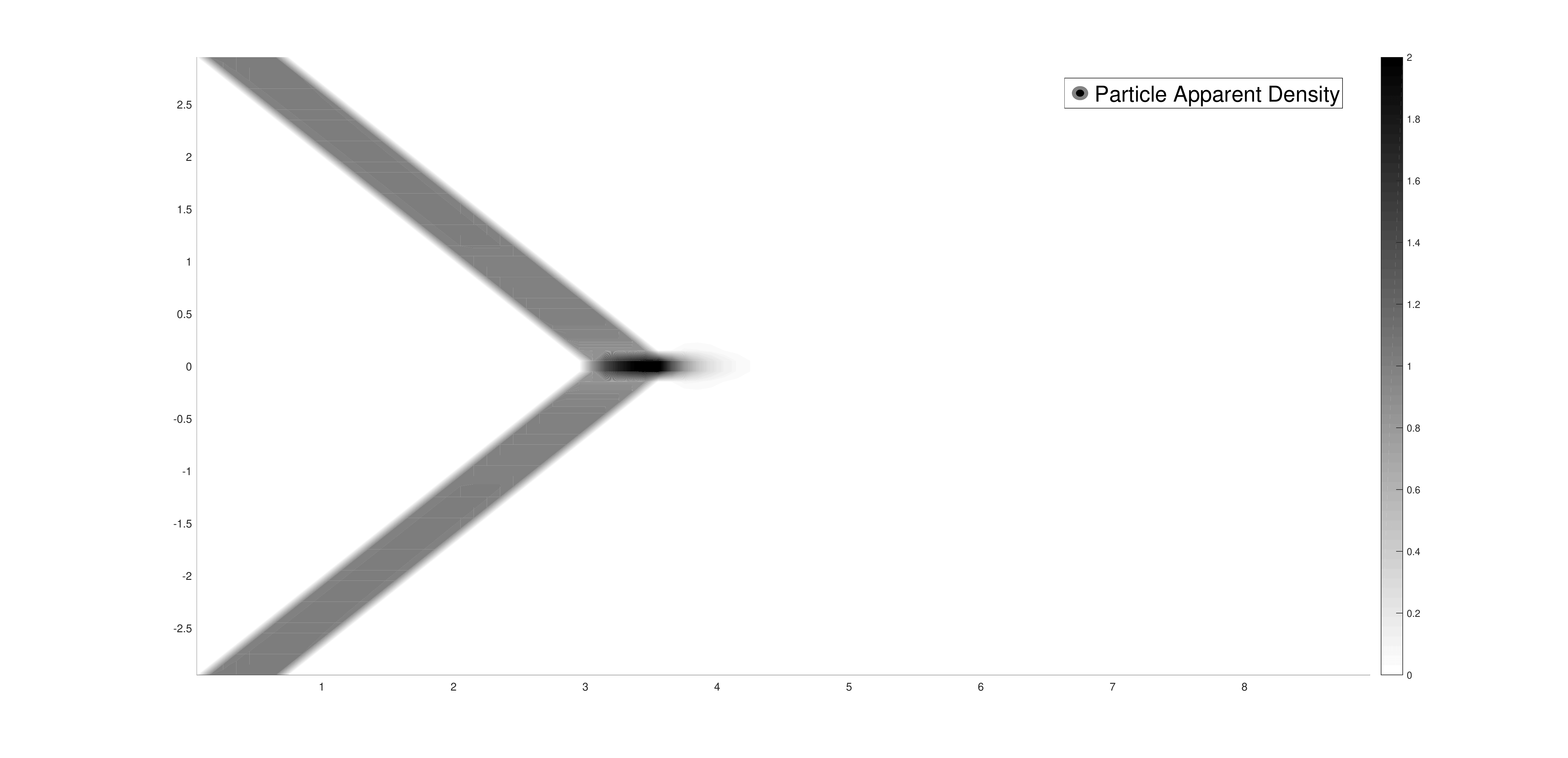}} \hspace{1em}
\subfloat[]{\includegraphics[width=0.44\textwidth]{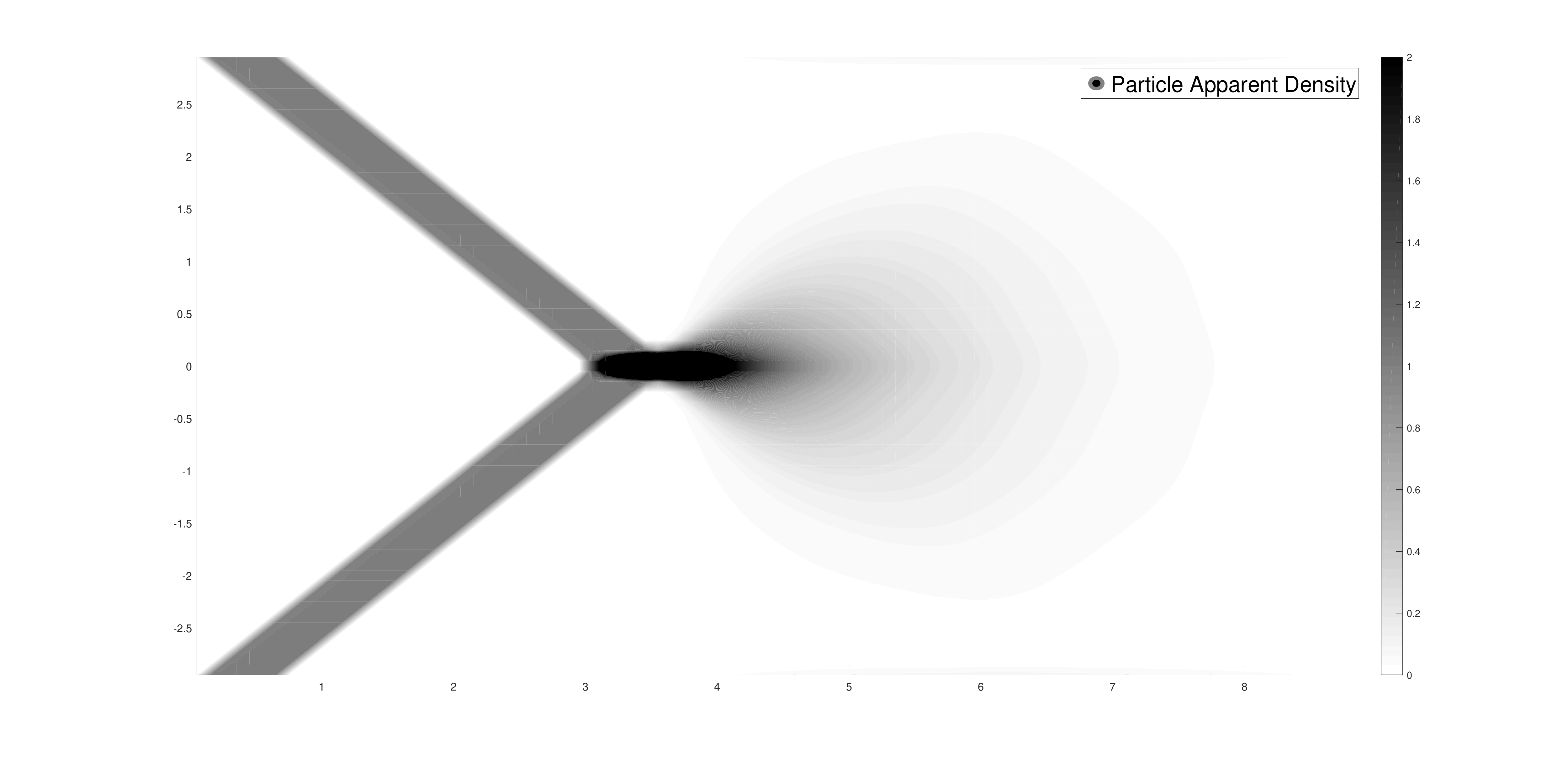}} \\
\subfloat[]{\includegraphics[width=0.44\textwidth]{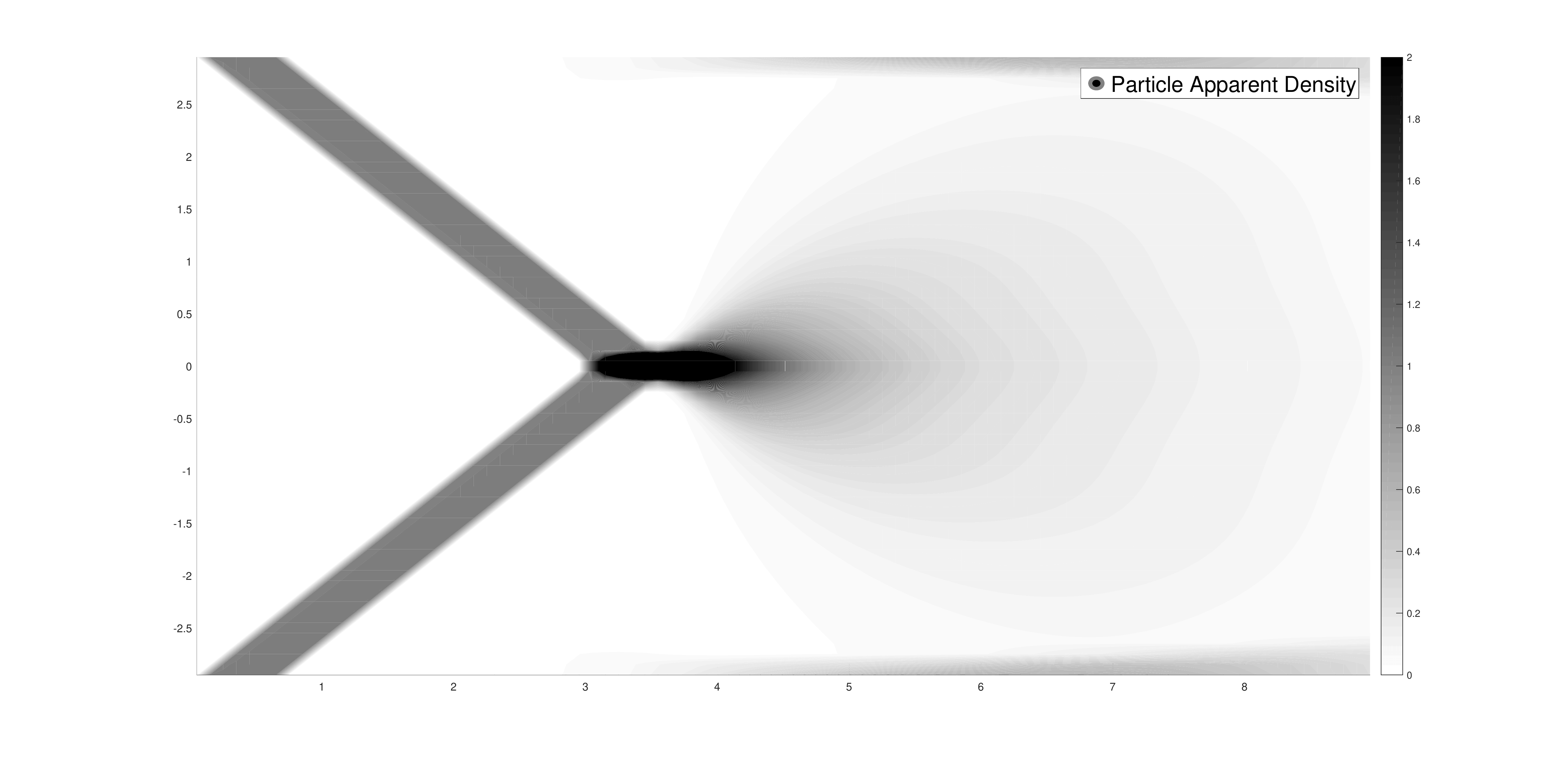}} \hspace{1em}
\subfloat[]{\includegraphics[width=0.44\textwidth]{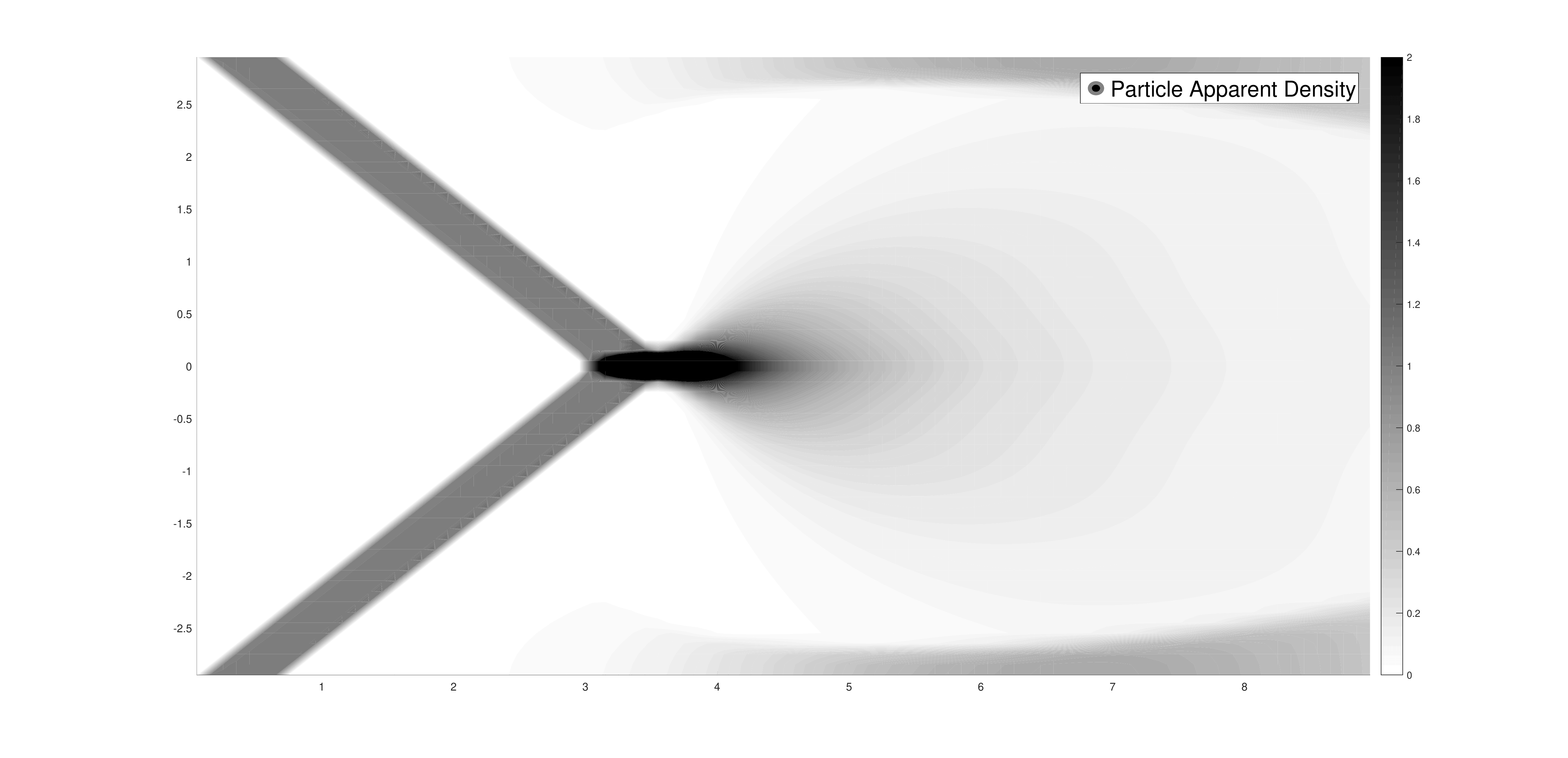}}
\caption{\label{fig:jet_r04}Particle number density at different output times with ${ Kn}_s=1.0\times 10^{-4}$ and $r=0.4$ \cite{liu2019granular}.}
\end{figure}

\begin{figure}[htbp]
\centering
\subfloat[]{\includegraphics[width=0.44\textwidth]{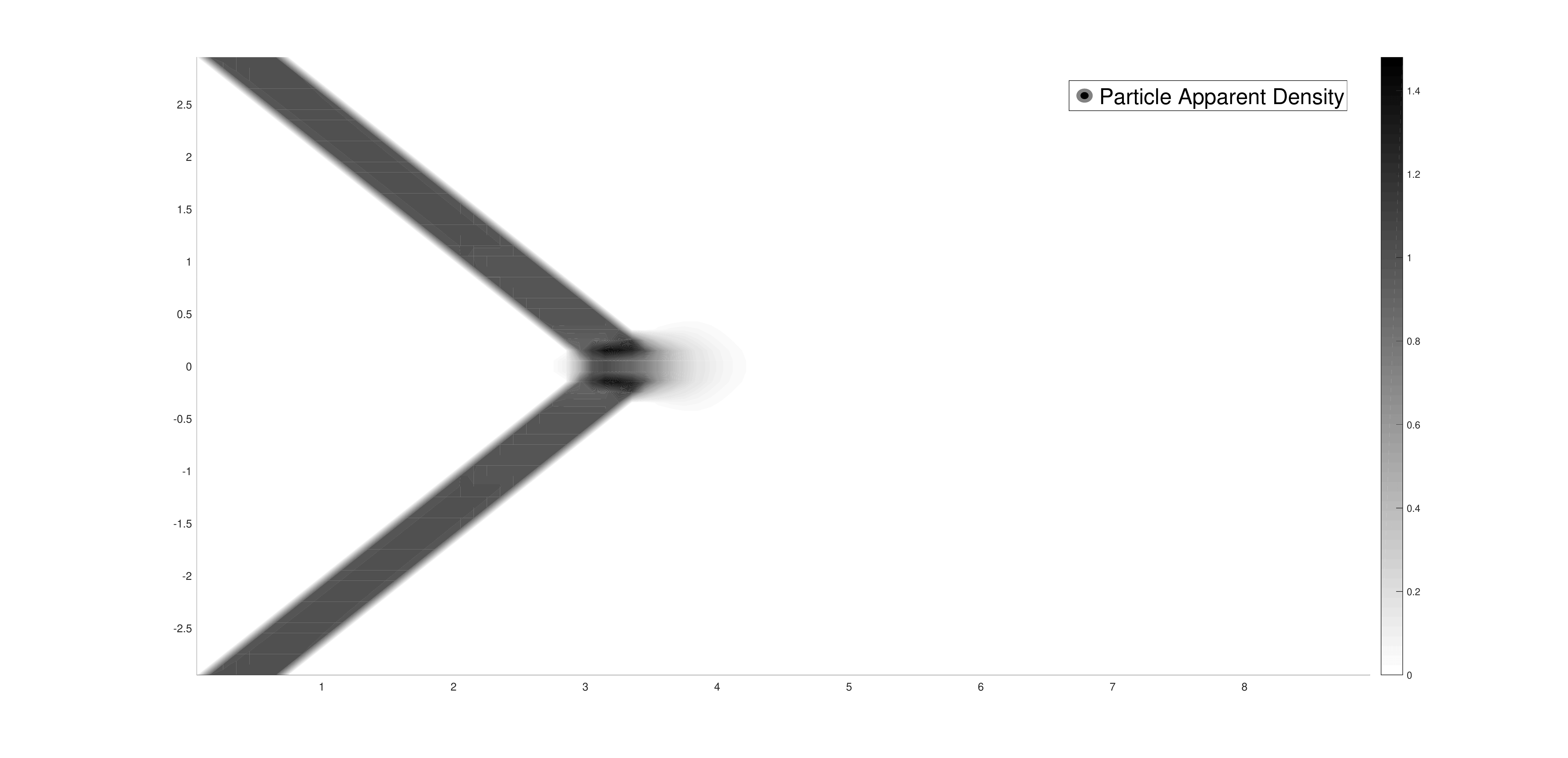}} \hspace{1em}
\subfloat[]{\includegraphics[width=0.44\textwidth]{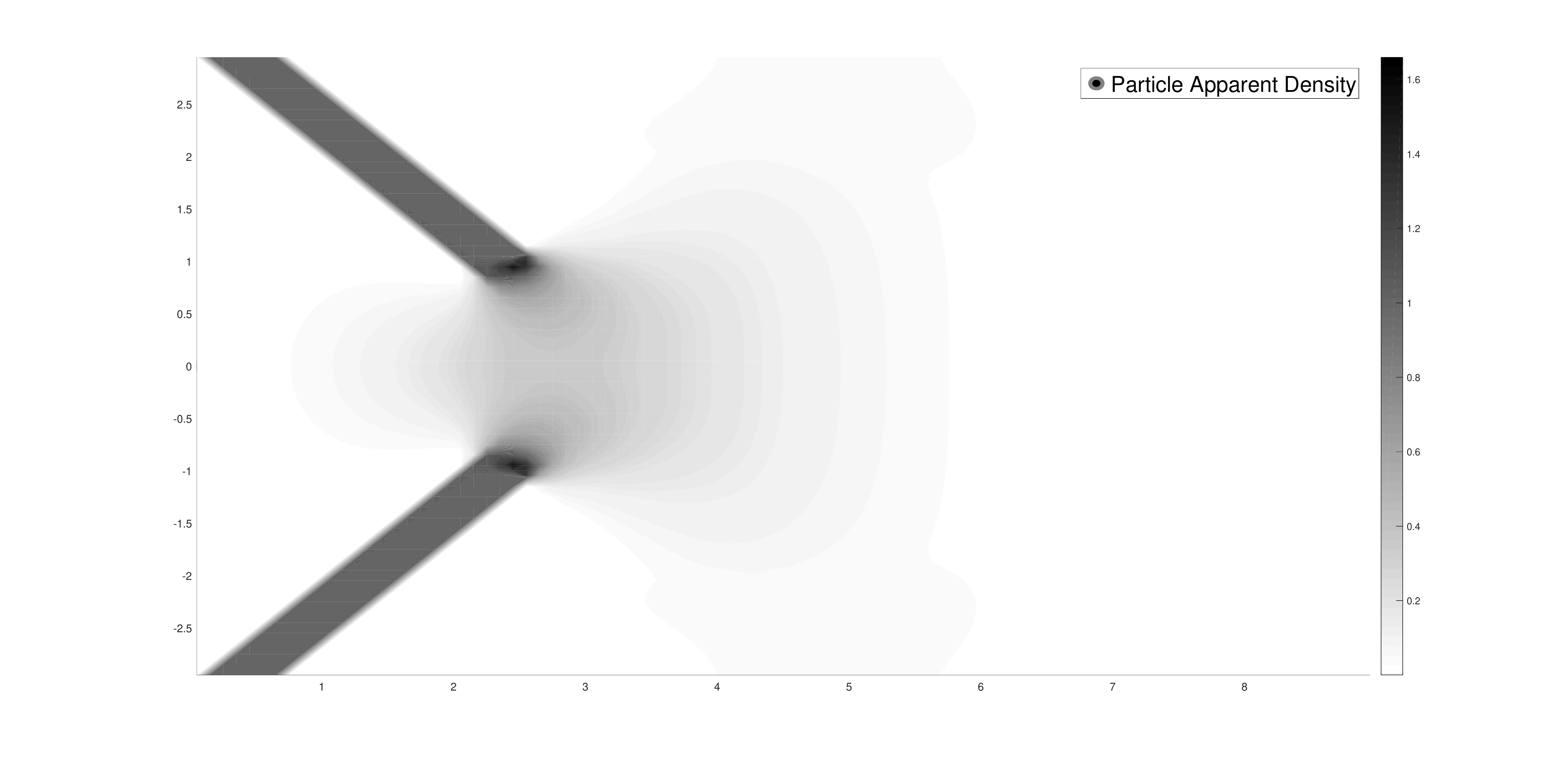}} \\
\subfloat[]{\includegraphics[width=0.44\textwidth]{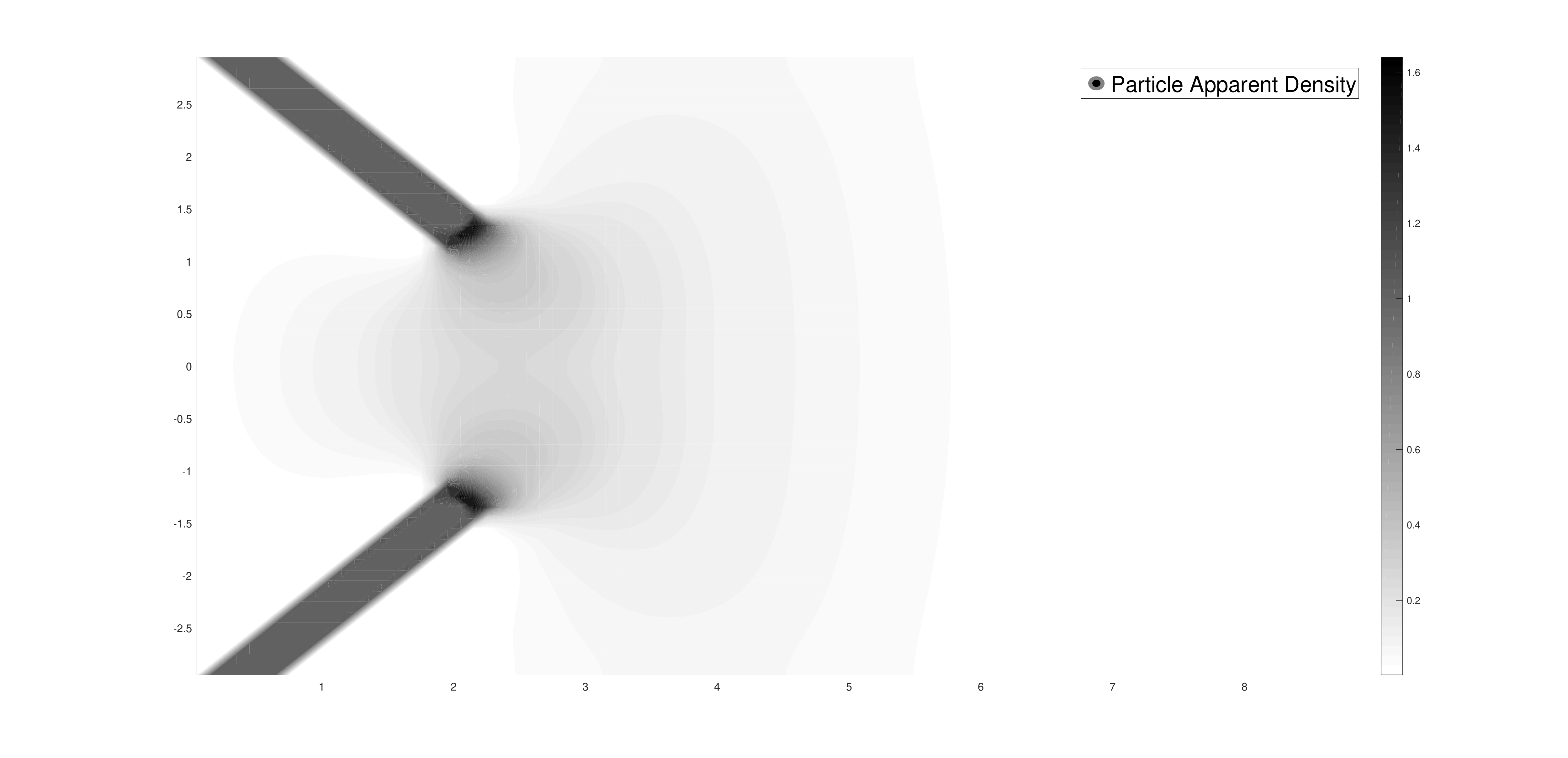}} \hspace{1em}
\subfloat[]{\includegraphics[width=0.44\textwidth]{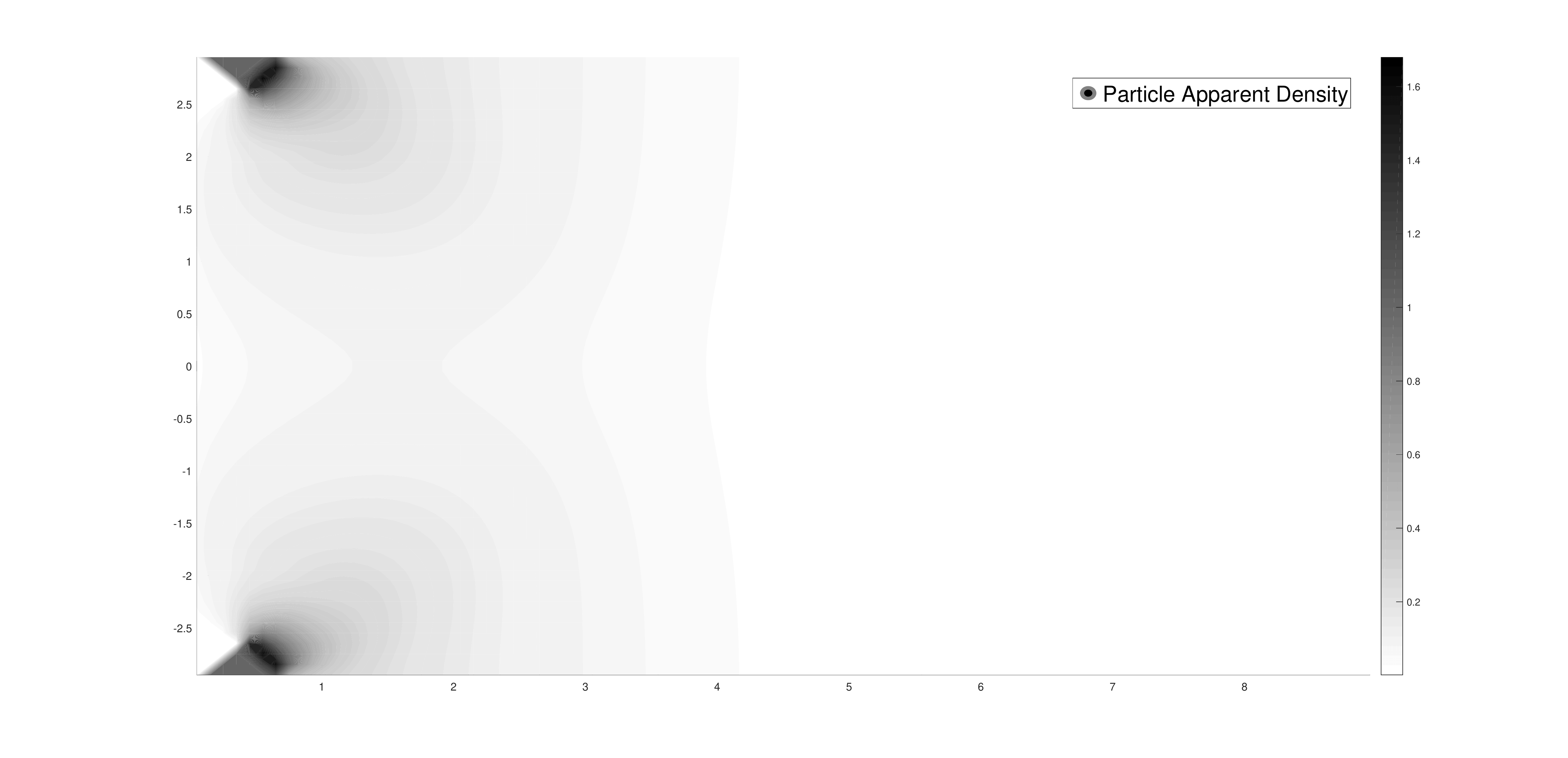}}
\caption{\label{fig:jet_r1}Particle number density at different output times with ${ Kn}_s=1.0\times 10^{-4}$ and $r=1.0$ \cite{liu2019granular}.}
\end{figure}

The particle trajectory crossing (PTC) and particle wall reflecting are two important tests to show the ability of the numerical scheme in capturing the rarefied particle flow.
The hydrodynamic models fail to capture these two phenomena and gives nonphysical $\delta$-shock \cite{marchisio2013computational}.
In this example, the problem of two particle jets impinging into a rectangular chamber is computed to demonstrate the ability of the UGKS to capture the PTC and particle wall reflecting in two-dimensional flows.
To omit the influence of gas phase, the Stokes number is set as infinity.
The channel geometry as well as the mesh geometry is shown in Fig.~\ref{fig:jet_initial}, the used mesh is an unstructured one with $\Delta x=0.1$.
Initially, two particle jets are injected from left top and left bottom corner of a rectangular chamber with adiabatic wall.
The apparent density of the jet flow is used as the reference apparent density and set $\epsilon_s \rho_s=1$, and the injection velocity is along $135^\text{o}$ and $225^\text{o}$ directions with respect to the positive x-axis.
The velocity magnitude of the jet is used as the reference velocity and set $|\vec{U}_s|=1.0$; and granular temperature is $T_s=0$.
The velocity space is $[-\sqrt{2},\sqrt{2}]$ divided into $17\times17$ cells.
Four sets of particle Knudsen number and restitution coefficient are calculated.
Firstly, the collisionless regime with infinite Knudsen number is considered and compared with the PIC result, the distribution of particle apparent density at $t=20$ is shown in Fig.~\ref{fig:jet_collisionless}.
It can be observed that UGKS recovers the physical consistent PTC and wall reflecting phenomena.
Then the case at particle Knudsen number $1.0\times 10^{-4}$ with restitution coefficient $r=0$ is simulated, and the distribution of particle apparent density at $t=20$ is shown in Fig.~\ref{fig:jet_collision}, compared with the PIC result.
In such a situation, two solid particles will share same speed after collision, and the two particle jets merge into a single one.
In addition, the evolution of particle density field for the cases with the restitution coefficient $r=0.4$ and $r=1.0$ is given in Figs.~\ref{fig:jet_r04} and \ref{fig:jet_r1}.
The particle scattering effect appears and the particles fill the chamber due to elastic collision.

\subsubsection{Particle segregation in Taylor-Green flow}

\begin{figure}[htbp]
\centering
\subfloat[]{\includegraphics[width=0.48\textwidth]{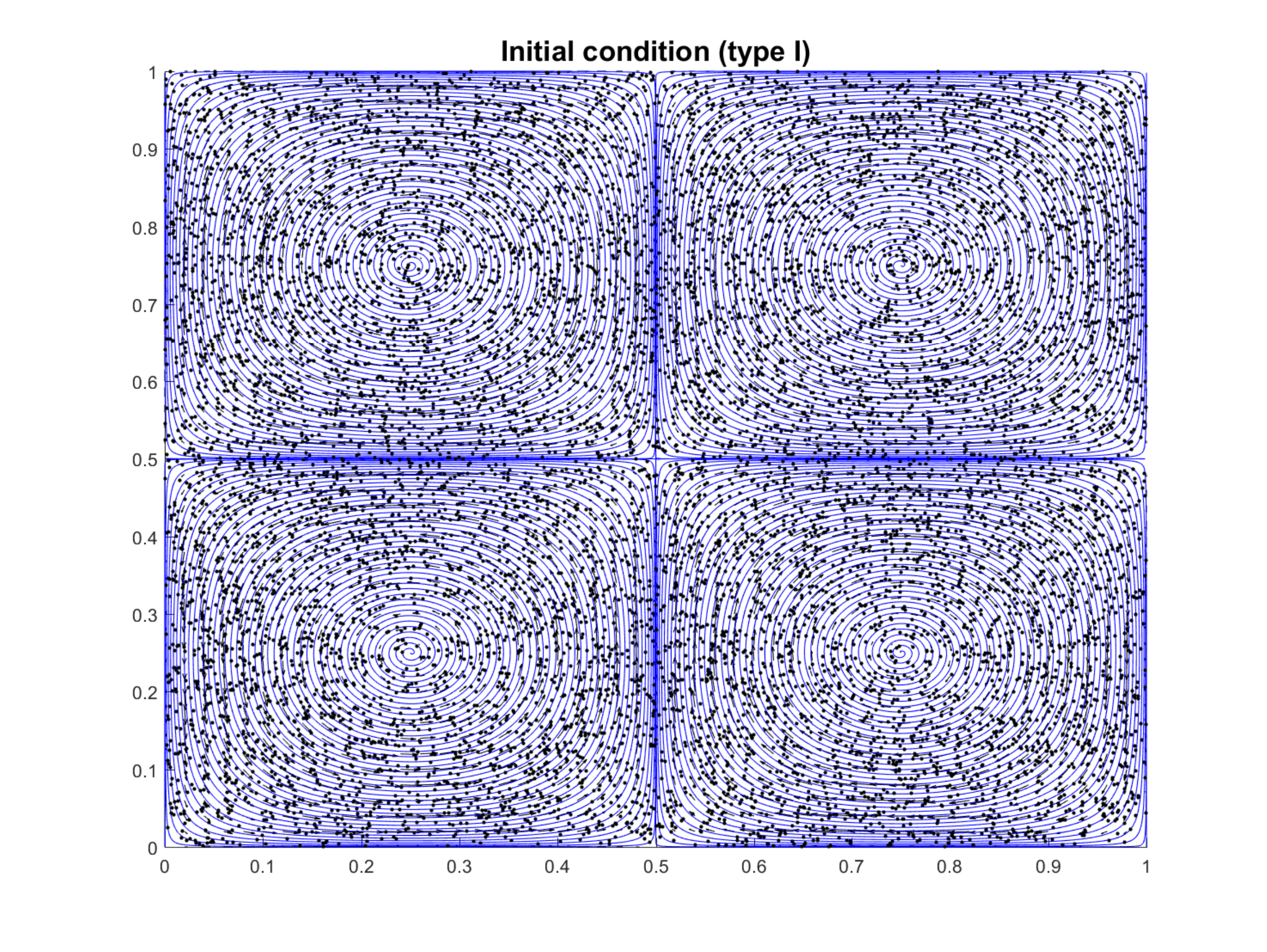}}
\subfloat[]{\includegraphics[width=0.48\textwidth]{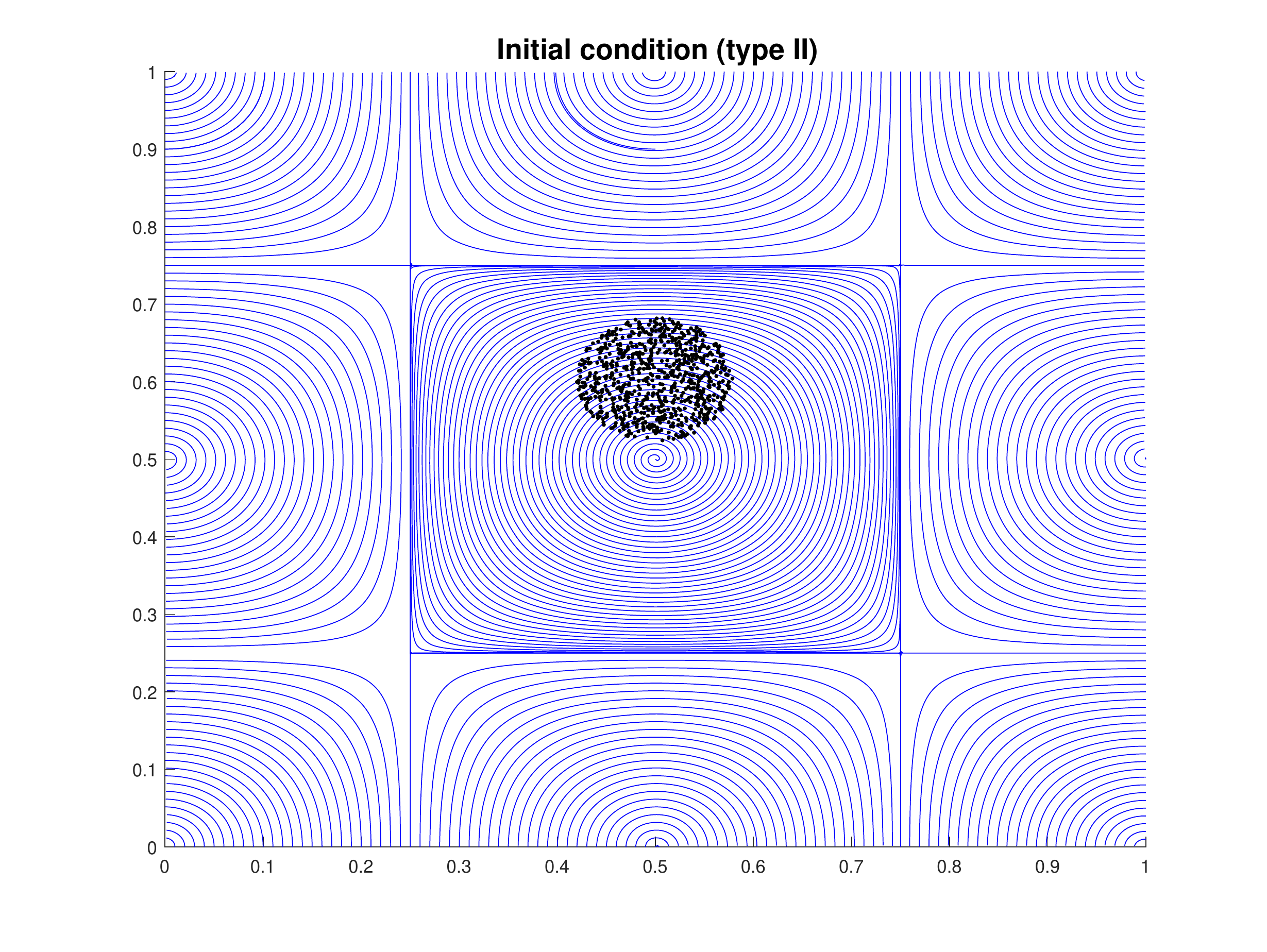}}
\caption{\label{fig:taylor_set}Two initial particle apparent density and streamline used for Taylor{-}Green flow test cases. \cite{liu2019granular}}
\end{figure}

Liu et al. \cite{liu2019granular} computed this case to study the particle segregation in Taylor-Green flow.
Two initial conditions are considered as shown in Fig.~\ref{fig:taylor_set}.
The gas field is assumed to be a two-dimensional Taylor-Green vortex with periodic boundary condition.
For the first initial condition, particles are uniformly distributed in space.
And for the second initial condition, the particles are set to be uniformly distributed in a circle centered in $(0.5,1-5/(4\pi))$ with radius $1/(4\pi)$ \cite{desjardins2008}.
For both test case, the initial particle velocity is the same as the initial gas flow velocity.
The physical domain is $[0,1]\times[0,1]$ divided equally into $200\times 200$ cells, and the velocity space is $[-1.2,1.2]\times[-1.2,1.2]$ with $42\times 42$ velocity cells.

\begin{figure}[htbp]
\centering
\subfloat[]{\includegraphics[width=0.44\textwidth]{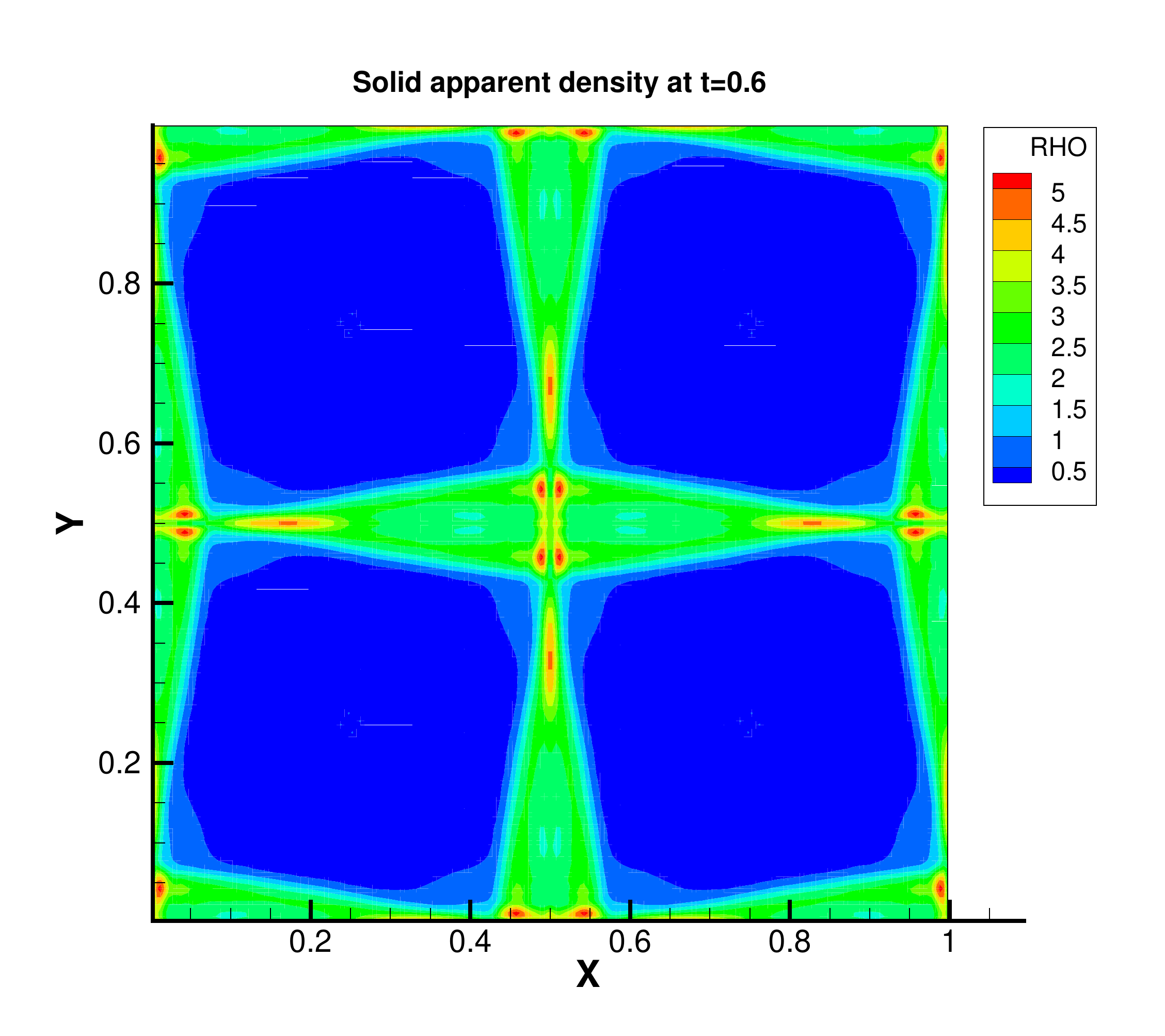}} \hspace{1em}
\subfloat[]{\includegraphics[width=0.44\textwidth]{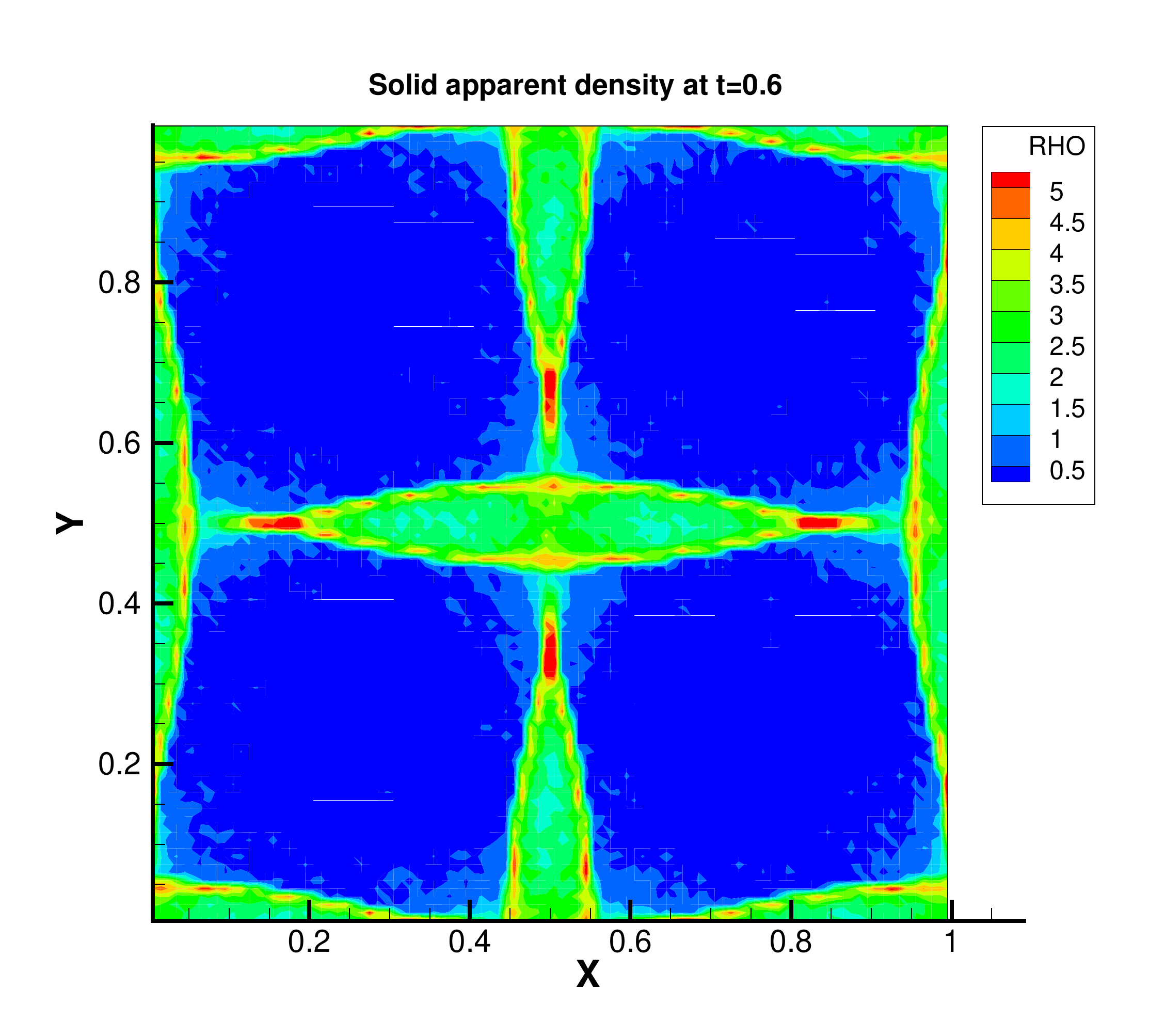}} \\
\subfloat[]{\includegraphics[width=0.44\textwidth]{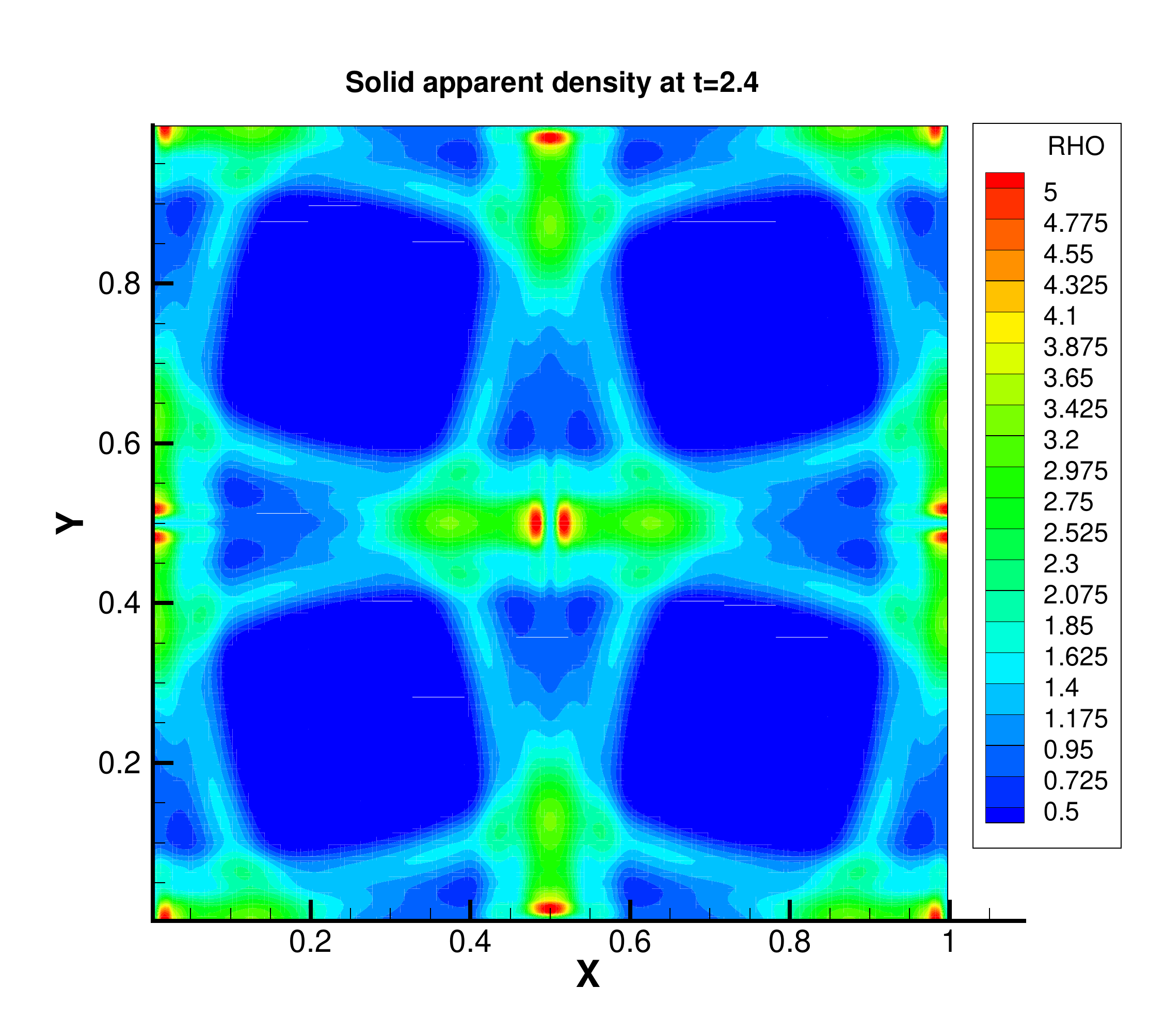}} \hspace{1em}
\subfloat[]{\includegraphics[width=0.44\textwidth]{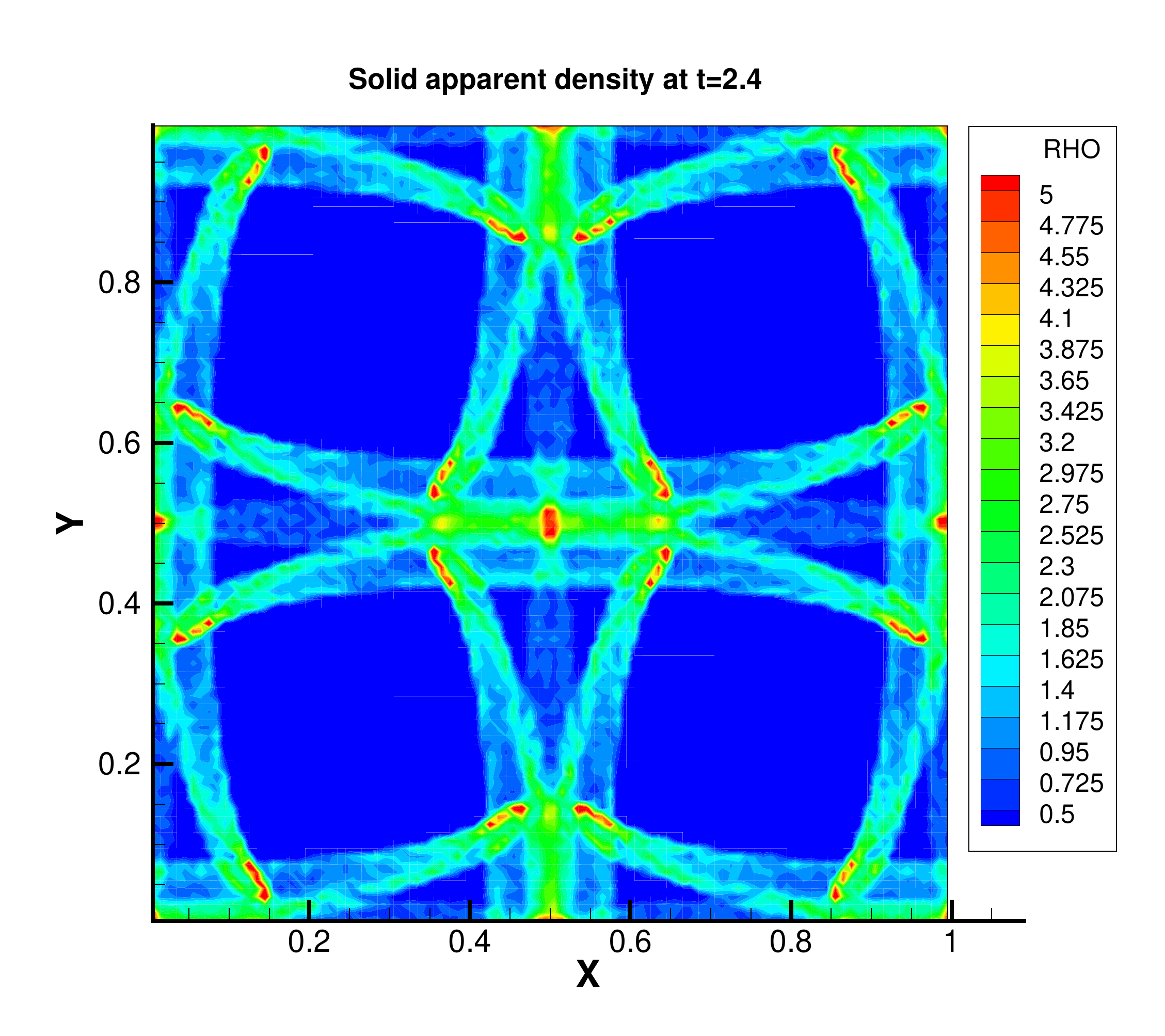}}
\caption{\label{fig:taylor_kn4}The UGKS and PIC result of particle apparent density distribution in {Taylor-Green} flow  with parameter ${ Kn}_s=10^{4}$ and ${ St}_s=0.3$ \cite{liu2019granular}. (a) UGKS at $t=0.6$, (b) PIC at $t=0.6$, (c) UGKS at $t=2.4$, and (d) PIC at $t=2.4$.}
\end{figure}

\begin{figure}[htbp]
\centering
\subfloat[]{\includegraphics[width=0.44\textwidth]{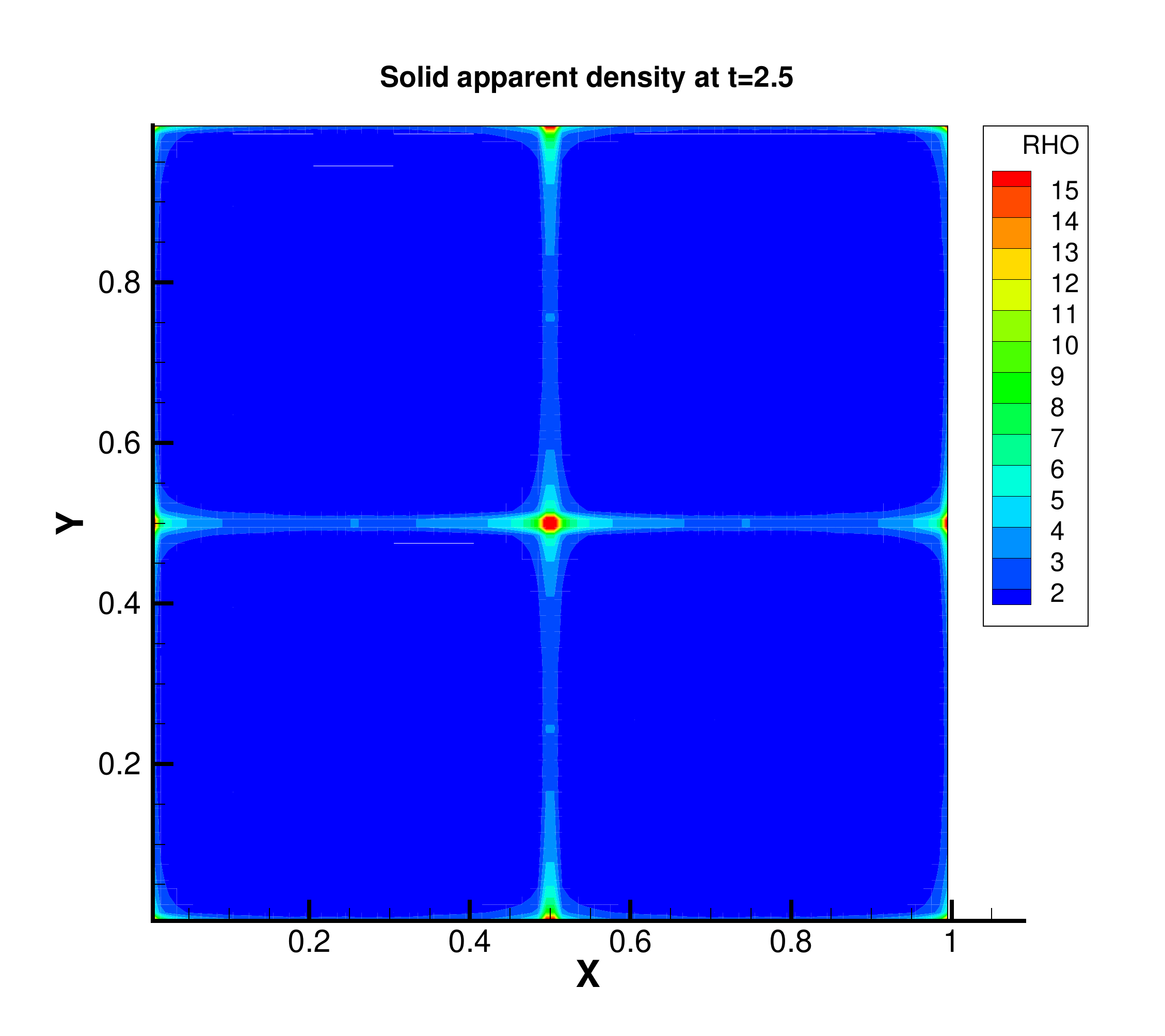}}
\subfloat[]{\includegraphics[width=0.44\textwidth]{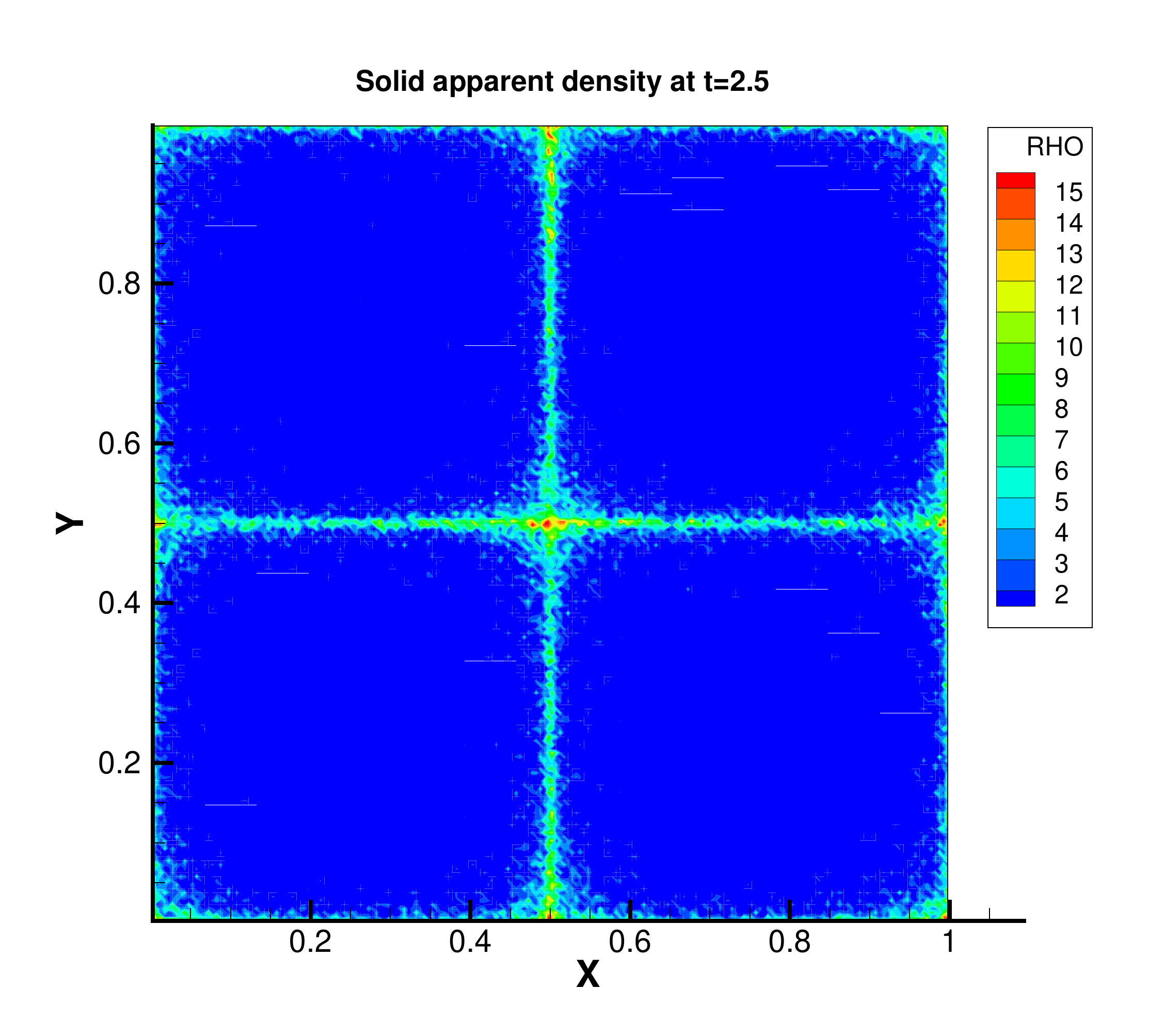}}
\caption{\label{fig:taylor_st003} Particle apparent density distribution in {Taylor-Green} flow with parameter ${ Kn}_s=10^{4}$ and ${ St}_s =0.03$ \cite{liu2019granular}. (a) UGKS solution, and (b) PIC data.}
\end{figure}

\begin{figure}[htbp]
\centering
\subfloat[]{\includegraphics[width=0.44\textwidth]{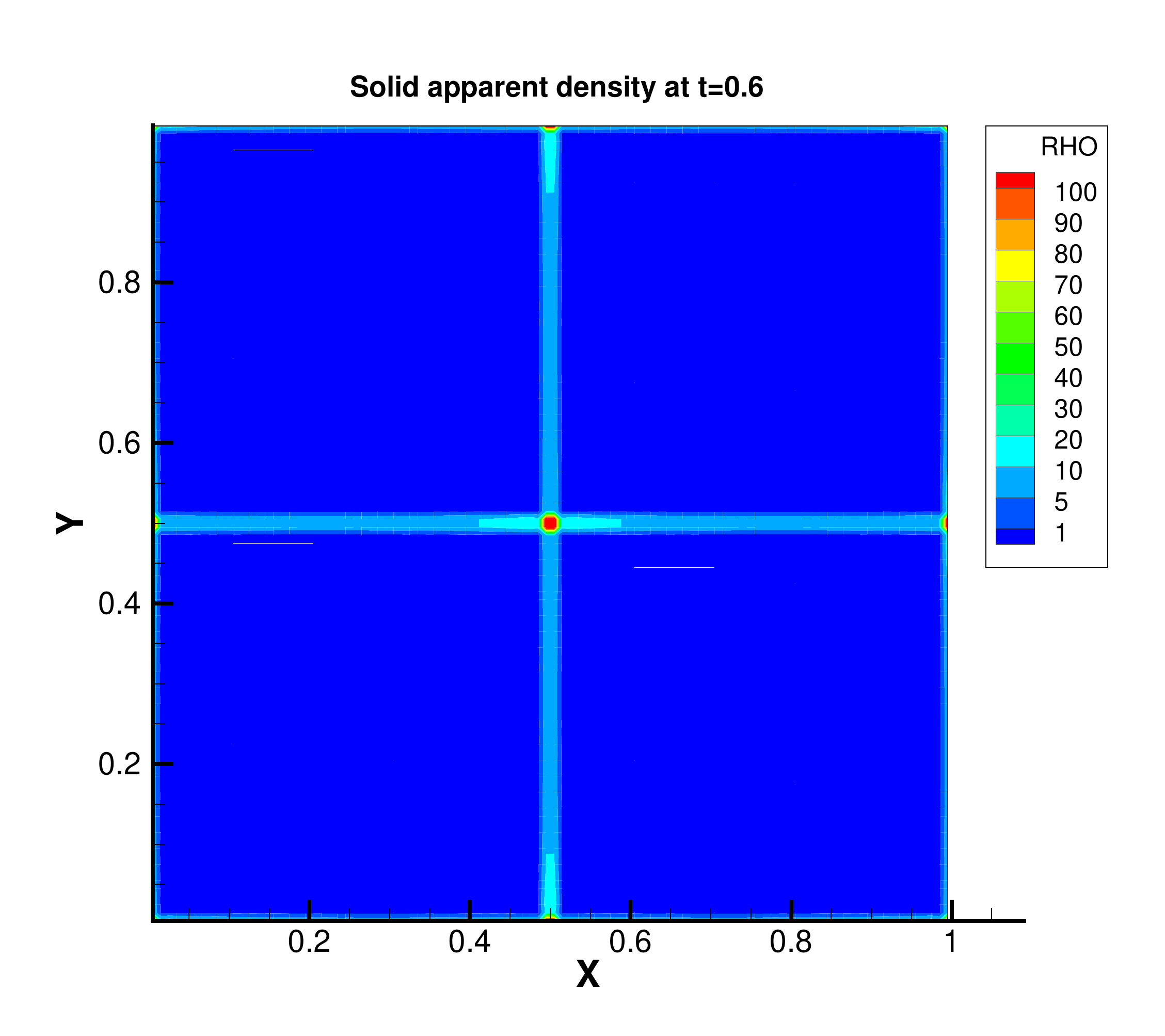}}
\subfloat[]{\includegraphics[width=0.44\textwidth]{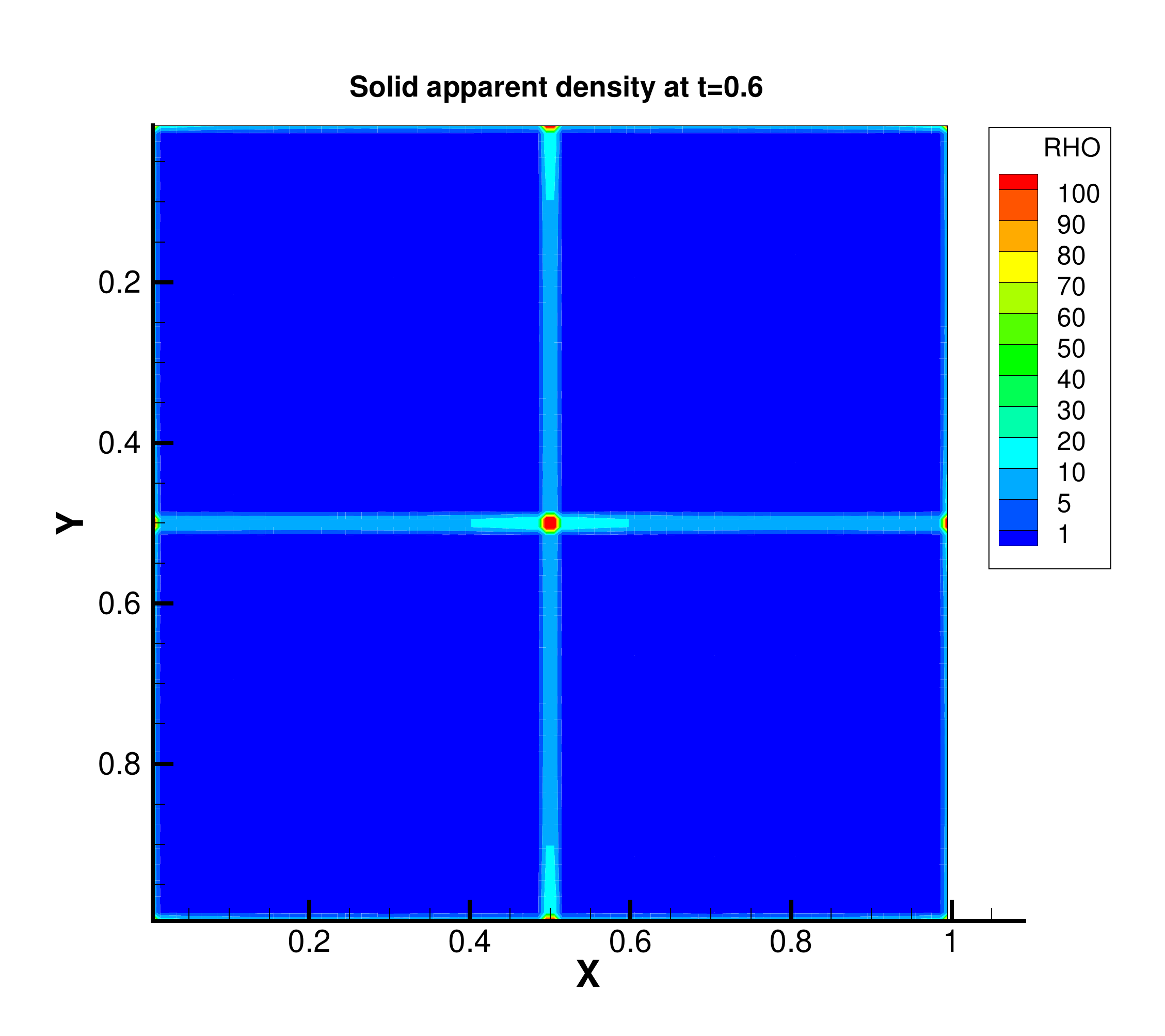}}
\caption{\label{fig:taylor_st03} Particle apparent density distribution in {Taylor-Green} flow at $t=0.6$ with parameter ${ Kn}_s = 10^{-4}$ and ${ St}_s=0.3$ \cite{liu2019granular}. (a) UGKS solution, and (b) pressureless Euler results.}
\end{figure}

\begin{figure}[htbp]
\centering
\subfloat[]{\includegraphics[width=0.44\textwidth]{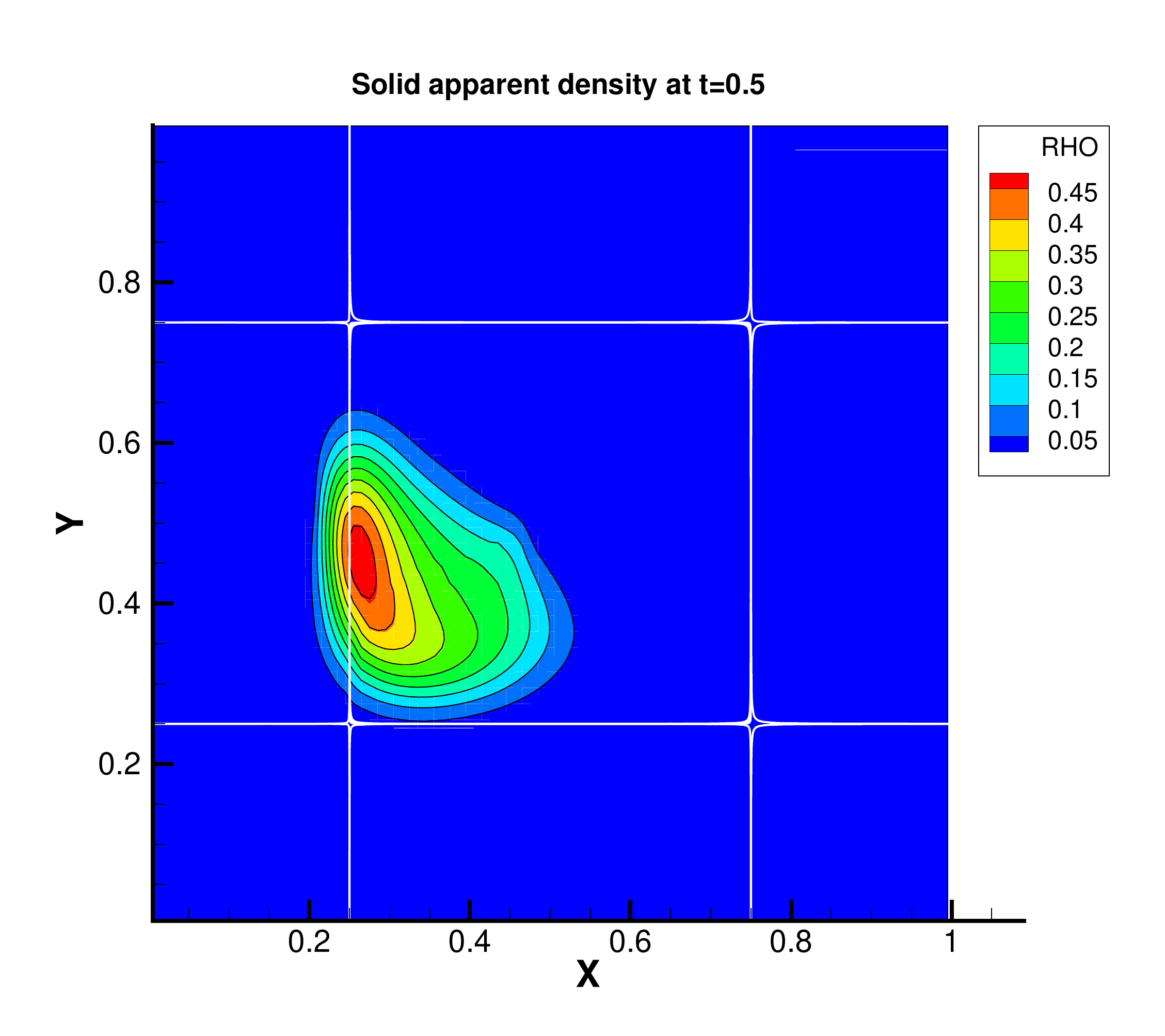}}
\subfloat[]{\includegraphics[width=0.44\textwidth]{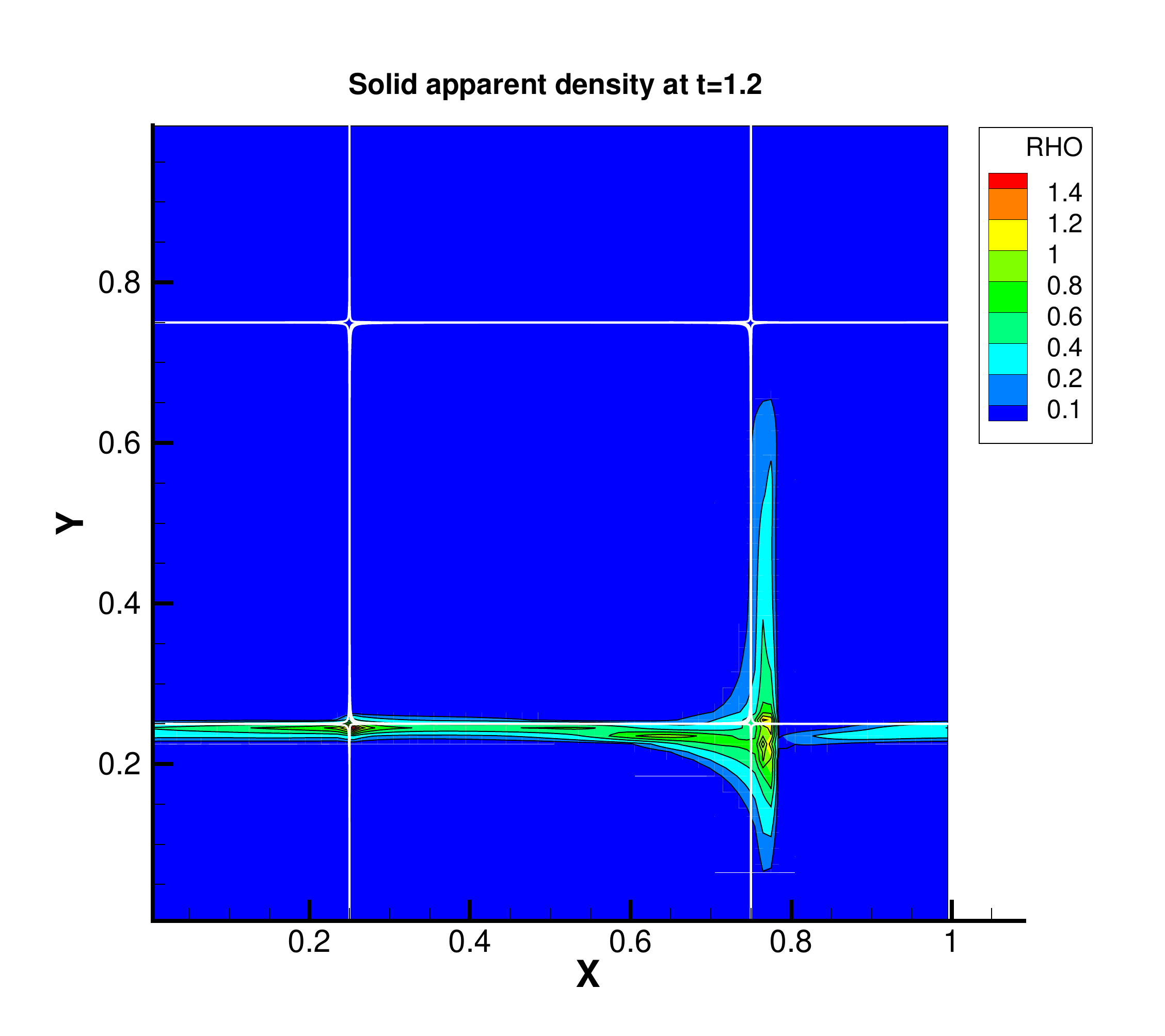}}
\caption{\label{fig:taylor_st01} The UGKS (contour) and pressureless Euler (lines) results of particle apparent density distributions in {Taylor-Green} flow with parameter ${ Kn}_s=10^{-4}$ and ${ St}_s=0.1$ \cite{liu2019granular}. (a) $t=0.5$, and (b) $t=1.2$. }
\end{figure}

\begin{figure}[htbp]
\centering
\subfloat[]{\includegraphics[width=0.44\textwidth]{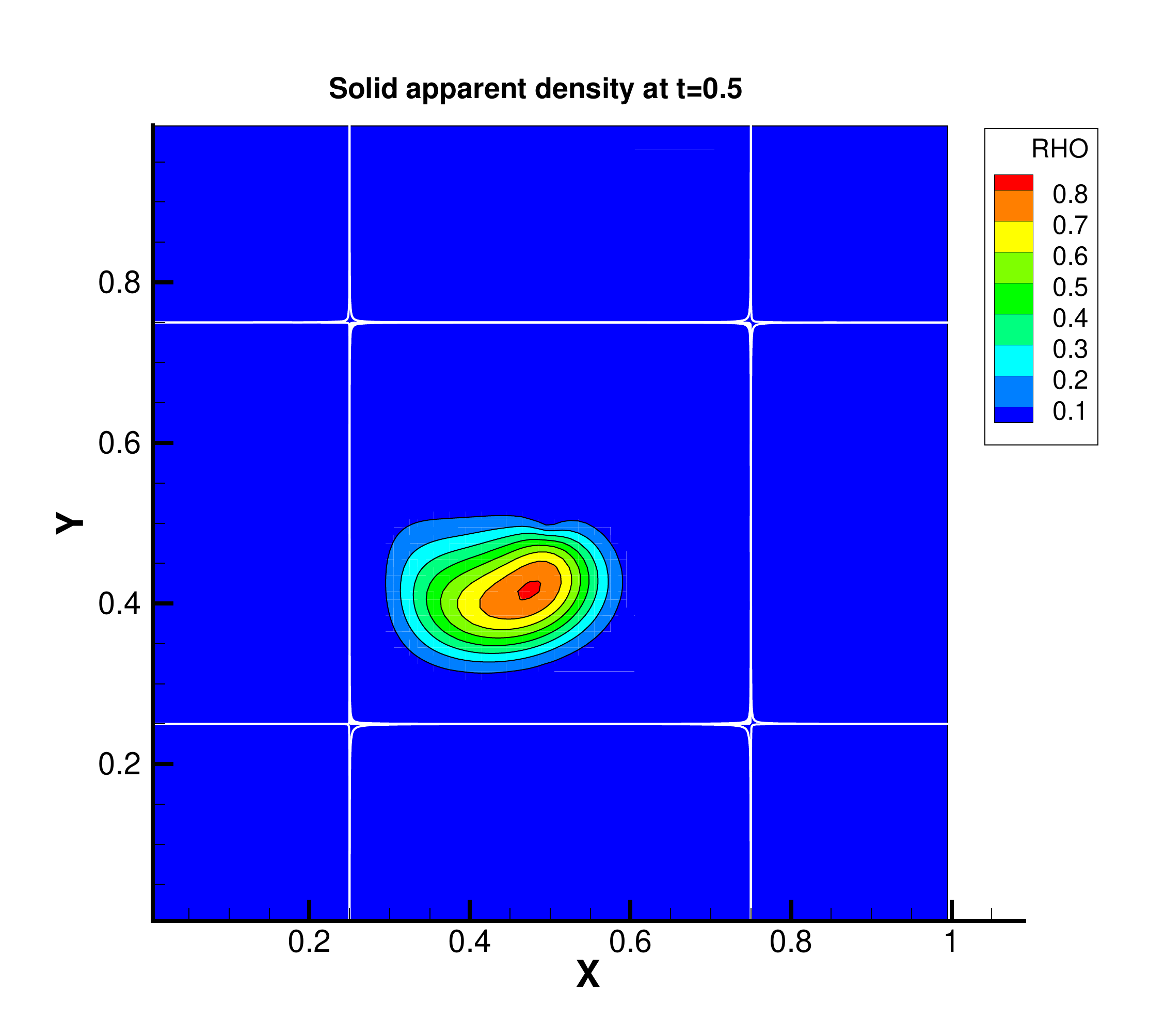}}
\subfloat[]{\includegraphics[width=0.44\textwidth]{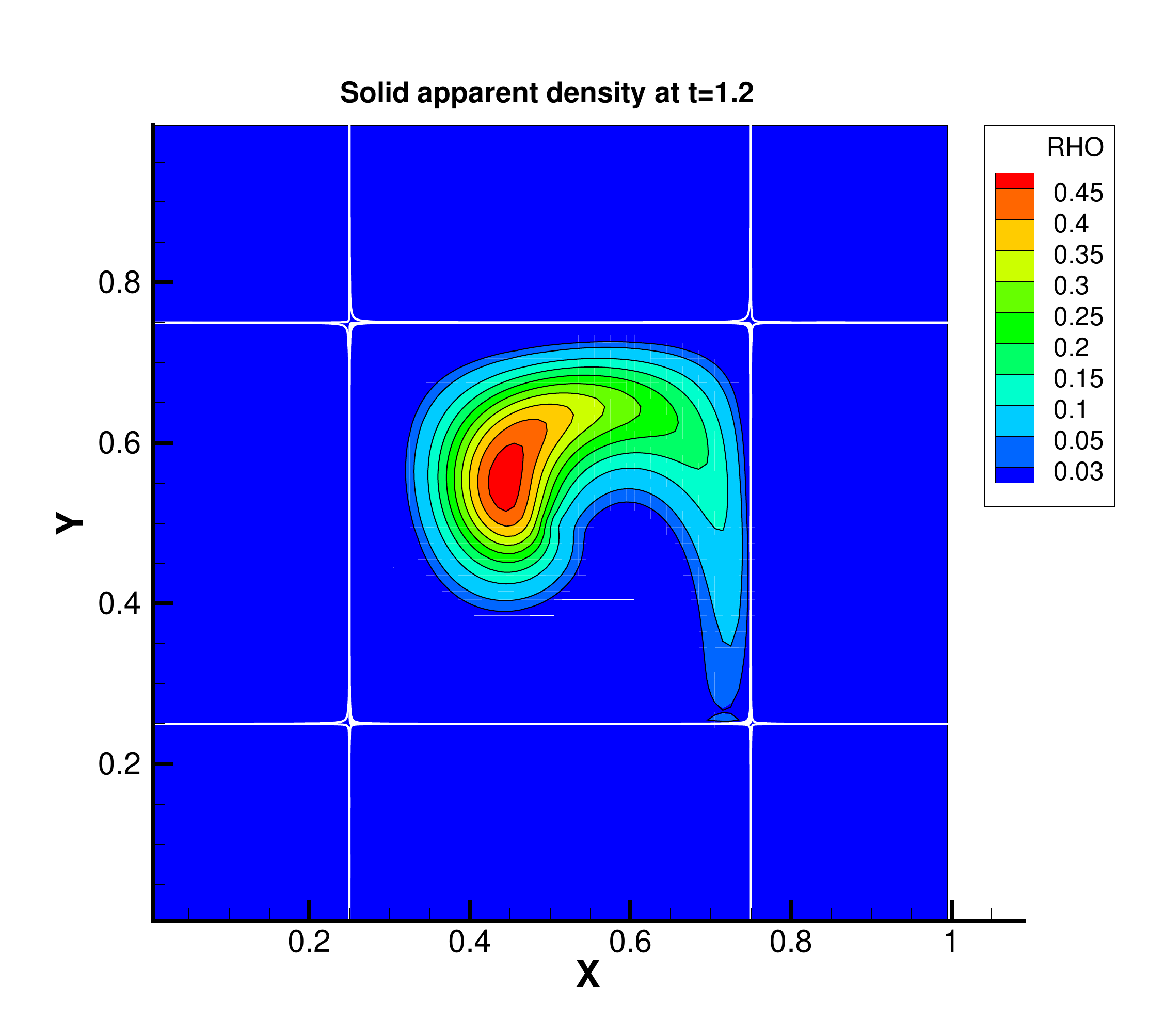}}
\caption{\label{fig:taylor_st-3} The UGKS (contour) and pressureless Euler (lines) results of particle apparent density distributions in {Taylor-Green} flow with parameter ${ Kn}_s=10^{-4}$ and ${St}_s=10^{-3}$ \cite{liu2019granular}. (a) $t=0.5$, and (b) $t=1.2$. }
\end{figure}

This problem is characterized by two important parameters, i.e., the Knudsen number and the Stokes number.
According to the analysis in \cite{chaisemartin2007}, the critical Stokes number is ${ St}_c =1/8\pi$, below which the kinetic number density function will keep mono-kinetic, and above which the particle trajectory crossing can occur.
For the first initial condition, the case of $\tau_{st}=0.3> { St}_c$ and ${ Kn}_s = 10^{4}$ is considered first.
The solution of the UGKS is compared with PIC solution at time $t=0.6$ and $t=2.4$ as shown in Fig.~\ref{fig:taylor_kn4}.
The physical consistent particle trajectory crossing is captured, and the UGKS gives satisfactory result comparing to PIC up to $t=0.6$, however, due to the numerical dissipation of finite volume scheme, the numerical resolution decreases for a long time calculation at $t=2.4$.
When decreasing the Stokes number to 0.03, which is less than the critical Stokes number, the solution of the UGKS and PIC results are shown in Fig.~\ref{fig:taylor_st003}.
Under this Stokes number, the velocity distribution will remain mono-kinetic and particles will concentrate on the edge of vortexes.
The UGKS solution agrees well with the PIC solution.
Furthermore, the case at ${ Kn}_s=10^{-4}$ is computed.
In such regime, the intense inelastic collision will dissipate the kinetic energy of particles and even for large Stokes number $\tau_{st}=0.3$, an efficient preferential concentration occurs.
The density distribution at $t=0.6$ of UGKS solution is shown in Fig.~\ref{fig:taylor_st03} comparing with the pressureless Euler solution.
For the second initial condition, if the Stokes number smaller than ${ St}_c$, the particles will remain inside of a {Taylor-Green} vortex forever and no particle trajectory crossing will appear, and eventually the particles will accumulate at four corners of the vortex where the flow velocity is small.
For the Stokes number larger than the critical stokes number, some particles will escape from the original vortex and enter into neighboring cells, and the particle trajectory crossing will appear \cite{desjardins2008,chaisemartin2007}.
The cases at ${ Kn}_s=10^{-4}$ with $\tau_{st}=0.1$ and $10^{-3}$ are calculated.
The solutions of UGKS at $t=0.6$ and $t=1.2$ are shown in Figs.~\ref{fig:taylor_st01} and \ref{fig:taylor_st-3}, comparing with the pressureless Euler equation.
For both Stokes numbers, the solutions of UGKS are consistent with the theoretical analysis, and agrees well with Euler solution.

\section{Summary}

In this paper, the unified gas kinetic scheme (UGKS) has been introduced in details.
The UGKS is a discrete velocity-based method under the finite volume framework, which is able to capture physical solution in all Knudsen number regimes.
There are several key components which make the UGKS a truly multiscale method.
\begin{itemize}
\item [(a)] The velocity space is discretized by discrete velocity points, which allows the UGKS to record and capture non-equilibrium flow physics through the evolution of discrete distribution function.
\item [(b)] The underlying methodology is the direct modeling of flow physics on the discretization scale, and following the basic and general conservation laws of the macroscopic flow variables and distribution function. The UGKS does not constrain the modeling scale on a specific physical scale, and provides a general framework for construction of multiscale methods through modeling on a flexible discretization scale.
\item [(c)] The multiscale flux function is constructed from the integral solution of the BGK-type kinetic model equation, which couples particles' free transport and collision in the evolution process, providing a scale transition from the kinetic particle transport to the hydrodynamic wave propagation. This is the key ingredient of the UGKS, which adapts the physical evolution on the discretization scale according to the numerical resolution and local physics, and makes the concept of direct modeling fully realizable.
\end{itemize}

Due to the multiscale property, the UGKS is more efficient than the discrete velocity method based on splitting treatment of particles' transport and collision.
For neutral gas flow, the asymptotic limits of the UGKS are  the Navier--Stokes equations and collisionless equation, respectively in the hydrodynamic and the free molecular regimes.
With such asymptotic property, the flow physics in hydrodynamic and kinetic scales is bridged through the integral solution of kinetic model equation in UGKS and the flow physics in the transition regime can be simulated reliably even without explicit PDEs.

In the past decade, many computational techniques have been developed and equipped in the UGKS, which significantly enhance the capability of the UGKS in numerical simulations.
The implicit scheme and multigrid method improve the convergence efficiency of the UGKS by one or two order of magnitudes for steady and unsteady state solutions.
The memory reduction and adaptive mesh techniques in the velocity space reduce the memory consumption of the UGKS significantly, which make it possible to run a three dimensional UGKS code on a small cluster or a personal computer.
With parallel algorithms for the decomposition of both physical space and velocity space, the UGKS is capable of simulating large-scale problems.
A novel multiscale method, the unified gas kinetic wave-particle (UGKWP) method, has been proposed using the wave-particle adaptation, which combines the advantages of the stochastic particle method and deterministic method.
The wave-particle adaption and decomposition concept reflects in two aspects, i.e., flow description and its dynamic evolution.
Specifically, on the one hand, during the transport process, the free streaming process is simulated by moving discrete particles and the collective effect of particles' collision and wave interaction is computed by the deterministic method from macroscopic flow variables.
  On the other hand, the equilibrium and non-equilibrium information are described by the analytic macroscopic flow variables and the stochastic discrete particles, respectively.
Since both the evolution and description employ the wave-particle formulation, the UGKWP method gains a superior adaptive property, which enables the UGKS to take numerical calculations of all Knudsen flows in an optimized efficient way.
Moreover, this adaptivity comes from the consistent treatment  between flow physics and the numerical representation.
After a decade of development and improvement, the advantage of the multiscale UGKS has been well explored.
The methodology of the UGKS has also been applied in many areas for constructing the corresponding multiscale methods, such as radiative transfer, phonon transport, multi-component and multi-phase flows, neutron transport, plasma, which show good performance in solving various multiscale transport problems.
Many numerical test cases have been carried out, which provides a reliable validation of the UGKS in different field.

In the first decade of the UGKS, great efforts have been focused on the fundamental research of the UGKS, including numerical scheme, computational techniques.
The UGKS is beginning to show its great potentials in solving real engineering problems.
Targeting on solving real engineering problems, the study in the following ten years may focus on
\begin{itemize}
\item [(a)] Large-scale computation using the UGKS or UGKWP methods for three dimensional problems of various multiscale transports.
\item [(b)] Incorporation of more physics, such as real gas effect including ionization and chemical reaction into the UGKS framework.
\item [(c)] Further extension to multi-resolution process and multiscale physics, such as turbulent flow.
\end{itemize}
We believe that the further development of the UGKS could make great contributions in the modeling and studying multiscale flow physics,
which is beyond the single scale partial differential equations in the kinetic and hydrodynamic scales alone.

Many thanks to all the researchers who have been and will be working on the unified gas kinetic related methods.

\bibliographystyle{apalike}
\bibliography{ugks_chapter}
\end{document}